\pdfoutput=1
\documentclass[12pt,preprint]{aastex}
\usepackage {morefloats}

\def\lessim{\mathrel{\hbox{\rlap{\hbox{\lower4pt\hbox{$\sim$}}}\hbox{$<$}}}}
\def\grtsim{\mathrel{\hbox{\rlap{\hbox{\lower4pt\hbox{$\sim$}}}\hbox{$>$}}}}

\shorttitle{Recurrent Novae in M31}
\shortauthors{Shafter et al.}

\begin{document}


\title{Recurrent Novae in M31}


\author{
A. W. Shafter\altaffilmark{1},
M. Henze\altaffilmark{2},
T. A. Rector\altaffilmark{3},
F. Schweizer\altaffilmark{4},
K. Hornoch\altaffilmark{5},
M. Orio\altaffilmark{6,7},
W. Pietsch\altaffilmark{8},
M. J. Darnley\altaffilmark{9},
S. C. Williams\altaffilmark{9},
M. F. Bode\altaffilmark{9},
J. Bryan\altaffilmark{10}}
\altaffiltext{1}{Department of Astronomy, San Diego State University, San Diego, CA 92182, USA}
\altaffiltext{2}{European Space Astronomy Centre, P.O. Box 78, 28692 Villanueva de la Ca\~{n}ada, Madrid, Spain}
\altaffiltext{3}{Department of Physics and Astronomy, University of Alaska Anchorage, 3211 Providence Dr., Anchorage, AK 99508}
\altaffiltext{4}{Carnegie Observatories, 813 Santa Barbara St., Pasadena, CA 91101, USA}
\altaffiltext{5}{Astronomical Institute, Academy of Sciences, CZ-251 65 Ondrejov, Czech Republic}
\altaffiltext{6}{Astronomical Observatory of Padova (INAF), 35122 Padova, Italy}
\altaffiltext{7}{Department of Astronomy, University of Wisconsin, 475 N. Charter St., Madison, WI 53706, USA}
\altaffiltext{8}{Max Planck Institute for extraterrestrial Physics, PO Box 1312, Giessenbachstr., 85741, Garching, Germany}
\altaffiltext{9}{Astrophysics Research Institute, Liverpool John Moores University, Liverpool L3 5RF, UK}
\altaffiltext{10}{McDonald Observatory, Austin, TX, 78712, USA}



\begin{abstract}
The reported positions of 964 suspected nova eruptions in M31 recorded
through the end of calendar year 2013 have been compared
in order to identify recurrent nova candidates.
To pass the initial screen and
qualify as a recurrent nova candidate
two or more
eruptions were required to be coincident
within $0.1'$, although this criterion was relaxed to $0.15'$
for novae discovered on early photographic patrols.
A total of 118 eruptions from 51 potential
recurrent nova systems satisfied the screening criterion. 
To determine what fraction of these novae are indeed recurrent
the original plates and published images of the relevant eruptions
have been carefully compared.
This procedure has resulted in the elimination of 27 of the 51 progenitor
candidates (61 eruptions) from further consideration as recurrent novae,
with another 8 systems (17 eruptions) deemed unlikely to be recurrent.
Of the remaining 16 systems, 12 candidates (32 eruptions)
were judged to be recurrent novae, with an additional
4 systems (8 eruptions) being possibly recurrent.
It is estimated that $\sim$4\% of the nova
eruptions seen in M31 over the past century are associated with
recurrent novae. A Monte Carlo analysis shows that
the discovery efficiency for recurrent novae
may be as low as 10\% that for novae in general, suggesting
that as many as one in three nova eruptions observed in
M31 arise from progenitor systems having recurrence times
$\lessim$100~yr.
For plausible system parameters,
it appears unlikely that recurrent novae can provide
a significant channel for the production of Type Ia supernovae.

\end{abstract}




\section{Introduction}

Classical novae are all semi-detached binary systems consisting of a late-type
star that fills its Roche lobe and transfers material to a white dwarf
companion \citep{war08}.
The transferred material slowly accumulates on the white dwarf's
surface until the temperature and density at the base of the accreted layer
become sufficiently high for a thermonuclear runaway (TNR) to ensue, leading to
a nova eruption \citep[][and references therein]{sta08}.
The progenitor binary is not disrupted in the process, and
following the eruption, the mass transfer process resumes, eventually
leading to another eruption. The time interval between successive
eruptions varies considerably depending on properties of the
progenitor system such as the mass and temperature of the white
dwarf component and the rate of accretion onto its surface.

Models of nova eruptions suggest that systems with massive
and luminous (hot) white dwarfs can trigger a TNR after accreting
a relatively small envelope. Thus, for a given accretion rate,
such systems are expected to have the shortest recurrence times
\citep[e.g.,][]{yar05,tow05,kat14}.
Systems that have been observed to have more than one eruption
(with recurrence times $\lessim$100~yr) have been traditionally
referred to as ``Recurrent Novae" (RNe), although the distinction
with Classical Novae (CNe) is arbitrary given that strictly
speaking {\it all\/} CNe are believed to be recurrent.

Given that RNe are believed to harbor the most massive white dwarfs
among all nova binaries, they are of particular interest as
potential Type Ia supernova (SNe~Ia) progenitors \citep[e.g.,][]{mao12}.
In particular,
in the so called ``single degenerate" progenitor model, SNe~Ia are
produced when a massive white dwarf in a close binary system
accretes material from its companion star, pushing it over the
Chandrashekhar limit. If the white dwarf has a Carbon-Oxygen (CO)
composition,
the result is the explosive burning of the CO white dwarf
(a deflagration) leading
to a Type~Ia supernova explosion. If, on the other hand,
the white dwarf is a ONeMg core, the result is believed to be an
accretion induced collapse leading to the production of a neutron star.

\citet{del96} estimated the frequency of RN outbursts relative to CN outbursts
in the Galaxy, the LMC and in M31 and concluded that RNe were not a major
channel for the production of SNe~Ia. Here, we present a comprehensive study
of the positions of the 964 nova eruptions reported in M31 from September 1909
to the end of 2013 with the goal of determining
the fraction of nova eruptions that are recurrent, and thus
associated with the massive white dwarf progenitors potentially capable
of producing Type~Ia supernovae.

\section{Identification of M31 Recurrent Nova Candidates}

In principle, searching for RN systems should be straightforward.
One simply looks for spatial
coincidences among the reported positions of nova candidates
observed in M31. The database assembled by W. Pietsch\footnote{\tt
http://www.mpe.mpg.de/$\sim$m31novae/opt/m31/index.php} is ideal for this
purpose. In practice, however, uncertainties in the reported nova
positions significantly complicate the process. The earliest recorded
novae in M31 were discovered on photographic plates and
reported primarily by \citet{hub29}, \citet{arp56}, \citet{ros64,ros73}, and
\citet{ros89}.
The equatorial coordinates were not reported,
but instead Cartesian offsets from the nucleus were given in a
system with the X-axis oriented along the major axis of the galaxy, with
$+$X pointing to the North-East. The Y axis passes through the
nucleus, with $+$Y pointing to the North West ($+$Y points South-East
in Arp's convention).
Most of these early positions were reported
to a precision of only 0.1$'$.

In order to account for differences of precision in the reported
positions of novae, we have adopted an initial screening criterion that
varies depending on the expected uncertainty of the nova position.
In all cases, to minimize the chance of missing a potential RN,
loose screening criteria have been adopted in the initial search.
For CCD surveys conducted after 1980, where the coordinates are
generally well determined, novae with reported positions
that differed by as much as $0.1'$ were considered to be
potential RNe.
We have also adopted this screening criterion for
novae discovered in the photographic surveys of \citet{ros64,ros73,ros89}
where revised astrometry is now available (Pietsch et al., in preparation).
To account for uncertainty in the published positions of novae from
earlier photographic patrols \citep[e.g.][]{hub29,arp56}, where the
reported nova positions are known to be less reliable, we have expanded
our initial screening for RNe to include novae
with reported positions that differed by as
much as 0.15$'$.

\begin{figure}
\includegraphics[angle=90,scale=.70]{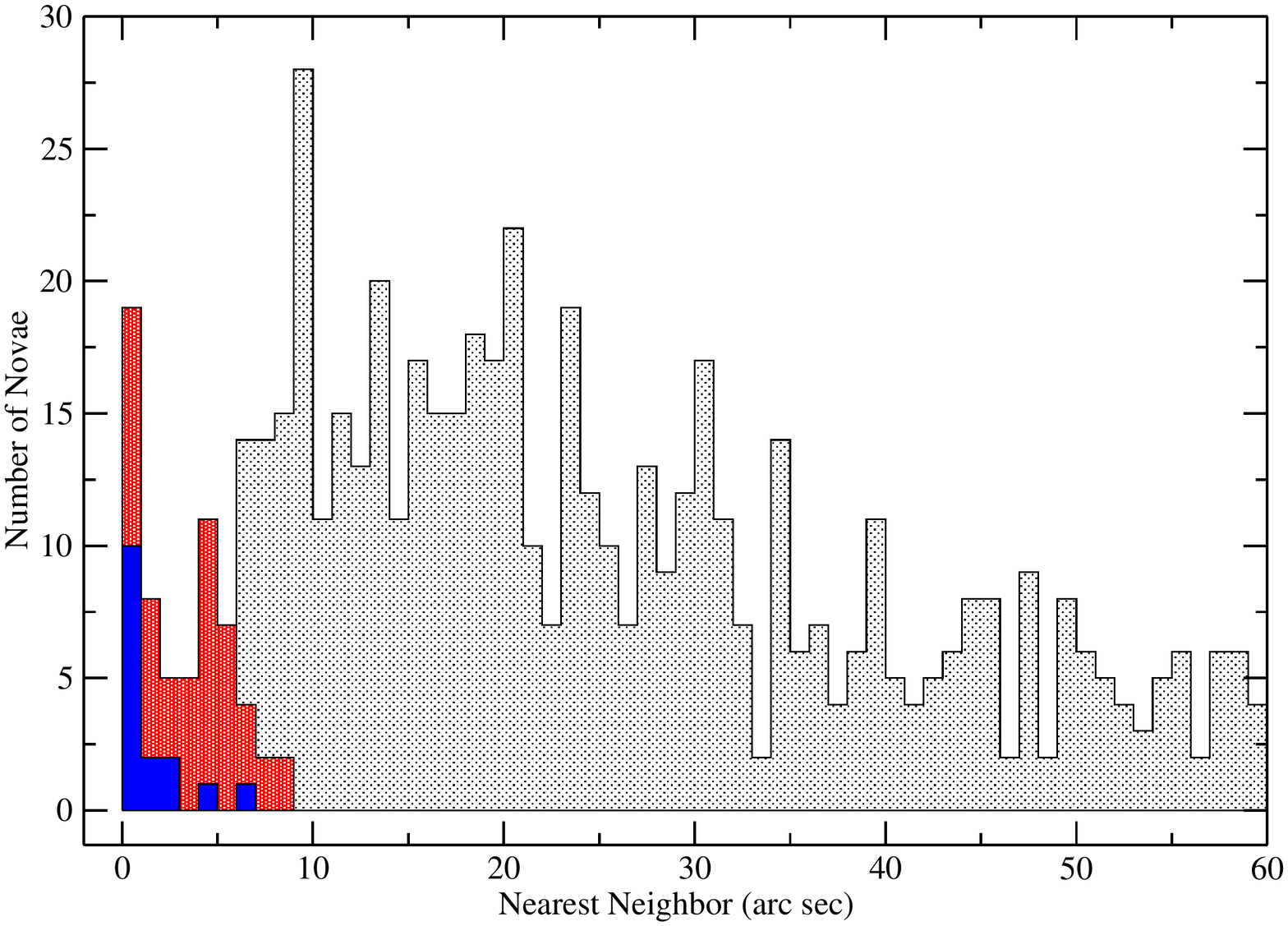}
\caption{The nearest neighbor distribution for novae in M31. The angular
separation between each of the 964 nova candidates in M31 and their
nearest neighbor have been divided into 1$''$ bins,
and the total number of novae
in each bin plotted as a function of angular separation. The shaded
red region shows the novae identified in our initial screen for RNe, while
the dark blue region represents systems that we have subsequently
identified as RNe or likely RN systems.
\label{fig1}}
\end{figure}

Clearly, such a coarse screening process will introduce significant
numbers of chance positional near-coincidences, or ``false positives"
in our search for RNe.
We can gain some insight into how likely chance positional
coincidences are to affect our screening process by considering a
``nearest neighbor nova" distribution.
Figure~\ref{fig1} shows the result of
binning the angular separations between each of the nova
eruptions reported through the end of calendar year 2013
and their closest neighbor nova. The plot, which has been truncated to
a maximum separation of $60''$, reveals that
there are a total of 55 nova candidates with a neighbor within $0.1'$.
All of these novae are included within our coarse screen.
An additional 8 novae from early photographic surveys have been included
in the expanded screen.

In an attempt to quantify the significance
of a given positional near-coincidence,
we have estimated the probability of its occurrence as a function of position
in M31. Specifically,
for a given observed angular separation, $s$,
the probability of a chance positional near-coincidence, $P_C$, can be
estimated by considering the observed nova surface density
(assumed to follow the background
$R$-band light of the galaxy\footnote{We ignore possible variations in the
(recurrent) nova density between bulge and disk populations, which are assumed
to be small.})
in the vicinity of the nova, which we estimate
by considering the number of novae in
an elliptical annulus centered on the nucleus of
M31 that contains the position of the nova. The width of the annulus is
taken to be 1$'$, with the inner and outer elliptical annuli located
0.5$'$ on either side of the
nova's position. Our estimate of the surface density is then simply
given by the number of novae observed in the annulus, $N$,
divided by the area of the annulus, $A$. Since we do not know {\it a priori\/}
which novae will be RN candidates,
the computation of $P_C$ must consider the probability
that any nova in the annulus lies within a distance $s$ of any other nova
in the annulus. In practice, $P_C$, for a given nova pair
is simply a function of the observed
separation, $s$, and the surface density of novae at the location of interest.
Specifically, we have

\begin{equation}
P_C \simeq 1 - \prod_{k=1}^{N-1}~(1 - kx),
\end{equation}

\noindent
where $x=\pi s^2/A$ is the overlap area for a chance positional coincidence
of separation less than or equal to $s$.
Not surprisingly, the probability of a chance positional near-coincidence
is particularly high close to the center of M31
where the nova density is the highest \citep[e.g.,][]{cia87}.

Out of a total of 964 nova eruptions
reported in M31 through the end of 2013\footnote{Of these,
25 nova candidates have been deemed to be non-novae.
However, we have included all nova candidates in our screen as some of
the systems, e.g. M31N 1957-10b, have possibly been misclassified.
We discuss the nature of all objects that pass our screen
in the following section.},
a total of 118 RN candidates have survived
our initial screening process and
are presented in Table~\ref{rntable}. The first column gives the first recorded
eruption of a RN candidate with possible subsequent eruptions
given in column 2.
The time interval between the initial outburst of a
nova and its possible subsequent
eruptions ($\Delta t$), the reported spatial separation ($s$),
the isophotal radius ($a$) of the nova
(i.e., the semimajor axis of an elliptical
isophote that passes through the position of the nova),
the number of novae ($N$) observed in the $1'$ wide annulus passing through the position
of the nova,
and the probability of
a chance positional coincidence ($P_C$), computed from Equation (1),
are also summarized in Table~\ref{rntable}.
For systems with multiple recurrences, we also include
screening data between all subsequent recurrences.

\section{Individual Systems}

Below we consider the individual RN candidates that have passed our
initial screen. For all RN candidates, we attempted to
locate the original plate or digital image data in order to make a
detailed comparison of the nova positions. We were largely
successful, with the exception of some images taken as part of the
\citet{cia87} survey that are no longer available. Photographic
plates, for example from the surveys of \citet{hub29} and \citet{arp56},
were scanned and converted into FITS images for comparison with
modern CCD images. Images of the eruptions
of a given RN candidate have been carefully registered using
the \texttt{geomap} and \texttt{geotran} routines in IRAF\footnote{
IRAF is distributed by the National Optical Astronomy Observatory,
which is operated by the Association for Research in Astronomy, Inc.
under cooperative agreement with the National Science Foundation.}.
Once aligned, the relative spatial positions of the novae
have been displayed through
a comparison image formed by the ratio of the images of the
individual eruptions.

In cases where the positions of the original nova eruption
appears to coincide within observational errors
with the subsequent eruption(s), we have
re-computed the astrometry and determined
revised coordinates for objects with poorly determined positions
(e.g., those systems discovered on photographic plates prior to $\sim$1980).
Our revised coordinates are given in Table~\ref{revcoord}.
For RN candidates that erupted in the early part of the 20th century
(e.g., novae from the Hubble survey), there can be significant proper motion
of the field stars used in the revised astrometry.
When known, we have excluded any field stars with measured proper motions in
excess of 10 mas/yr in our astrometric solutions.
In most cases we estimate our revised astrometry to produce nova positions
accurate to 1$''$.

Given the limitations of the data (e.g., finite temporal and
spatial resolution of the images), it is never possible to establish with
absolute certainty that a given candidate is in fact a RN system.
For example, there is always
the possibility of a chance positional coincidence within observational
uncertainty of two (or more)
unrelated nova eruptions. In other cases, given limited temporal sampling,
there is the possibility for long period
variable (LPV) stars to masquerade as novae \citep[e.g.,][]{sha08}.
Despite these challenges, in the discussion to follow we have
brought to bear all available evidence in judging the likelihood that
a given candidate is a RN. This process has led us to
place the RN candidates from our initial screen into one of four categories:
(1) Recurrent Novae -- systems where the weight of the evidence strongly
supports the spatial coincidence of the eruptions, and where the probability
of a chance positional coincidence is low ($P_C\lessim0.1$),
(2) Possible Recurrent Novae -- systems where there remains some doubt as to 
the precise spatial coincidence of the eruptions or where
the nature of the object is in doubt (e.g., a possible LPV
or foreground Galactic dwarf nova),
(3) Unlikely Recurrent Novae -- often systems without available finding charts
where the probability of chance positional near coincidence is relatively high
($P_C\grtsim0.4$), and
(4) Rejected Recurrent Nova Candidates -- systems where the archival images
clearly establish that the eruptions are not spatially coincident.
We begin by discussing the systems judged to be RNe.

\subsection{Recurrent Novae}

\subsubsection{M31N 1919-09a}

\begin{figure}
\includegraphics[angle=0,scale=.28]{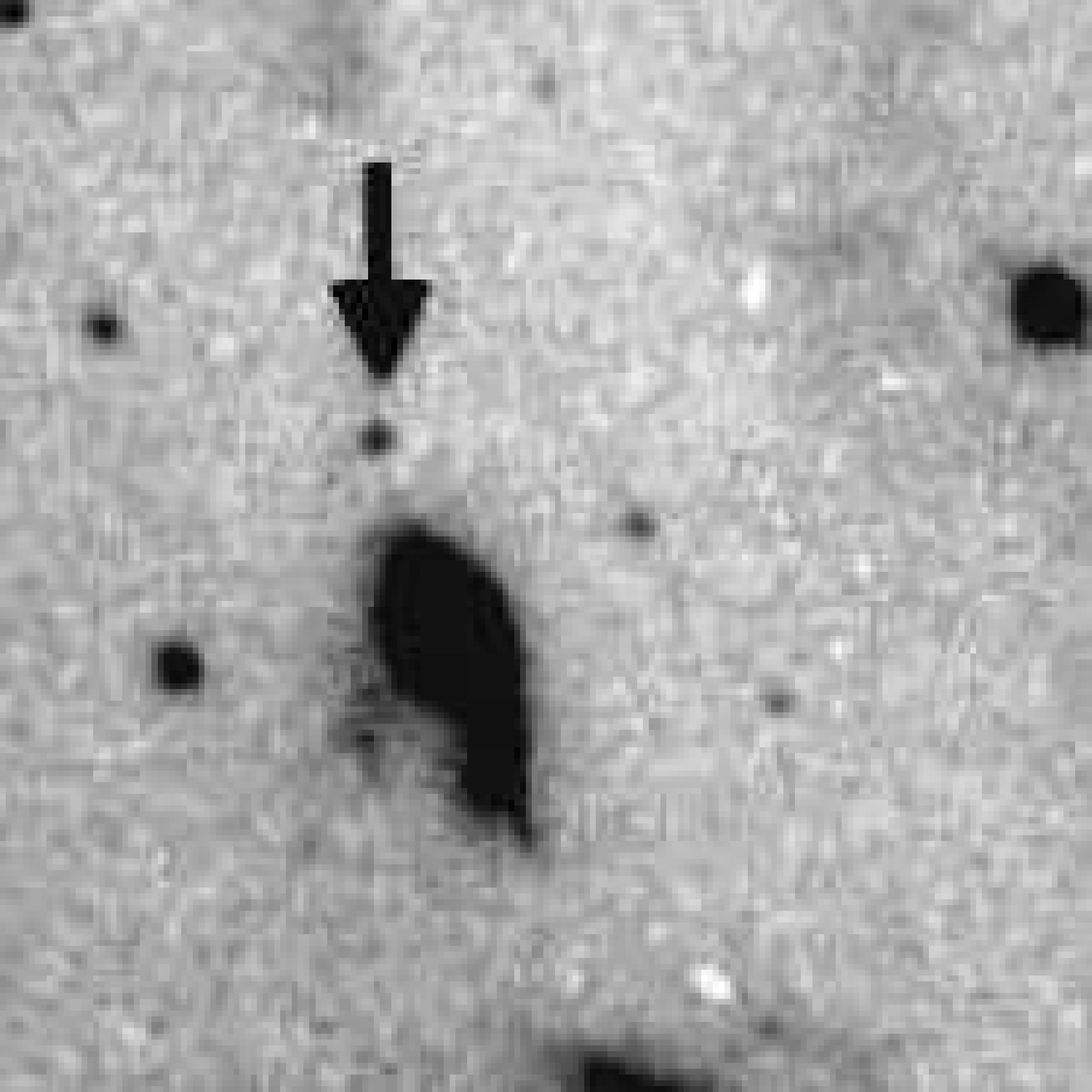}
\includegraphics[angle=0,scale=.28]{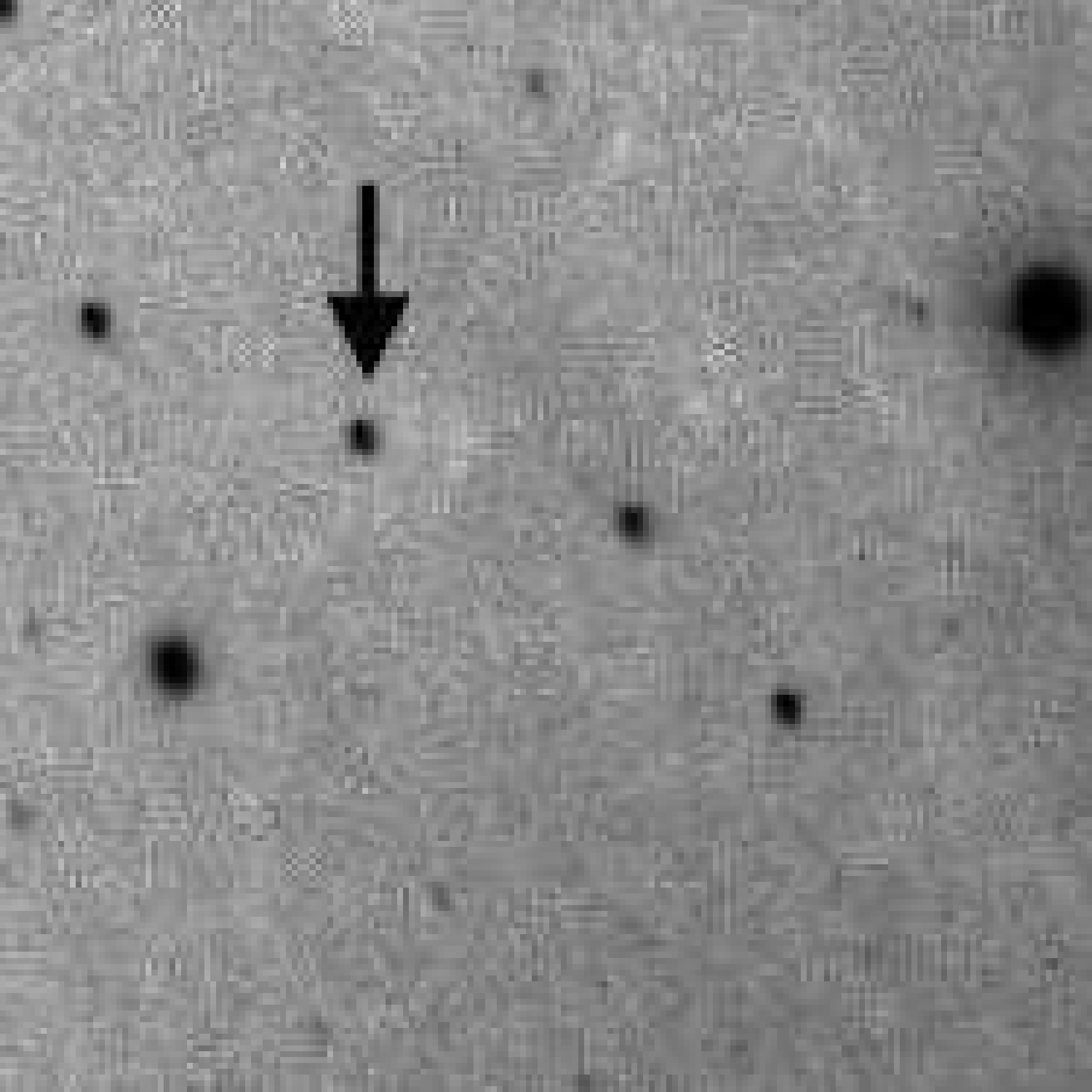}
\includegraphics[angle=0,scale=.28]{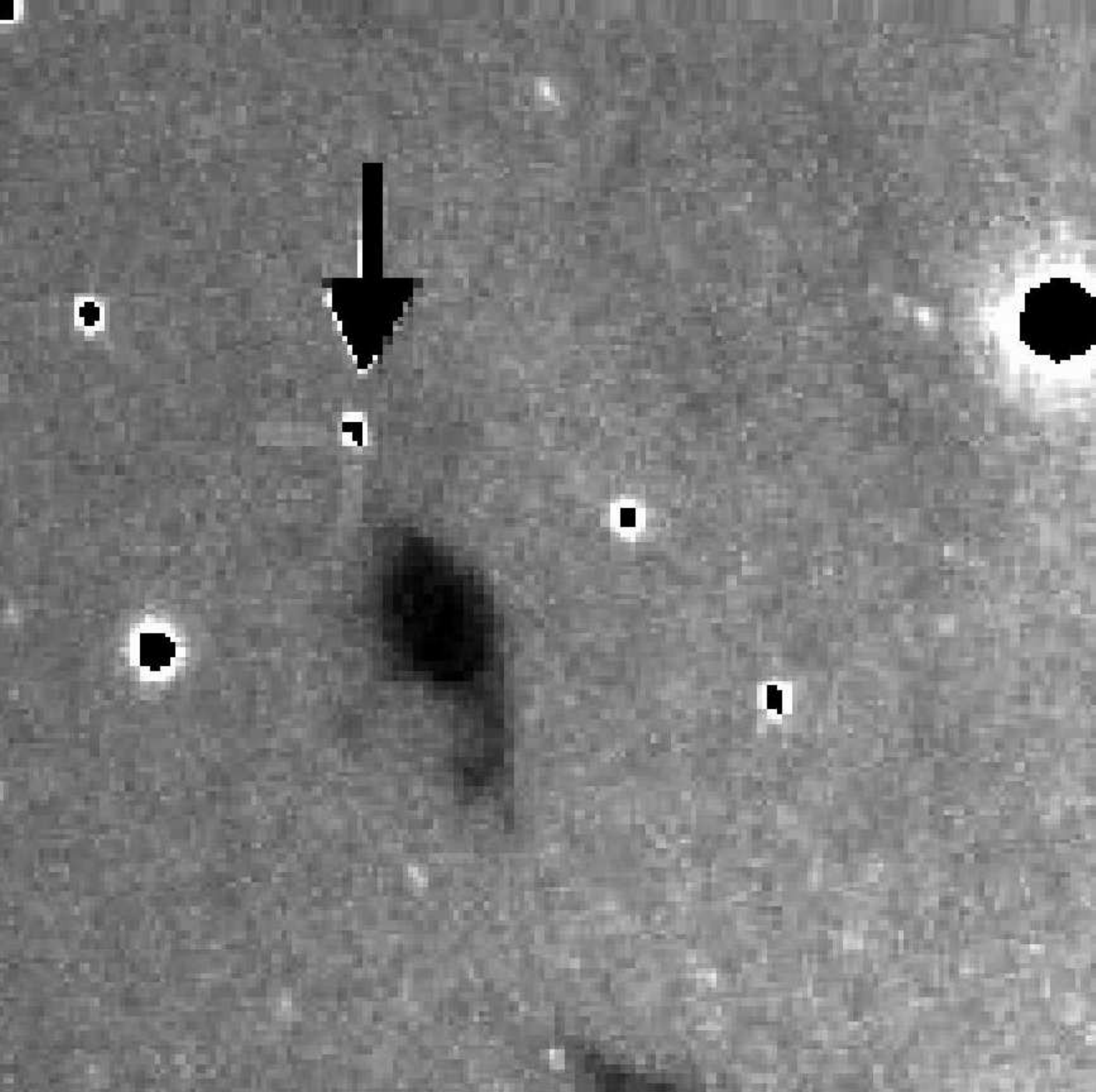}
\caption{Images of M31N 1919-09a, 1998-06a, and their comparison
(left, center, and right, respectively).
The image for M31N 1919-09a is from plate \#5054 in the Carnegie archives,
while that for 1998-06a is from the RBSE program
at Kitt Peak National Observatory \citep{rec99}.
The comparison image reveals that the two novae are spatially
coincident to within the resolution afforded by the images ($\sim 1''$).
The black ``smear" is part of ink marks on the glass side of the plate.
North is up and East to the left, with a scale of $\sim2'$ on a side.
\label{fig2}}
\end{figure}

M31N 1998-06a was discovered at $m_{H\alpha}=16.3$ on 1998 June 06 UT
as part of the Kitt Peak National Observatory's
Research-Based Science Education
(RBSE) program \citep{rec99}.
The object lies approximately $10.5'$ from the nucleus of M31
and just $1.8''$ from the nominal position of M31N 1919-09a \citep{hub29}.
The probability of a chance positional coincidence
with $s\leq1.8''$ at this location is estimated to be 0.030, making the object
a strong RN candidate. To confirm this possibility we have located the
original discovery plate (\#5054) in the Carnegie archives. The plate was taken
on 1919 September 21, and shows the nova at $m_{pg}=17.6$.
Figure~\ref{fig2} shows
a scan of the plate compared with the RBSE image of 1998-06a. The novae
clearly appear to be spatially coincident. Revised astrometry
of M31N 1919-09a (see Table~\ref{revcoord})
shows that the nova lies $\sim1.5''$ from the measured position
of 1998-06a, dropping the probability of a chance coincidence to 0.016.
To within the uncertainty of the measured positions, the novae are
consistent with being spatially coincident, and
we conclude that M31N 1919-09a is very likely a RN.
We note that M31N 1998-06a was detected as a faint supersoft X-ray source
(SSS) $1028\pm92$ days after the optical outburst \citep{pie05}.
Owing to the low number of detected photons, no estimate of the
effective temperature was possible.
The SSS disappeared $1773\pm463$ days after outburst \citep{pie07a}.

\subsubsection{M31N 1923-12c}

\begin{figure}
\includegraphics[angle=0,scale=0.28]{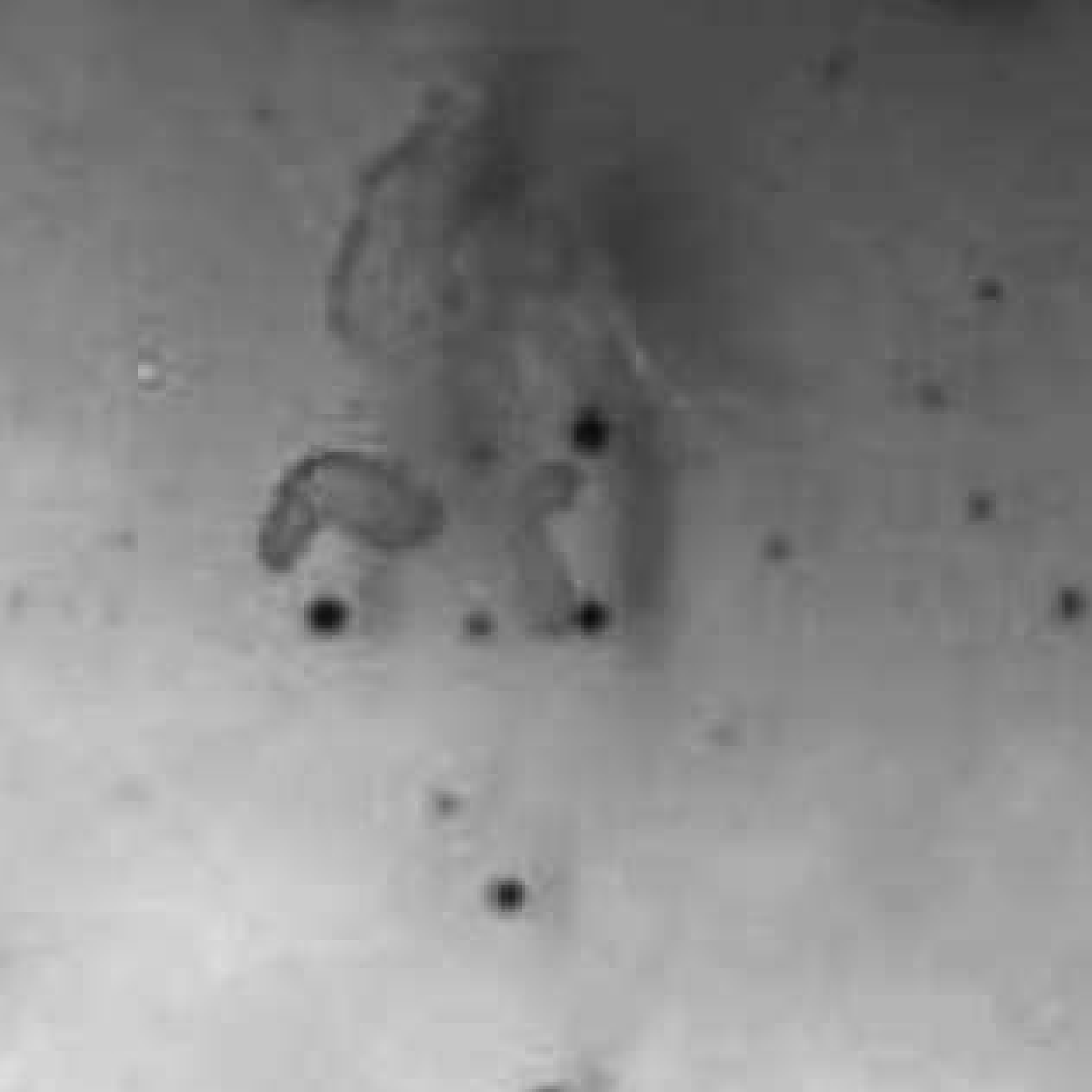}
\includegraphics[angle=0,scale=0.28]{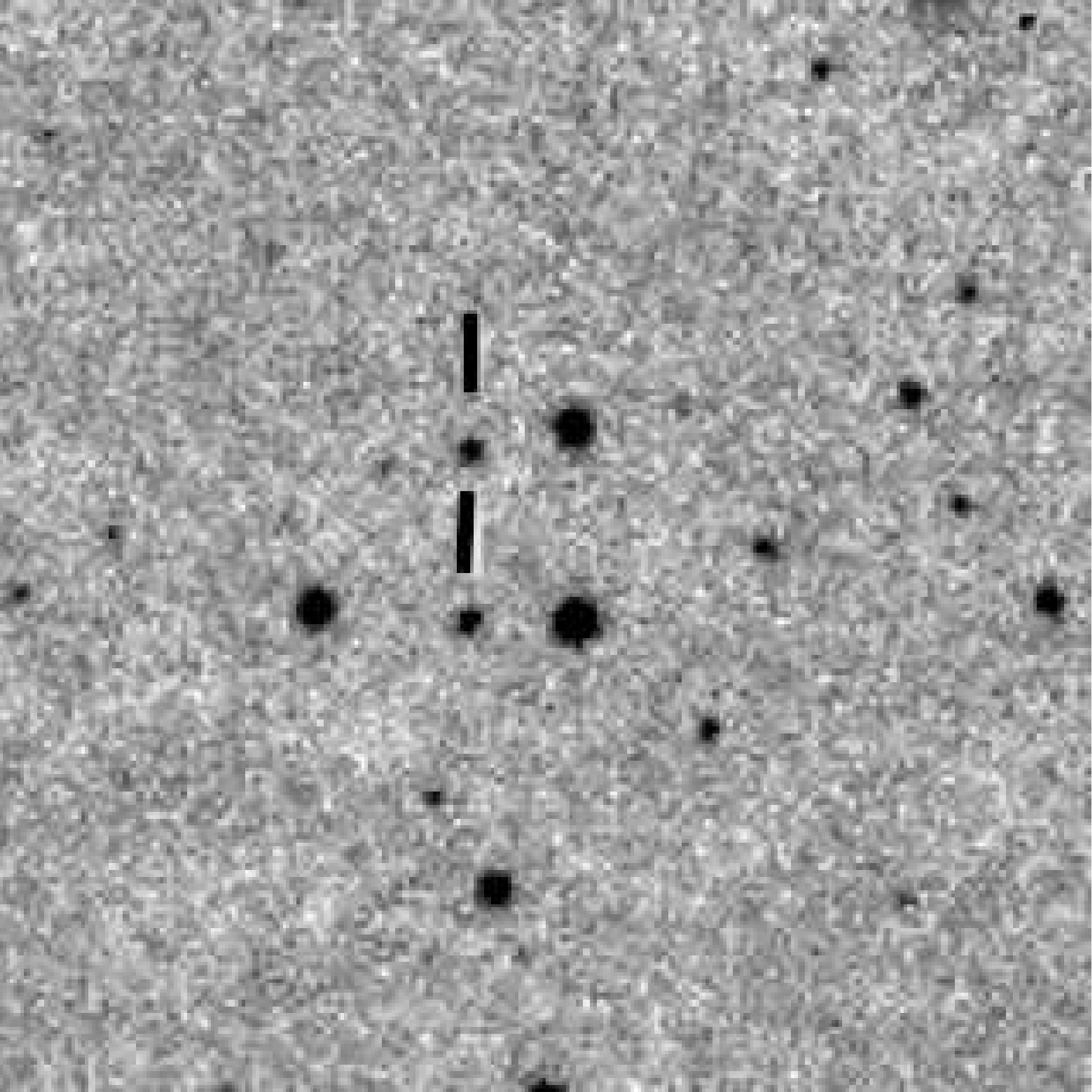}
\includegraphics[angle=0,scale=0.28]{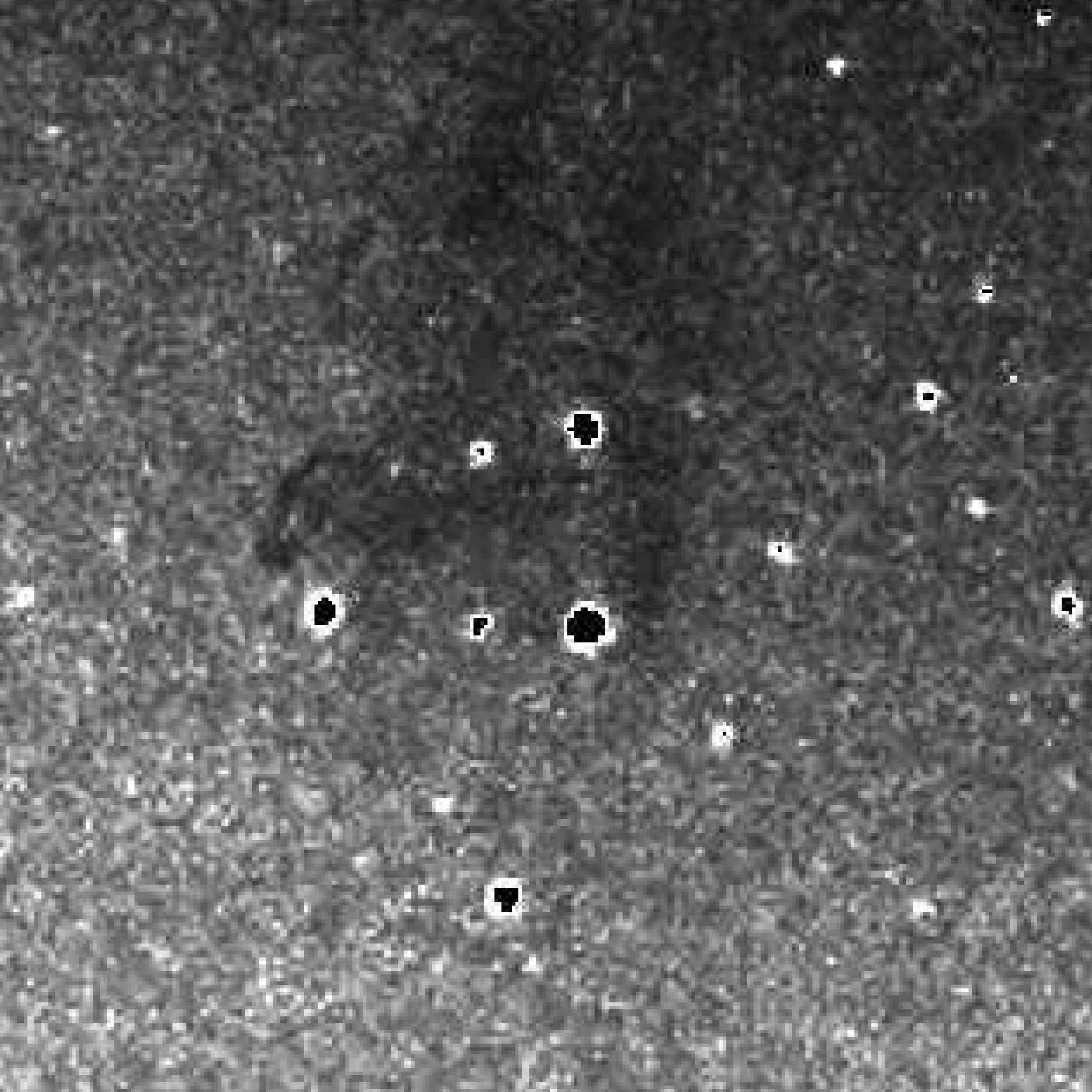}
\caption{Images of M31N 1923-12c, 2012-01b, and their comparison
(left, center, and right, respectively).
The image for M31N 1923-12c is from plate H348H in the Carnegie archives,
while the chart of 2012-01b is courtesy of K. Nishiyama and
F. Kabashima (Miyaki-Argenteus Observatory, Japan).
As revealed by the comparison image, the novae are spatially coincident
to within measurement uncertainties ($\sim 1''$).
North is up and East to the left, with a scale of $\sim3'$ on a side.
\label{fig3}}
\end{figure}

M31N 2012-01b was a rapidly-fading, He/N nova that was discovered on
2012 January 21.419 UT by K. Nishiyama and F. Kabashima
approximately $6''$ from the nominal position of
M31N 1923-12c, a nova discovered by Hubble on 1923 December 11.
We located Hubble's plate (H348H) in the Carnegie archives and
produced a scan of the plate, a portion of which is reproduced in
Figure~\ref{fig3} along with a chart of 2012-01b courtesy of K. Nishiyama and
F. Kabashima (Miyaki-Argenteus Observatory, Japan). Nova 1923-12c,
which is indicated by Hubble's hand-drawn ink marks, appears to lie
at the same position as 2012-01b. The revised astrometry of 1923-12c
(see Table~\ref{revcoord})
confirms that the nova lies within $0.45''$ of 2012-01b. The probability
of a chance coincidence at this location in the galaxy is estimated
to be just 0.0033. We conclude, as initially reported in \cite{sha12b}, that
1923-12c is a RN, and that 2012-01b is a subsequent eruption of
Hubble's 1923 nova.

\subsubsection{M31N 1926-06a}

\begin{figure}
\includegraphics[angle=0,scale=.28]{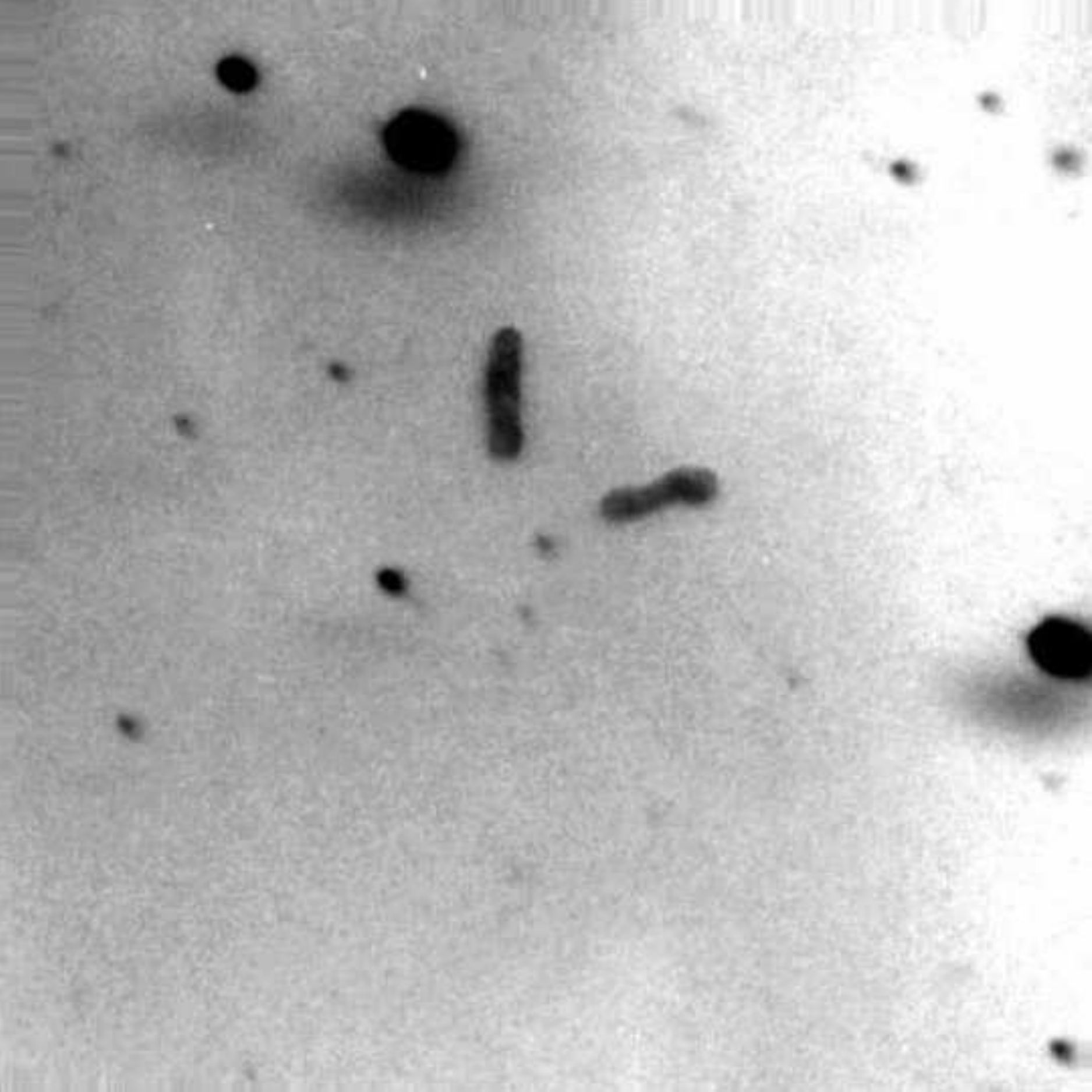}
\includegraphics[angle=0,scale=.28]{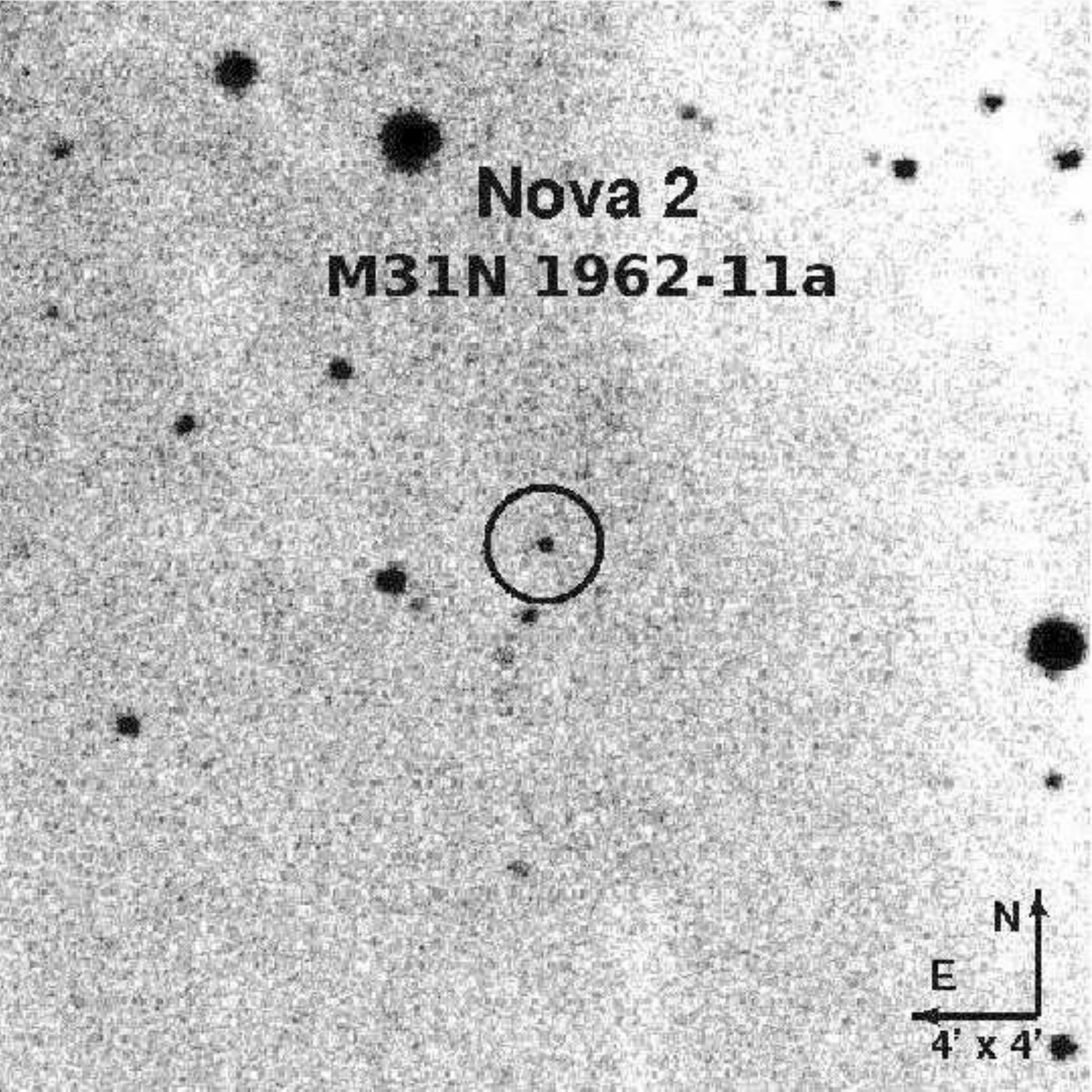}
\includegraphics[angle=0,scale=.28]{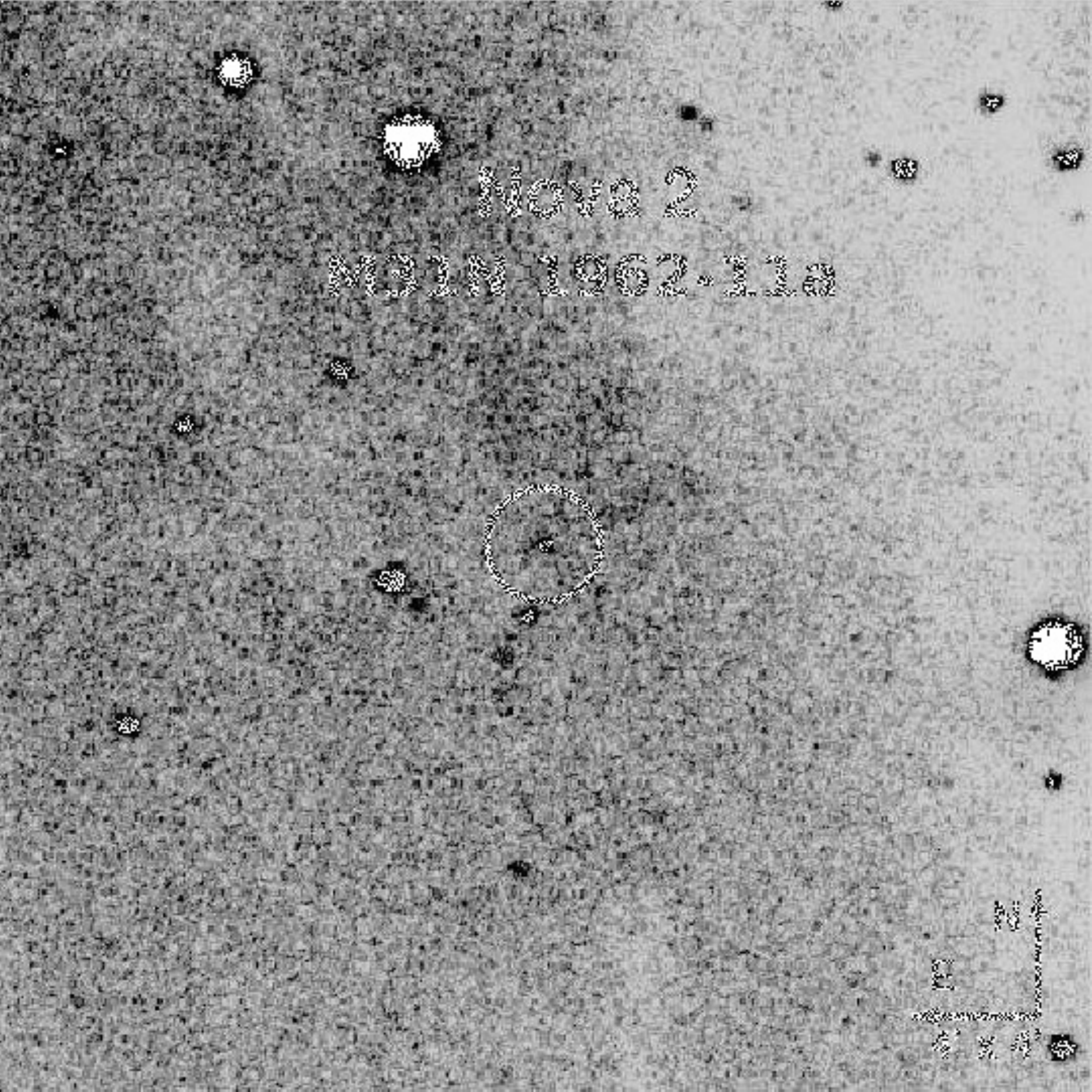}
\caption{Images of M31N 1926-06a, 1962-11a, and their comparison
(left, center, and right, respectively).
The image for M31N 1926-06a was produced from plate H304D in the
Carnegie archives, while the chart for 1962-11a is from \citet{hen08a}.
A comparison of the images reveals that the novae are spatially coincident
to within measurement uncertainties ($\sim 1''$).
North is up and East to the left, with a scale of $\sim4'$ on a side.
\label{fig4}}
\end{figure}

A possible recurrence of M31N 1926-06a, 1962-11a, was
discovered by \citet{ros64} as part of his
multi-year nova survey. As pointed out by \citet{hen08a}, the recurrence was
discovered independently
by \citet{bor68}. Our expanded screen shows that the published position
of M31N 1962-11a lies $\sim6.3''$ from the reported position
of 1926-06a (nova \#62 in \citet{hub29}). We were
successful in locating the original discovery plates (H304D and H309D) in the
Carnegie archives.
Figure~\ref{fig4} shows a comparison of the position of M31N 1926-06a and that
of 1962-11a from \cite{hen08a}. The novae appear coincident. Revised
astrometry of M31N 1926-06a (see Table~\ref{revcoord})
reveals that Hubble's nova lies within
$1.40''$ of the position of 1962-11a. The probability of a chance coincidence
at this location in the galaxy is just 0.010,
and we conclude that M31N 1962-11a is almost certainly a recurrence of 1926-06a.

\subsubsection{M31N 1926-07c}

\begin{figure}
\includegraphics[angle=0,scale=.28]{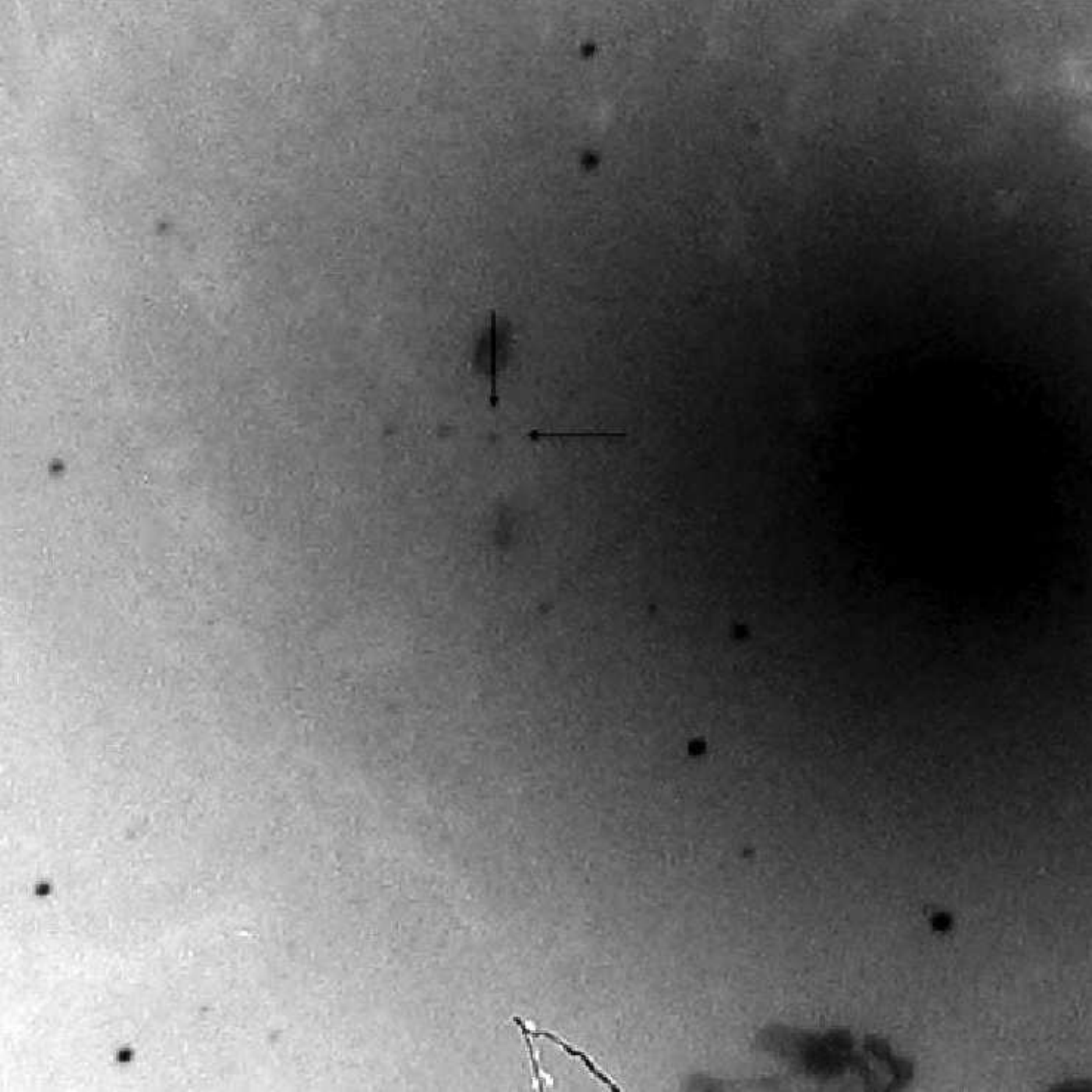}
\includegraphics[angle=0,scale=.28]{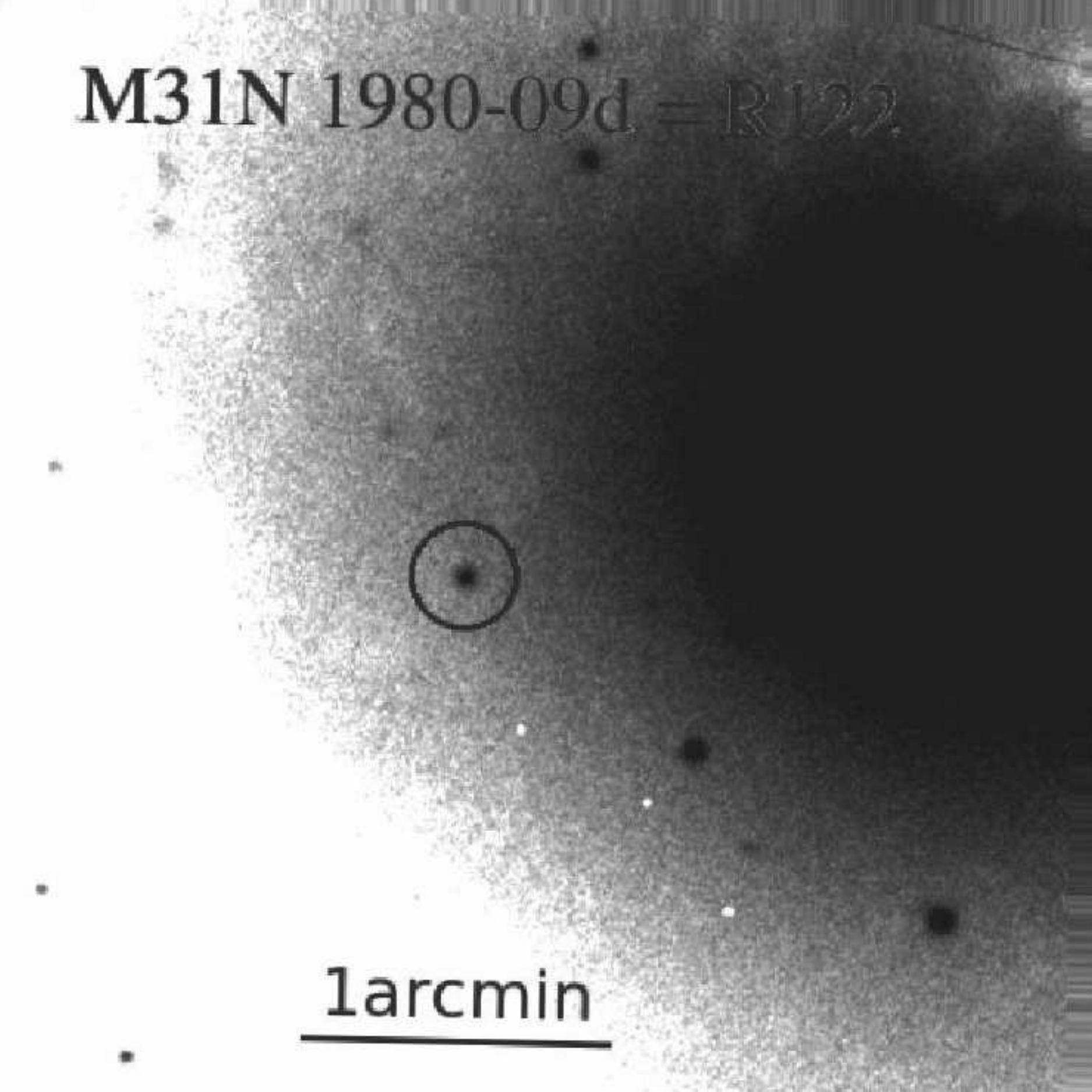}
\includegraphics[angle=0,scale=.28]{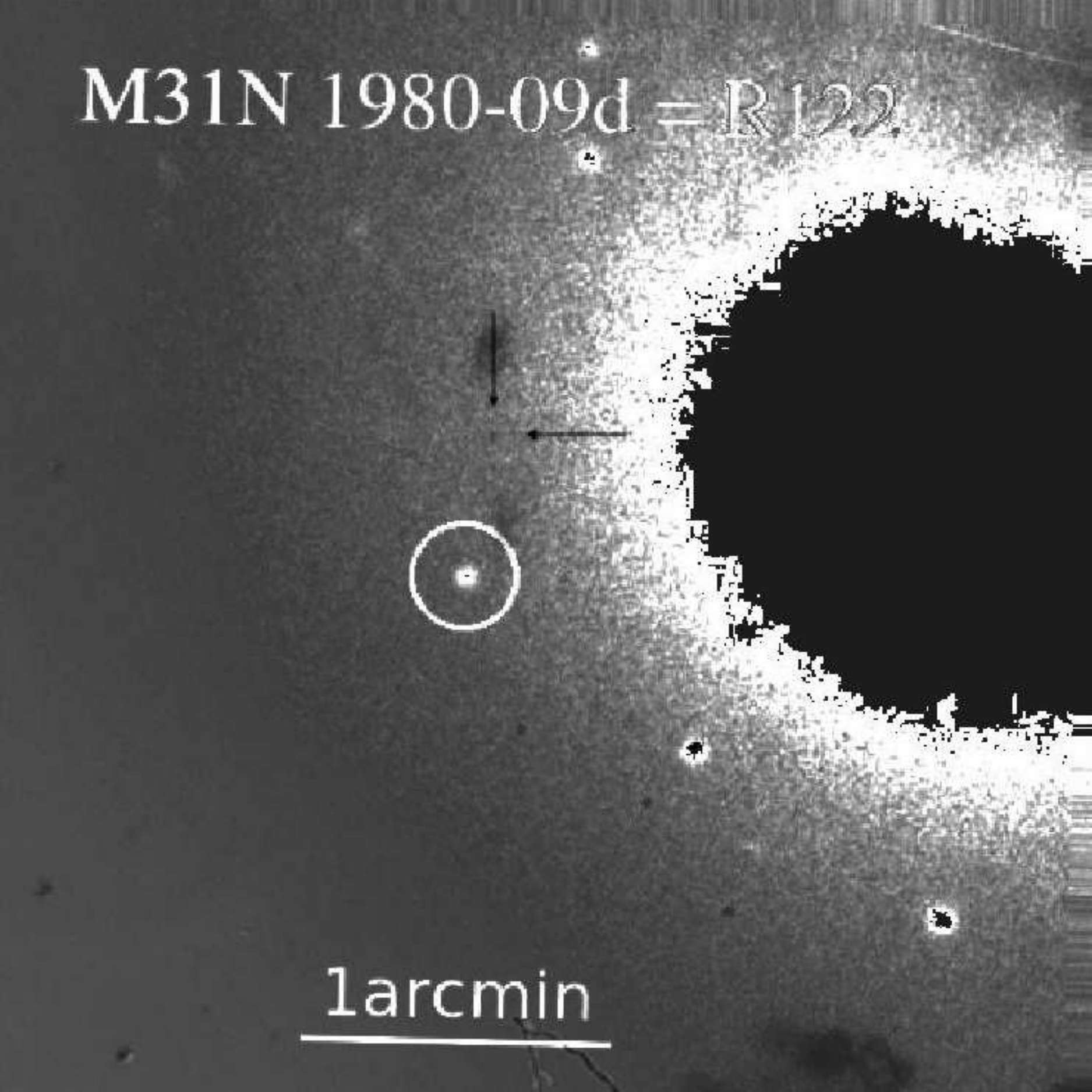}
\caption{Images of M31N 1926-07c, 1980-09d and their comparison
(left, center, and right, respectively).
The comparison image shows that
M31N 1980-09d (shown in white) is clearly not a recurrence of 1926-07c, with
the former nova being located $\sim27''$ to the S of 1926-07c.
However, as seen in Figure~\ref{fig6}, M31N 1926-07c
turns out to be coincident with M31N 1997-10f and 2008-08b, and thus
is in fact a RN.
The finding chart for M31N 1926-07c was reproduced from plate
H668H from the
Carnegie archives, while the chart for 1980-09d is taken from
the survey of \citet{ros89}.
North is up and East to the left, with a scale of $\sim3.5'$ on a side.
\label{fig5}}
\end{figure}

Our coarse screen indicated that M31N 1926-07c, nova \#65 in \cite{hub29},
might be coincident with
a nova that erupted 54 years later, M31N 1980-09d \citep{ros89}.
Examination of the
finding charts for the two novae
shown in Figure~\ref{fig5}, clearly demonstrate that the novae are not
spatially coincident. However, revised astrometry for M31N 1926-07c
(see Table~\ref{revcoord}) and further analysis of it's precise position
unexpectedly revealed that the nova was coincident with both
M31N 1997-10f and 2008-08b, a nova pair also identified as a RN candidate
in our coarse screen.

M31N 1997-10f was discovered by \citet{sha01}, and confirmed by \citet{lee12}.
A possible recurrence was discovered by \citet{hen08a}, who noted that the
position of M31N 2008-08b was nearly coincident with 1997-10f
($P_C\simeq0.045$), making the latter
object likely a RN. We have located the original image from \citet{sha01} and
re-measured the coordinates of M31N 1997-10f (see Table~\ref{revcoord}),
confirming the association with 2008-08b.

\begin{figure}
\includegraphics[angle=0,scale=.28]{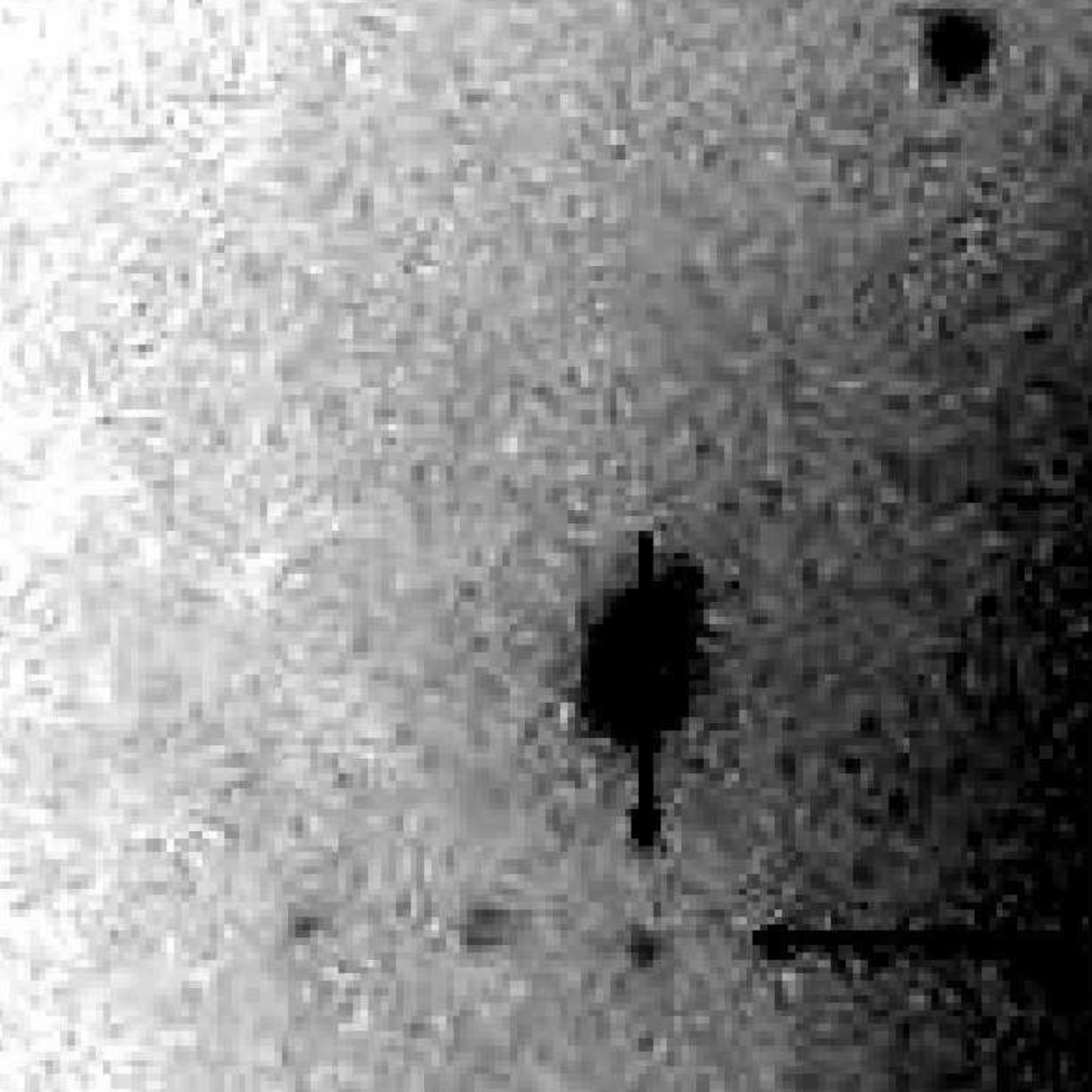}
\includegraphics[angle=0,scale=.28]{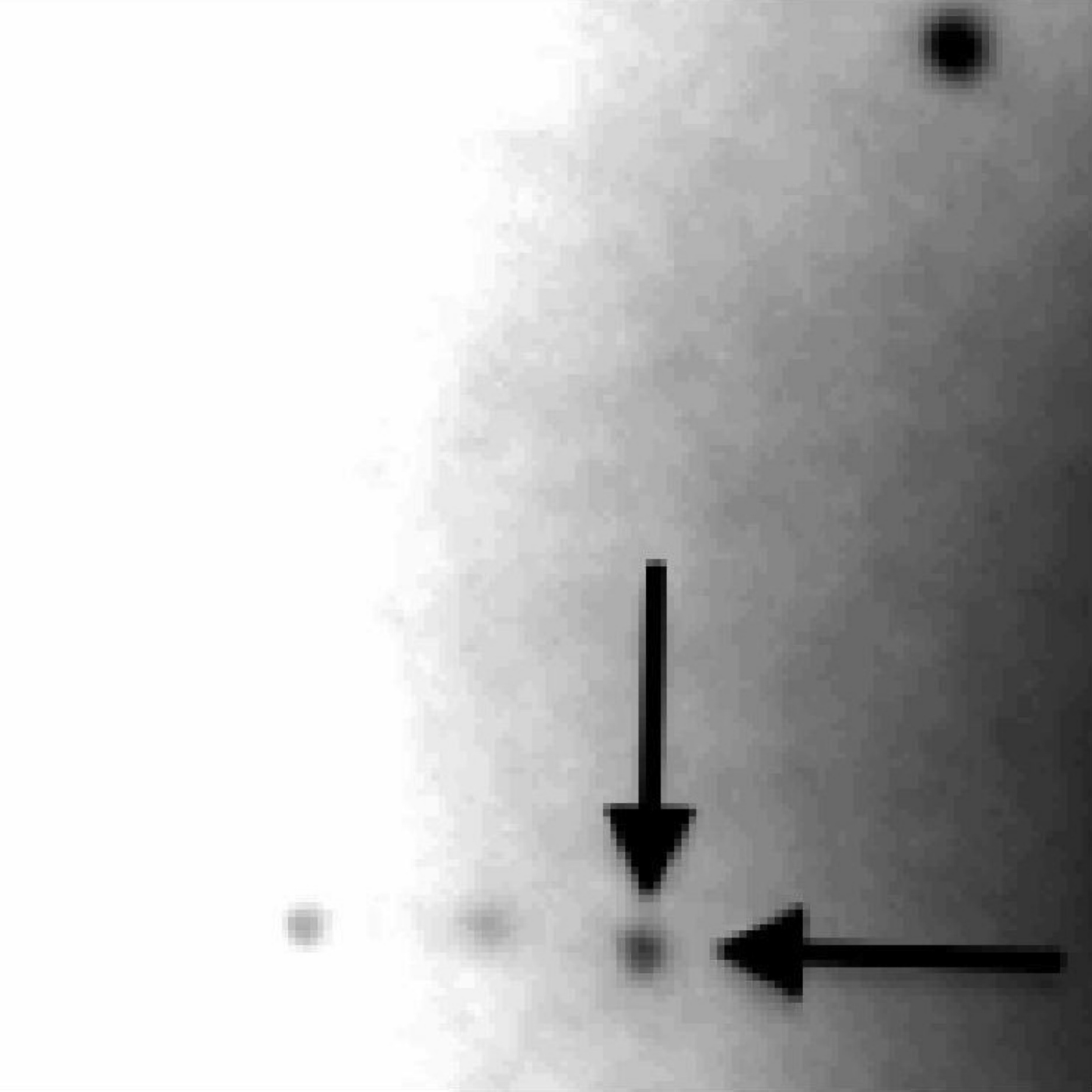}
\includegraphics[angle=0,scale=.28]{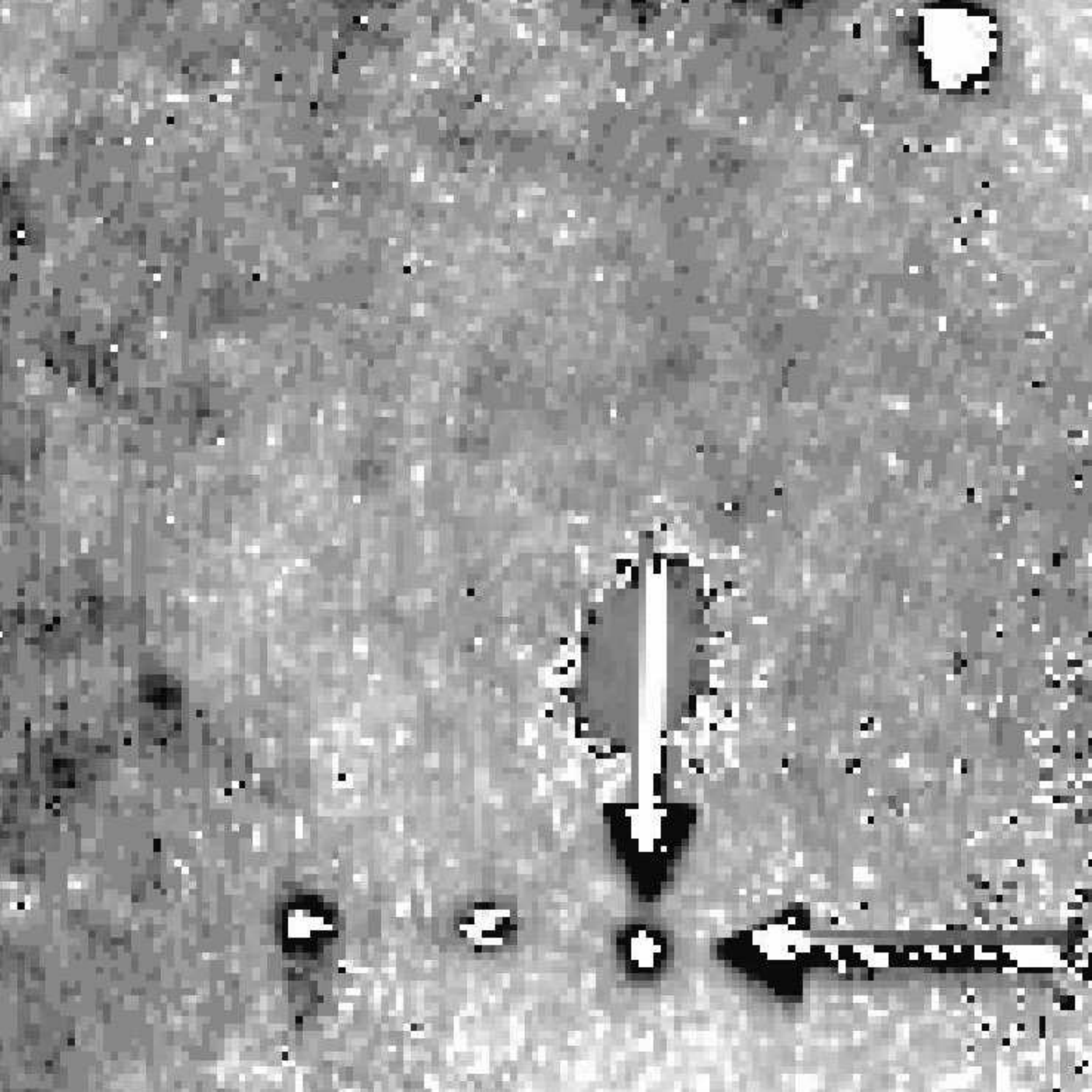}

\includegraphics[angle=0,scale=.28]{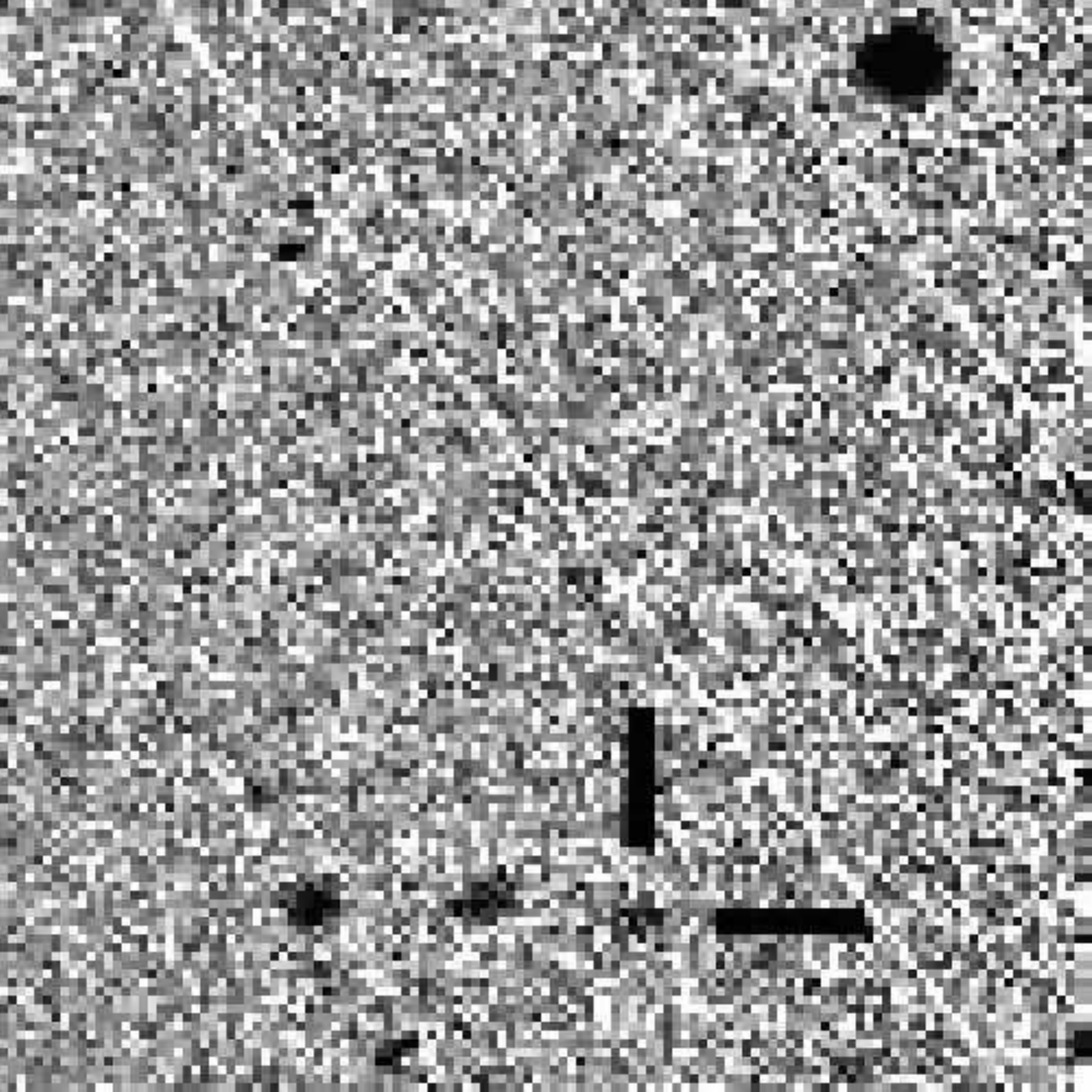}
\includegraphics[angle=0,scale=.28]{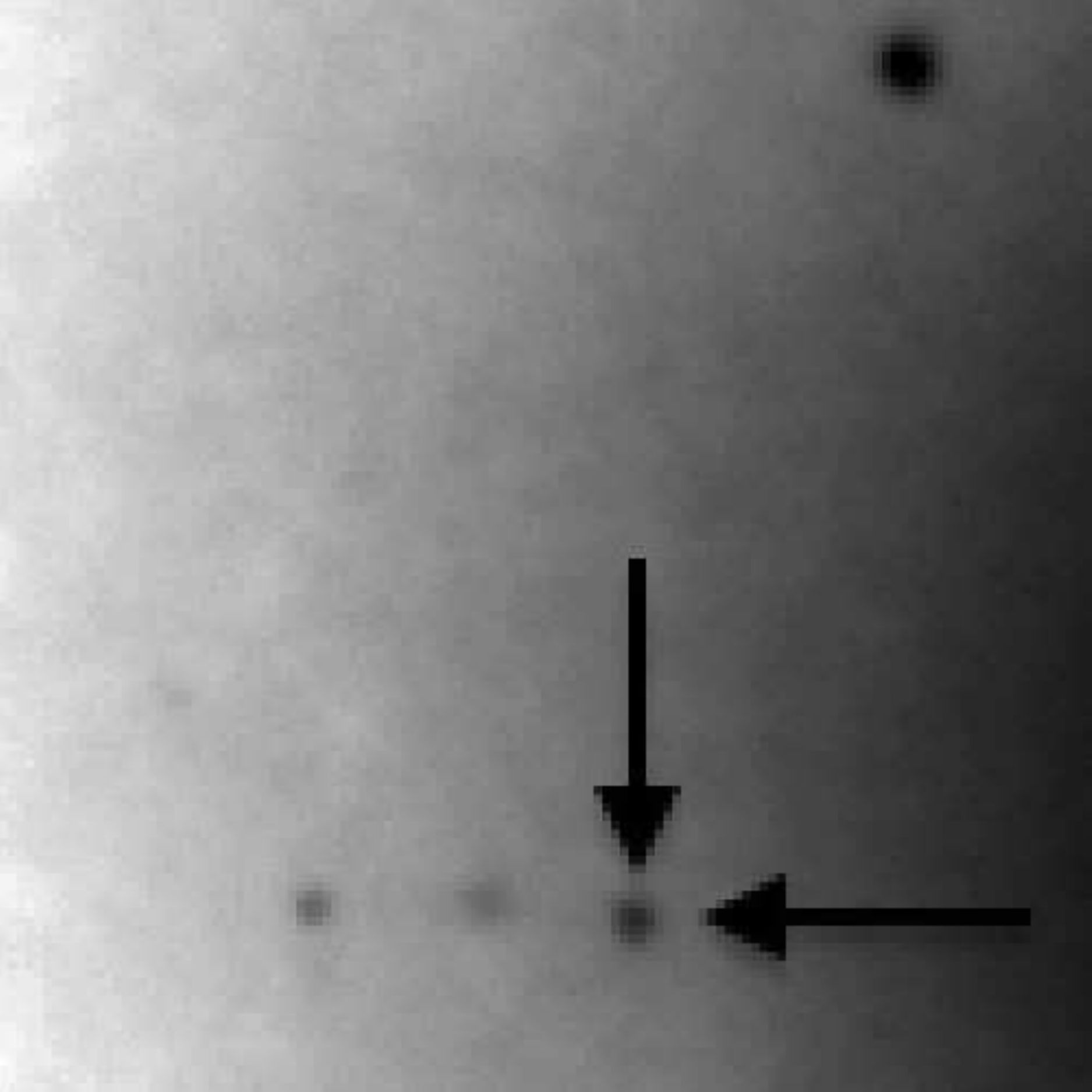}
\includegraphics[angle=0,scale=.28]{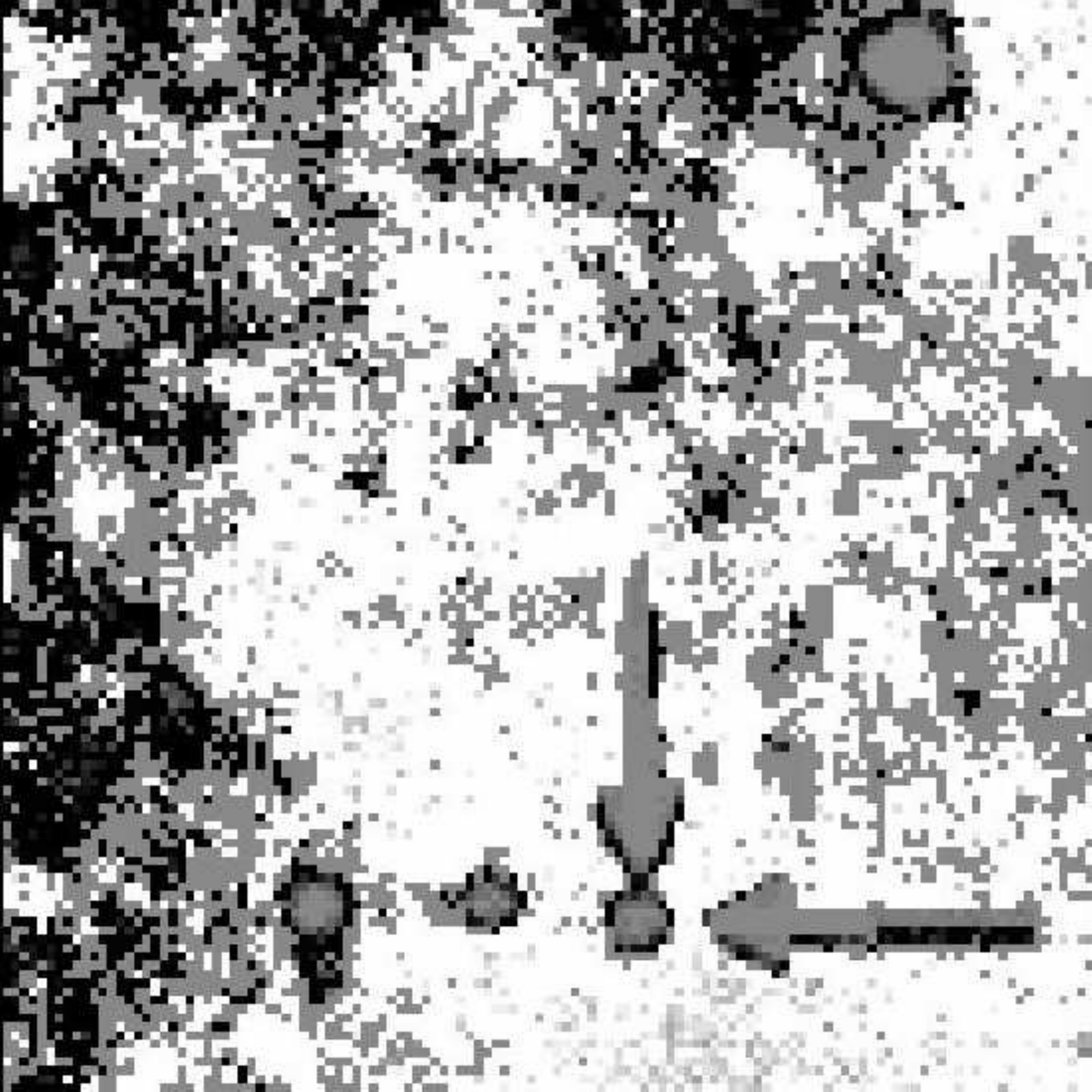}
\caption{Images of M31N 1926-07c, 2008-08b, and their comparison
(top left, center, and right), and
M31N 1997-10f, 2008-08b, and their comparison
(bottom left, center, and right, respectively).
Although the novae erupted relatively close to the nucleus of M31, the
comparison images reveal that the novae are all coincident to within
measurement uncertainties estimated to be $\sim 1''$. Thus,
M31N 1926-07c, although not coincident with 1980-09d (see Fig.~\ref{fig5}),
appears to be a RN clearly associated with both 1997-10f and 2008-08b.
The chart for 1997-10f was produced from data taken in the survey
of \citet{sha01}, while that for
2008-08b is taken from data reported in \citet{hen08a}.
North is up and East to the left, with a scale of $\sim1'$ on a side.
\label{fig6}}
\end{figure}

The positions of M31N 1926-07c, 1997-10f and 2008-08b are compared
in Figure~\ref{fig6}, which shows that all 3 novae are indeed spatially
coincident to within the resolution of the images
($P_C\simeq0.024$). We also note that
M31N 2008-08b was spectroscopically classified as a possible He/N nova
(with a quite narrow H$\alpha$ FWHM) by \citet{dim08}.
We conclude that M31N 1926-07c is a RN with 1997-10f
and 2008-08b representing subsequent outbursts.

\subsubsection{M31N 1945-09c}

\begin{figure}
\includegraphics[angle=0,scale=.28]{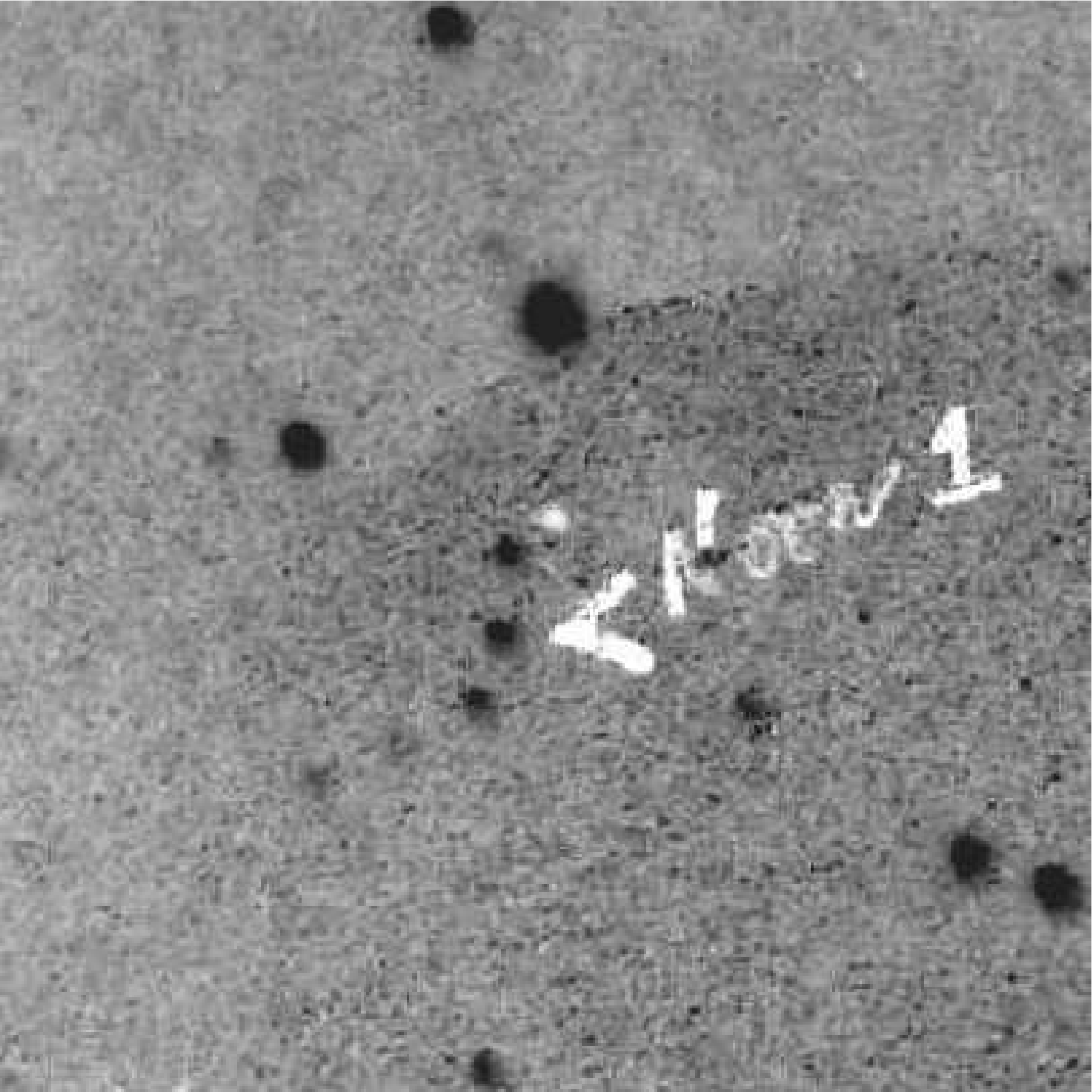}
\includegraphics[angle=0,scale=.28]{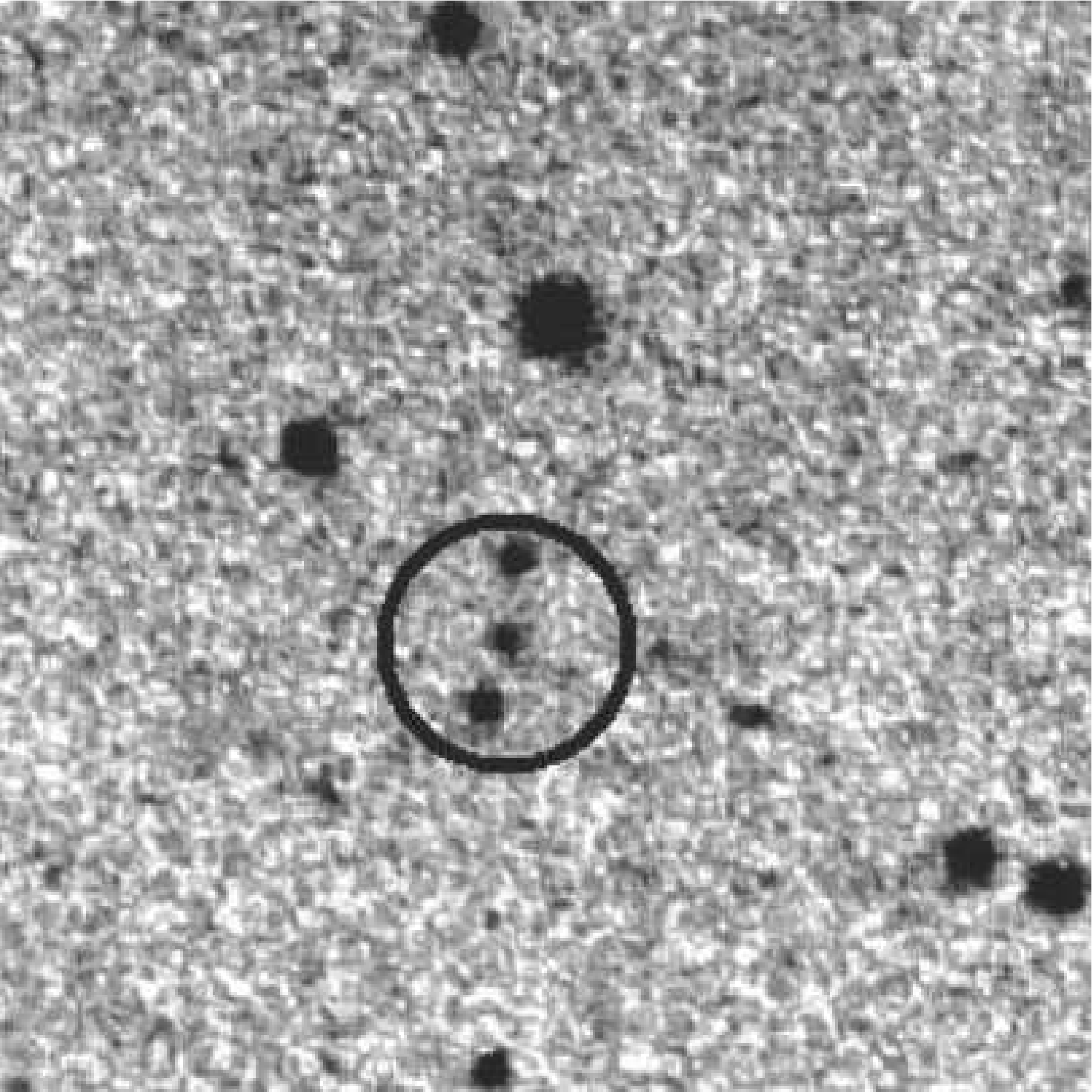}
\includegraphics[angle=0,scale=.28]{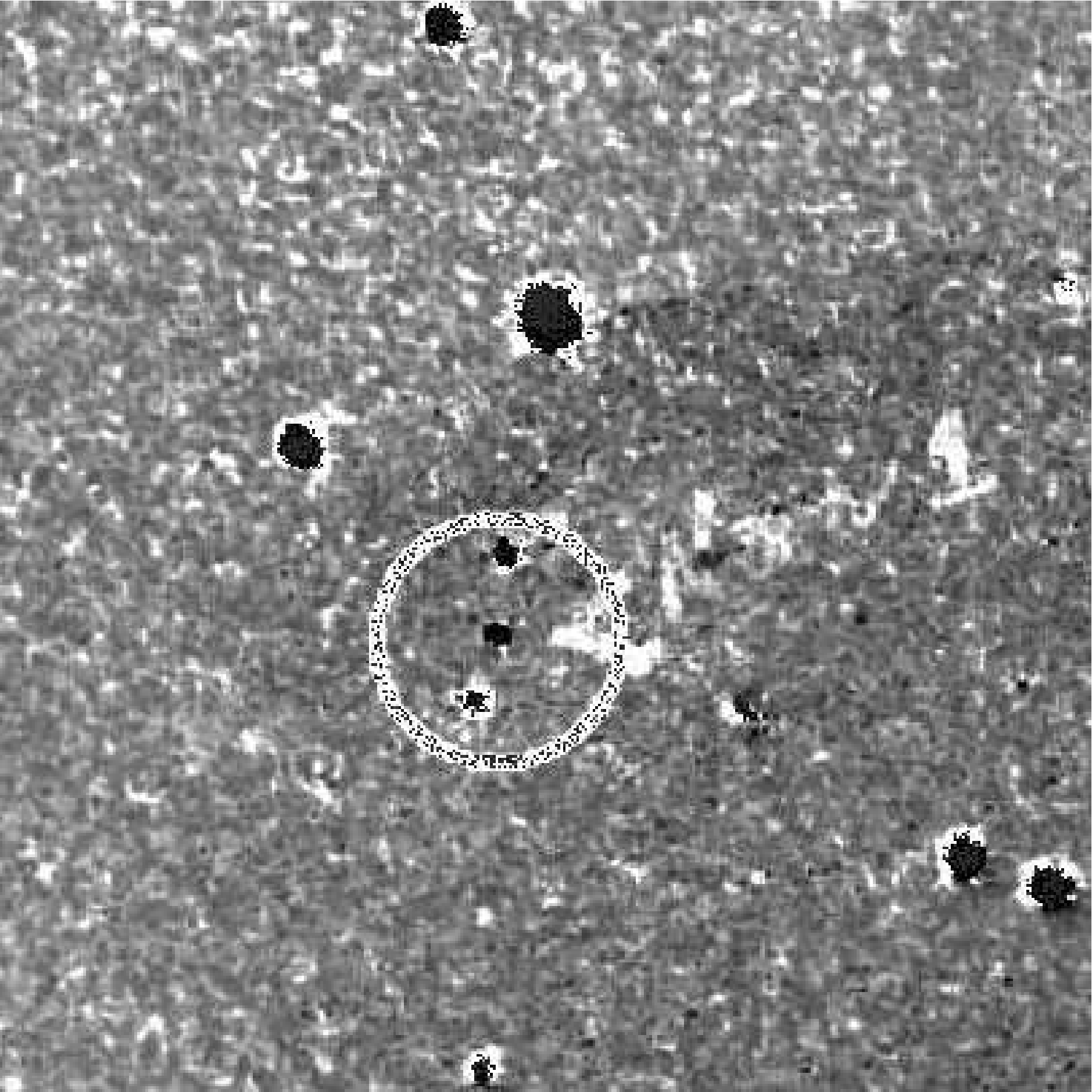}
\caption{Images of M31N 1945-09c, 1975-11a, and their comparison
(left, center, and right, respectively).
The image for M31N 1945-09c is reproduced from print B1677B in
the Huntington Library's Baade collection, while the chart
for 1975-11a is taken from \citet{hen08a}.
The comparison image clearly establishes that the novae are spatially
coincident to within less than an arcsec.
North is up and East to the left, with a scale of $\sim2'$ on a side.
\label{fig7}}
\end{figure}

M31N 1975-11a was observed to erupt in the outskirts of M31
at an isophotal radius,
$a\sim 29'$, and less than $1''$ from the
reported position of 1945-09c \citep{baa64}. At this position in the
galaxy, the probability of a chance positional coincidence is negligible
($\sim 0.0004$). We conclude, as did \citet{hen08a}, that M31N 1945-09c
is a RN. Finding charts for the novae are compared in Figure~\ref{fig7},
and confirm that the novae are indeed spatially coincident.
The revised coordinates for M31N 1945-09c are given in Table~\ref{revcoord}
lie only 0.41$''$ from the position of 1975-11a leading to an even
smaller probability of a chance positional near coincidence, $P_C\simeq0.0001$.

\subsubsection{M31N 1960-12a}

\begin{figure}
\includegraphics[angle=0,scale=.28]{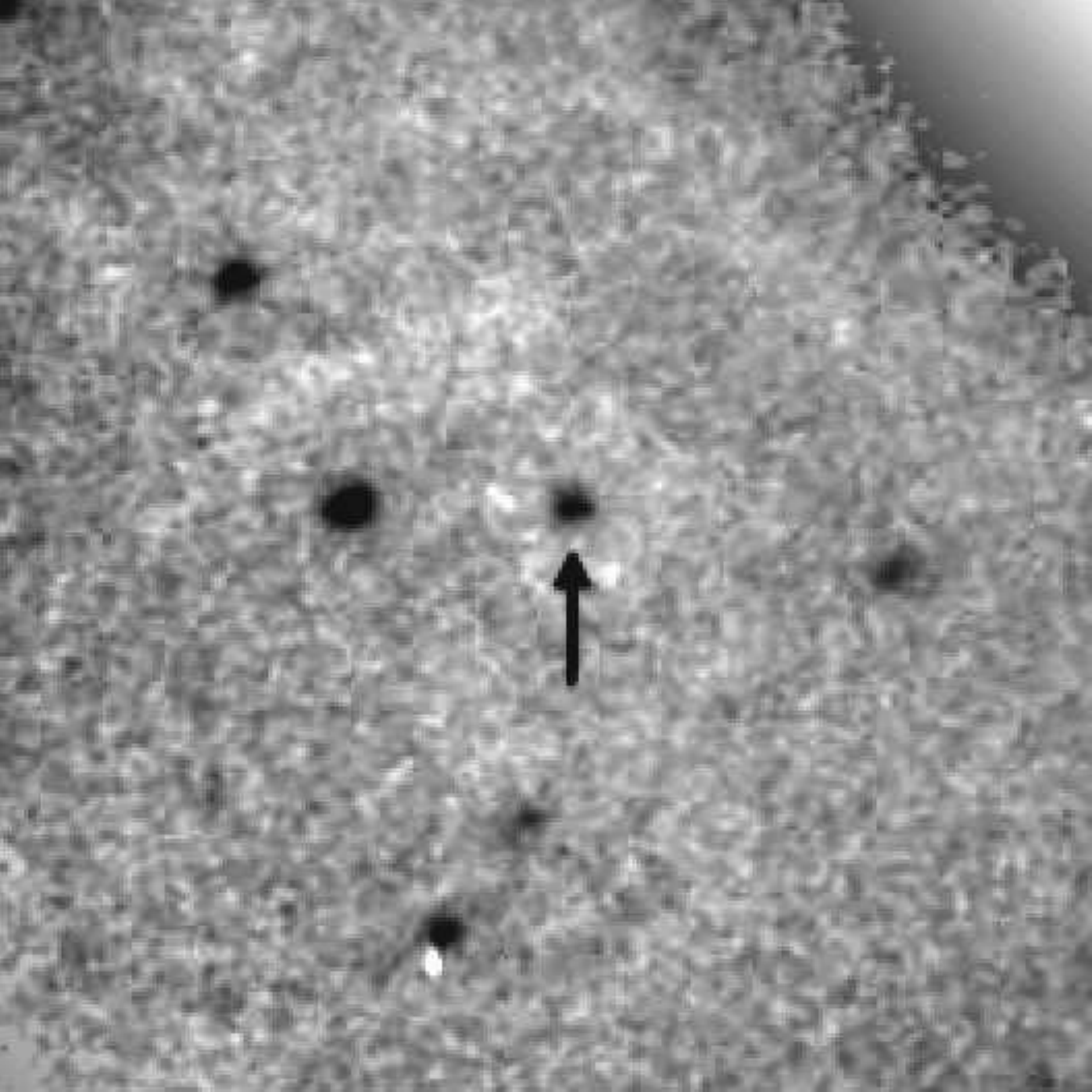}
\includegraphics[angle=0,scale=.28]{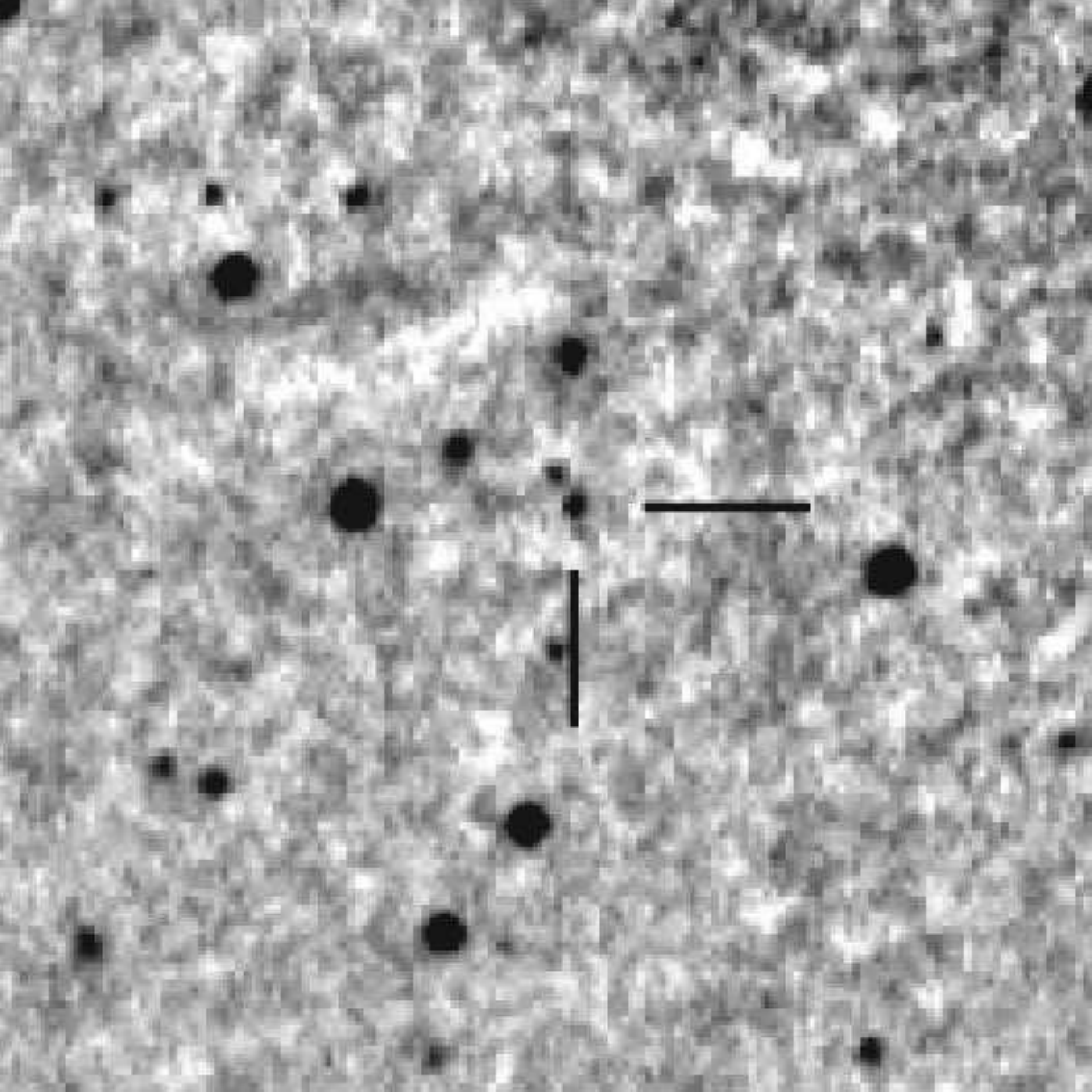}
\includegraphics[angle=0,scale=.28]{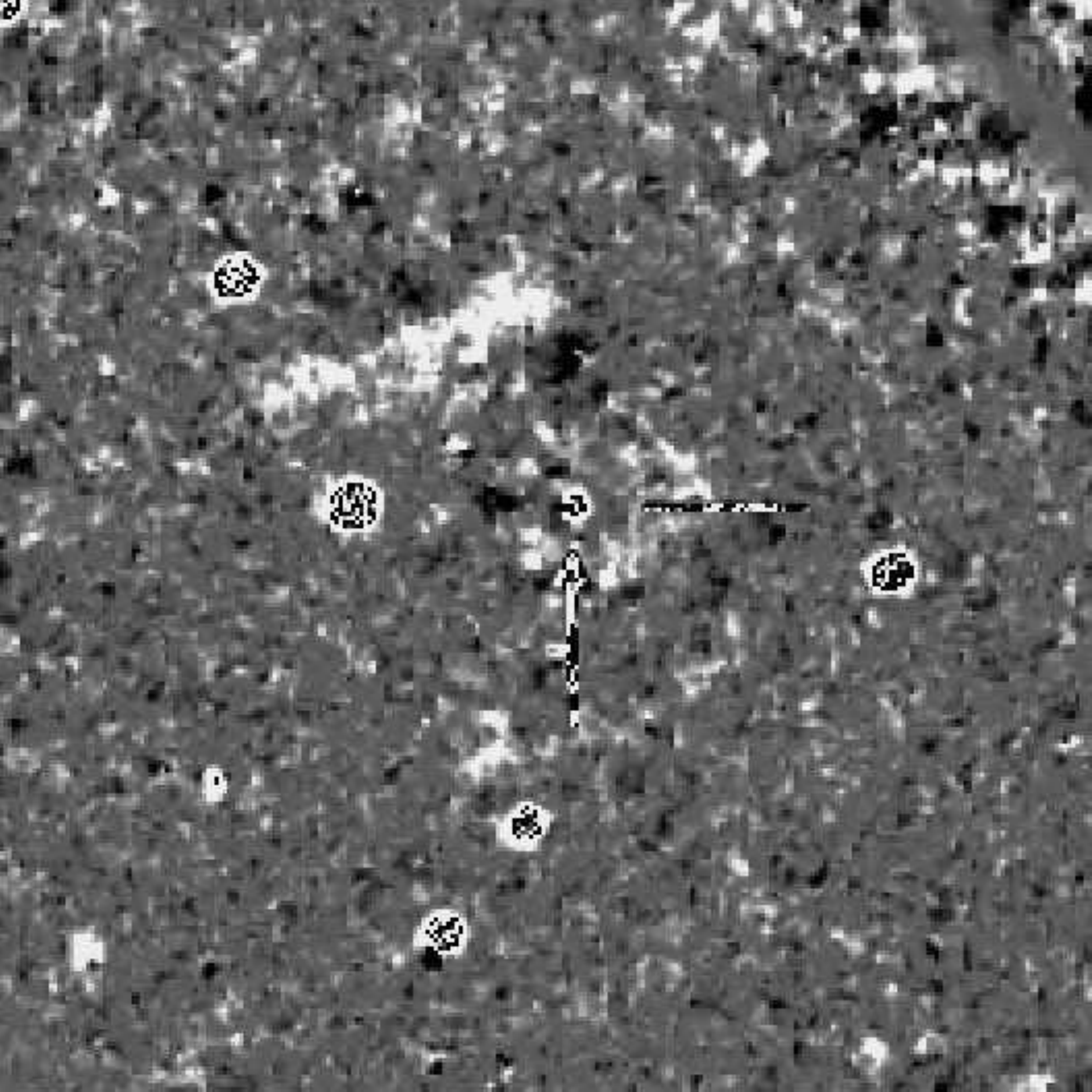}
\caption{Images of M31N 1960-12a, 2013-05b, and their comparison
(left, center, and right, respectively).
The image for M31N 1960-12a was produced from data taken as part of the
\citet{ros64} survey, while
the chart for 2013-05b is from \citet{hor13a}.
The comparison image establishes that the novae are spatially
coincident to within measurement uncertainties ($\sim$ 1$''$).
North is up and East to the left, with a scale
of $\sim2.5'$ on a side.
\label{fig8}}
\end{figure}

M31N 1960-12a erupted relatively close to the nucleus of M31
($a\sim4.4'$) and our initial screen revealed two possible recurrences:
1962-11b and 2013-05b.
Figure~\ref{fig8} shows the
position of M31N 1960-12a compared with 2013-05b.
Revised astrometry
of M31N 1960-12a yields the coordinates
given in Table~\ref{revcoord}, and shows that
the nova is coincident with the published position of 2013-05b to
within $\sim0.9''$. We estimate the probability of a chance coincidence
at this location in M31 is $\sim0.091$, and thus we consider it very likely
that M31N 2013-05b is a recurrence of 1960-12a.

As will be discussed further below, M31N 1960-12a, 1962-11b, and 2013-05b were
also flagged as a possible recurrences of 1953-11a. Of these, M31N 1962-11b
was the most likely recurrence with a reported separation $s\sim2.9''$.
A comparison of the finding charts
shown in both Figures~\ref{fig8} and later in \ref{fig27}
reveals that none of these novae are coincident
with M31N 1953-11a.

\subsubsection{M31N 1963-09c}

\begin{figure}
\includegraphics[angle=0,scale=.28]{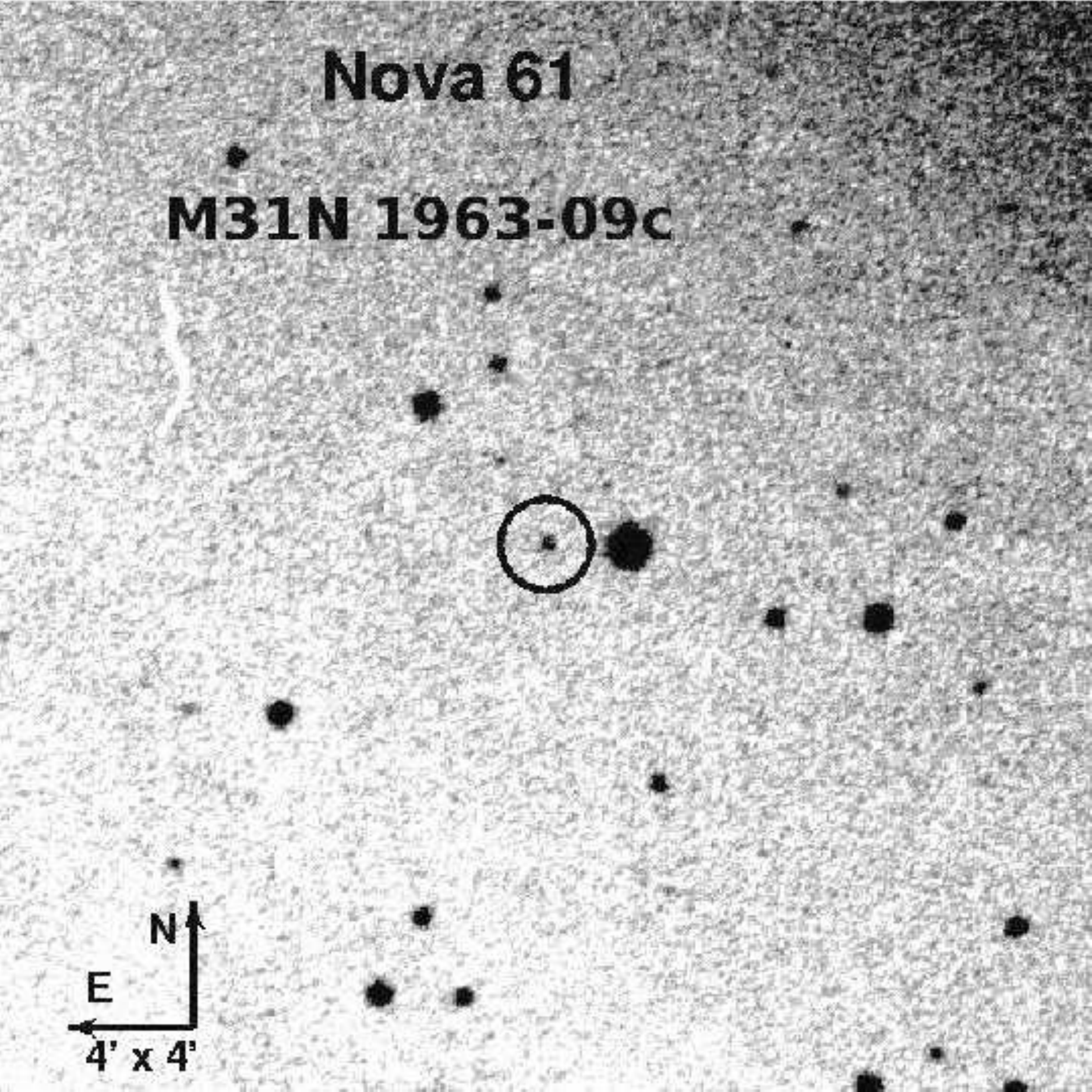}
\includegraphics[angle=0,scale=.28]{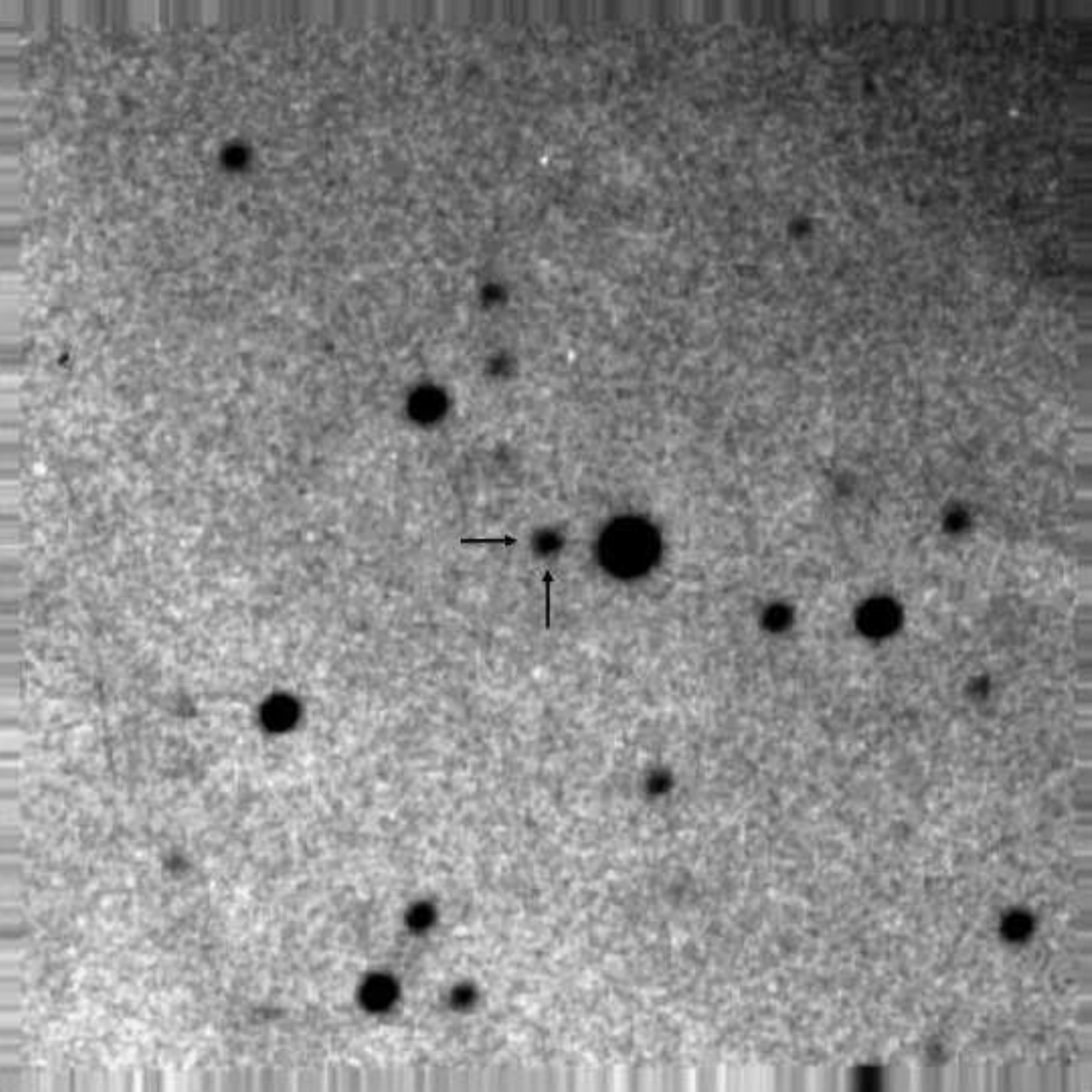}
\includegraphics[angle=0,scale=.28]{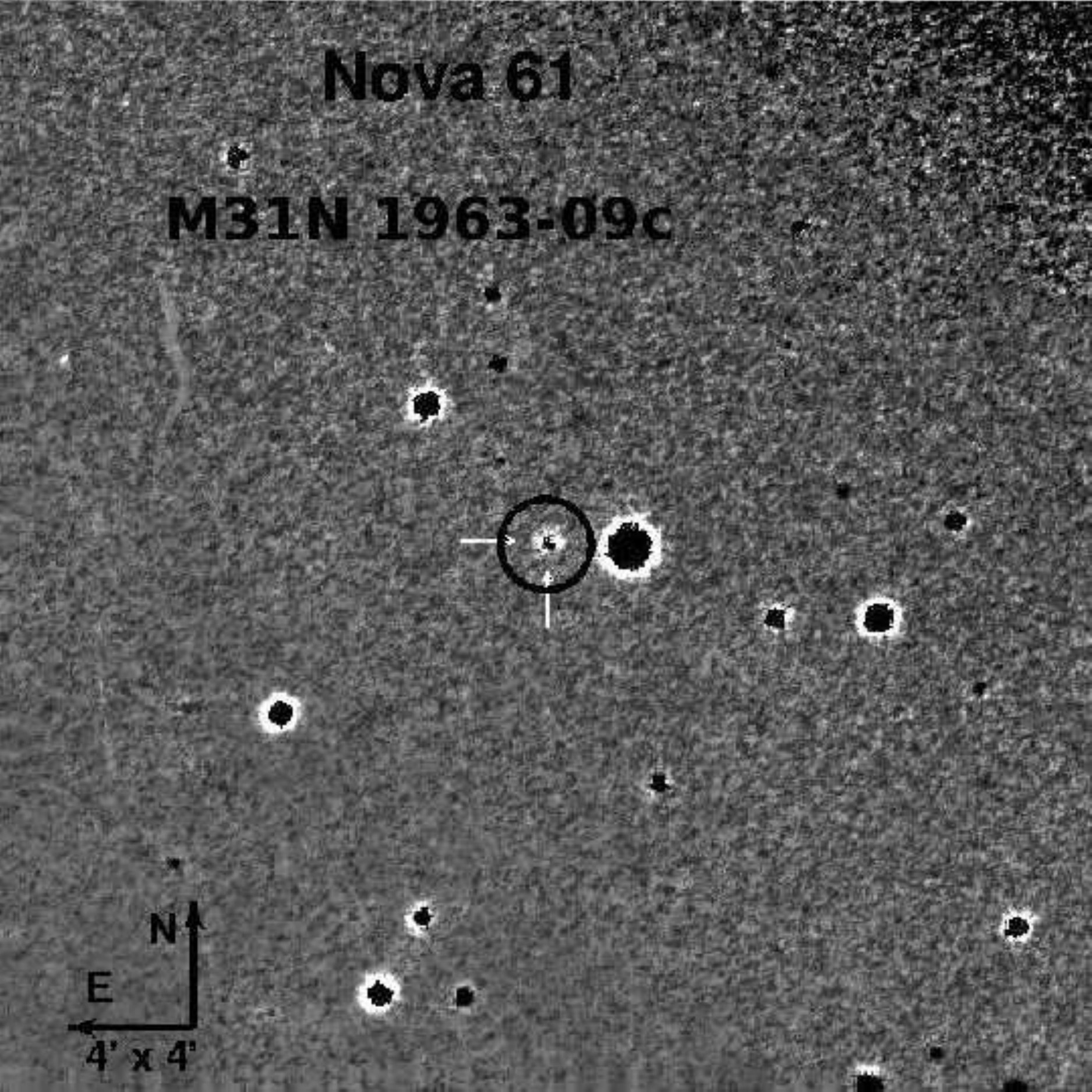}

\includegraphics[angle=0,scale=.28]{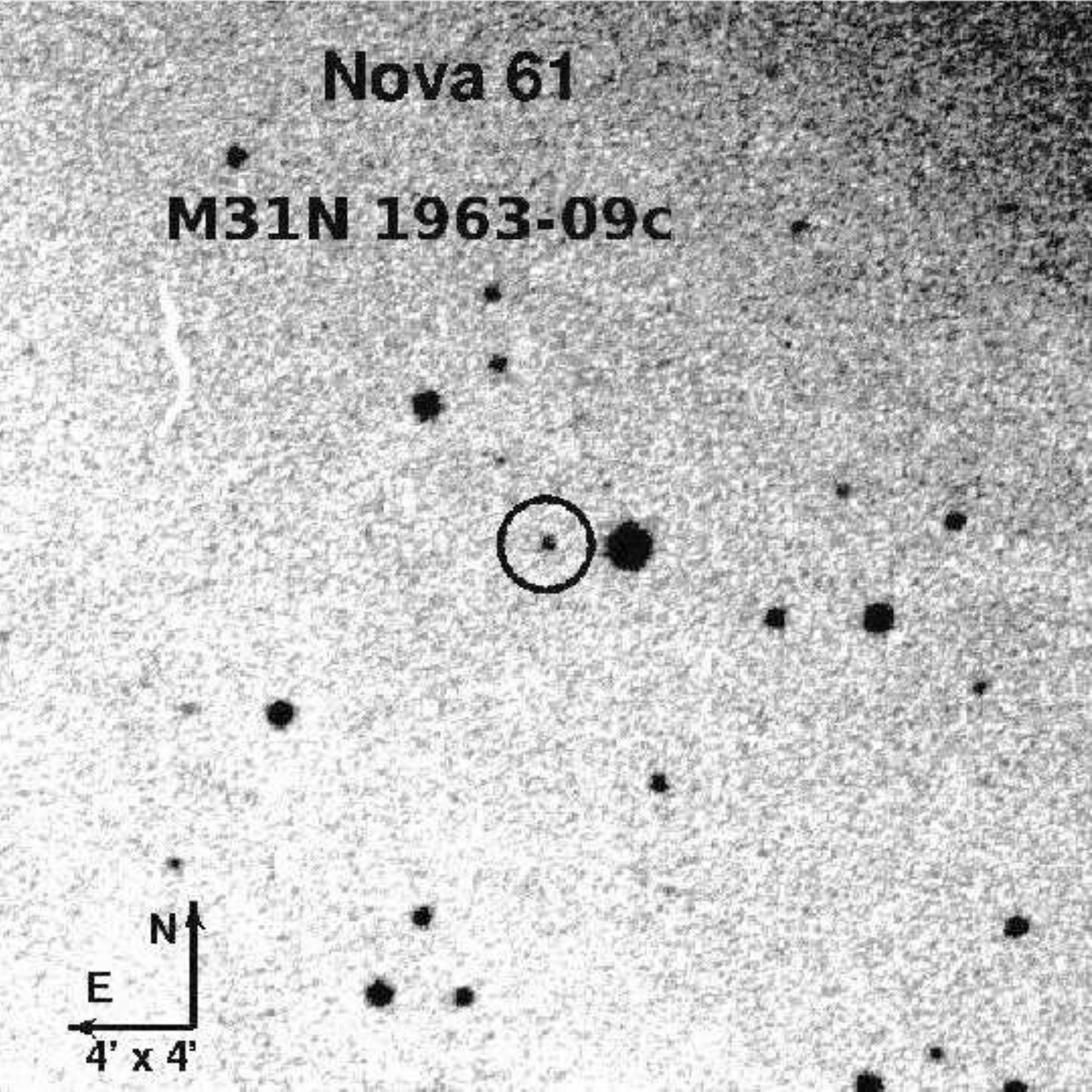}
\includegraphics[angle=0,scale=.28]{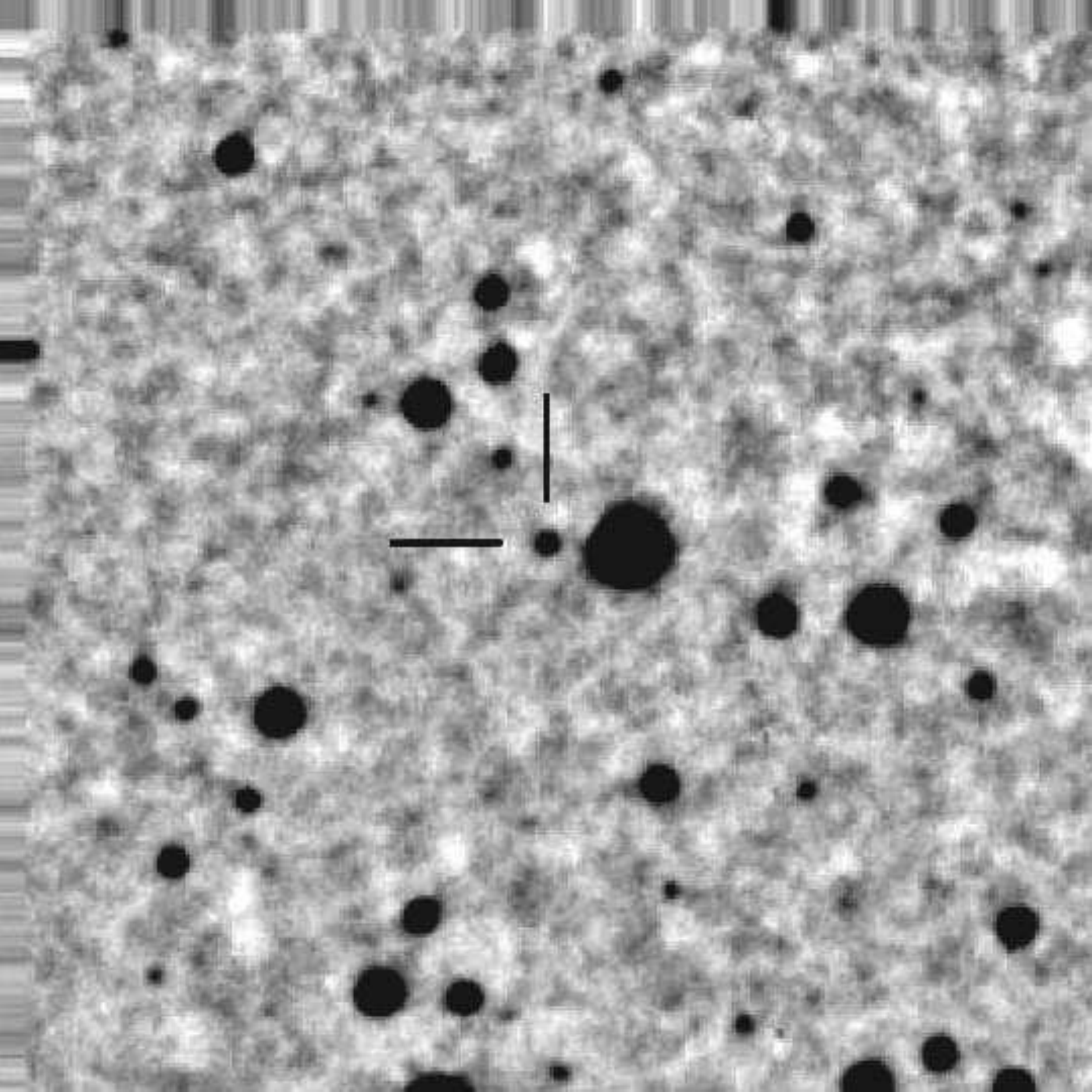}
\includegraphics[angle=0,scale=.28]{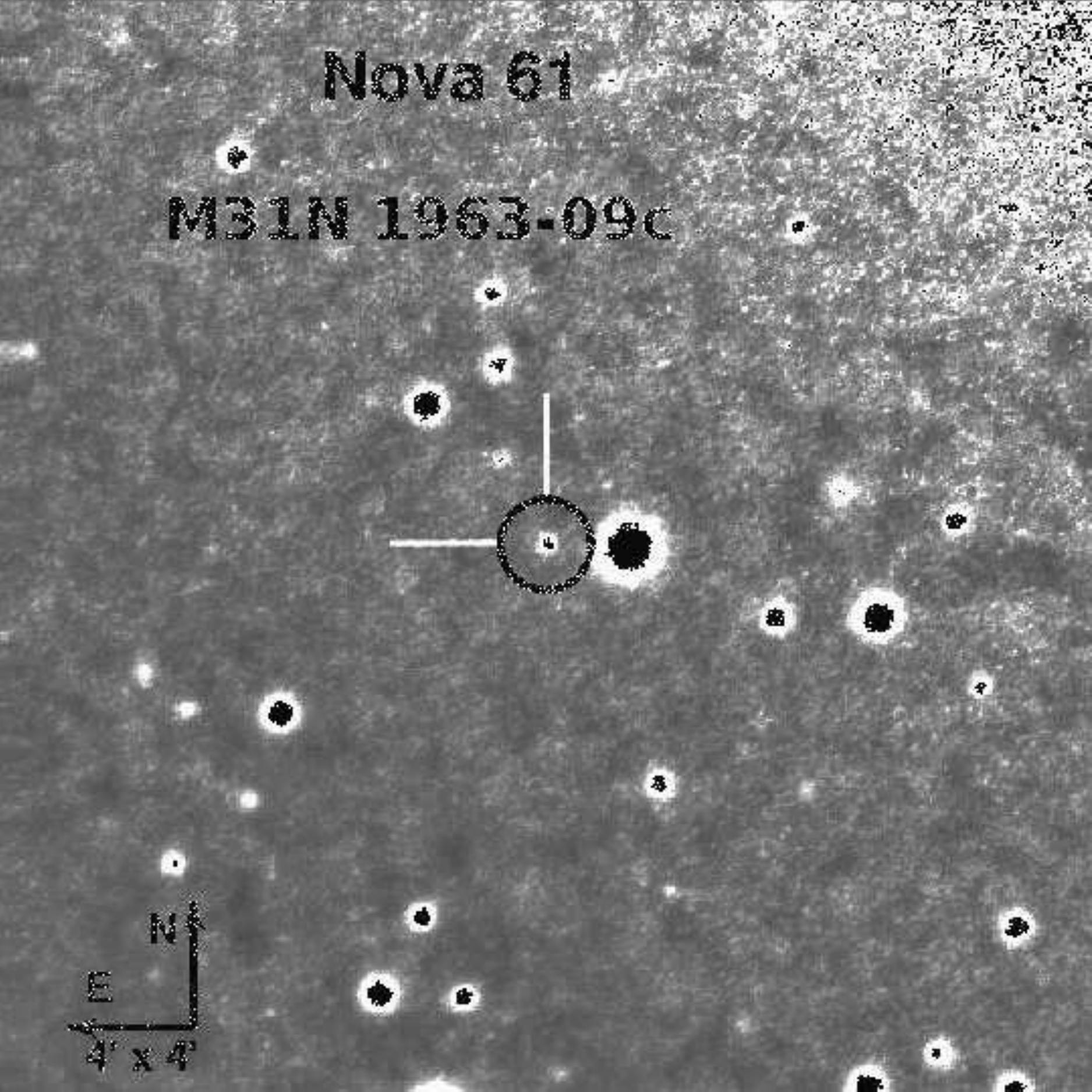}
\caption{Images of
M31N 1963-09c, 1968-09a, and their comparison (top left, center, and right),
and M31N 1963-09c, 2010-10e, and their comparison
(bottom left, center, and right, respectively).
The images for M31N 1963-09c, 1968-09a and 2010-10e are from
\citet{hen08a}, \citet{ros73} and \citet{hor10b}, respectively.
As is clear from the comparison images
M31N 1963-09c, 1968-09a and 2010-10e are spatially
coincident to within measurement uncertainties ($\sim$ 1$''$).
North is up and East to the left, with a scale of $\sim4'$ on a side.
\label{fig9}}
\end{figure}

M31N 1963-09c, which was discovered by \citet{ros73},
has been observed to have three possible recurrences.
1968-09a \citep{ros73}, 2001-07b \citep{lee12},
and 2010-10e \citep{hor10b}
have all erupted within 1.32$''$
of the reported position of 1963-09c. \citet{ros73} was first
to note that M31N 1963-09c and 1968-09a appeared to be spatially
coincident. Given that the novae are
located rather far ($a\sim17.6'$) from the nucleus of M31, the probability
of a chance positional coincidence with $s<1.32''$ is $\sim0.01$. Finding
charts for the eruptions of M31N 1963-09c and 1968-09a from the
survey of \citet{ros73} are compared with the chart of
2010-10e \citep{hor10b} in Figure~\ref{fig9}.
The novae are clearly coincident, and
we conclude, as did \citet{sha10}, that M31N 1963-09c is a RN
in M31. Consistent with this finding,
spectroscopic observations of the most recent eruption
(2010-10e) by \citet{sha10} revealed that the nova
was a member of the He/N class. In addition,
M31N 2010-10e was first detected as a bright SSS only $14\pm1$ days
after outburst by \citet{pie10b}. \citet{hen14a} described an X-ray
spectrum characterized by a relatively high effective (blackbody)
temperature of $61^{+6}_{-3}$~eV and reported that the SSS
had disappeared on day $92\pm5$ after the optical outburst.
Additionally, during one of the observations of \citet{hen14a}
the X-ray light curve of M31N 2010-10e showed strong,
aperiodic variability on time scales of hours.

This RN is noteworthy in that the first two recorded
eruptions occurred within a time span of just 5 years. This was the shortest
recurrence time measured for any RN until the discovery that
M31N 2008-12a has a recurrence time of just 1 year (see below).
\citet{wil14} has studied the field of M31N 2010-10e in search of 
a potential red giant companion for the nova. Archival
{\it Hubble Space Telescope\/} ({\it HST\/}) images
reveal a resolved source $0.80\sigma$ away from the nominal
position of the nova. According to \citet{wil14},
the density of stars at this location
suggests there is a 15.6\% probability of such an
alignment occurring by chance.

\subsubsection{M31N 1966-09e}

\begin{figure}
\includegraphics[angle=0,scale=.28]{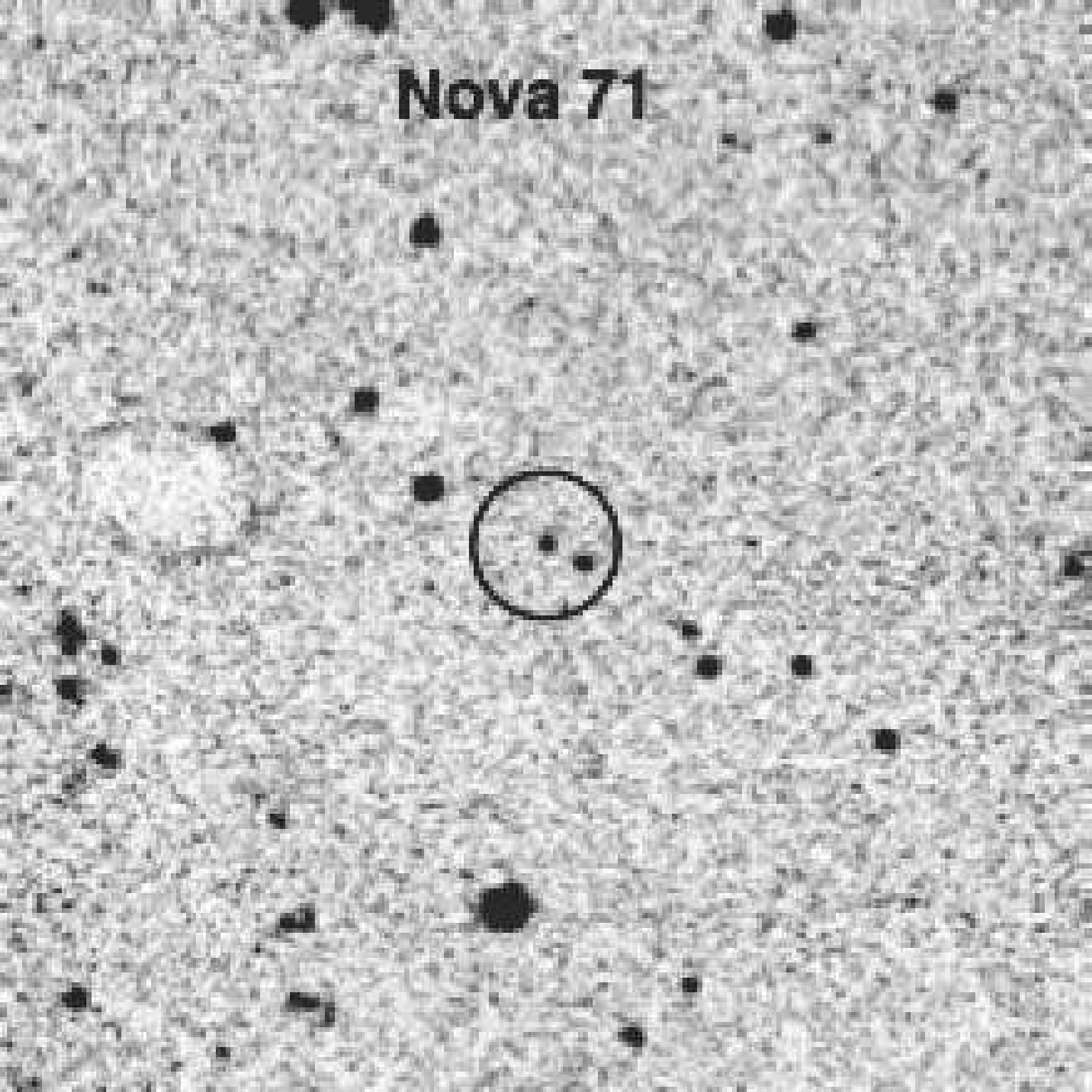}
\includegraphics[angle=0,scale=.28]{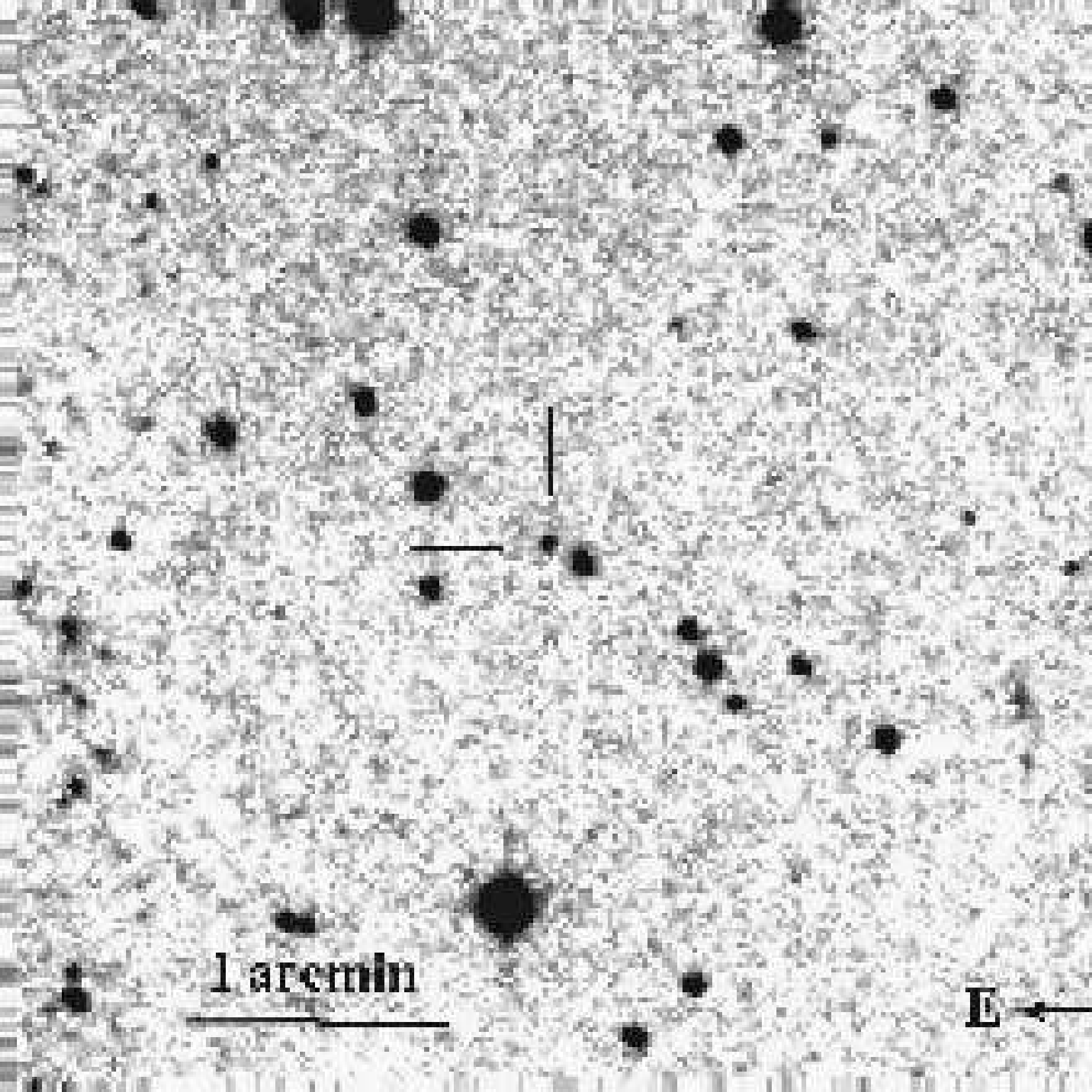}
\includegraphics[angle=0,scale=.28]{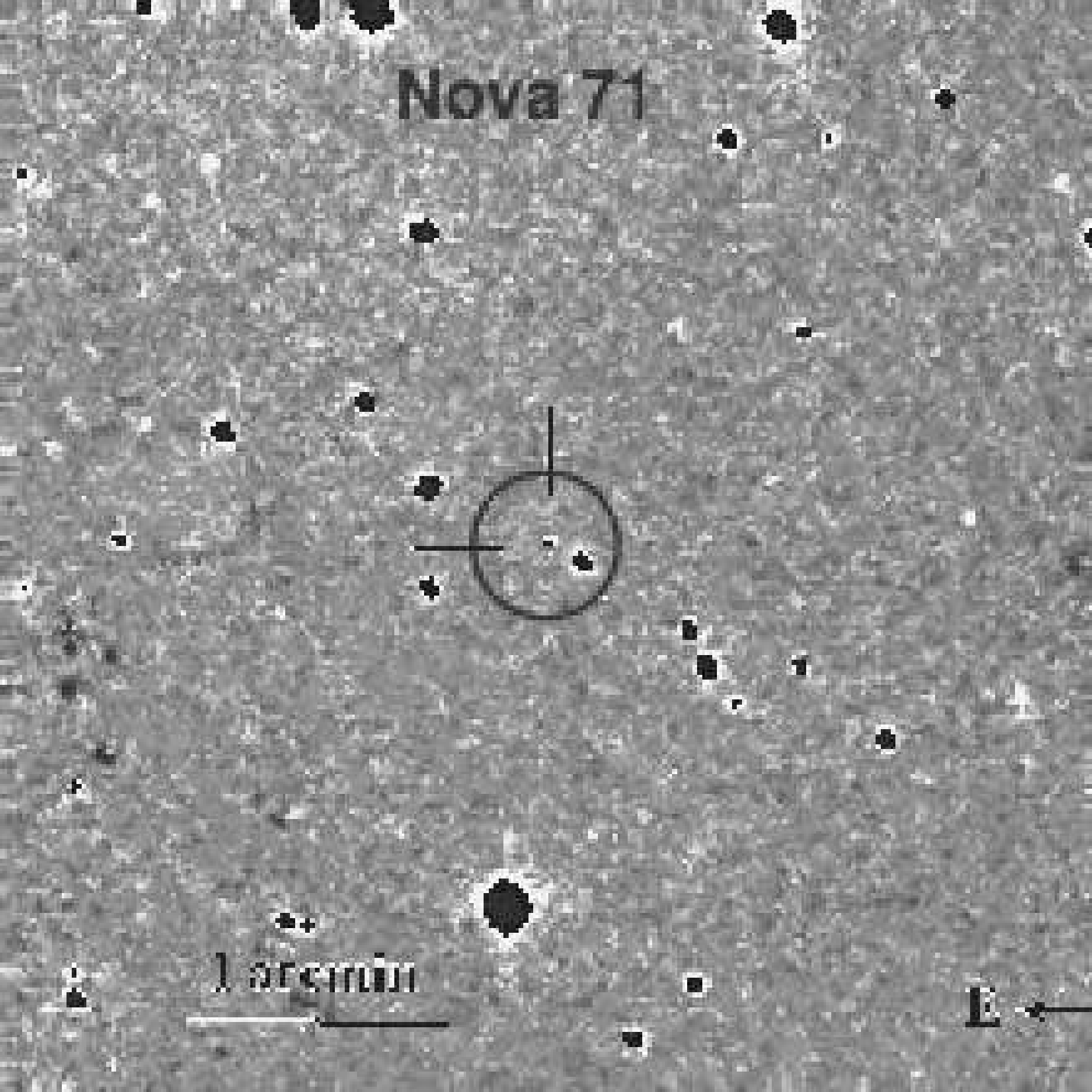}
\caption{Images of M31N 1966-09e, 2007-08d, and their comparison
(left, center, and right, respectively).
The chart for M31N 1966-09e is from
\citet{hen08a}, while that for 2007-08d is from \citet{pie07b}.
The comparison image established that the novae are spatially coincident
to within $\sim 1''$.
North is up and East to the left, with a scale of $\sim4'$ on a side.
\label{fig10}}
\end{figure}

M31N 1966-09e, discovered as nova \#71 by \citet{hen08a} on
archival Tautenburg Schmidt plates,
is noteworthy in that it was observed to erupt in the
outskirts of M31 (almost a full degree from the nucleus of the galaxy).
A recurrence was observed as M31N 2007-08d \citep{pie07b}. The
reported positions differ by just $0.36''$, and at this position
in the galaxy the probability of a chance coincidence is negligible.
Astrometry of M31N 1966-09e yields the coordinates
given in Table~\ref{revcoord}.
Figure~\ref{fig10} confirms that the two eruptions are indeed spatially
coincident, and we conclude that M31N 1966-09e is a RN.
The spectroscopic type and light curve properties of M31N 2007-08d
was measured by \citet{sha11b} as a relatively slow ($t_2\sim81\pm11$~d)
Fe~II system. The recurrence was also observed in the infrared by the
{\it Spitzer Space Telescope\/} but did not reveal an infrared excess
that could be attributed to dust formation \citep{sha11a}.

\subsubsection{M31N 1982-08b}

\begin{figure}
\includegraphics[angle=0,scale=.28]{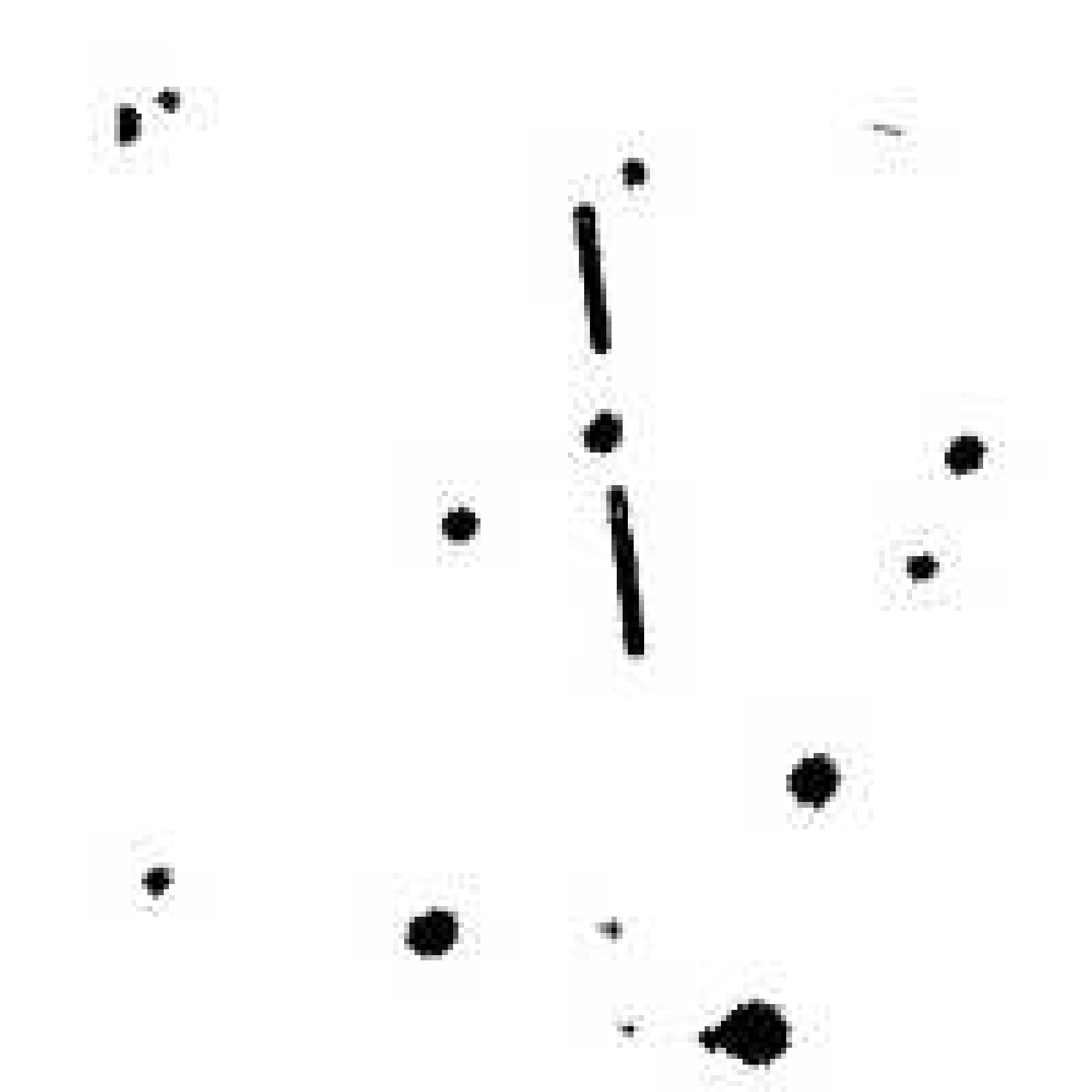}
\includegraphics[angle=0,scale=.28]{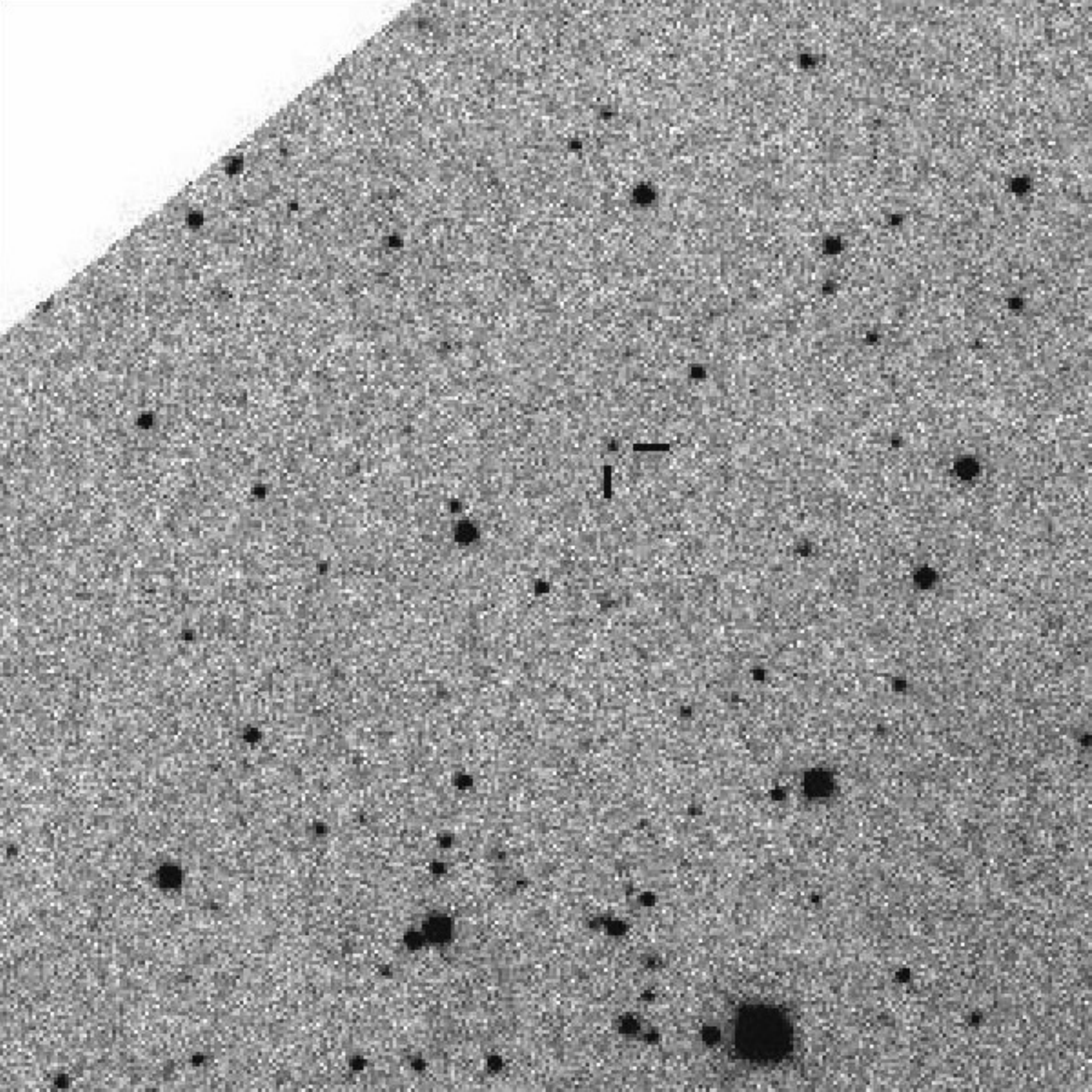}
\includegraphics[angle=0,scale=.28]{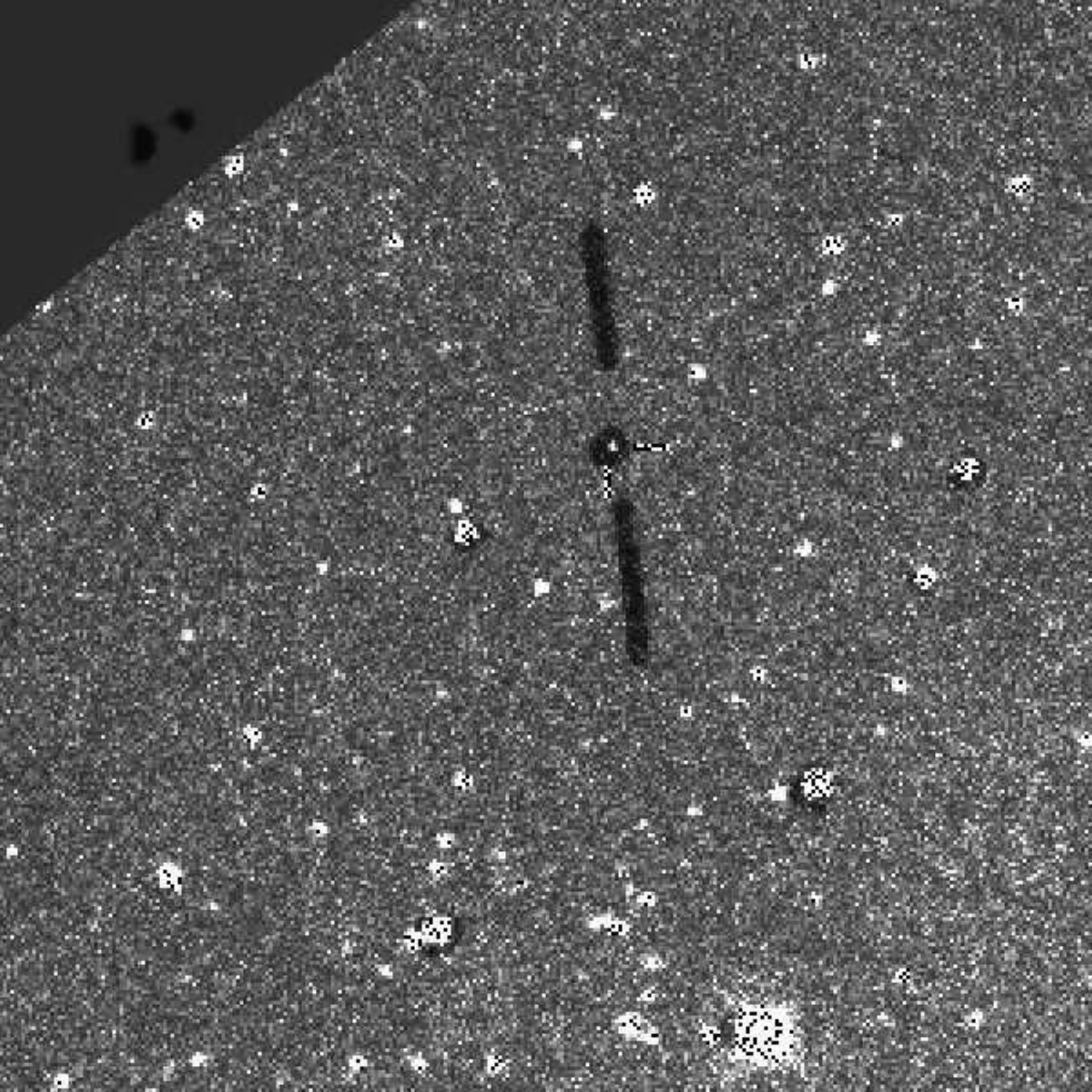}
\caption{Images of M31N 1982-08b, 1996-08c, and their comparison
(left, center, and right, respectively).
The chart for M31N 1982-08b is taken from \citet{sha92}, with that for
1996-08c taken from the survey of \citet{sha01}. Although the finding chart
for M31N 1982-08b has poor spatial resolution, the novae appear spatially
coincident to within measurement uncertainties ($\sim 2''$). At the
location of the novae in the outskirts of M31 ($a\sim60'$), the
probability of a chance positional coincidence is negligible.
North is up and East to the left, with a scale of $\sim5'$ on a side.
\label{fig11}}
\end{figure}

M31N 1982-08b was discovered in the outskirts of M31
as nova \#27 in the survey of \citet{sha92}. A likely
recurrence (nominally located $\sim3''$ away)
was discovered by \citet{sha01}, who reported
a nova discovered on 1996 August 12. Subsequently, the nova acquired the
designation M31N 1996-08c. In reviewing the original data
to produce a finding chart, we have determined that the
nova was actually discovered a year later, on 1997 August 01 UT.
Revised astrometry
given in Table~\ref{revcoord} shows that the novae are spatially coincident to
less than $1''$, which is less than the uncertainty in the absolute positions.
Finding charts for the images are shown in Figure~\ref{fig11}, confirming
that the novae are spatially coincident.
The probability of a chance positional coincidence is negligible
($P_C\simeq0.0001$),
and we conclude that M31N 1982-08b is recurrent.

\subsubsection{M31N 1984-07a}

\begin{figure}
\includegraphics[angle=0,scale=.28]{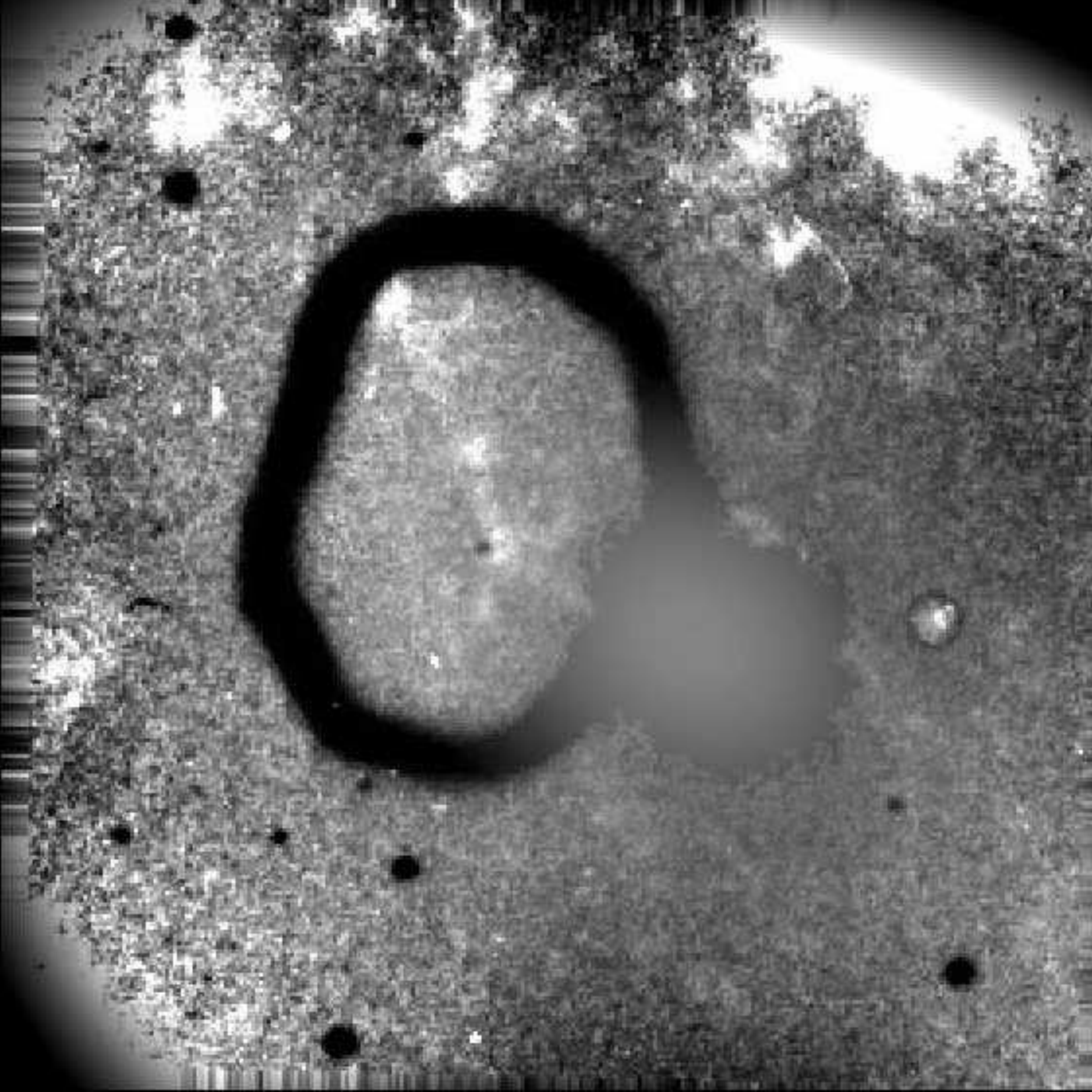}
\includegraphics[angle=0,scale=.28]{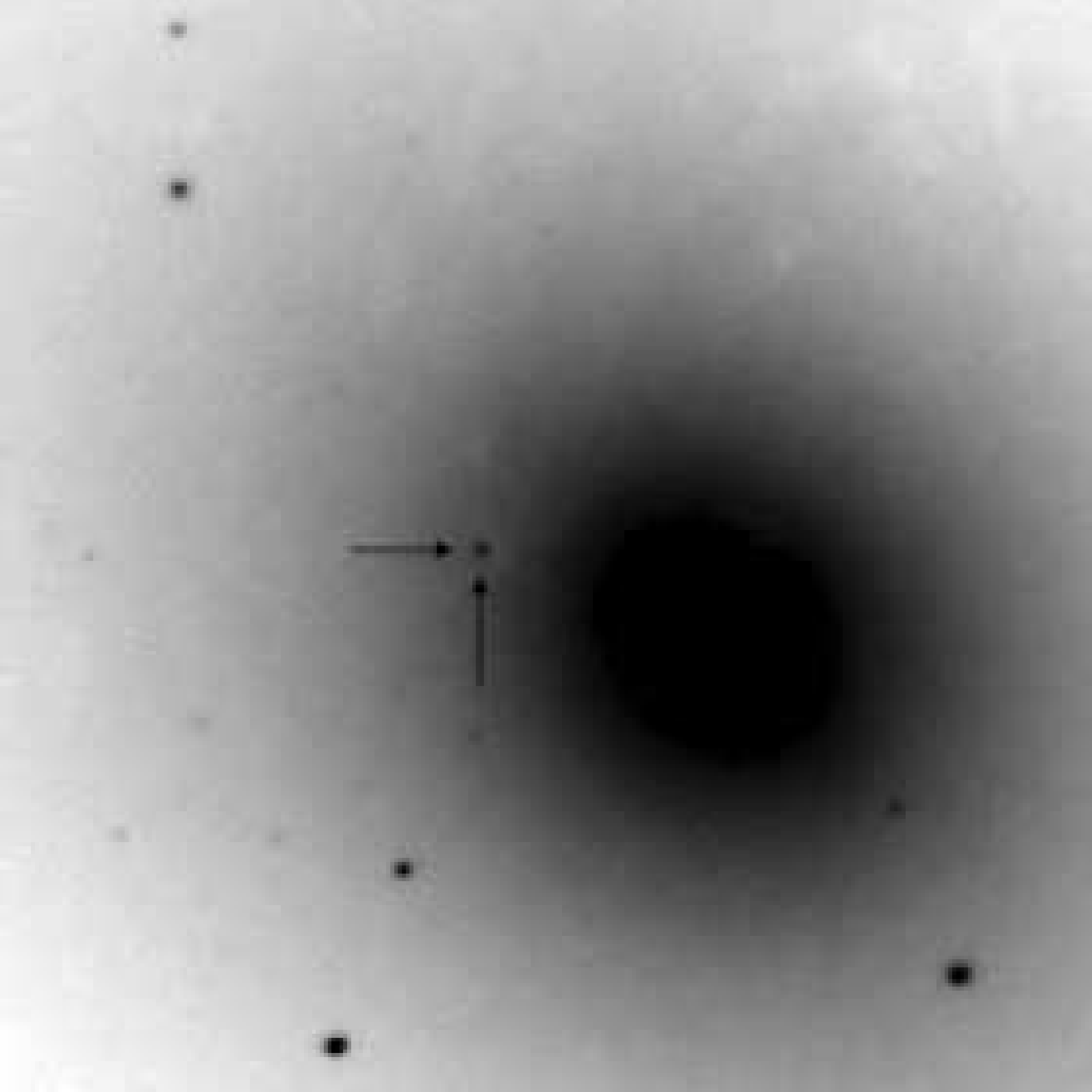}
\includegraphics[angle=0,scale=.28]{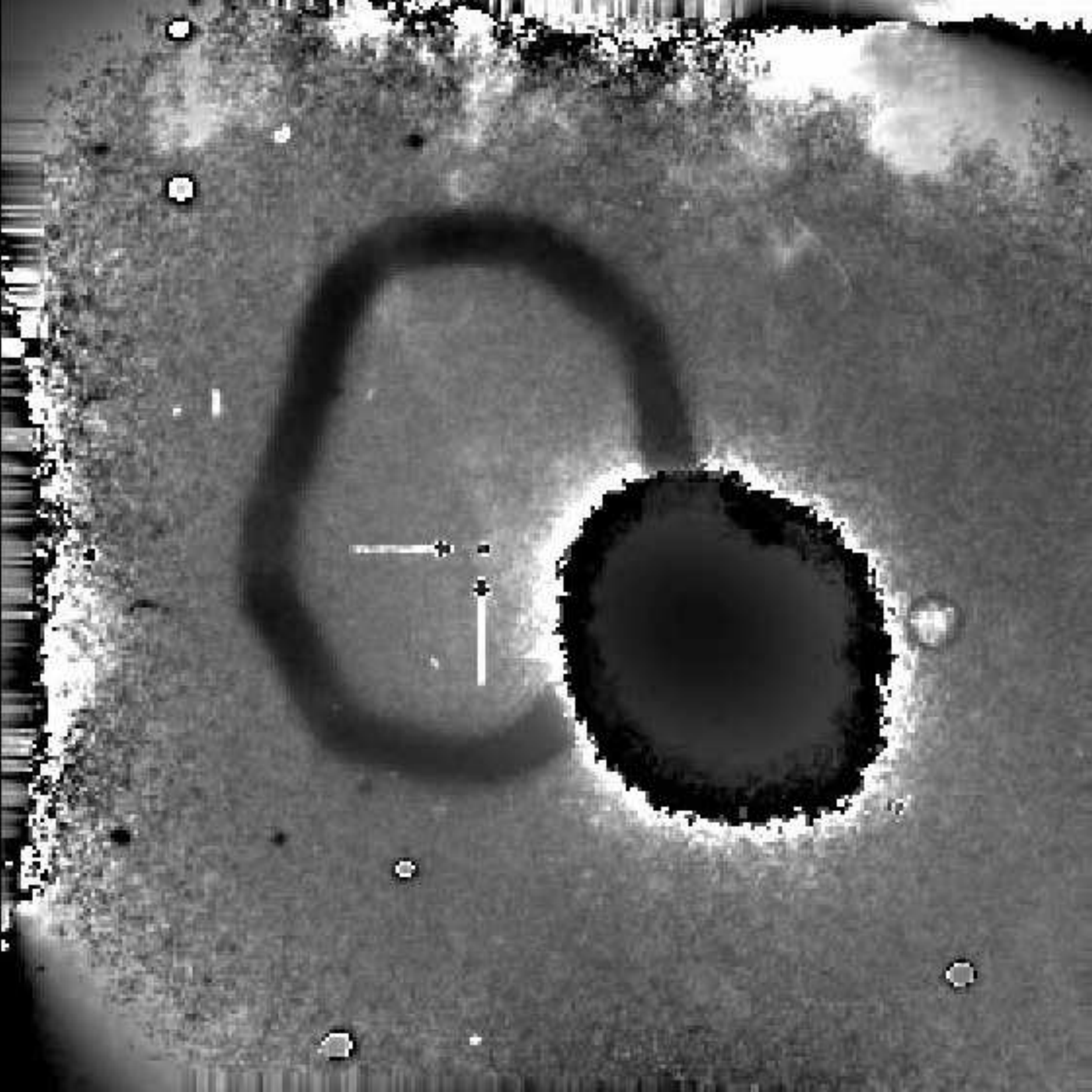}

\includegraphics[angle=0,scale=.28]{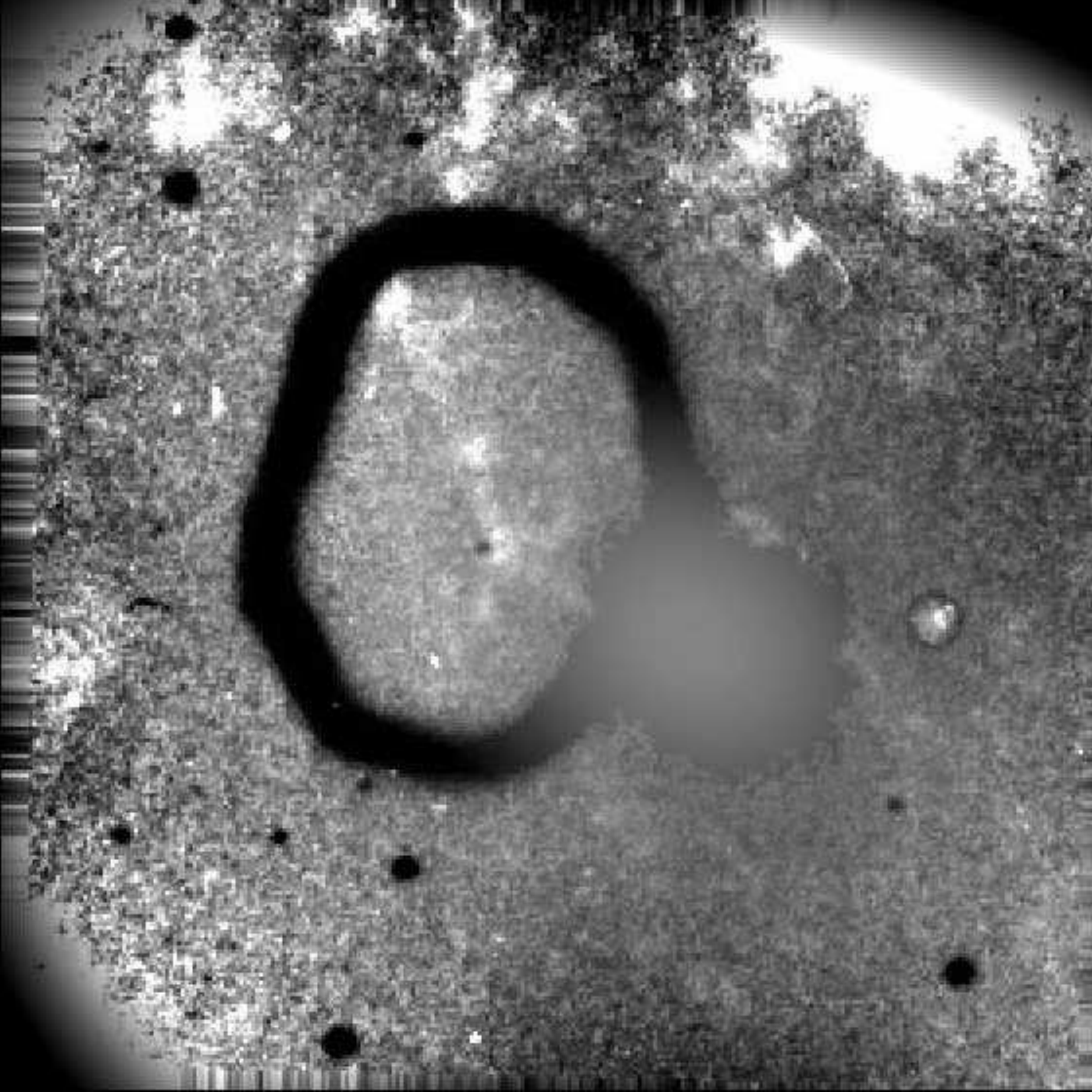}
\includegraphics[angle=0,scale=.28]{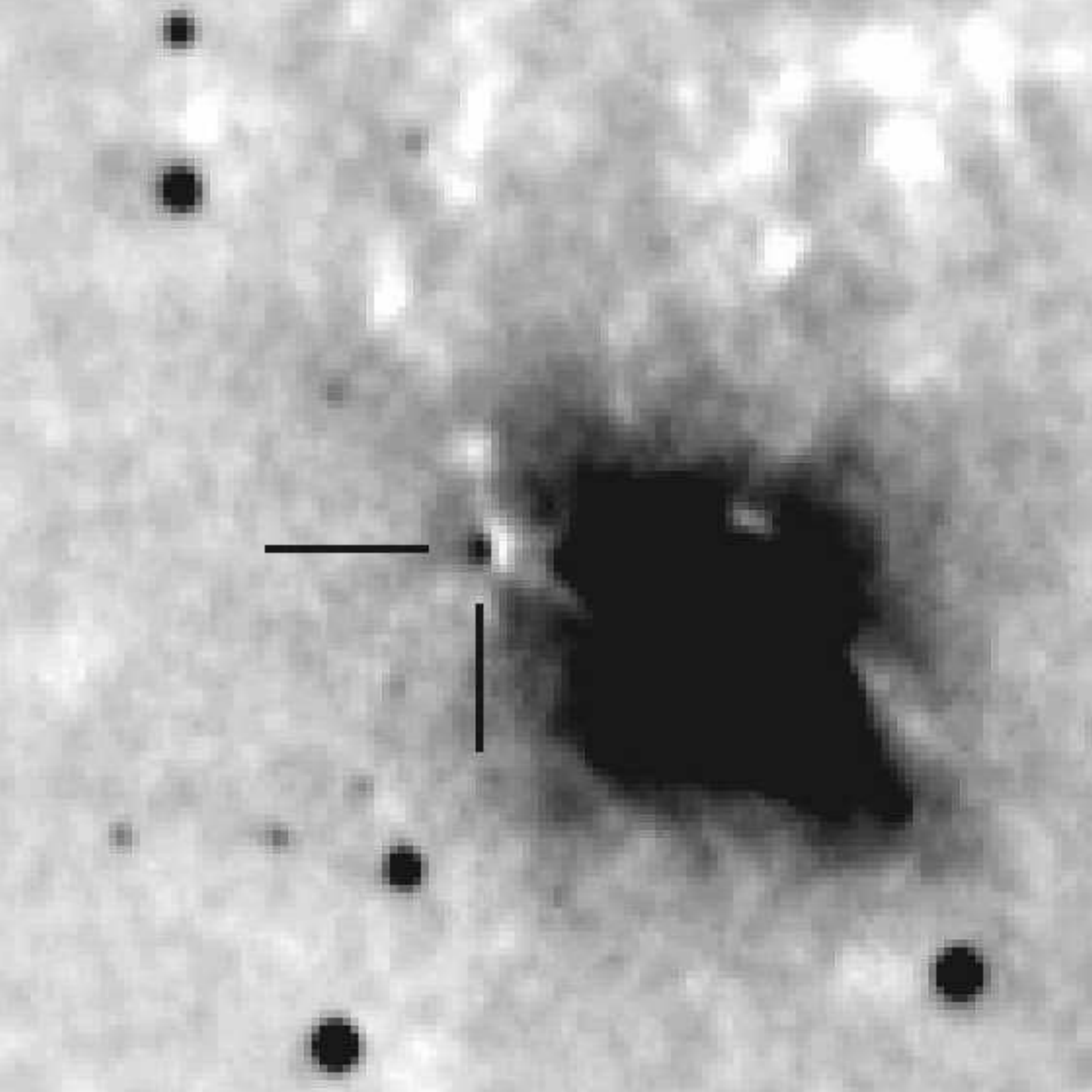}
\includegraphics[angle=0,scale=.28]{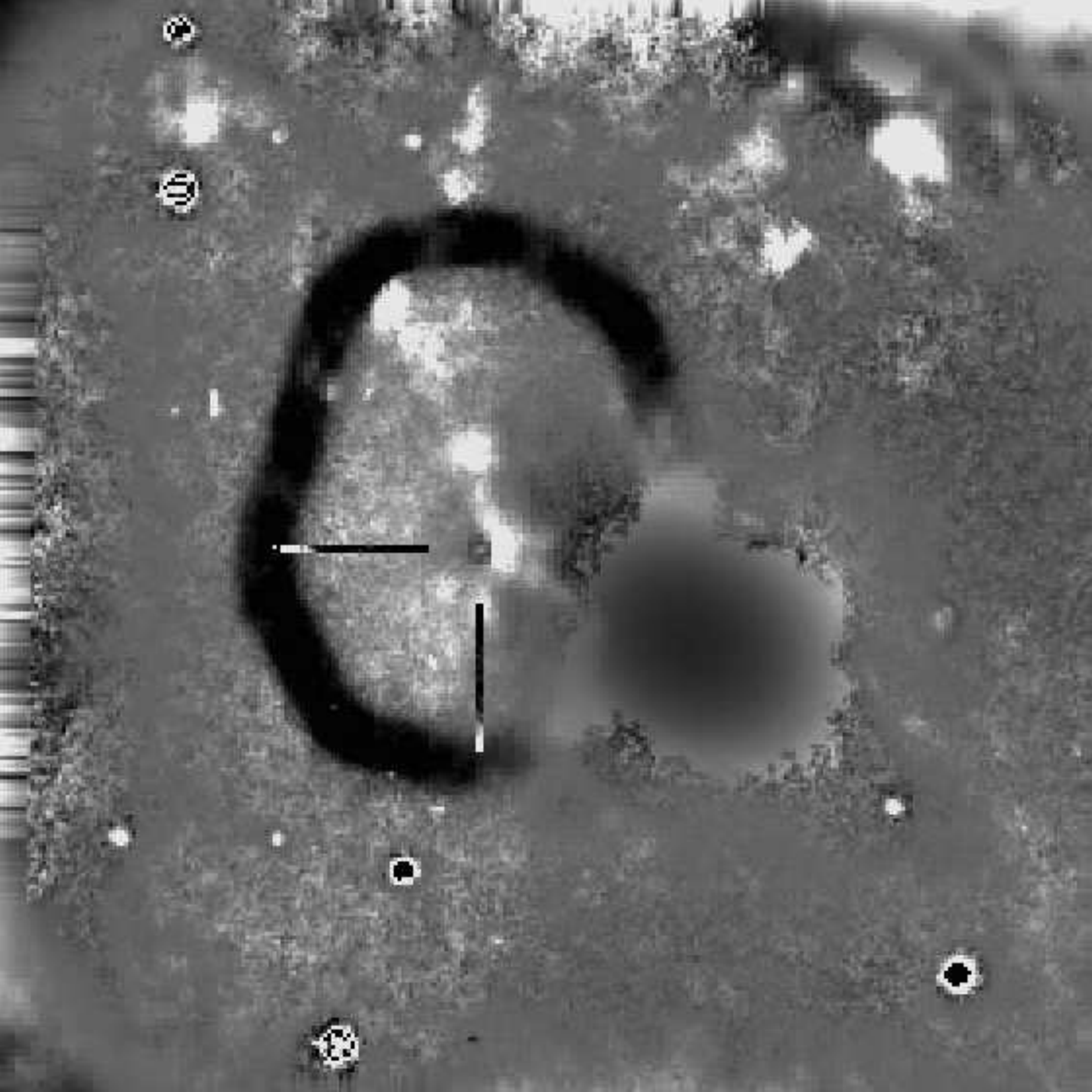}
\caption{Images of M31N 1984-07a, 2004-11f and their comparison
(top left, center, and right), and
M31N 1984-07a, 2012-09a, and their comparison
(bottom left, center, and right, respectively).
The chart for M31N 1984-07a is from the survey of \citet{ros89},
that for 2004-11f is from the RBSE project \citep{rec99},
and that for 2012-09a is from the data reported in \citet{hor12a}.
Although the novae erupted very close to the nucleus of M31 making a detailed
comparison of the positions challenging, the comparison images
show that the novae appear to be spatially coincident to less than 1 arcsec.
North is up and East to the left, with a scale of $\sim2.5'$ on a side.
\label{fig12}}
\end{figure}

M31N 1984-07a was discovered by \citet{ros89} very close
to the nucleus of M31. There have been several subsequent novae seen
to erupt within $2''$ of M31N 1984-07a, including 2001-10c, 2004-02a, 2004-11f,
and 2012-09a.
We have remeasured the position of M31N 1984-07a using the image
from the survey of \citet{ros89}, and determined the revised coordinates
given in Table~\ref{revcoord}. The revised coordinates are within
1.1$''$, 2.3$''$, 0.3$''$, and 0.6$''$ of M31N 2001-10c, 2004-02a, 2004-11f, and
2012-09a, respectively. Thus, it
appears likely that M31N 2004-11f and 2012-09a are recurrences, with
2001-10c and 2004-02a being less likely to be recurrences.
\citet{pie07a} have studied
the positions of M31N 2001-10c, 2004-02a, and 2004-11f, and concluded
that these three novae are not spatially coincident. They were not able,
however, to rule out the possibility that M31N 2004-11f
was a recurrence of 1984-07a.
The positions of M31N 1984-07a, 2004-11f, and 2012-09a are shown
in Figure~\ref{fig12}.
It appears very likely that both M31N 2004-11f and 2012-09a are recurrences
of 1984-07a.

M31N 2004-11f was detected as a bright and fast SSS by \citet{pie07a}.
The X-ray emission was already present 34 days after the optical
outburst and had started to decline in luminosity soon afterwards,
by day 55 after outburst.
\citet{pie07a} reported strong indications that the X-ray
emission was supersoft, but unfortunately did not detect sufficient
photons to estimate the effective SSS temperature.
Notably, $\sim$10 months prior to the discovery
of M31N 2004-11f, \citet{pie07a} identified what they believed to be
the ``pre-nova" in archival {\it HST\/} images. If so, and if M31N 2012-09a
is also a recurrence of 1984-07a,
it is noteworthy that
archival {\it HST\/} images of the field of 2012-09a were analyzed
by \citet{wil14} who found no compelling evidence for a red giant companion.

\subsubsection{M31N 1997-11k}

\begin{figure}
\includegraphics[angle=0,scale=.28]{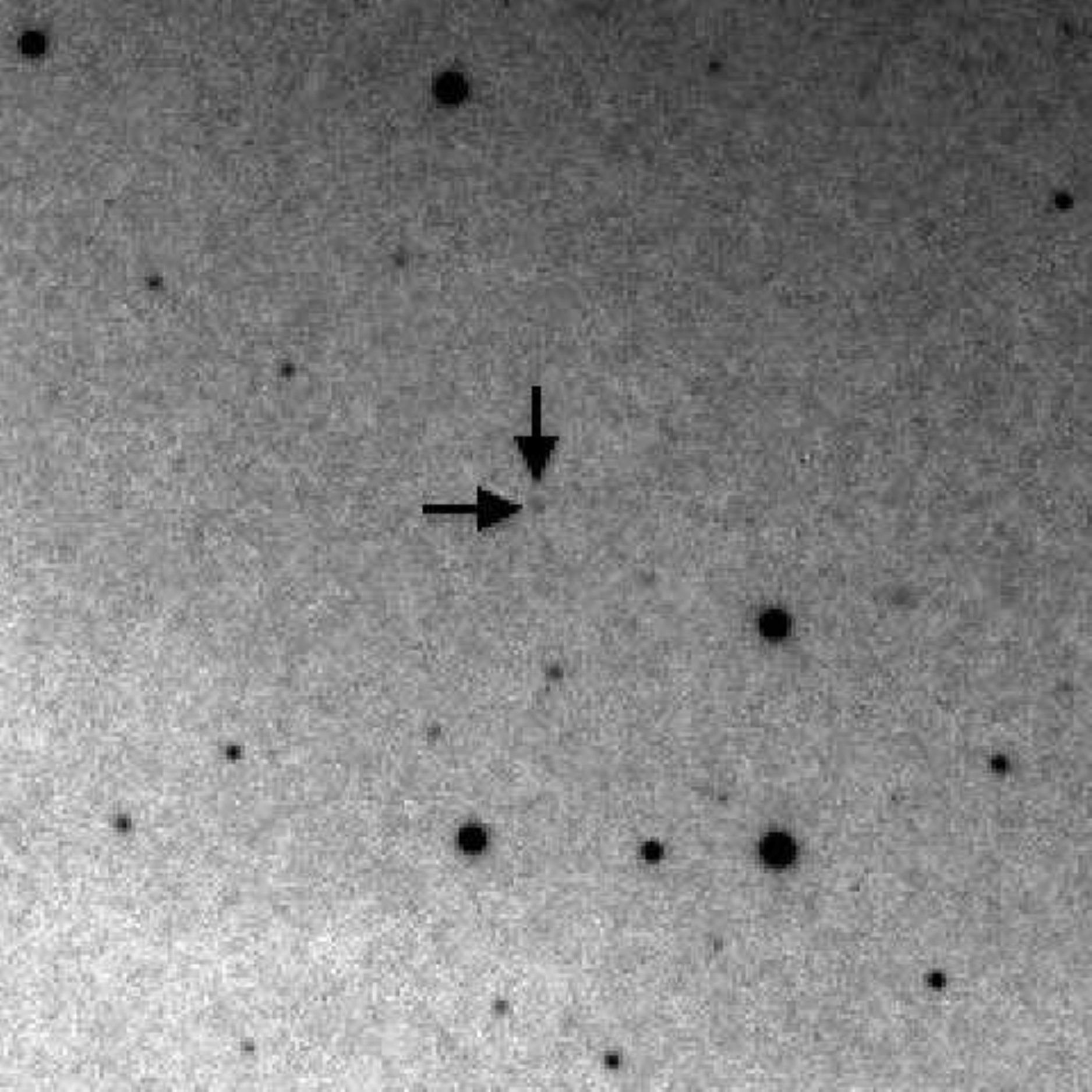}
\includegraphics[angle=0,scale=.28]{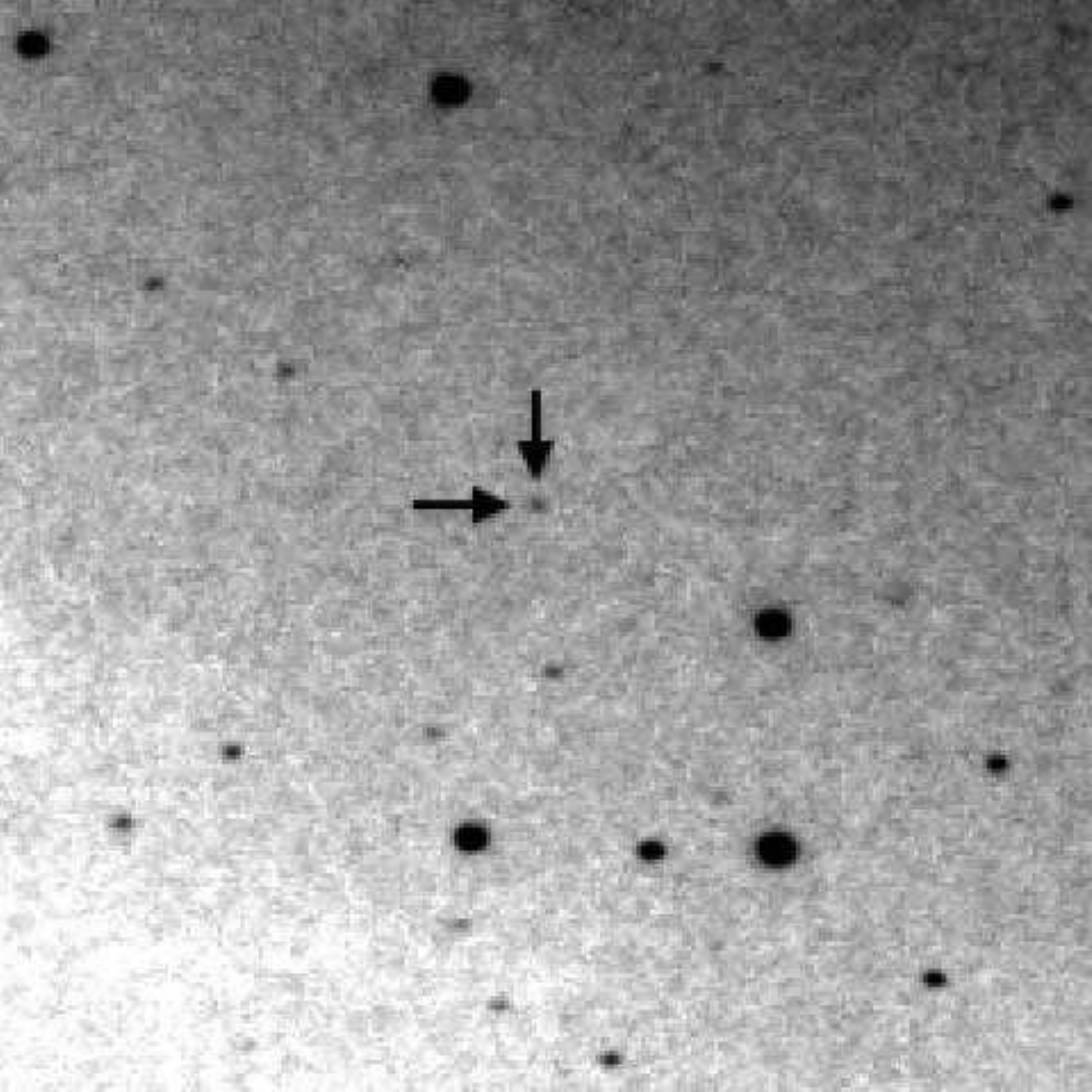}
\includegraphics[angle=0,scale=.28]{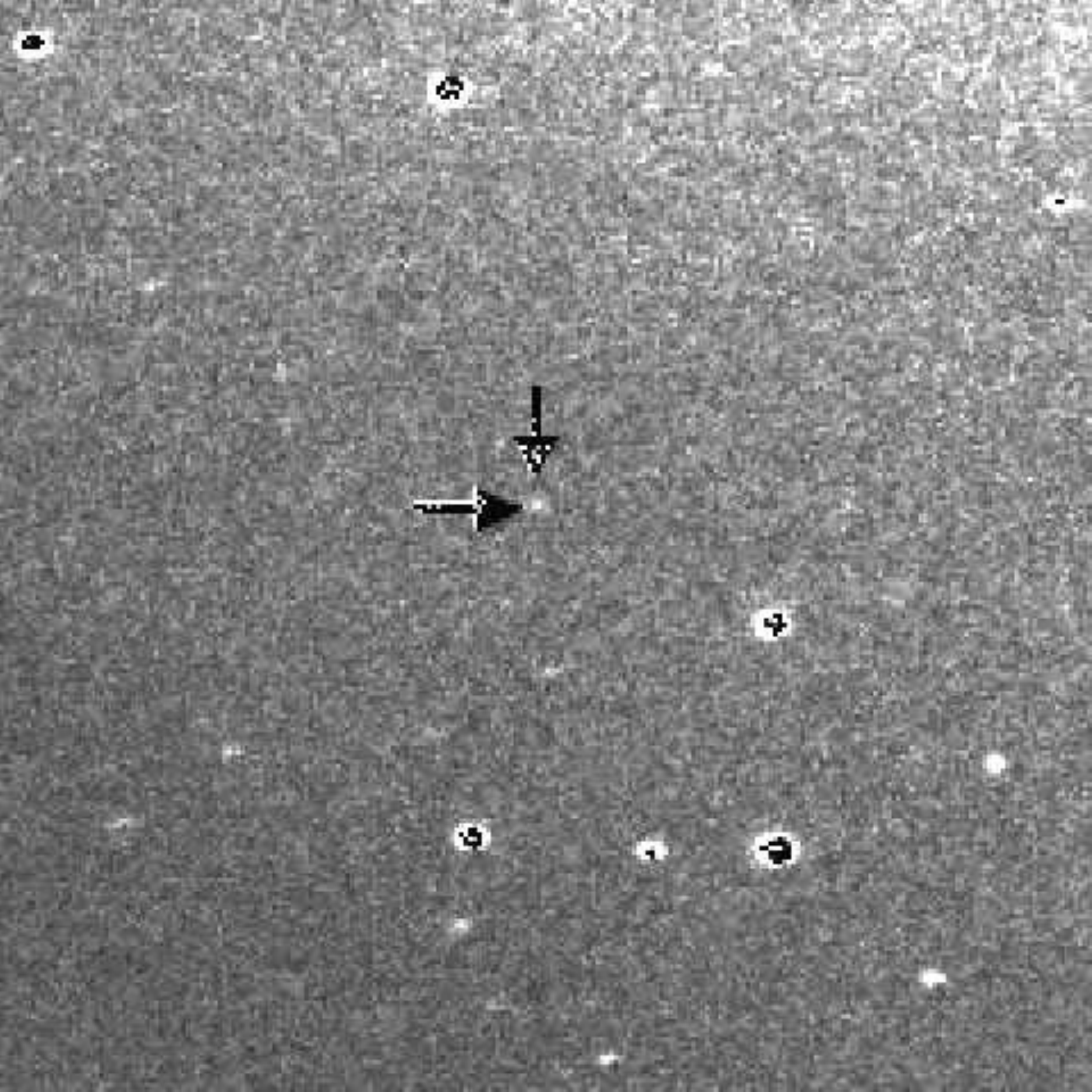}

\includegraphics[angle=0,scale=.28]{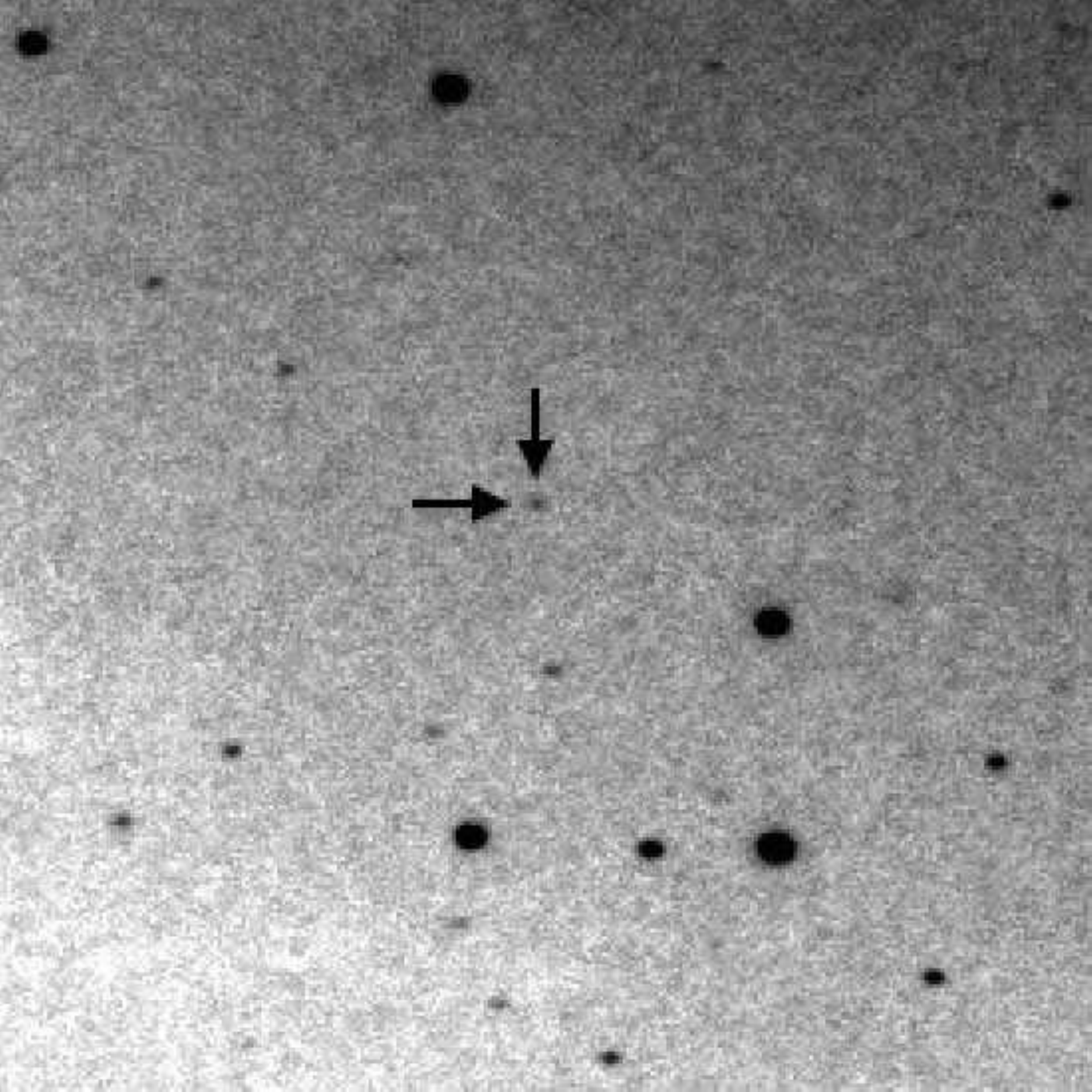}
\includegraphics[angle=0,scale=.28]{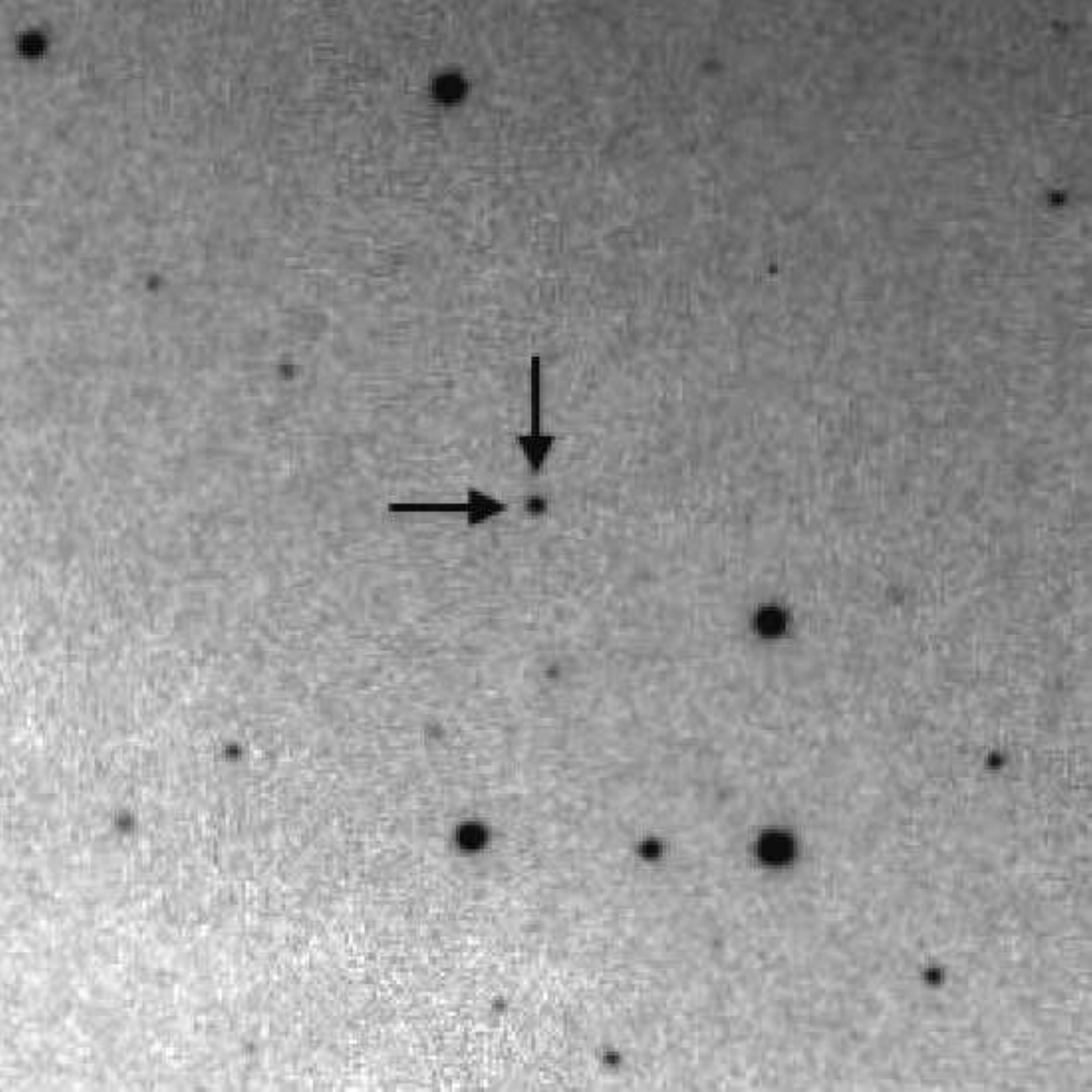}
\includegraphics[angle=0,scale=.28]{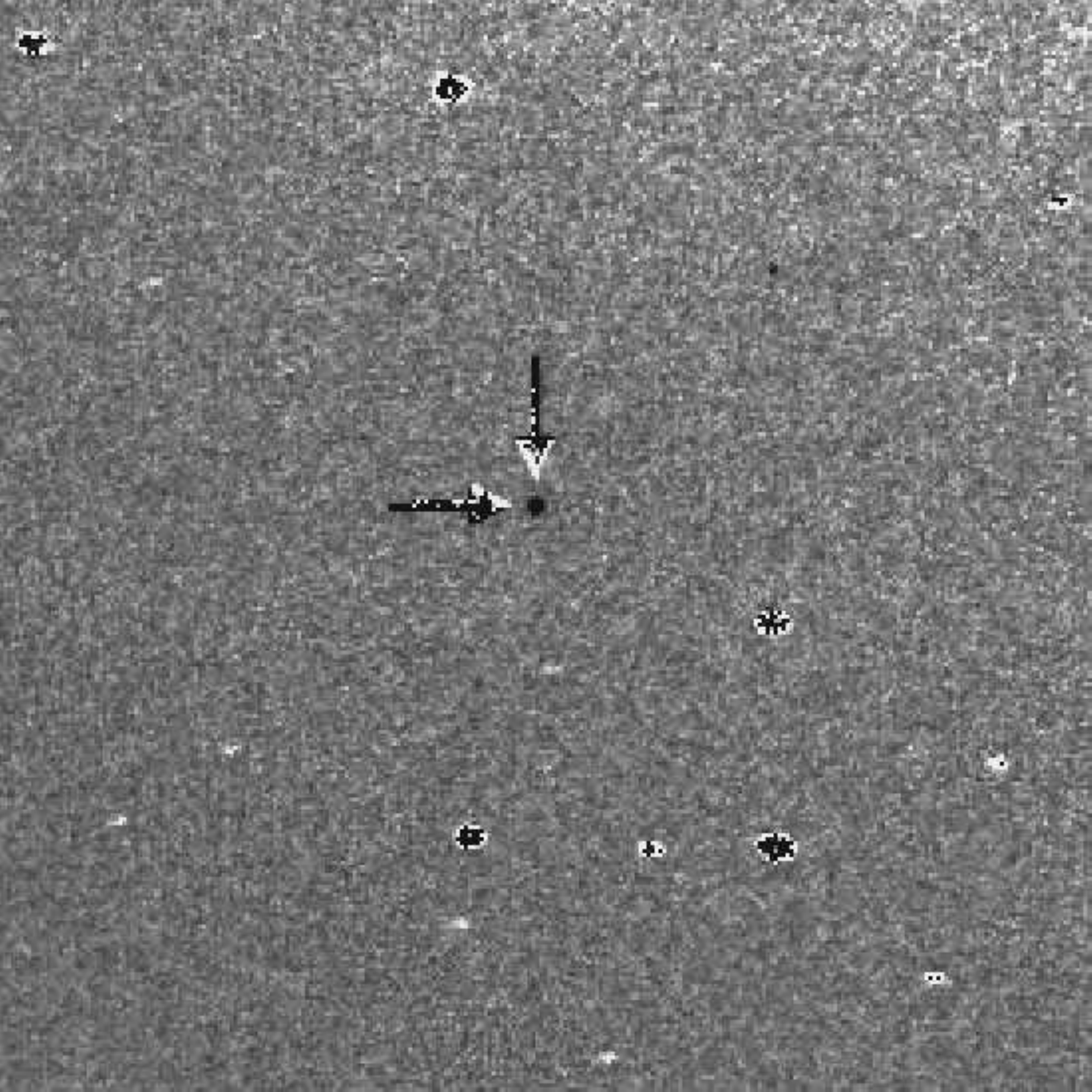}
\caption{Images of M31N 1997-11k, 2001-12b and their comparison
(top left, center, and right), and
M31N 2001-12b, 2009-11b and their comparison
(bottom left, center, and right, respectively).
Charts for all three novae
are taken from images obtained as part of the RBSE
program at KPNO \citep{rec99}.
Although, not shown clearly
in the comparison image (due to the dissimilar brightness of the novae
at the time the images were obtained),
a careful inspection of the positions reveal that all three novae are
spatially coincident to within the estimated
measurement uncertainties of $\sim 1''$.
North is up and East to the left, with a scale of $\sim2.8'$ on a side.
\label{fig13}}
\end{figure}

M31N 1997-11k was discovered as part of the RBSE program \citep{rec99} and
confirmed by \citet{lee12}. Two likely recurrences have been
observed: M31N 2001-12b \citep{lee12} and 2009-11b \citep{hen09b}.
\citet{hen09b} discussed the possibility that M31N 1997-11k might
either be a RN in M31 or a dwarf nova in the Galaxy, and urged
spectroscopic observations of 2009-11b to resolve the issue.
\citet{kas09} obtained a spectrum clearly establishing that
M31N 2009-11b was an Fe~II class nova in M31. Finding charts for the three
objects from the RBSE program
are shown in Figure~\ref{fig13}. We conclude, in
agreement with earlier suggestions, that M31N 1997-11k is a RN.
Given the relatively short recurrence time, it would appear that
the mass accretion rate in this system must be relatively high.
\citet{wil14} ruled out the presence of a luminous red giant secondary in the
progenitor system of M31N 2009-11b.

\subsubsection{M31N 2008-12a}

\begin{figure}
\includegraphics[angle=0,scale=.28]{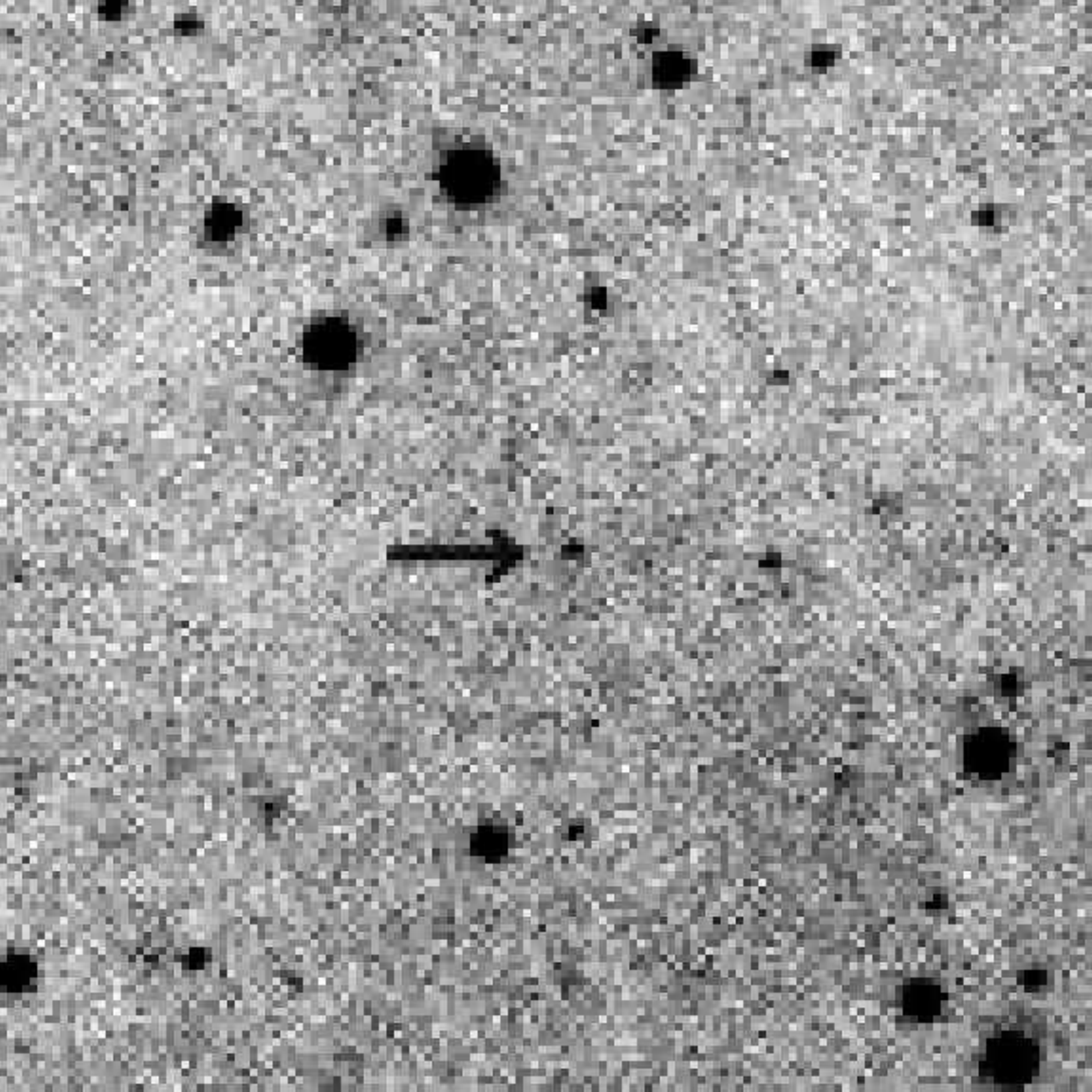}
\includegraphics[angle=0,scale=.28]{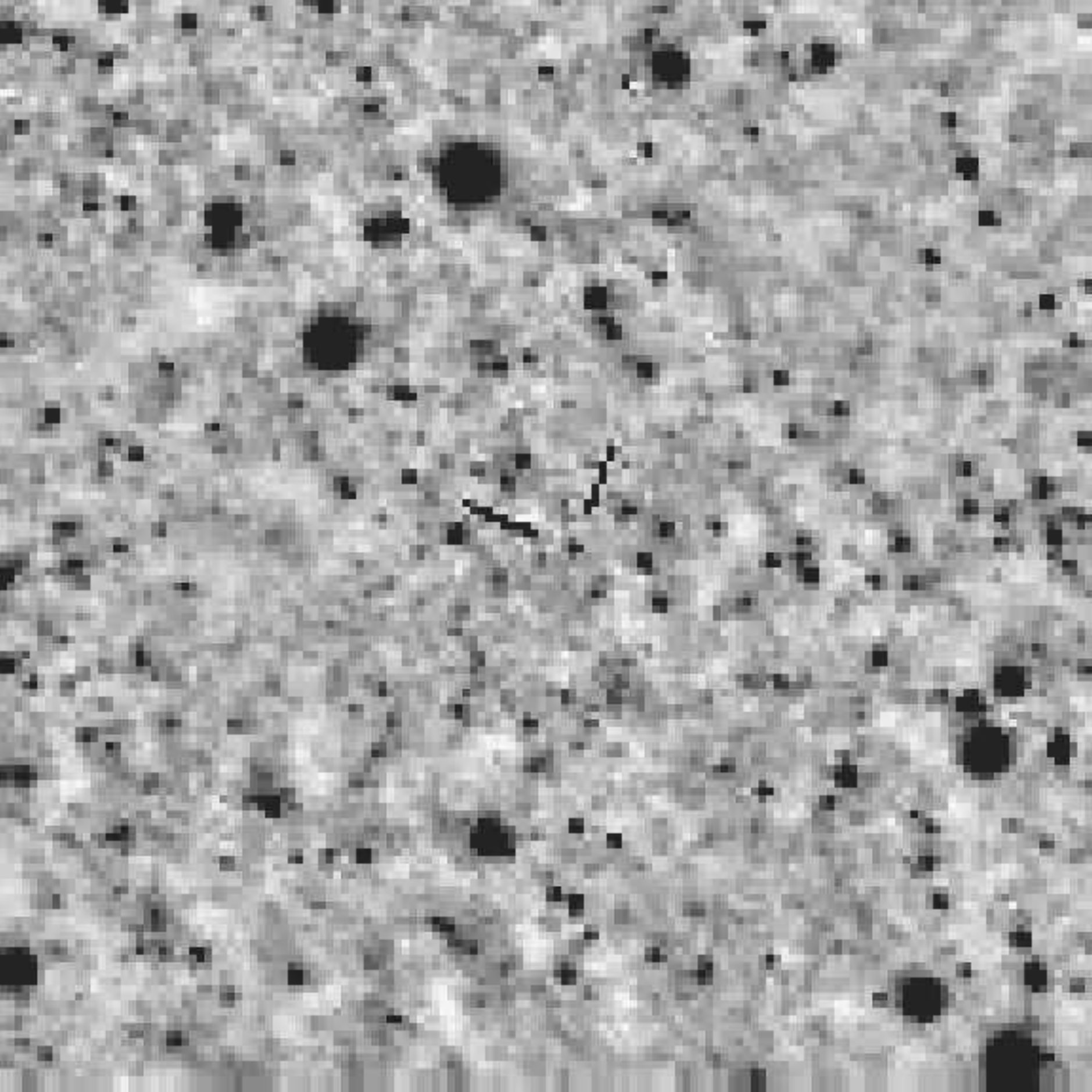}
\includegraphics[angle=0,scale=.28]{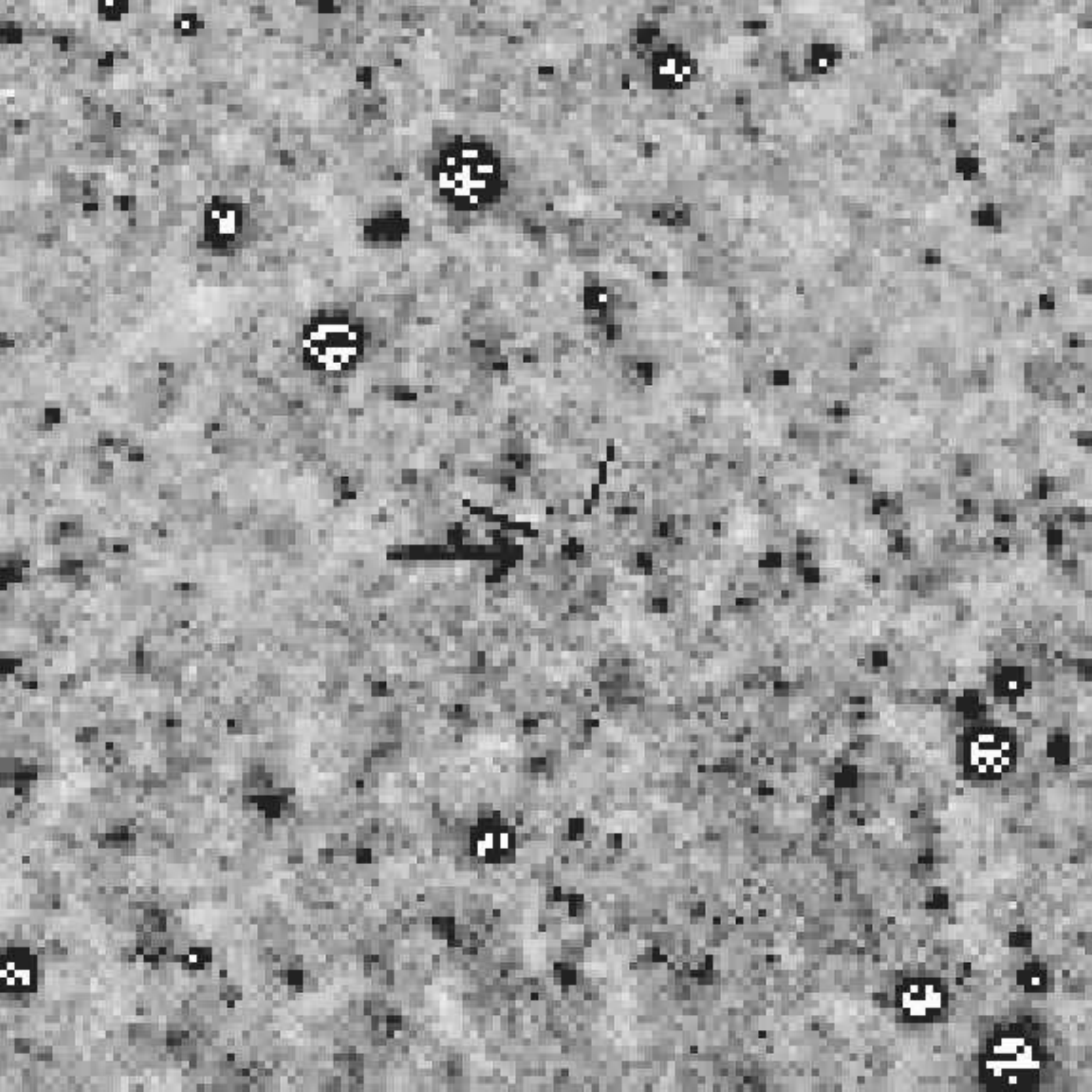}

\includegraphics[angle=0,scale=.28]{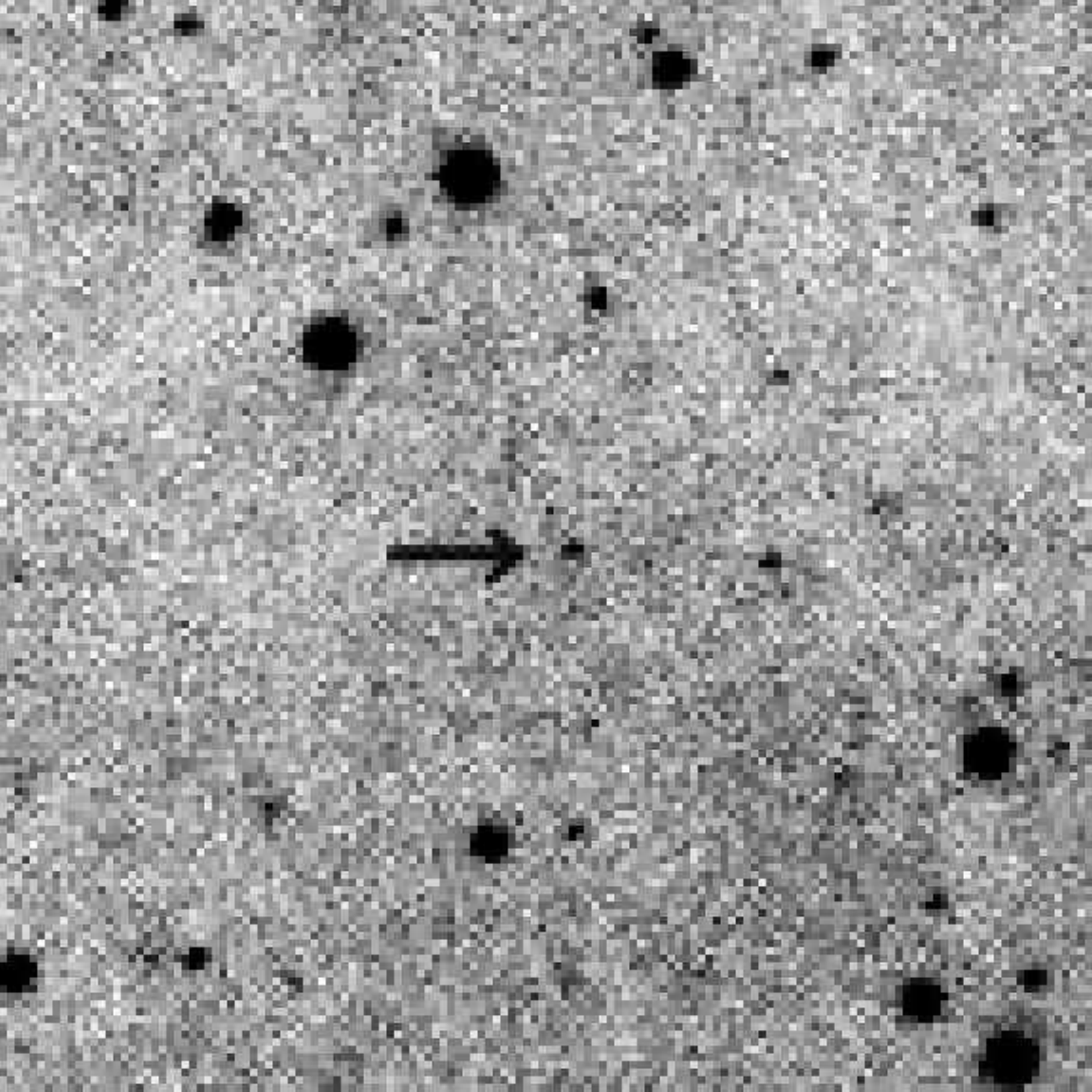}
\includegraphics[angle=0,scale=.28]{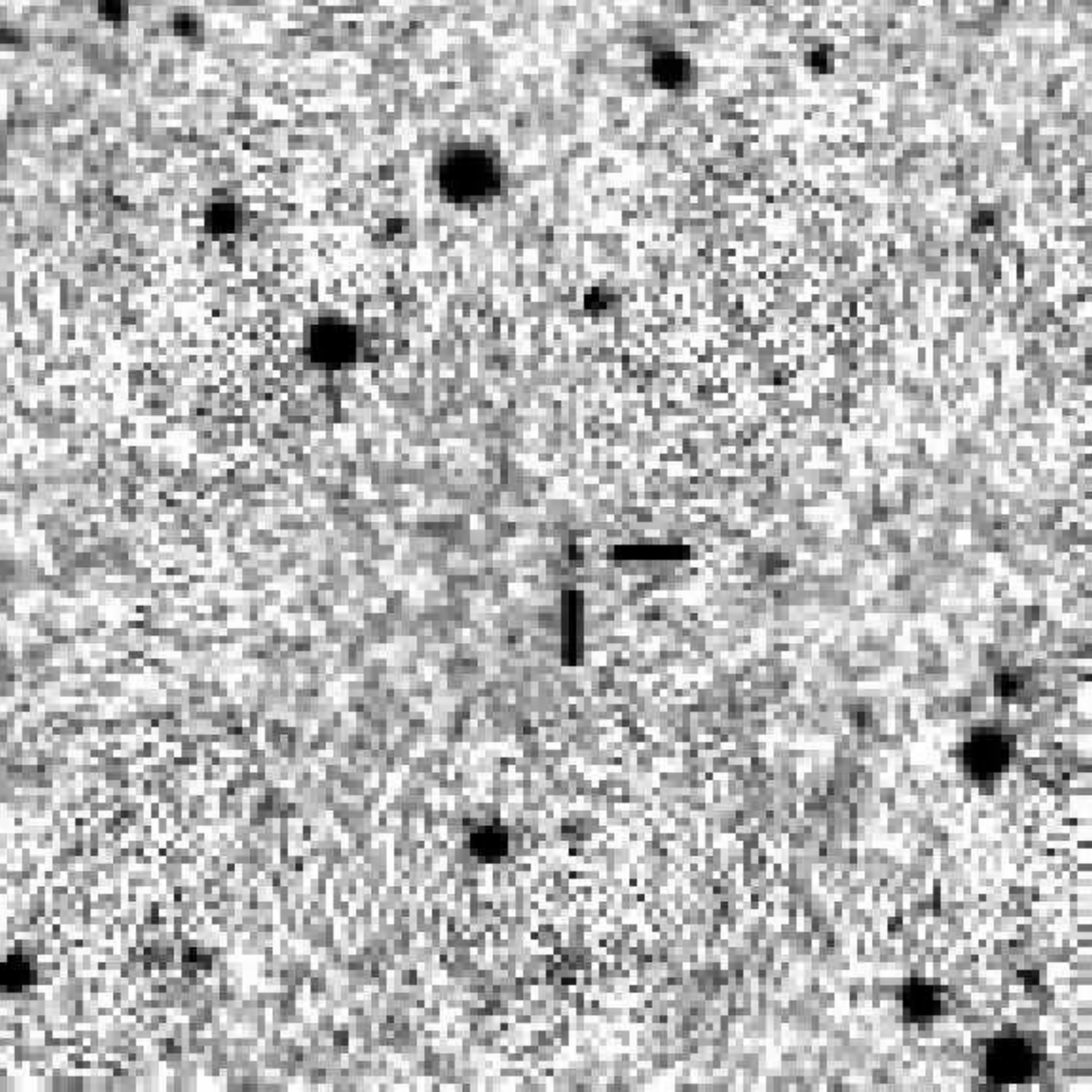}
\includegraphics[angle=0,scale=.28]{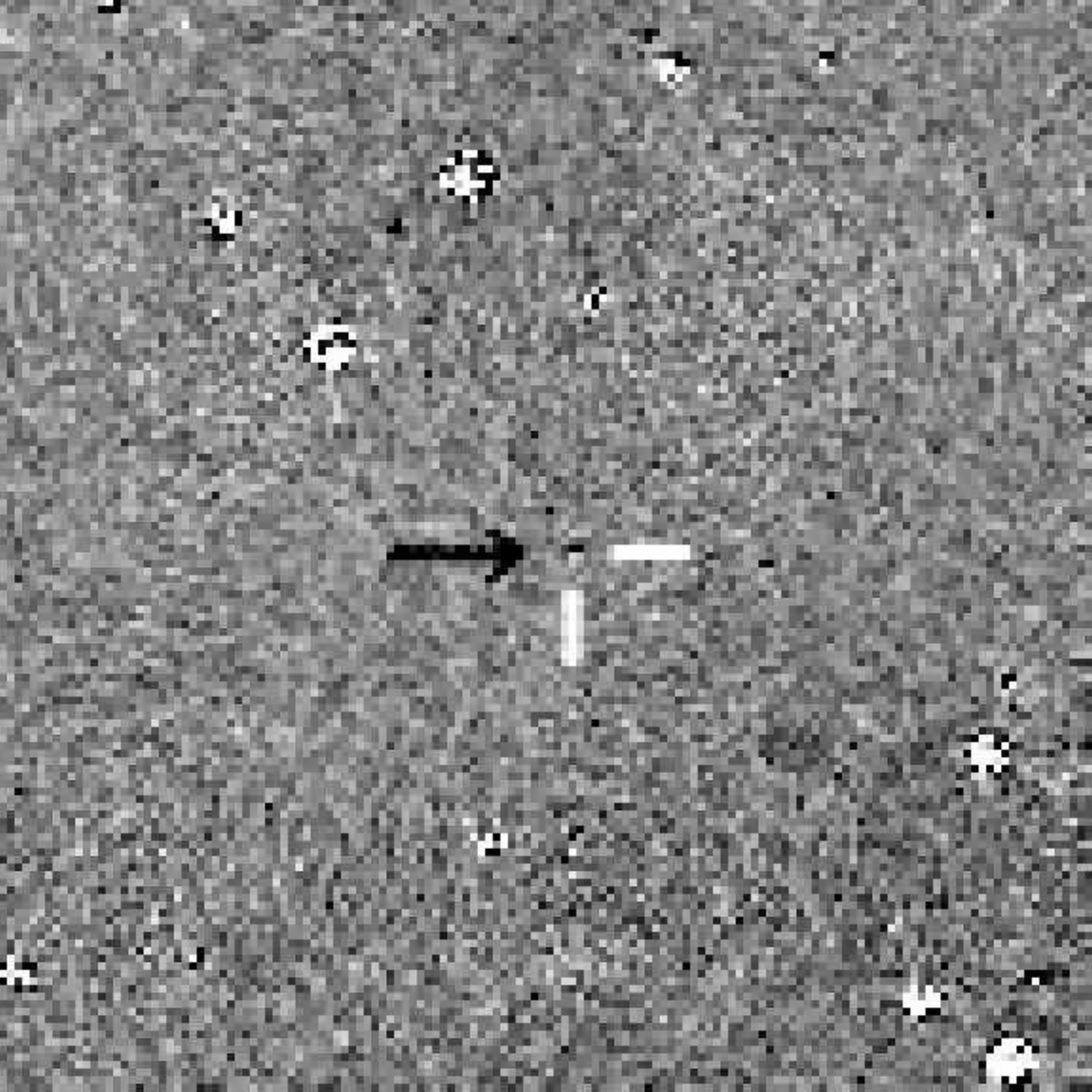}
\caption{Images of M31N 2008-12a, 2011-10e and their comparison
(top left, center, and right), and M31N 2008-12a, M31N 2012-10a,
and their comparison (bottom left, center, and right, respectively).
The charts for 2008-12a and 2012-10a are courtesy of K. Nishiyama and
F. Kabashima (Miyaki-Argenteus Observatory, Japan), while that for
2011-10e is from
\citep{bar11}. Although not obvious from the comparison images,
a careful inspection of the positions of the novae show
that all three are in fact
spatially coincident to within measurement uncertainties of
$\sim 1''$. At the nova's position in the outskirts of M31, the probability
of a chance positional coincidence is negligible.
North is up and East to the left, with a scale of $\sim3'$ on a side.
\label{fig14}}
\end{figure}

M31N 2008-12a, which was discovered by K. Nishiyama and F. Kabashima
(Miyaki-Argenteus Observatory, Japan), was first seen to erupt
in the outskirts of M31 ($a\simeq49'$). The nova has subsequently been
seen to have had four additional outbursts:
M31N 2009-12b \citep{tan14},
2011-10e \citep{bar11}, 2012-10a \citep{sha12a}, and
2013-11f \citep{tan13}, all 4
occurring within $1''$ of the measured position of 2008-12a\footnote{
While this paper was under review another outburst of
M31N 2008-12a, designated 2014-10c,
was discovered on 2014 Oct 02.903 UT \citep{dar14b}.}.
At this relatively large
galactocentric radius, the probability of a chance positional coincidence
is negligible, and it seems clear that the novae represent brightenings of the
same progenitor system (see Fig.~\ref{fig14}).
It would appear that M31N 2008-12a must be either
a RN with an extremely short interval between eruptions ($\sim 1$~yr), or
that the 2009, 2011, 2012 and 2013 events are simply rebrightenings of an
unusually slow nova.
Recent observations have shown that the latter possibility is untenable.
In particular, the He/N \citep{sha12a} spectroscopic type along with the
super-soft X-ray behavior strongly suggests
that the object is in fact a short recurrence time RN
\citep{dar14a,hen14b,tan14}. Further, as shown by \citet{dar14a}, the spectral
energy distribution of the quiescent counterpart of M31N 2008-12a is
consistent with a bright accretion disk (and thus a high accretion rate)
in this system.

A literature search has revealed three additional X-ray detections of
M31N 2008-12a in February 1992 and January 1993 \citep{whi95}
as well as in September 2001 \citep{wil04}.
The super-soft X-ray properties of the two earlier outbursts
(short duration, relatively high effective temperature)
are in agreement with the 2013 X-ray detection \citep{hen14b}.
These authors reported a very short SSS phase,
appearing on day $6\pm1$ and disappearing on day $19\pm1$
after the optical outburst, with an exceptionally
high effective (blackbody) temperature of $97^{+5}_{-4}$~eV.
With eight recorded outbursts, M31N 2008-12a is a very unusual object
that merits continued attention.
An archival study on potential previous outbursts is in preparation
(Henze et al. 2015).

\subsection{Possible Recurrent Novae}

\subsubsection{M31N 1953-09b}

\begin{figure}
\includegraphics[angle=0,scale=.28]{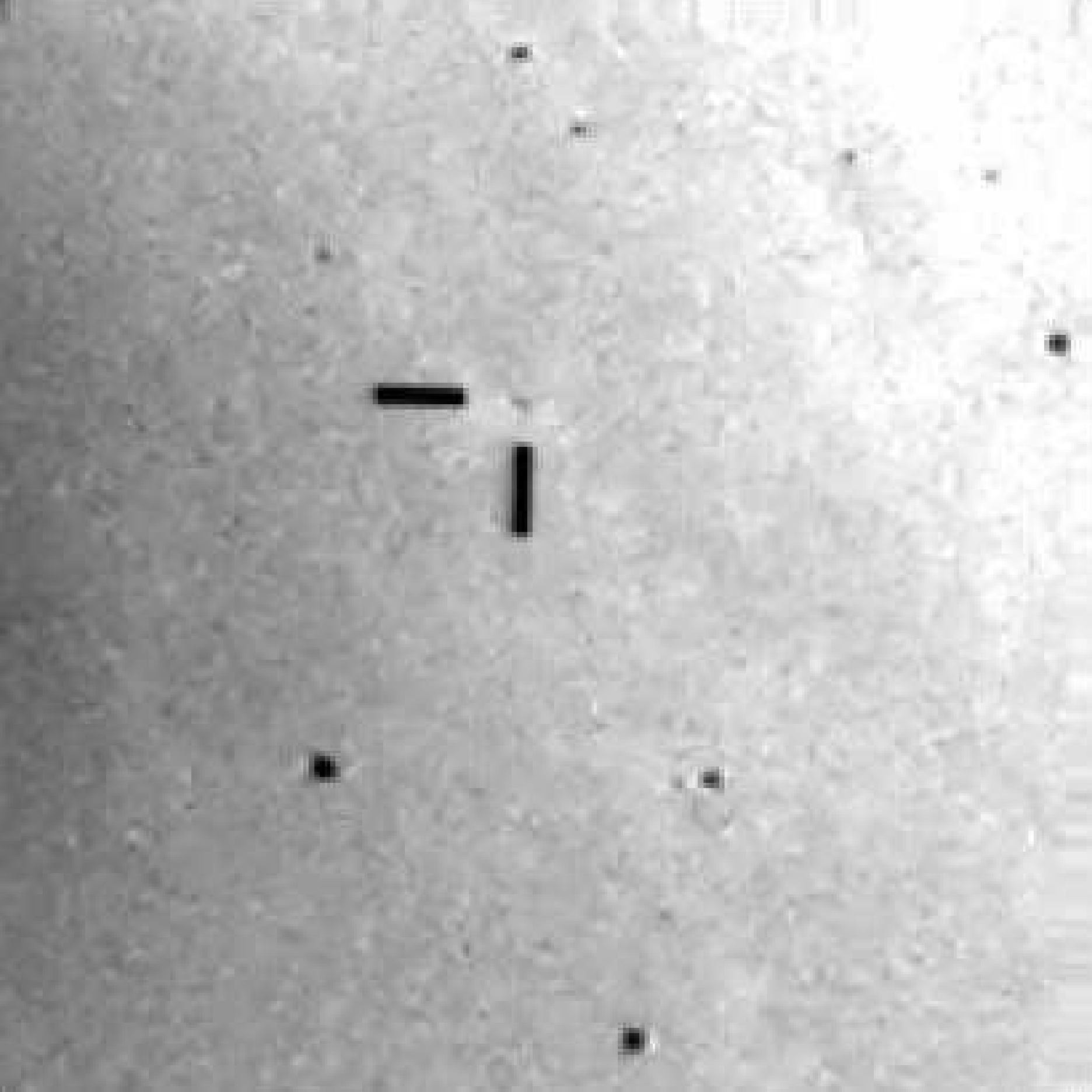}
\includegraphics[angle=0,scale=.28]{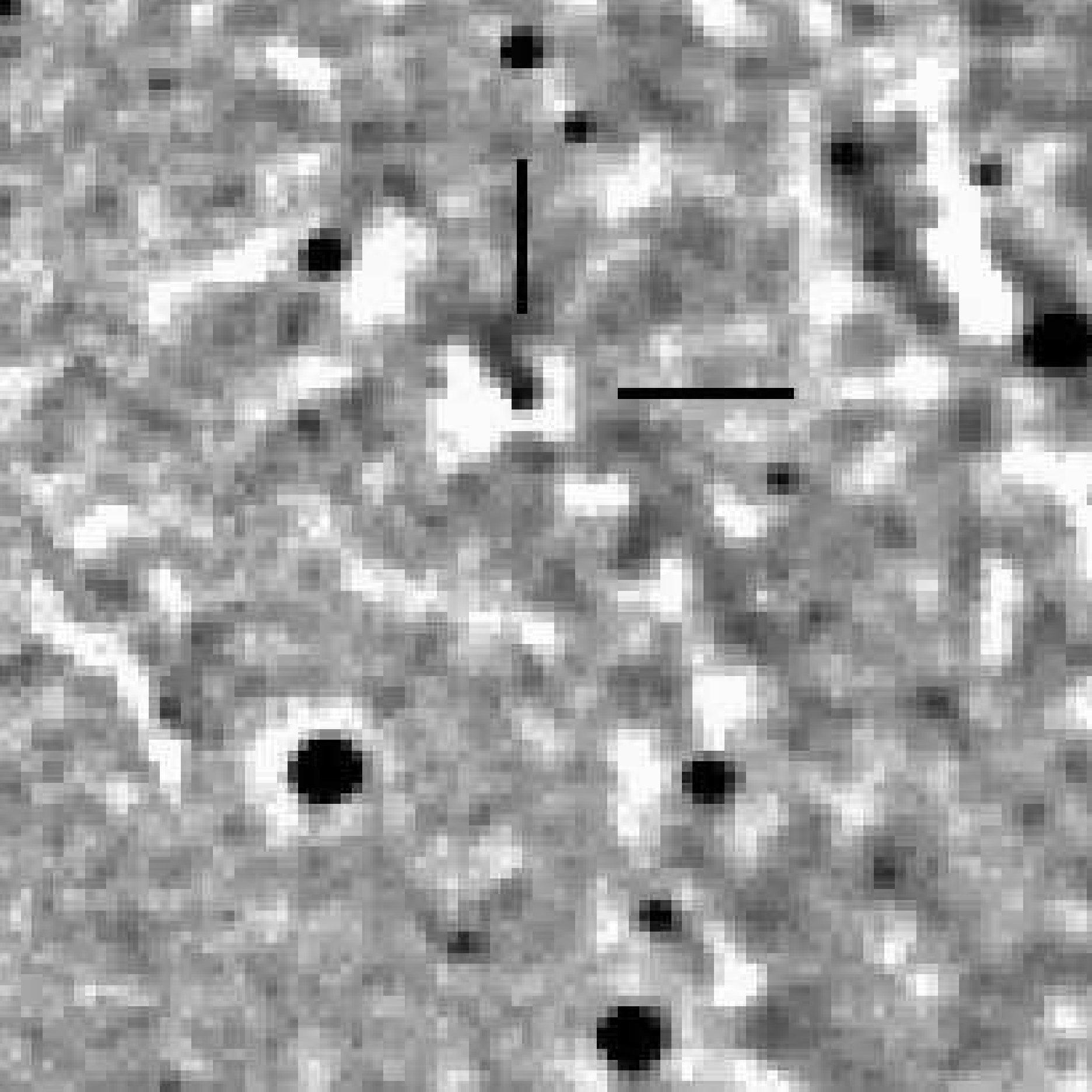}
\includegraphics[angle=0,scale=.28]{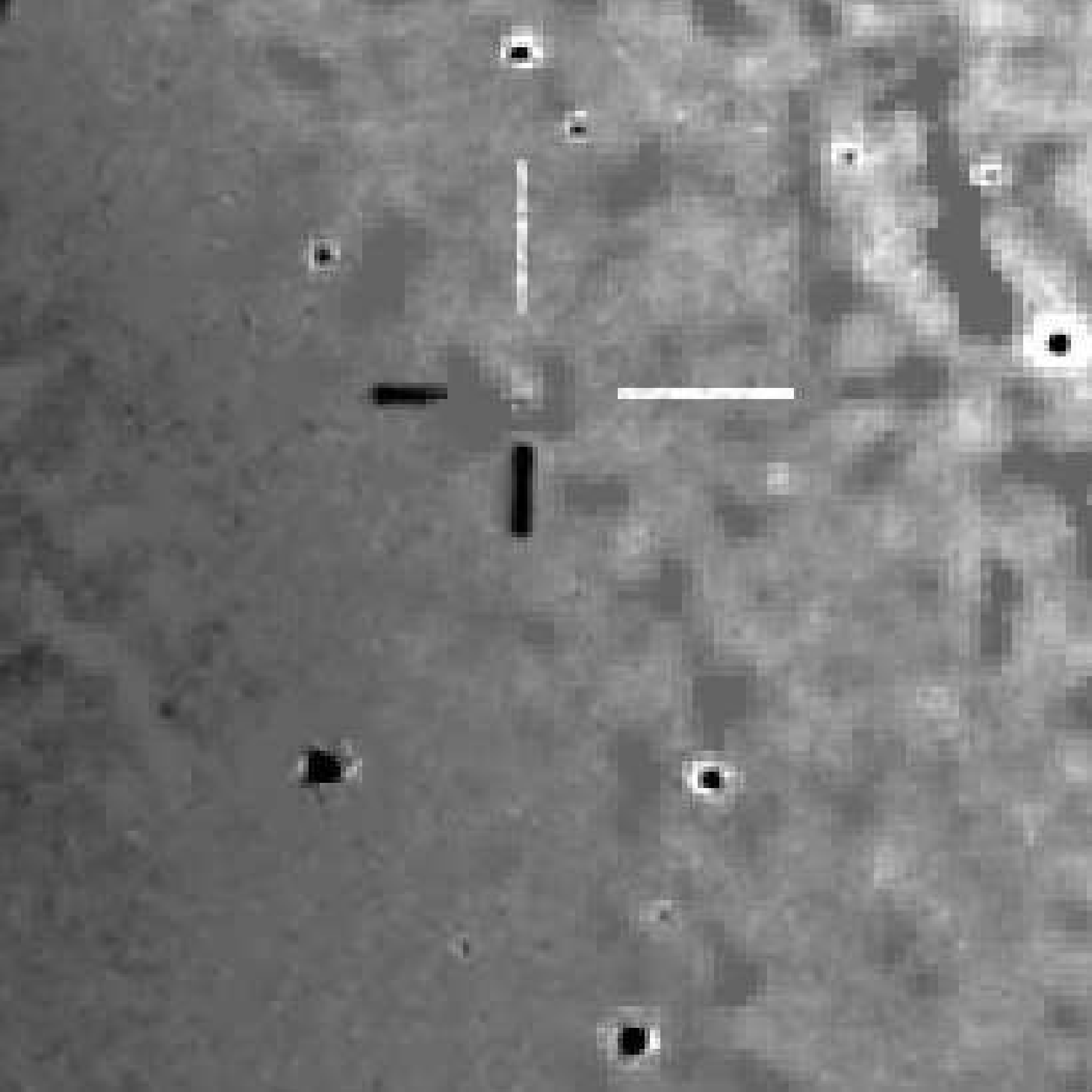}
\caption{Images of M31N 1953-09b, 2004-08a, and their comparison
(left, center, and right, respectively).
The image of M31N 1953-09b is a scan of the plate S967A from \citet{arp56},
while that of 2004-08a is from K. Hornoch (2004, unpublished).
The comparison image suggests that M31N 1953-09b may be slightly south
of the position of 2004-08a, but given the measurement
uncertainties, the objects are possibly coincident.
North is up and East to the left, with a scale of $\sim3.5'$ on a side.
\label{fig15}}
\end{figure}

M31N 1953-09b was discovered by \citet{arp56} as part of his
classic M31 nova study (Arp \#3). A subsequent nova, M31N 2004-08a, discovered
by K. Hornoch \citep{pie07a}, was seen to erupt $\sim4.8''$ away
from the reported position of 1953-09b. At the location of the nova
($a\sim5.8'$), the
probability of a chance coincidence is quite high ($P_C\simeq0.640$).
Nevertheless, we have located the original plate from Arp's survey (S967A) and
compared the position of M31N 1953-09b with that
of 2004-08a in Figure~\ref{fig15}.
A careful inspection of the charts shows that M31N 1953-09b may be
just slightly south of the position of 2004-08a, but
to within the limits of the seeing disks ($\sim1.5''$), it is possible
that the novae could be spatially coincident.
Revised coordinates for M31N 1953-09b are given in Table~\ref{revcoord},
lowering the probability of a chance positional coincidence to $P_C\simeq0.131$.

It is worth noting that \citet{lee12} judge the light curve of M31N 2004-08a
to be questionable for a nova, so there remains some doubt
regarding the nature of this object. That said,
M31N 2004-08a was detected as a SSS by \citet{pie07a}.
The nova showed a very short SSS phase and was only detected around
day 60 after outburst without being visible about 30 days before and after.
\citet{pie07a} also estimated a high effective (blackbody)
temperature of around 80~eV. In view of the available evidence,
the nature of M31N 1953-09b remains uncertain, with the
possibility that the object is a RN remaining viable.

\subsubsection{M31N 1961-11a}

\begin{figure}
\includegraphics[angle=0,scale=.28]{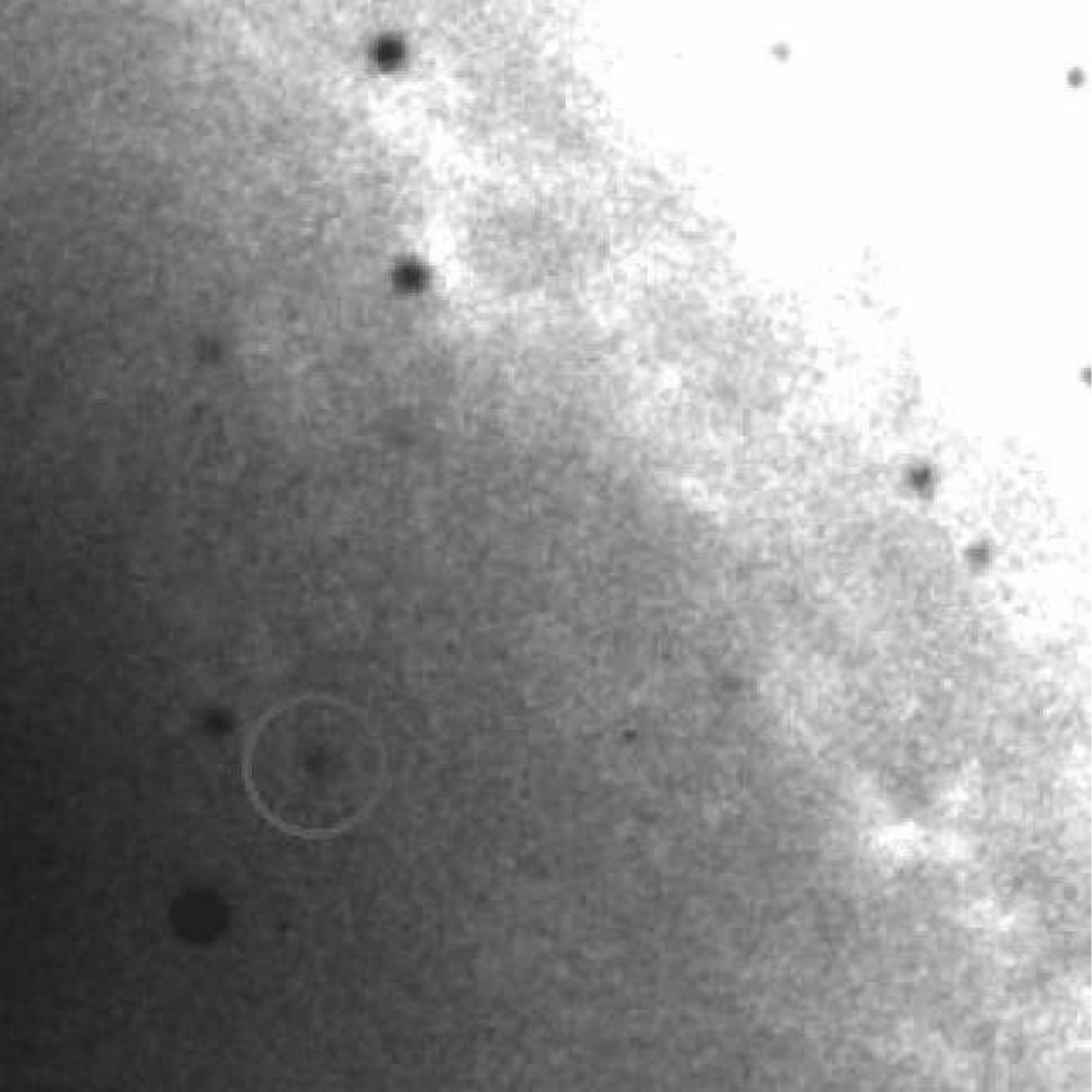}
\includegraphics[angle=0,scale=.28]{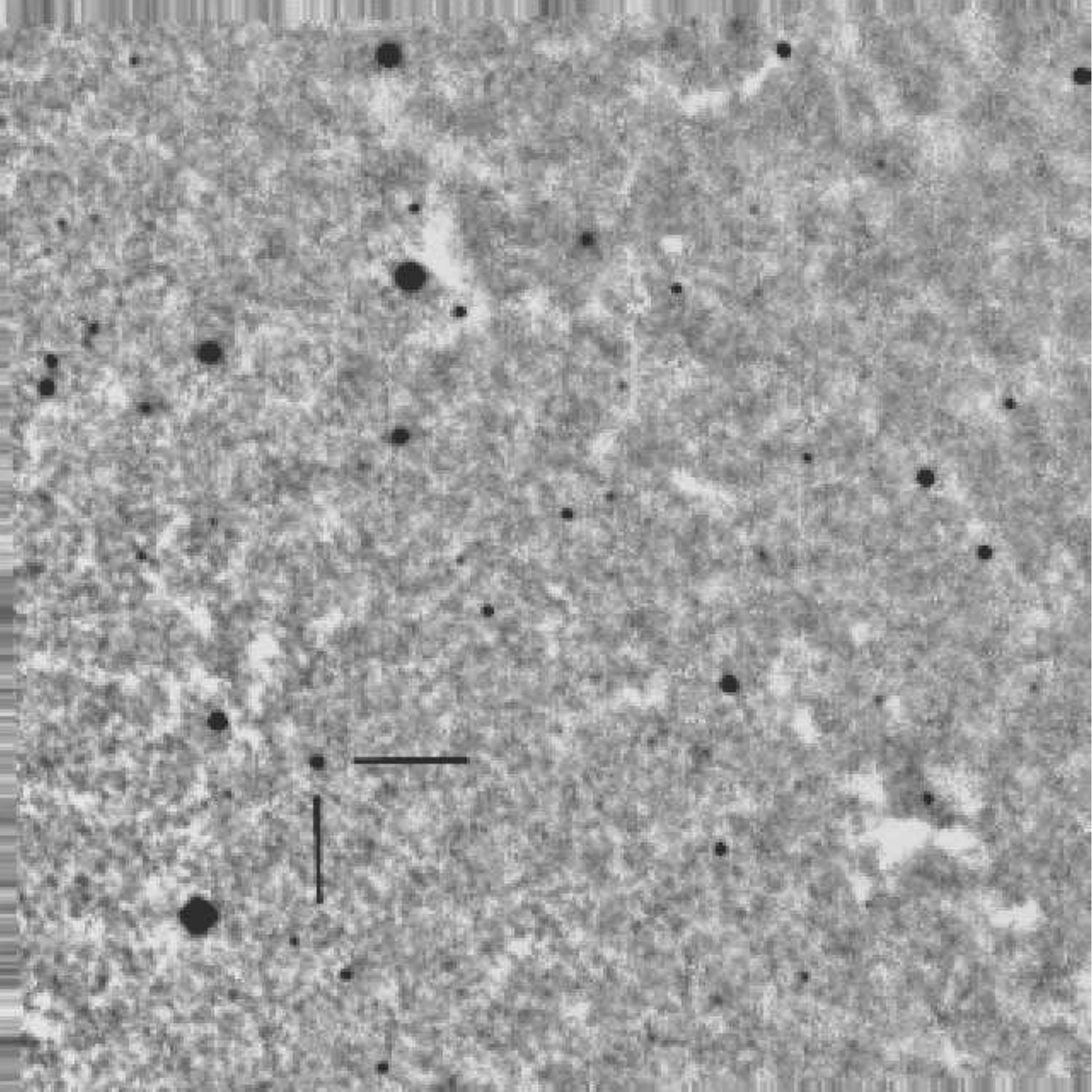}
\includegraphics[angle=0,scale=.28]{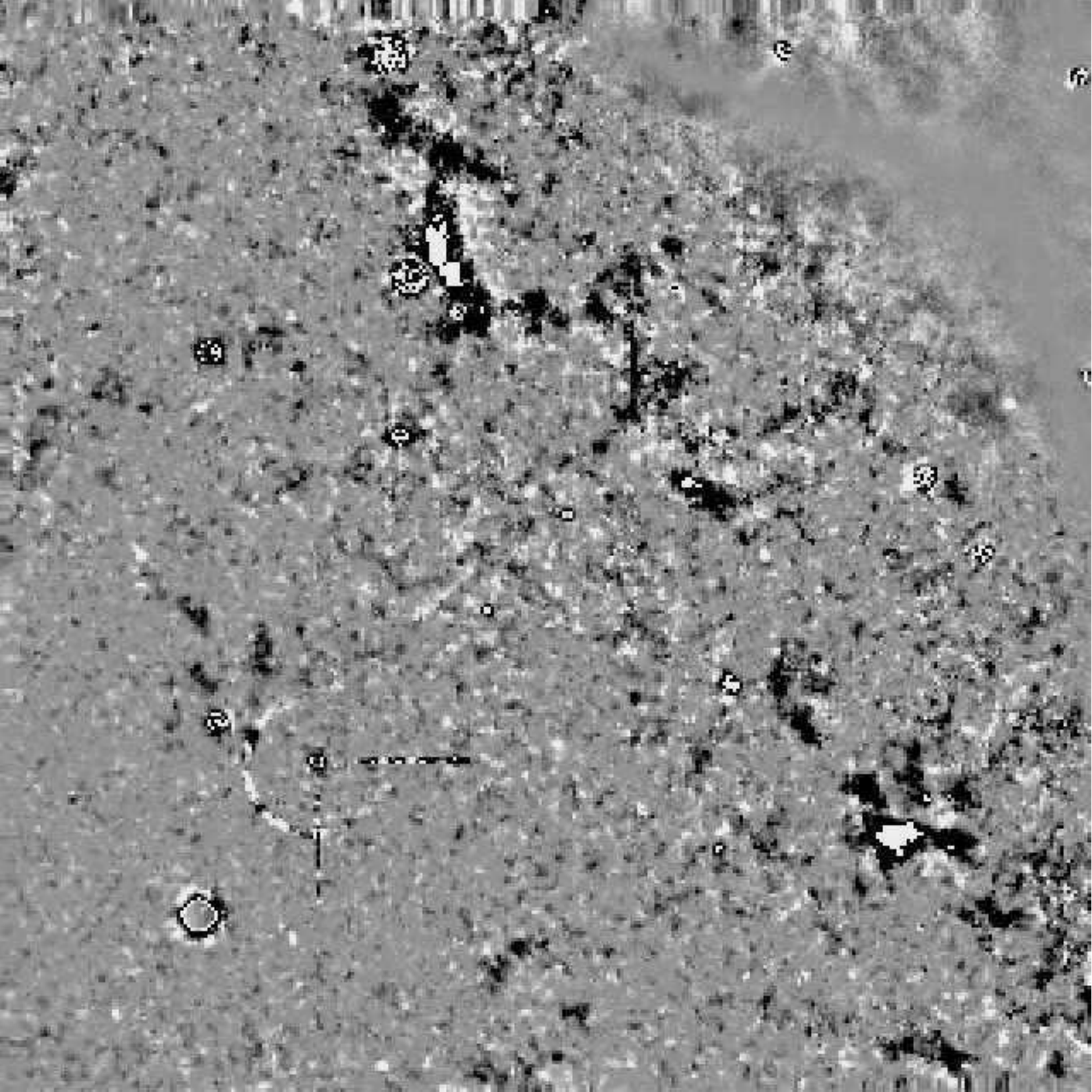}
\caption{Images of M31N 1961-11a, 2005-06c, and their comparison
(left, center, and right, respectively).
The image of M31N 1961-11a has been produced from data
taken as part of the \citet{ros64,ros73} survey,
while that of
2005-06c is from K. Hornoch (previously unpublished).
The comparison image confirms that the novae are coincident to 
within measurement uncertainties ($\sim 1''$).
North is up and East to the left, with a scale of $\sim3.5'$ on a side.
\label{fig16}}
\end{figure}

M31N 1961-11a erupted near the nucleus of M31 ($a\sim3'$) and
was discovered as nova \#35 in the survey of \citet{ros73}.
A possible recurrence, M31N 2005-06c, discovered by K. Hornoch \citep{pie07a},
erupted $\sim$2.4$''$ from the reported position of 1961-11a and
thus passed our initial RN screening. Given the nominal separation
of the two novae, and their close proximity to the nucleus,
the probability of a chance
positional coincidence is quite high ($P_C\simeq0.691$). However,
a careful comparison of the finding charts for the two novae shown
in Figure~\ref{fig16} reveals that they are in fact spatially coincident
to within measurement uncertainties.
Revised astrometry of M31N 1961-11a
based on the published chart (see Table~\ref{revcoord}) shows that the two
novae appear to be separated by just 0.45$''$, with a probability of a
chance coincidence, $P_C\simeq0.042$.
Despite the spatial coincidence, 
\citet{lee12} have identified a variable star near the position
of M31N 2005-06c, and have called into question the nature of the object.
In view of their findings, we hesitate to definitively classify
the object as a RN, and have instead included it in our list of uncertain RNe.

\subsubsection{M31N 1966-08a}

\begin{figure}
\includegraphics[angle=0,scale=.28]{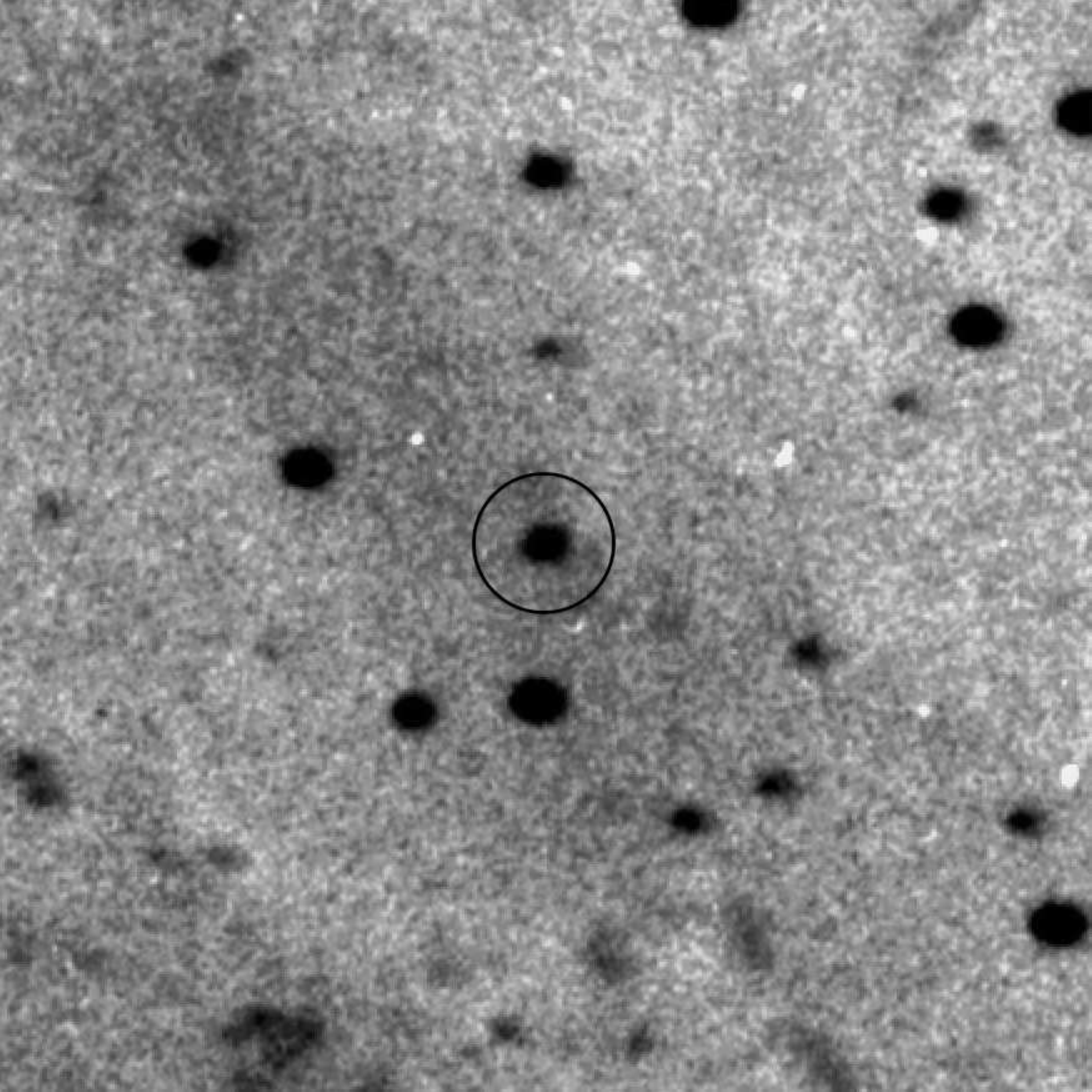}
\includegraphics[angle=0,scale=.28]{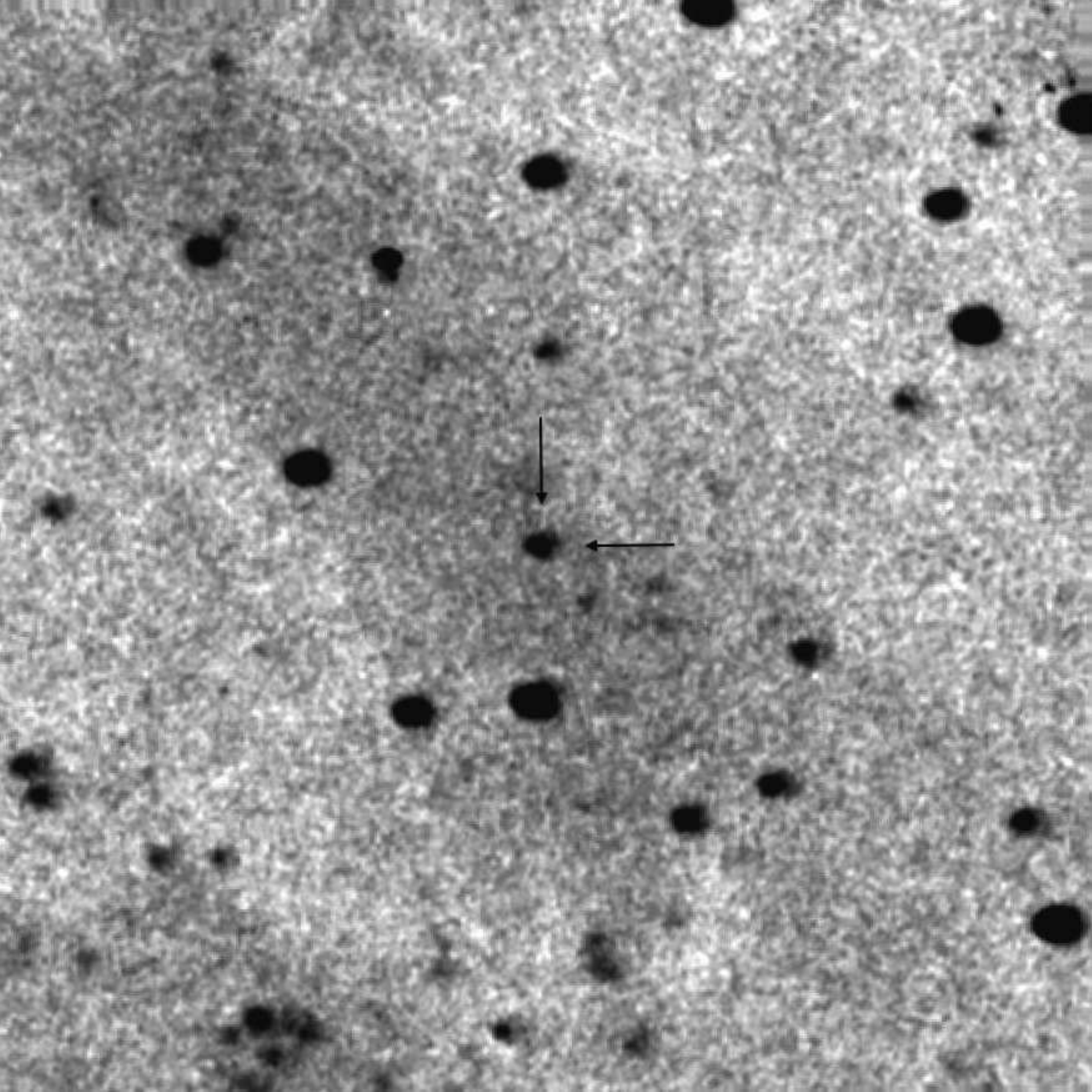}
\includegraphics[angle=0,scale=.28]{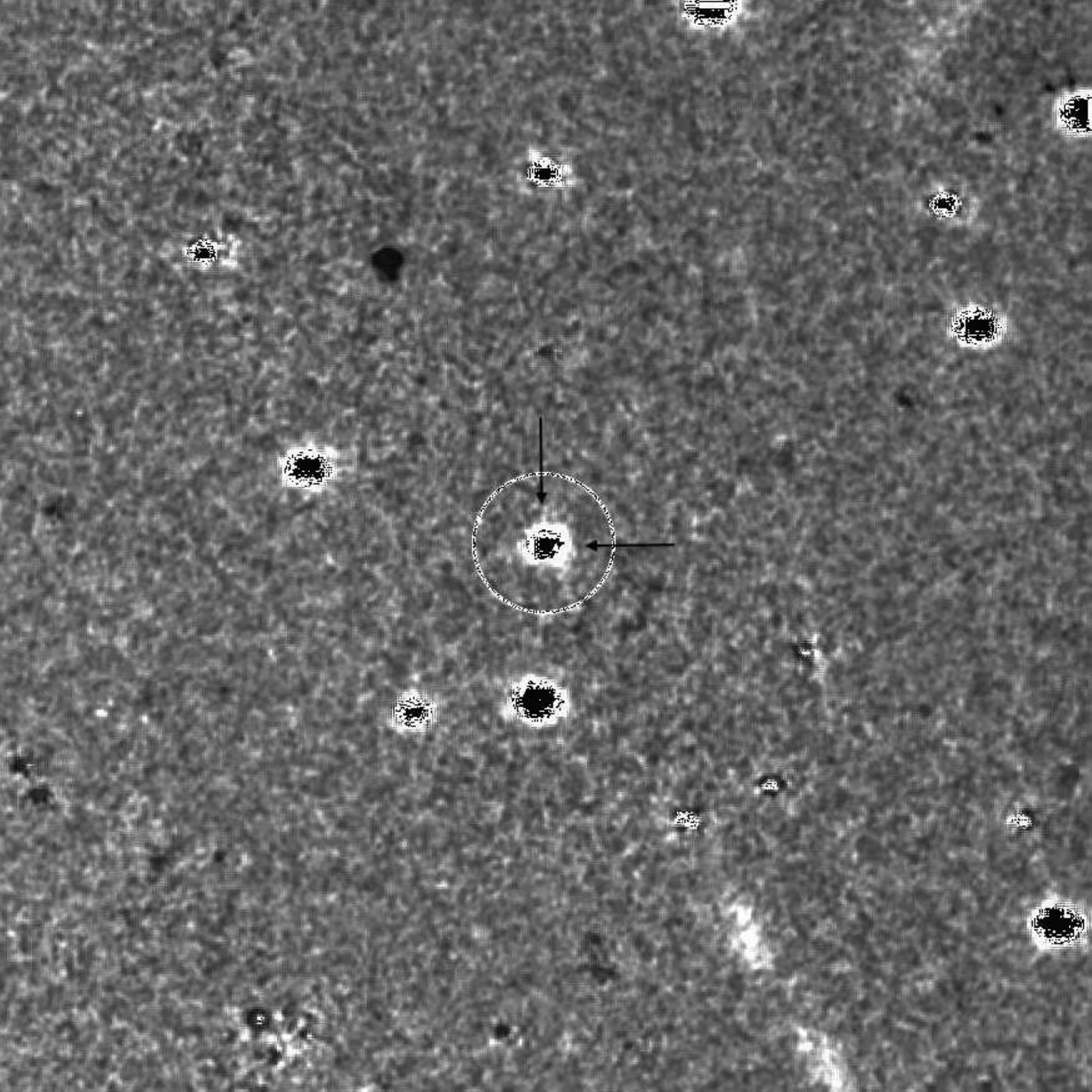}
\caption{Images of M31N 1966-08a, 1968-10c, and their comparison
(left, center, and right, respectively).
Both images are from the survey of \citet{ros73}.
The comparison image confirms that both novae are spatially coincident.
North is up and East to the left, with a scale of $\sim4'$ on a side.
\label{fig17}}
\end{figure}

M31N 1966-08a and 1968-10c were discovered as novae \#66 and \#81 in
the survey of \citet{ros73}, who gives the coordinates of the novae
as being identical.
The object is located
at a relatively large galactocentric radius ($a\simeq31.2'$) and there
is very little chance that the objects could be distinct systems.
The positions of the two novae are
shown in Figure~\ref{fig17}, confirming that the objects are
in fact spatially coincident.
\citet{sha89} argue that given the short interval
between eruptions, the system is likely a
Galactic foreground U Gem star (dwarf nova).
That said, the possibility that the object is a RN with a very short
recurrence time cannot be ruled out, and we have classified the
object as an uncertain RN. Whether or not the
system is a short recurrence time RN or a Galactic dwarf nova,
it is quite surprising that more
eruptions of the star have not been observed!

\subsubsection{M31N 1990-10a}

\begin{figure}
\includegraphics[angle=0,scale=.28]{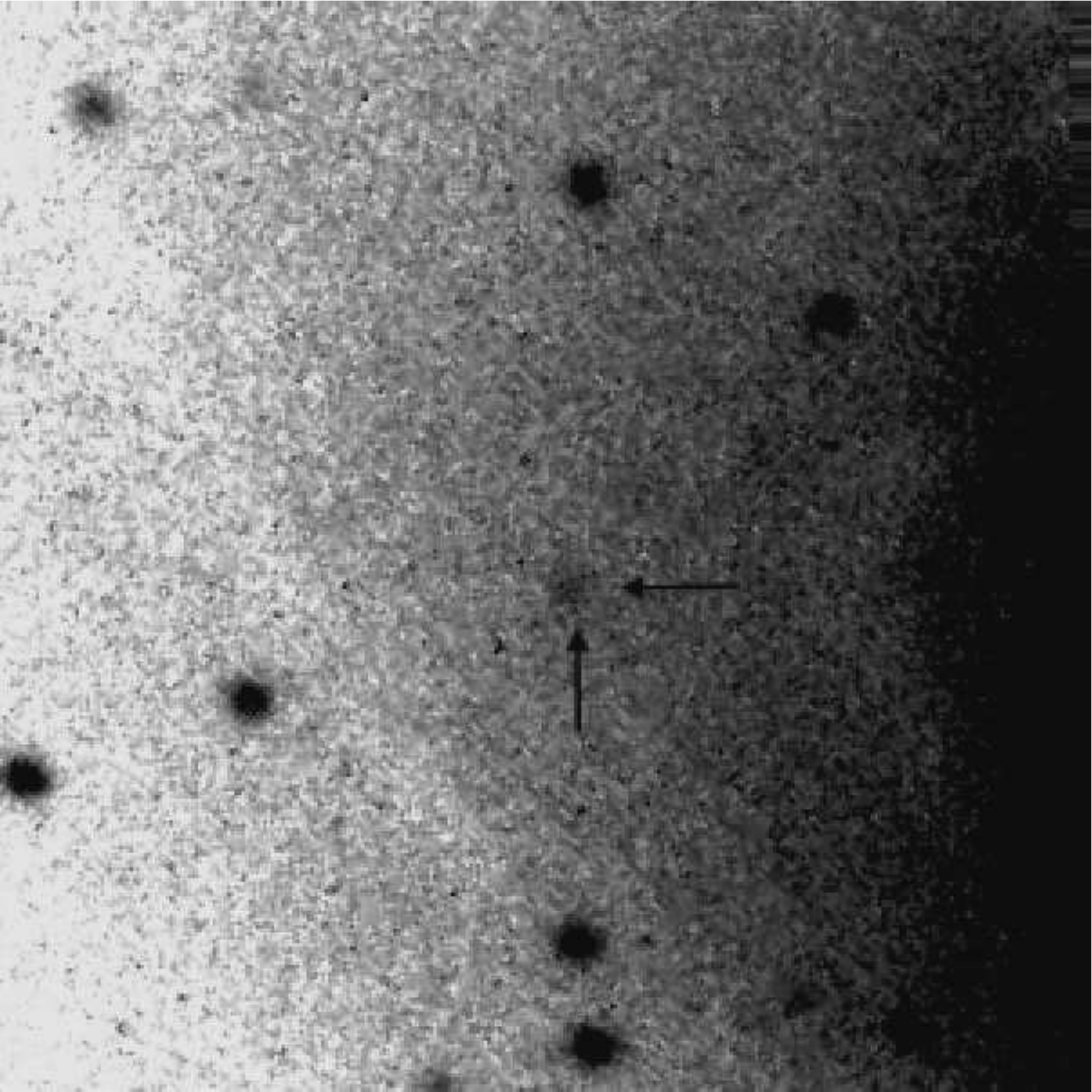}
\includegraphics[angle=0,scale=.28]{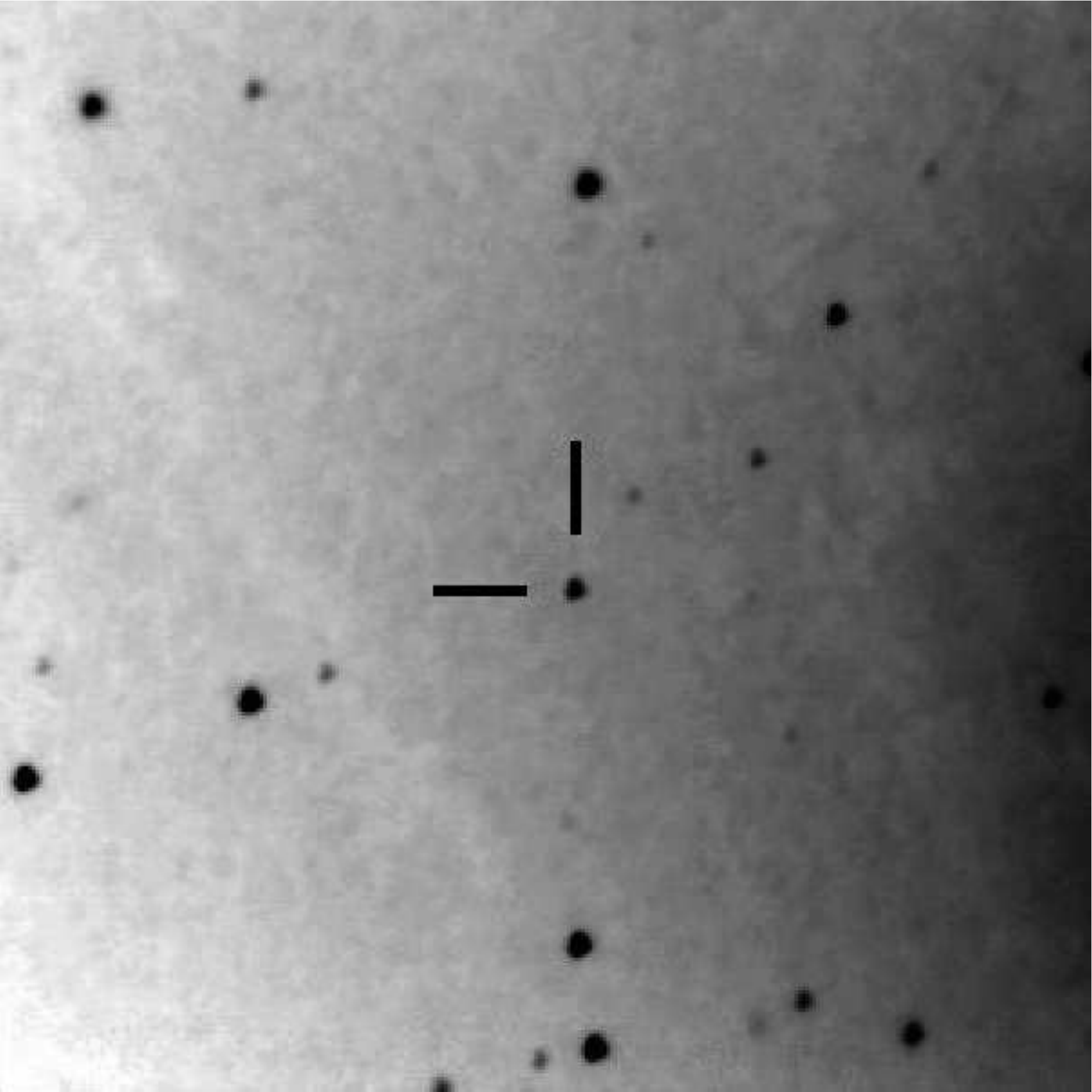}
\includegraphics[angle=0,scale=.28]{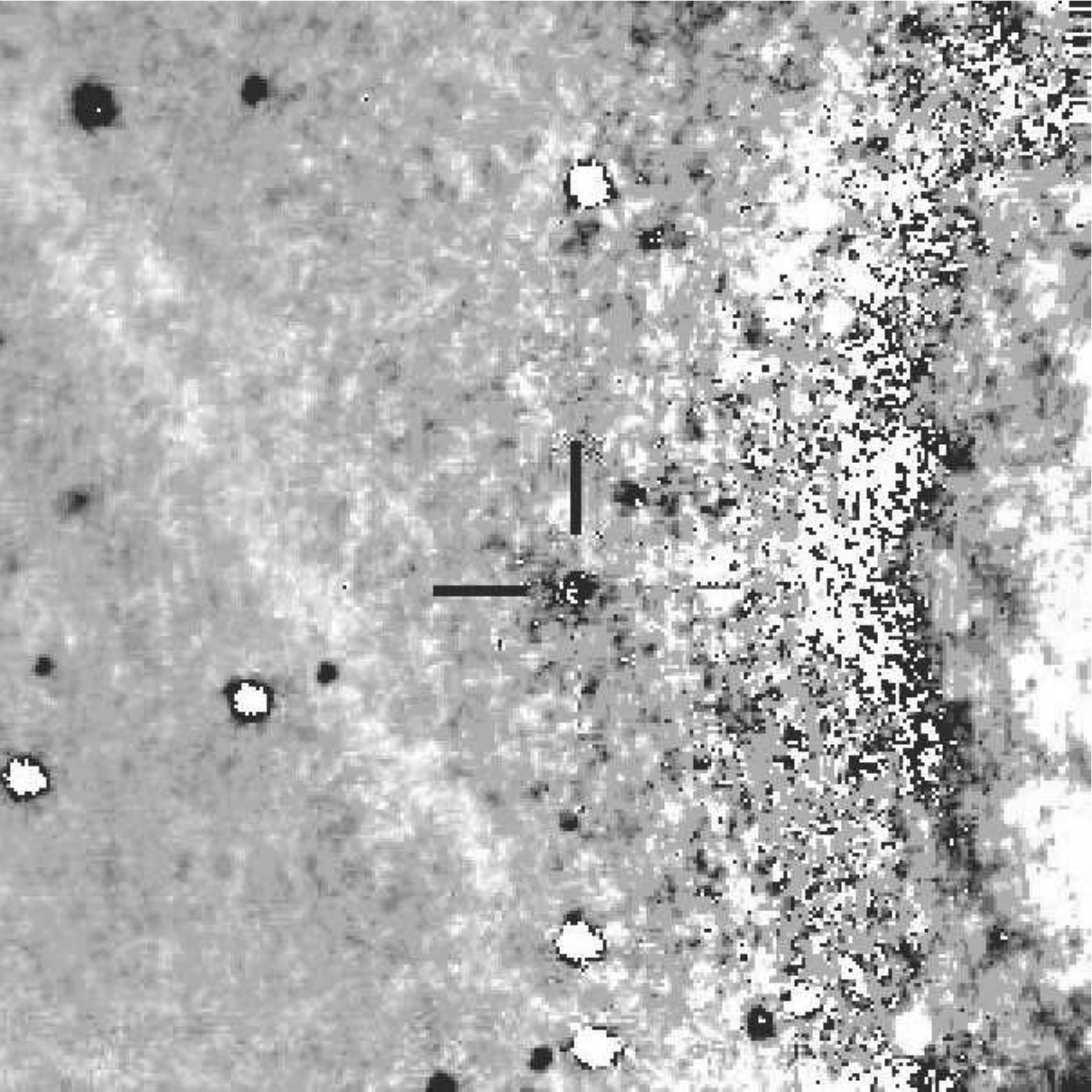}
\caption{Images of M31N 1990-10a, 2007-07a, and their comparison
(left, center, and right, respectively).
The chart for M31N 1990-10a is based on a 35-mm photographic negative
obtained by one of us (J. B.) on 1990 Oct 13.16, while that for 2007-07a
is taken from observations at the Skinakas Observatory \citep{hat07a}.
M31N 2007-07a is coincident to within $\sim 1''$ of the position of
M31N 1990-10a, although it appears that the former nova may be
just slightly NW of 1990-10a, and the relatively poor
image quality of the photographic negative does not allow us to make a
definitive judgment as to whether the two novae are in fact
spatially coincident.
North is up and East to the left, with a scale of $\sim2.5'$ on a side.
\label{fig18}}
\end{figure}

M31N 1990-10a was discovered on
1990 October 13.16 UT during a routine photographic patrol of M31 \citep{bry90}.
Our initial screen reveals two possible recurrences. The first, M31N 1997-10b,
was a faint nova candidate reported
by \citet{sha01}. A re-examination of the original image shows that
the object was likely a CCD artifact, and not a nova, as noted by \cite{lee12}.
Another possible recurrence was M31N 2007-07a \citep{hat07a}, which erupted
$\sim0.8''$ away from the reported position of 1990-10a.
Revised astrometry of M31N 1990-10a (see Table~\ref{revcoord})
reveals that the two novae
are spatially coincident to within $0.59''$ ($P_C\simeq0.041$).
However, as can be seen in Figure~\ref{fig18}, it appears that
M31N 2007-07a may lie just slightly to the NW of 1990-10a. The relatively
poor seeing of the M31N 1990-10a discovery image makes it
impossible to establish conclusively whether or not
the two novae are in fact coincident.
We conclude that M31N 1990-10a is a possible RN.

\subsection{Unlikely Recurrent Novae}

\subsubsection{M31N 1932-09d}

M31N 1932-09d was discovered by \citet{str36} and lies $\sim3.6''$ from
the position of a more recent nova, 2001-07d, which was discovered
by \citet{wli01} and independently by
the Wendelstein Calar Alto Pixellensing Project \citep{lee12}
and in data from the POINT-AGAPE survey \citep{dar04}.
Unfortunately, we have not been able to locate a
finding chart for M31N 1932-09d,
so we are unable to definitely establish whether or not the object is a RN.
We note that eruptions with a nominal separation of $3.6''$ or less
at a projected distance of $\sim3'$ from the nucleus of M31 have
probability of a chance coincidence that is quite high ($P_C\simeq0.939$).
We conclude that M31N 2001-07d is unlikely to be a recurrence of 1932-09d.

\subsubsection{M31N 1957-10b}

As noted by \citet{alk00}, M31N 1957-10b (PT~And) has exhibited several
recurrences (1983, 1986, 1988, and 1998) over the past 30 years
in addition to its most recent
outburst in 2010, M31N 2010-12a \citep{rua10a,zhe10}. \citet{sha89} and
\citet{alk00} have argued based on its light curve properties
that PT~And is a Galactic dwarf nova system, while
\citet{cao12} suggest based on an optical spectrum
that the object may indeed be a RN in M31.
We have not shown finding charts for the outbursts of PT~And
because it is clear that
they are spatially coincident and the light curves are
consistent between eruptions \citep{alk00}.
The only question is whether the object is a RN
in M31 or a Galactic dwarf nova.
After reviewing available data, including a spectrum published
online by Kao\footnote{
\tt http://www.cc.kyoto-su.ac.jp/$\sim$kao/blog/index.php/view/20},
which shows a blue continuum with no prominent emission features,
we conclude that the object is most likely a foreground Galactic dwarf nova
and not a RN.

\subsubsection{M31N 1982-09a}

M31N 1982-09a is the first of 6 novae to be discussed below
that were all discovered as part of the \citet{cia87} H$\alpha$ survey
for novae in M31.
Unfortunately, the original CCD data for these novae have been lost, making
it impossible to create the finding charts necessary to
definitively test whether these novae are recurrent.
Most of these novae erupted quite close to the nucleus of M31
and their putative recurrences
have high probabilities of chance positional coincidence.

M31N 1982-09a was one of the \citet{cia87} novae that
erupted very close ($a\sim1.08'$) to the nucleus of M31.
A possible recurrence, M31N 2011-02b, was seen to erupt
$\sim5.46''$ from the reported position of 1982-09a. The estimated probability
of a chance positional coincidence is extremely high, with $P_C\simeq0.998$.
Despite the fact that a direct comparison of the finding charts is not
possible, the coordinates are unlikely to be
in error by as much as $5''$ making it is extremely
unlikely that M31N 2011-02b is a recurrence of 1982-09a.

\subsubsection{M31N 1983-09c}

M31N 1983-09c was also discovered as part of the \cite{cia87} nova survey.
A possible recurrence, M31N 1997-11c \citep{rec99}, erupted
$\sim6''$ from the reported position of 1983-09c, and thus just
passed our RN screening criterion.
The novae erupted with $a\sim8'$ from the nucleus, and
probability of a chance positional coincidence is estimated to be
$\sim0.5$.
Since a finding chart for M31N 1983-09c
is not available, we cannot
test whether these novae might be spatially coincident. However,
given that both novae were registered with CCD detectors and
their positions computed directly from the images, it is highly unlikely that
a positional error of $\sim6''$ would be possible if M31N
1997-11c was a recurrence
of 1983-09c.

\subsubsection{M31N 1984-09b}

M31N 1984-09b was another nova found in the survey by \citet{cia87},
and like 1982-09a quite close to the nucleus of the galaxy
($a\simeq1.27'$). A possible recurrence,
M31N 2012-12a, was flagged by our screen. The
latter nova was discovered by \citet{hor12c}, and found to be a member
of the Fe~II spectroscopic class by \citet{sha12b}.
Despite the fact that
M31N 2012-12a has a reported position that is only $1.5''$ from that
of 1984-09b, the probability of a chance positional coincidence
so close to the nucleus of M31 is relatively high ($P_C\simeq0.450$).
Unfortunately, since a
finding chart for M31N 1984-09b no longer exists, once again
we cannot evaluate whether
the objects are spatially coincident.
\citet{wil14} studied the field of M31N 2012-12a and found that there is a
source within $3\sigma$ (1.532 ACS/WFC pixels or 0.077$''$)
of the nova's position. According to \citet{wil14}, the
local stellar density suggests a probability of chance coincidence
with this separation is 19.9\%, and it is unclear whether the source
might be associated with the nova progenitor.

\subsubsection{M31N 1985-09d}

As with two of the three the previous \citet{cia87} novae,
M31N 1985-09d was discovered close to the nucleus of M31
($a\simeq0.81'$). A possible recurrence, M31N 2009-08d, located $\sim4.12''$
away was discovered by K. Hornoch \citep{hen09a}.
Despite the fact that without a finding chart
we cannot definitively rule out the possibility that M31N 1985-09d
is recurrent,
it seems highly unlikely given that the coordinates differ by more than
4$''$ and the probability
of a chance positional coincidence with this separation
at this location in the galaxy is very high ($P_C\simeq0.993$).
We note that another nova,
M31N 2005-05b was also flagged by our screen, and lies 4.6$''$ away
from the nominal position of 2009-08d. As shown below M31N 2005-05b and 2009-08d
are not spatially coincident. Given that the reported
position of M31N 1985-09d is even further from 2005-05b than it is from 2009-08d,
it is even less likely that 2005-05b could be a recurrence of 1985-09d.
Finally, we note that M31N 2009-08d was determined to be an Fe~II nova by
\citet{dim09}, and no evidence for a red giant companion often associated with
RNe was found in the search for M31 nova progenitors by \citet{wil14}.

\subsubsection{M31N 1985-10c}

M31N 1985-10c, another nova discovered by \citet{cia87}, was again found
extremely close to
the nucleus of M31 ($a\simeq0.30'$). Two possible recurrences have been
subsequently noted, M31N 1995-12a and 2003-10b.
In addition to lacking a finding chart for M31N 1985-10c,
we were also unable to locate a chart for M31N 1995-12a despite
re-examining an image from the survey of \citet{sha01} taken
on 1996 January 14, 27 days after the nova was discovered by \citet{ans04}.
A chart does exits for M31N 2003-10b \citep{fia03}, but
given the extremely close proximity to the nucleus, it is unlikely that finding
charts, even if available for all three novae, would be useful
in definitively establishing the spatial coincidence of the objects.
Based on the published coordinates,
M31N 1995-12a and 2003-10b have probabilities of chance coincidence
with 1985-10c of $P_C\simeq0.708$ and $P_C\simeq0.422$, respectively, while
1995-12a and 2003-10b have a probability of chance coincidence
with respect to each other of $P_C\simeq0.953$.
We conclude that there is no compelling evidence that M31N 1985-10c
is a RN.

\subsubsection{M31N 1986-09a}

M31N 1986-09a, the final nova from the \citet{cia87} survey to be discussed,
was also discovered relatively close
to the nucleus of M31 ($a\simeq1.63'$). A possible recurrence
is M31N 2006-09b, which was seen to erupt $\sim4.8''$ away.
Despite the fact that a comparison of the finding charts for these novae
is not possible, as with the previous two systems, it is unlikely that
the coordinates would be in error by as much as 4.8$''$ if the system was a RN.
At this separation, the probability of a chance positional near coincidence,
$P_C\simeq0.994$.

\subsection{Rejected Recurrent Nova Candidates}

\subsubsection{M31N 1909-09b}

\begin{figure}
\includegraphics[angle=0,scale=.28]{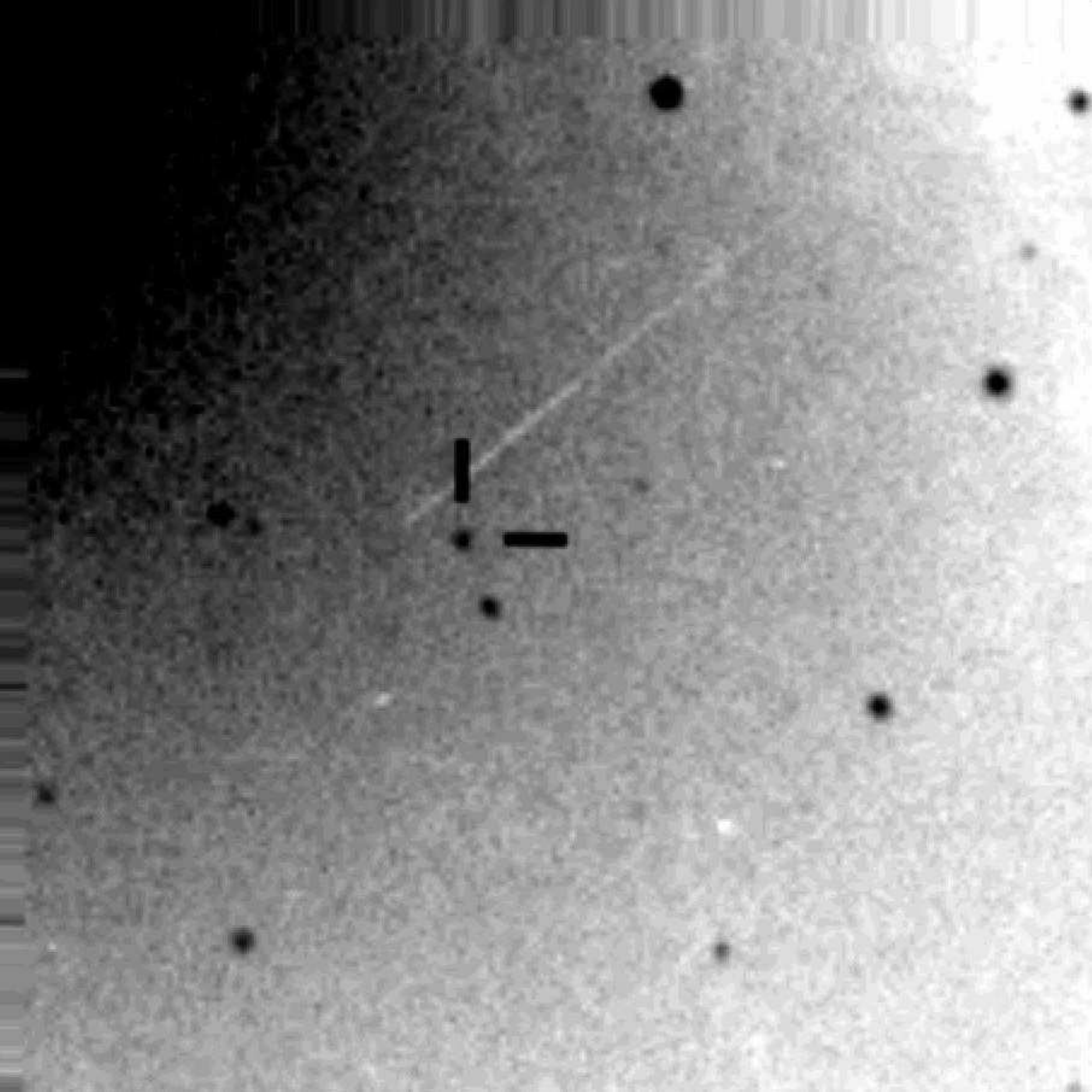}
\includegraphics[angle=0,scale=.28]{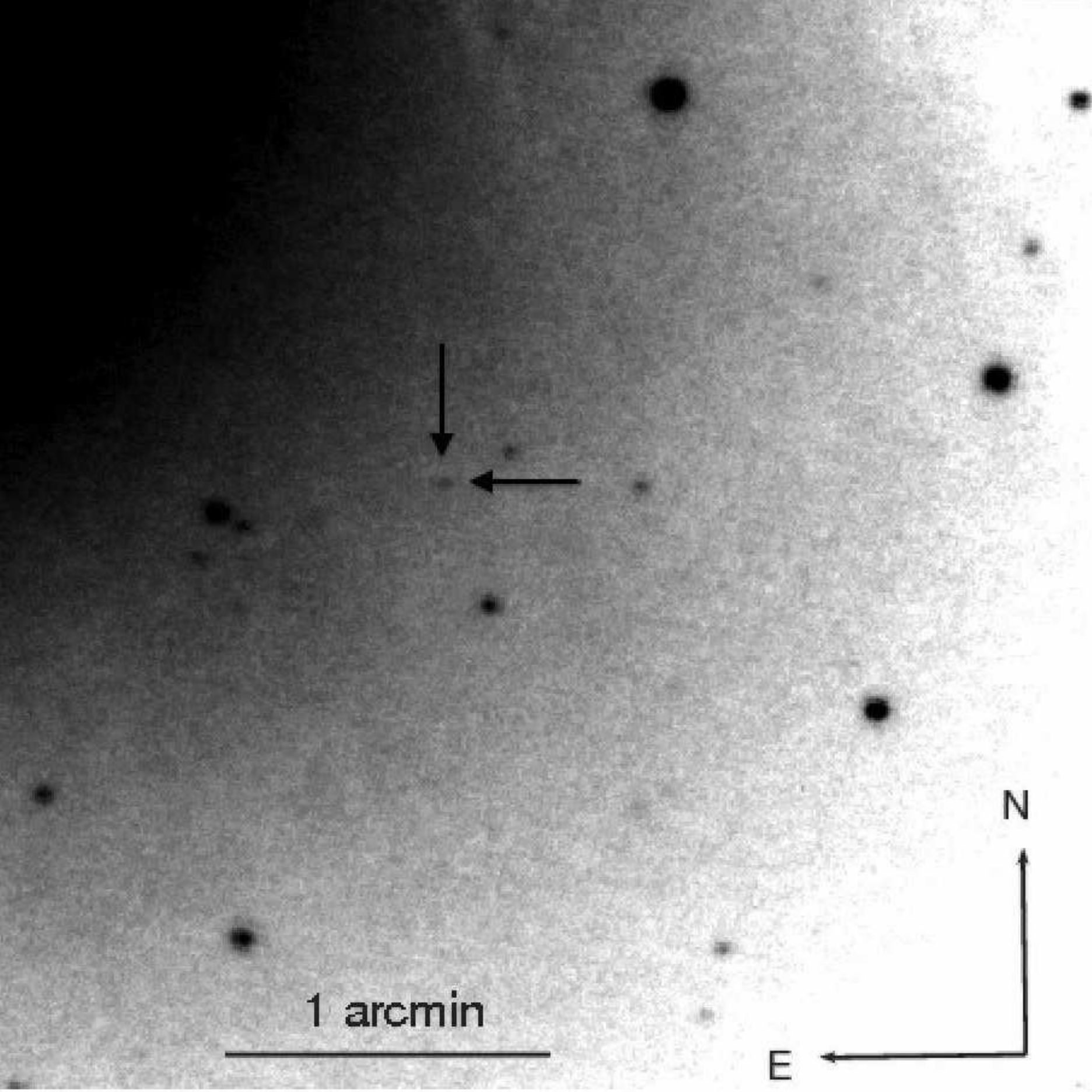}
\includegraphics[angle=0,scale=.28]{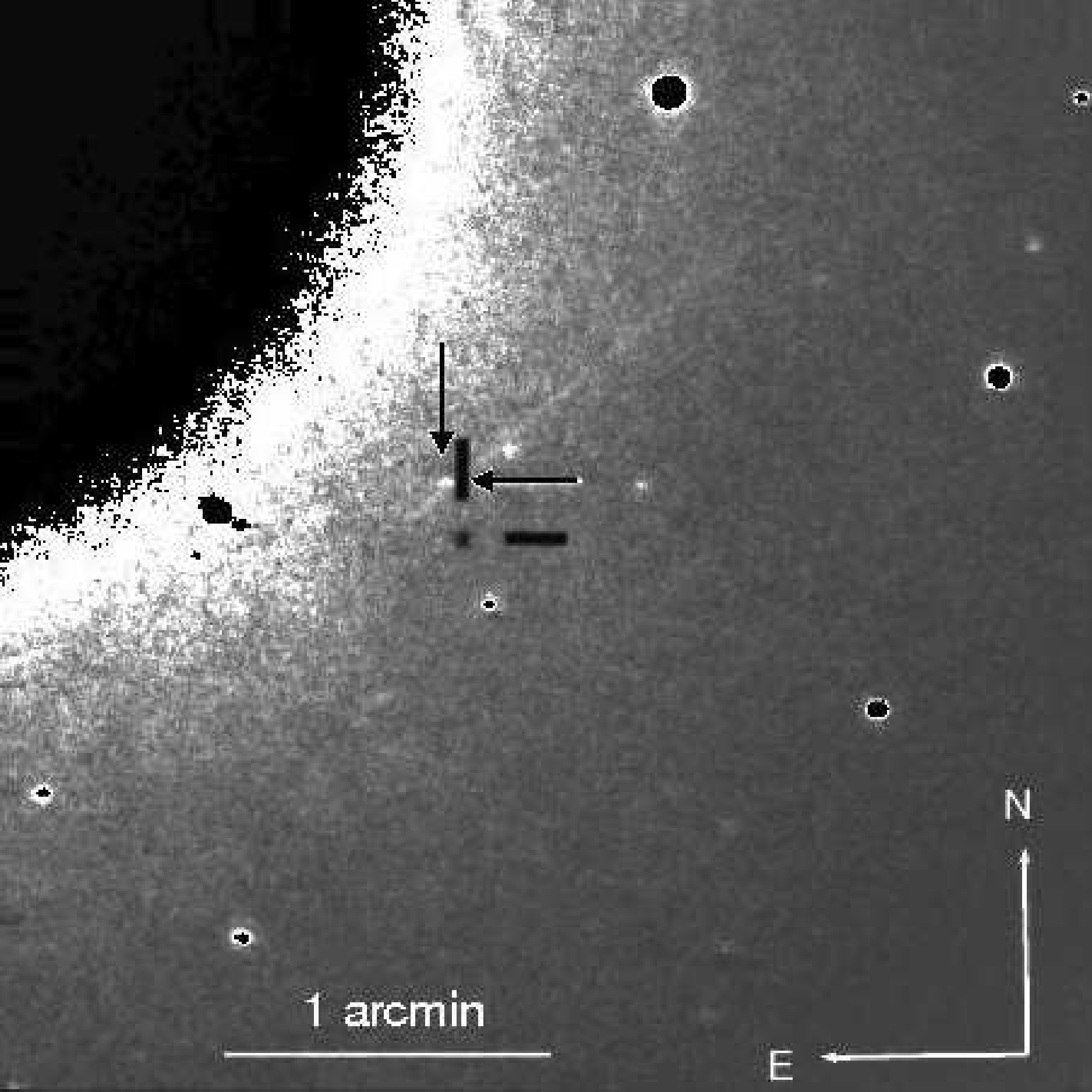}
\caption{Images of M31N 1909-09b, M31N 2009-02b, and their comparison
(left, center, and right, respectively).
The comparison image shows that
the positions of the two novae are clearly not coincident, with
M31N 2009-02b (in white) being located $\sim10.5''$ NNE
of the position of 1909-09b.
The image for M31N 1909-09b is a reproduction of plate S19-Ri from
the Carnegie archives, while that for 2009-02b is from the SuperLOTIS
project \citep{pie09a}.
North is up and East to the left, with a scale of $\sim3.4'$ on a side.
\label{fig19}}
\end{figure}

The relatively recent nova M31N~2009-02b \citep{pie09a}
erupted $\sim4.2''$ SSW of the
nominal position of M31N 1909-09b.
M31N 1909-09b is the second nova listed by \citet{hub29},
and was discovered on the rise on
1909 September 12 on a plate taken by Ritchey
using the Mount Wilson 60-in reflector \citep{rit17}.
The nova reached a maximum brightness
of $m_{pg}=16.7$ on September 15, and faded slowly to $m_{pg}=18.0$
on November 07. The available photometry places a lower limit on the
$t_2$ time of at least 55 days, making M31N 1909-09b a very slow nova.
A comparison of the position of M31N 1909-09b from plate S19-Ri
(1.5 hr exposure) taken by Ritchey
on 13 September 1909 and the position of 2009-02b from the SuperLOTIS
project clearly shows that the two novae are in fact
distinct objects separated by $10.5''$ (see Fig.~\ref{fig19}).
Astrometry of M31N 1909-09b based on a scan
of the Ritchey plate yields the revised coordinates given in
Table~\ref{revcoord}.

\subsubsection{M31N 1918-02b}

\begin{figure}
\includegraphics[angle=0,scale=.28]{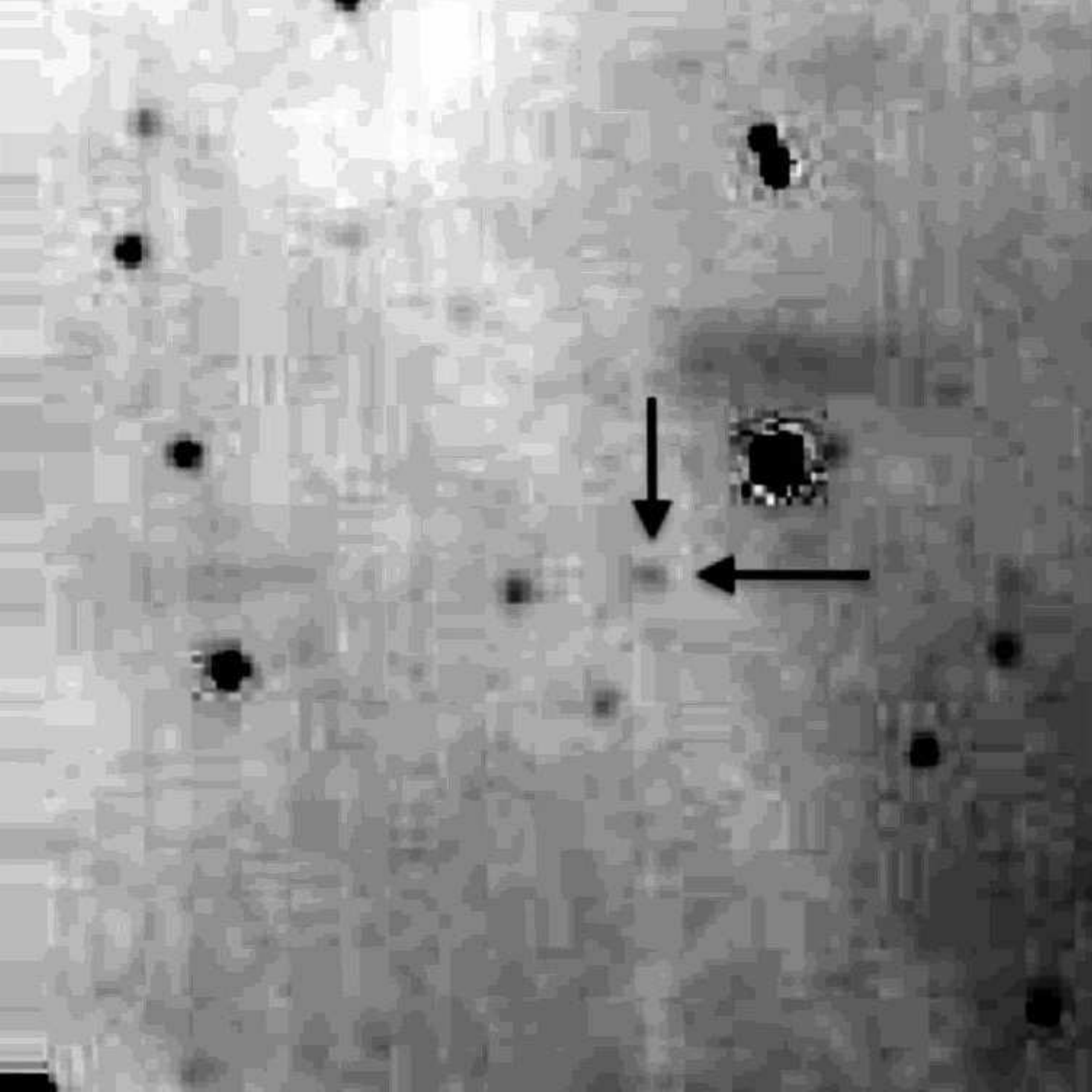}
\includegraphics[angle=0,scale=.28]{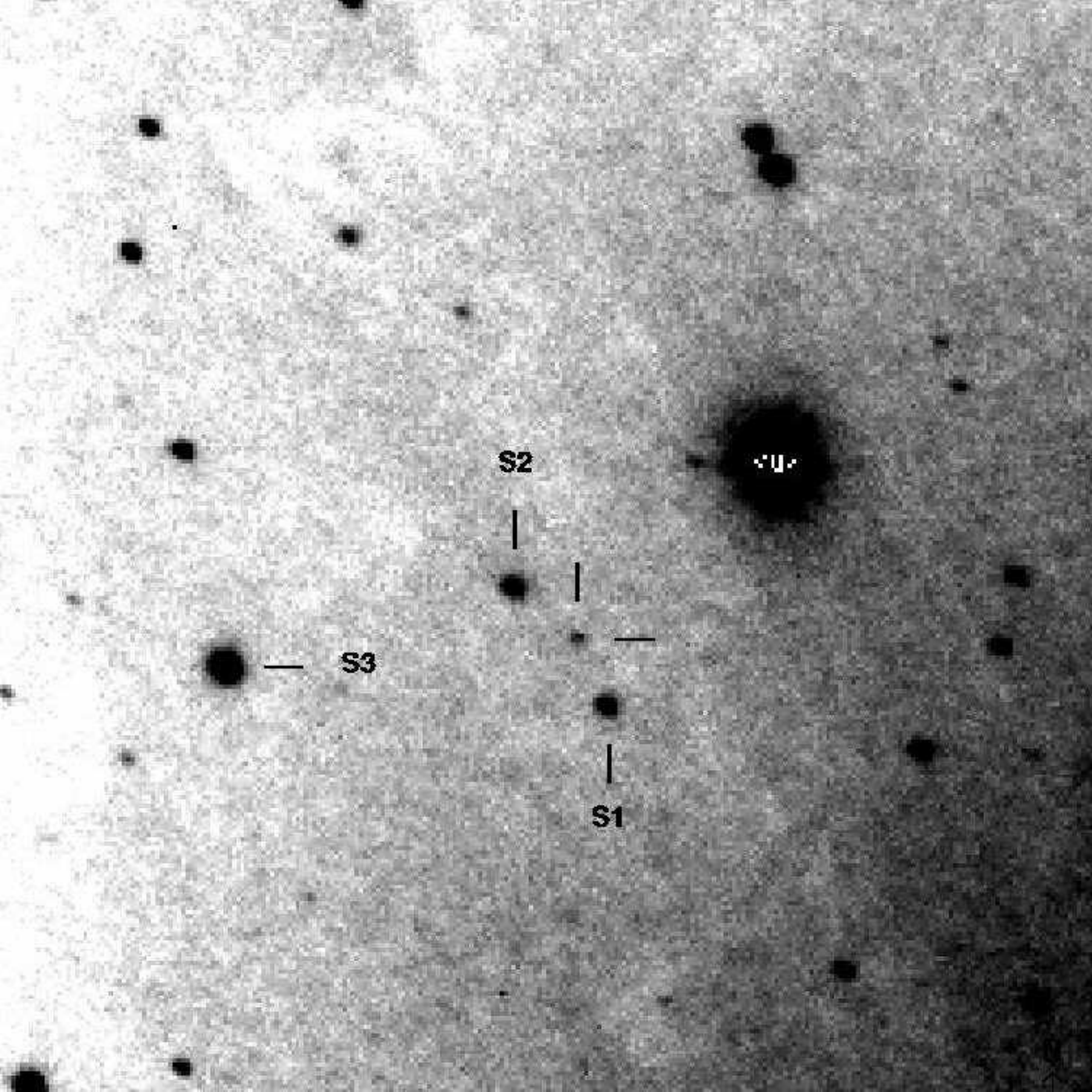}
\includegraphics[angle=0,scale=.28]{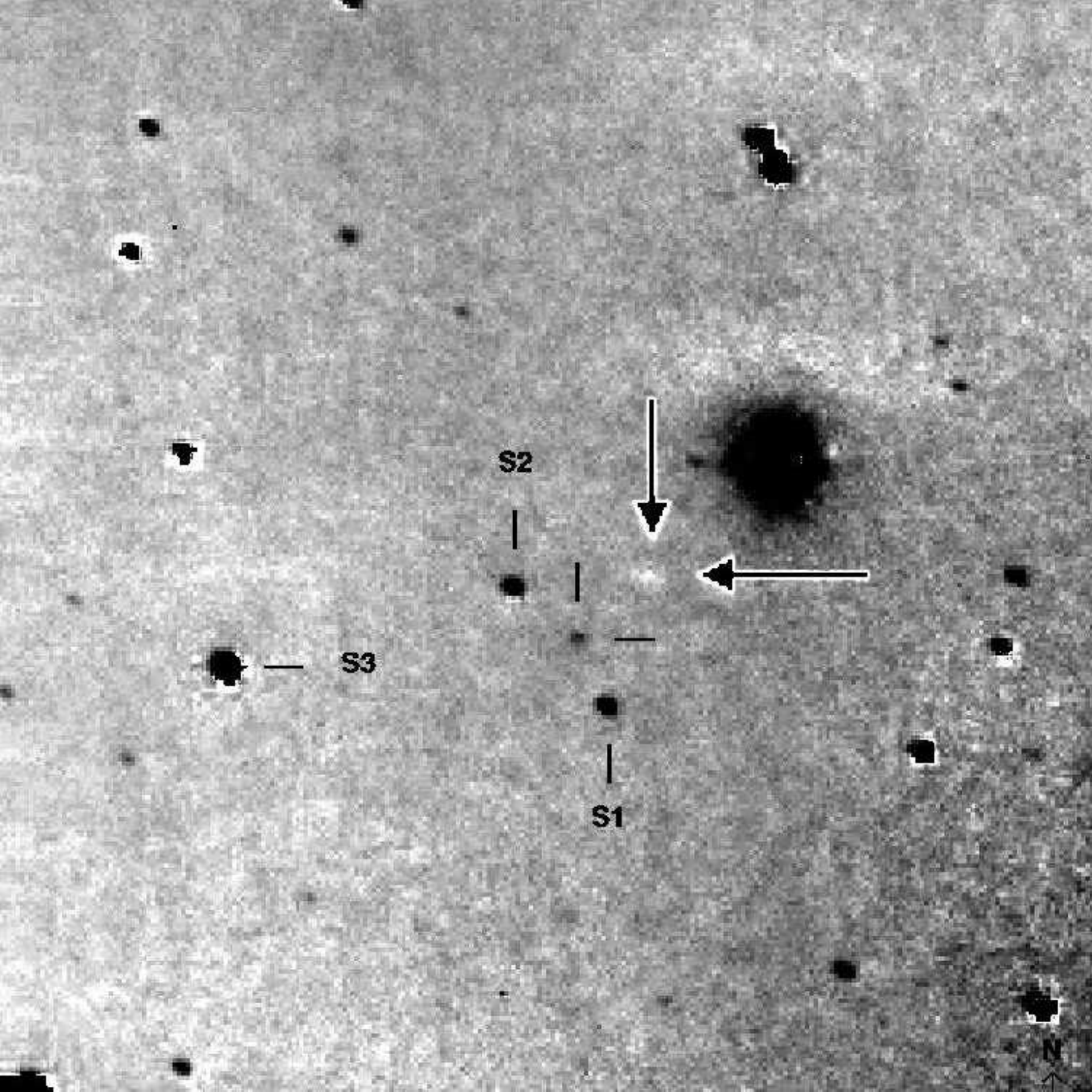}
\caption{Images of M31N 1918-02b, M31N 2013-10g, and their comparison
(left, center, and right, respectively).
The comparison image reveals that
the two novae are clearly not spatially coincident, with
M31N 1918-02b (in white)
being located $\sim18''$ NW of the position of 2013-10g.
The image for M31N 1918-02b is a reproduction of plate S162-Ri from the
Carnegie archives, while that of 2013-10g is from the Lick Observatory
Supernova Search.
North is up and East to the left, with a scale of $\sim3.5'$ on a side.
\label{fig20}}
\end{figure}

Our screen has revealed that
the recent Fe~II nova M31N 2013-10g, which reached a peak brightness
of $R=17.5$ \citep{fab13},
erupted within $\sim12''$
of nova \#9 in Hubble's survey \citep{hub29}. The latter nova was
discovered on 1918 February 10 UT (plate S162-Ri), and reached a similar peak
brightness ($m_{pg}=17.5$).
Despite the relatively large nominal separation,
given the uncertainty in the coordinates
derived from these early photographic surveys, we took a closer look
at the positions of these two novae. As can be seen
in Figure~\ref{fig20}, the novae clearly represent
eruptions from distinct progenitors,
and we conclude that M31N 2013-10g is not a recurrence of 1918-02b.
Updated astrometry for M31N 1918-02b yields the revised
coordinates given in Table~\ref{revcoord}.

\subsubsection{M31N 1923-02a}

\begin{figure}
\includegraphics[angle=0,scale=.28]{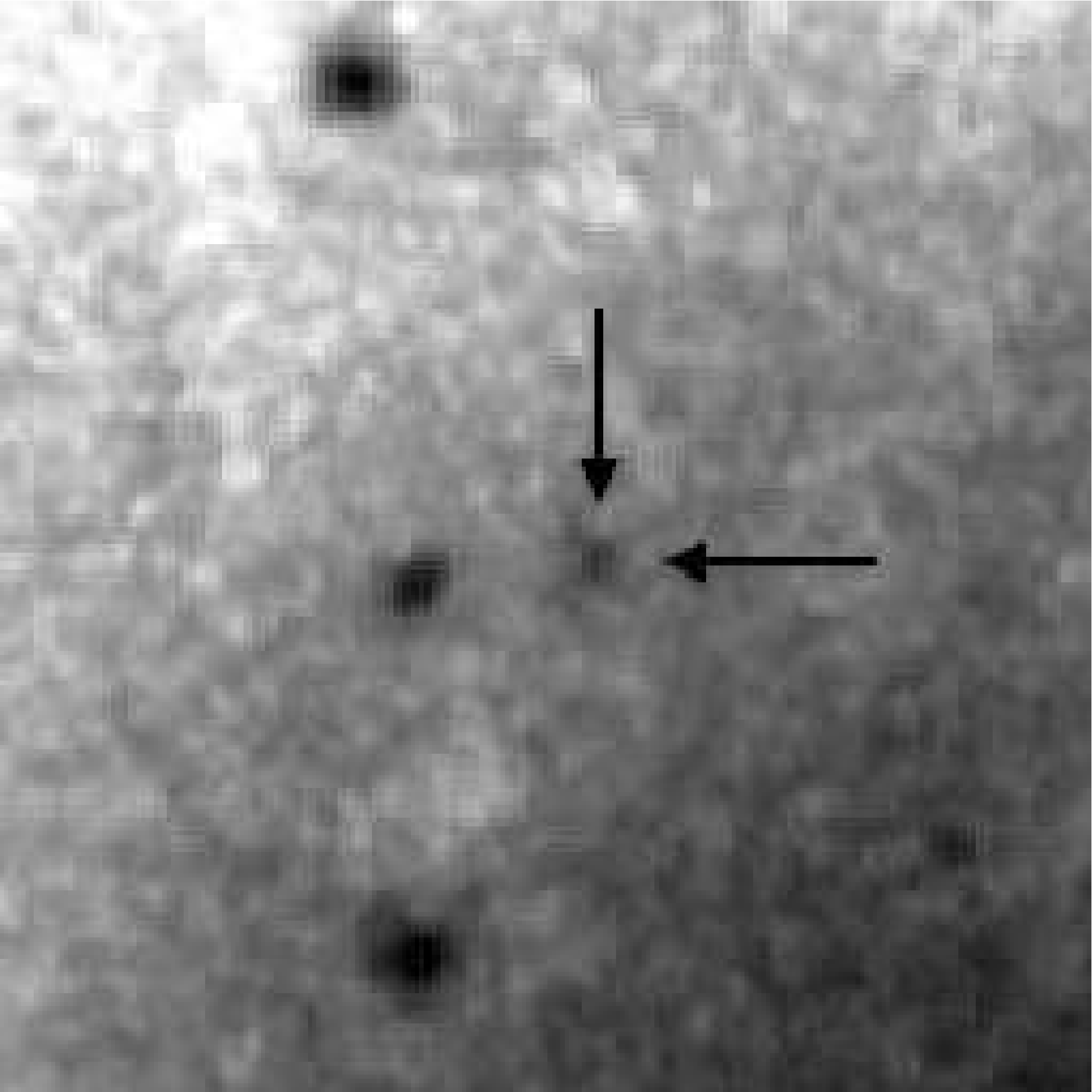}
\includegraphics[angle=0,scale=.28]{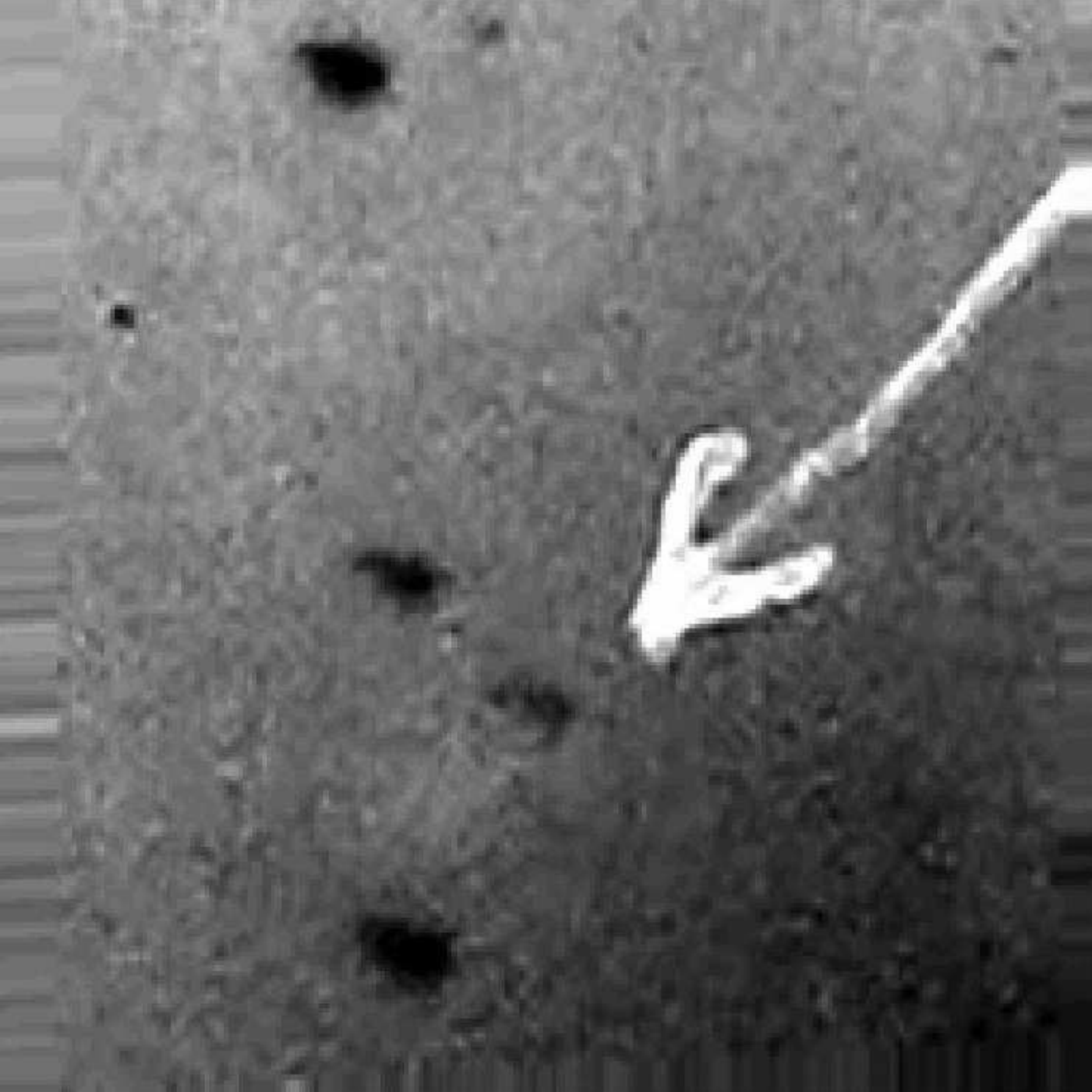}
\includegraphics[angle=0,scale=.28]{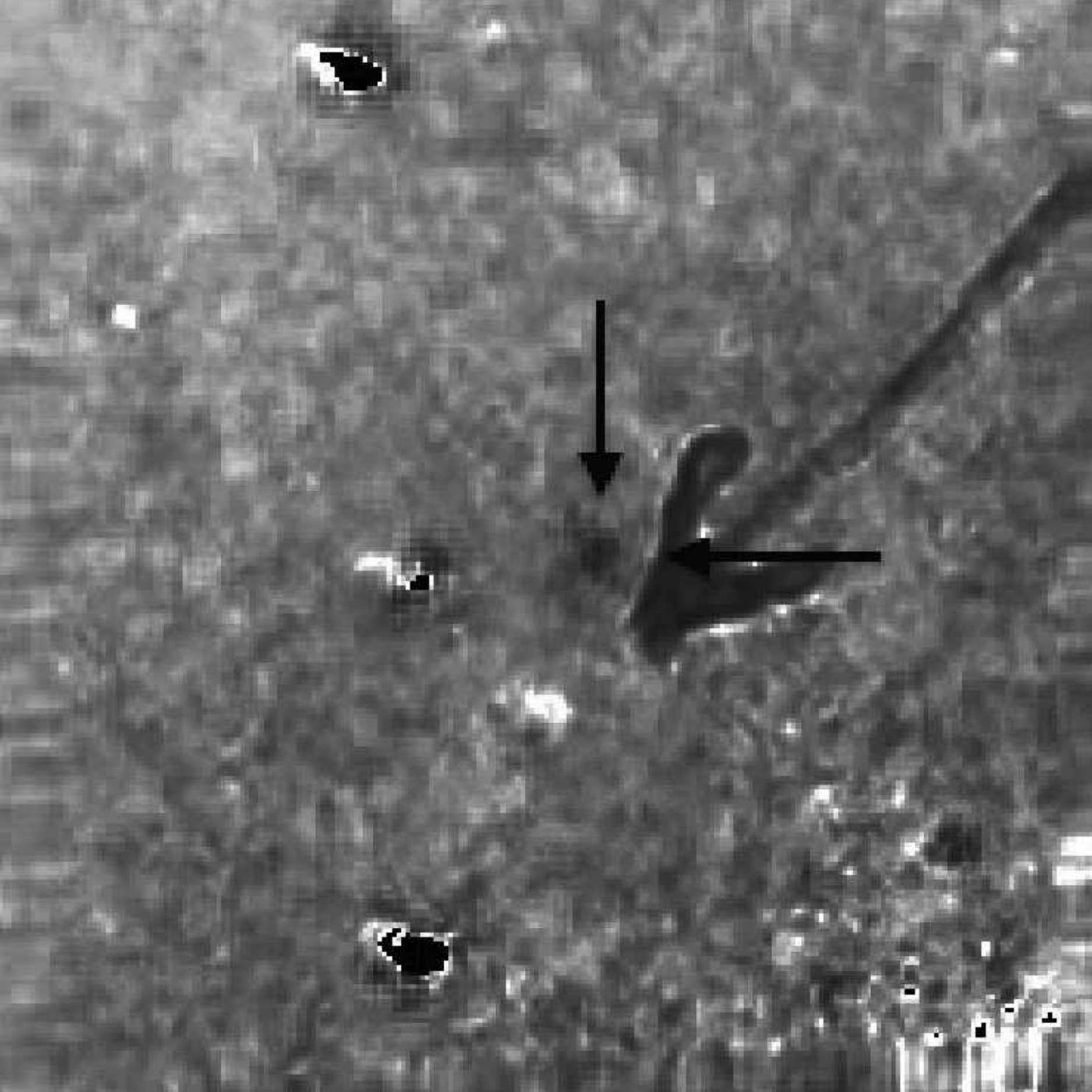}

\includegraphics[angle=0,scale=.28]{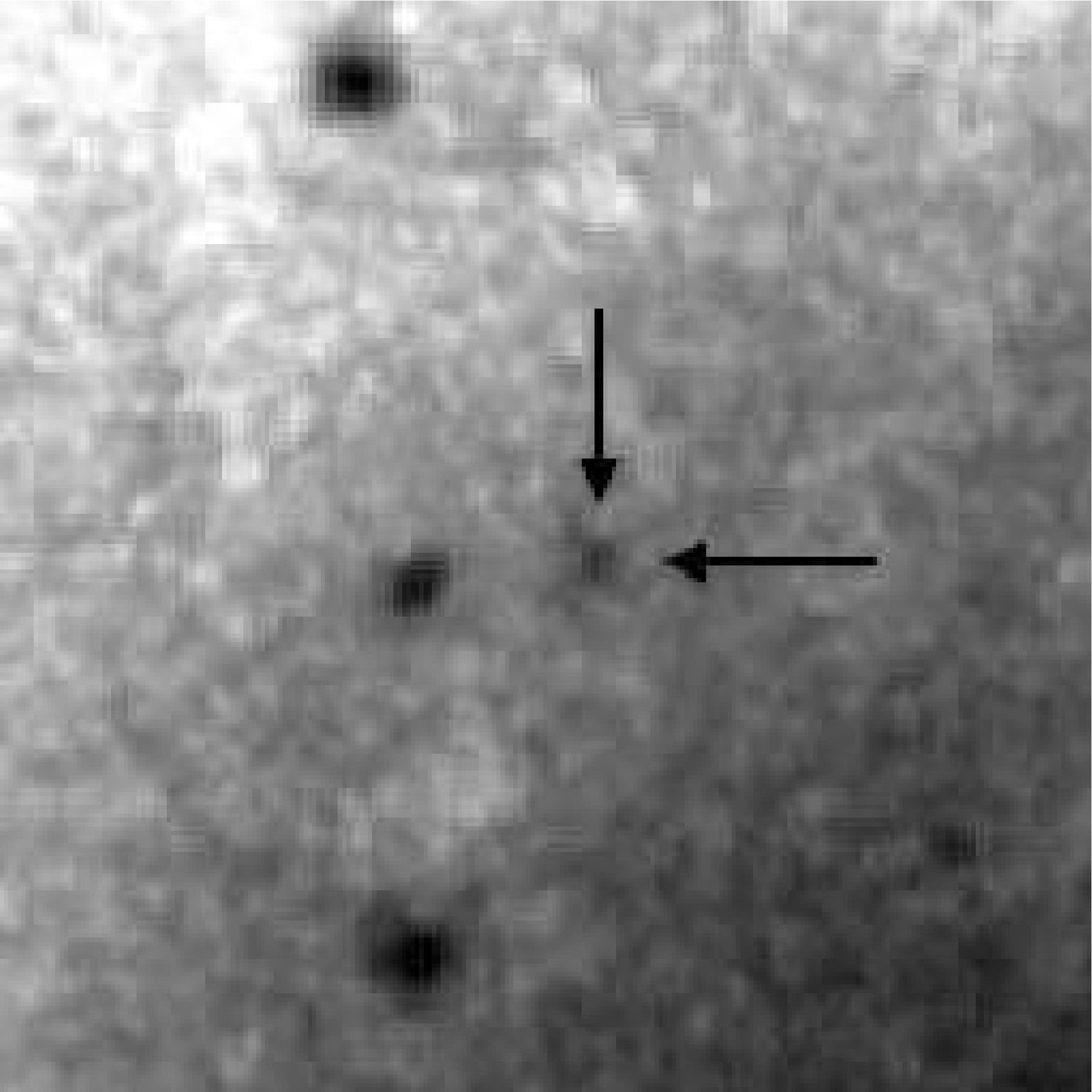}
\includegraphics[angle=0,scale=.28]{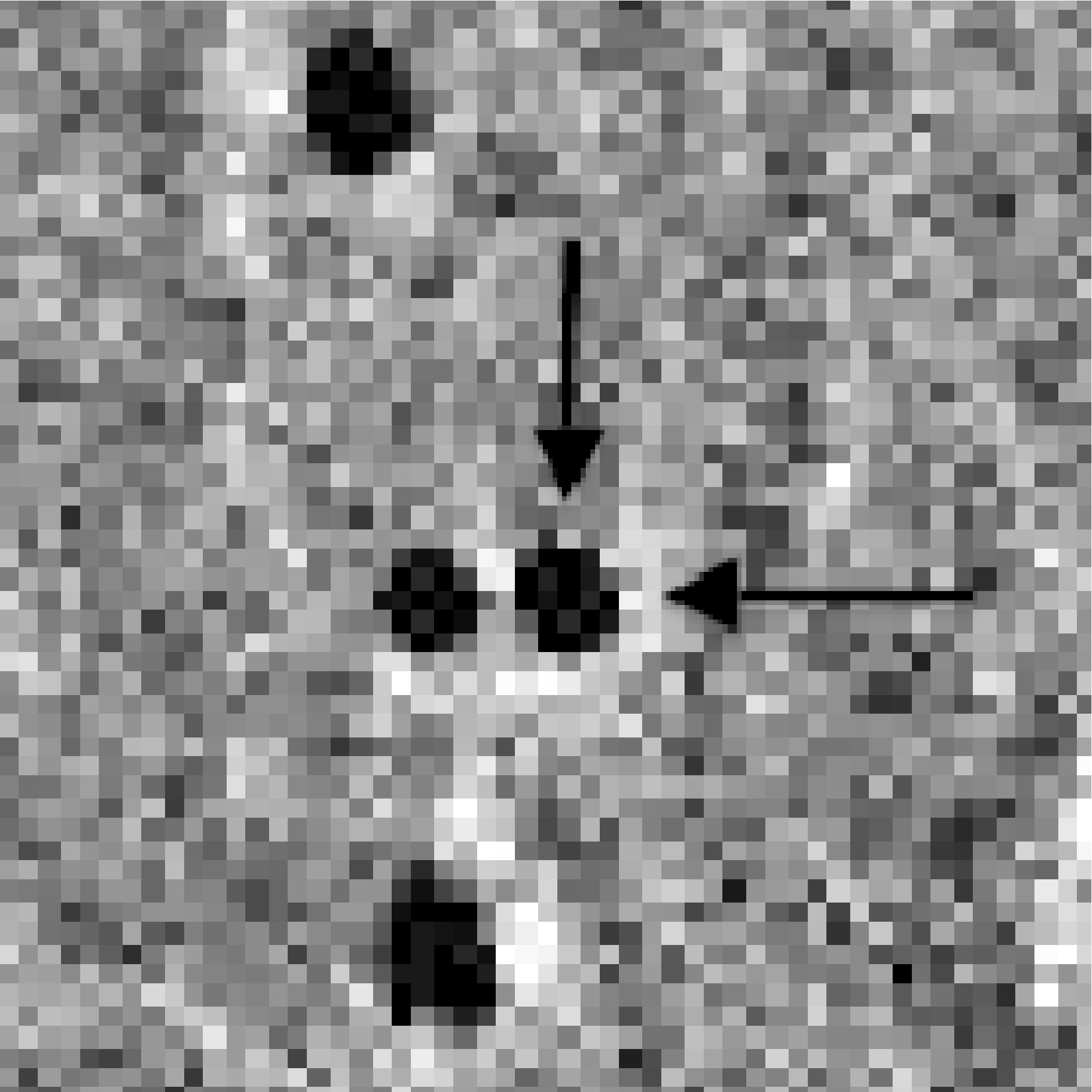}
\includegraphics[angle=0,scale=.28]{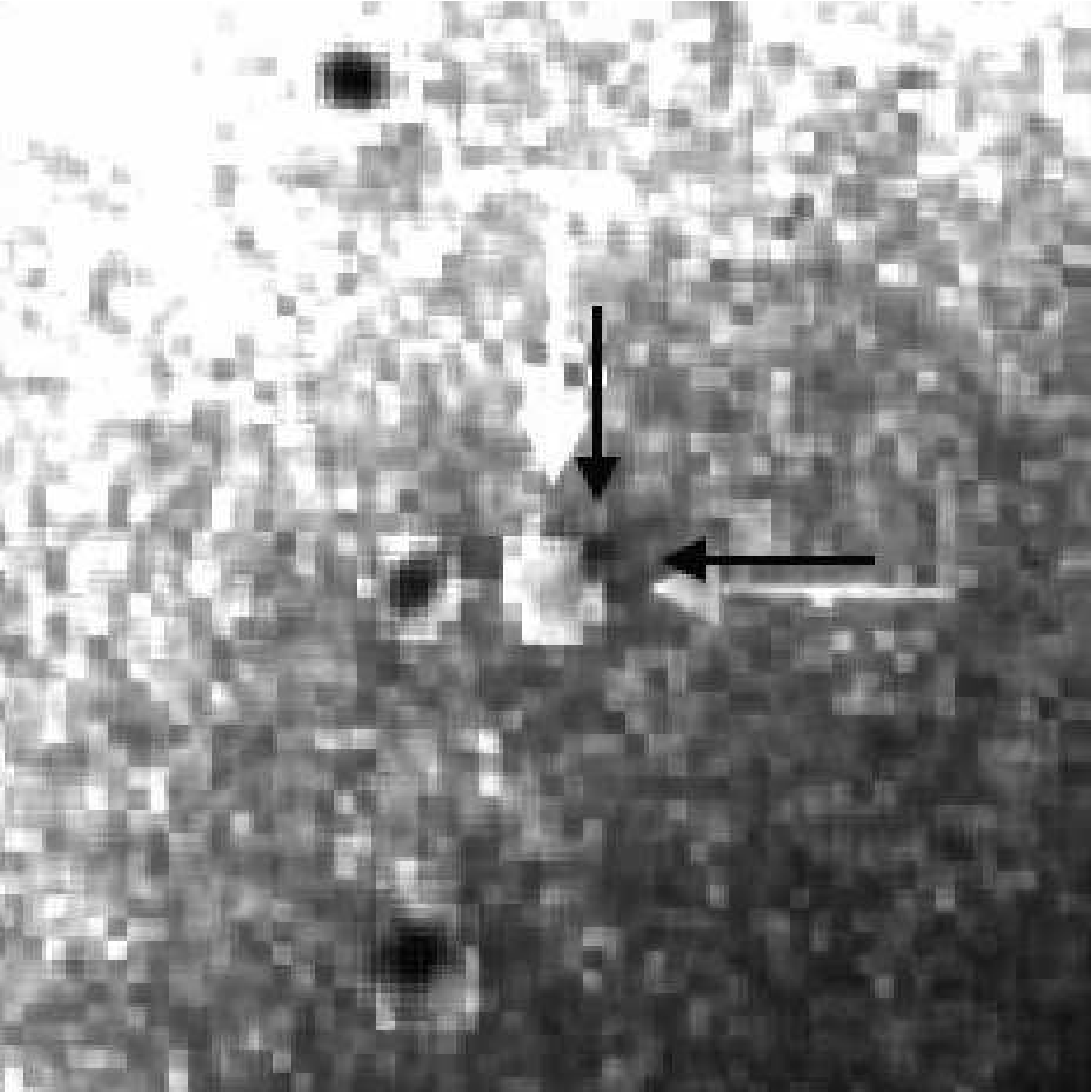}

\includegraphics[angle=0,scale=.28]{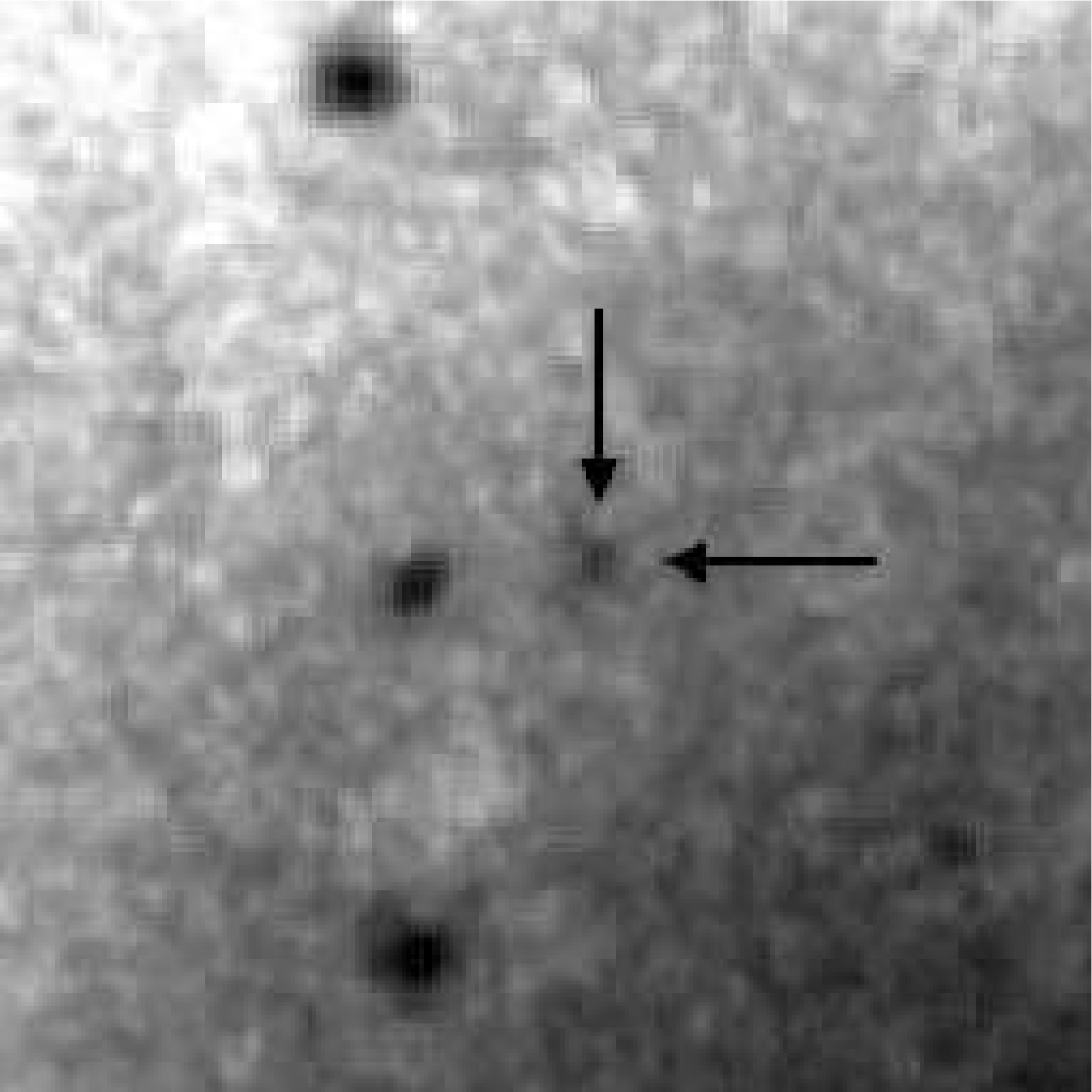}
\includegraphics[angle=0,scale=.28]{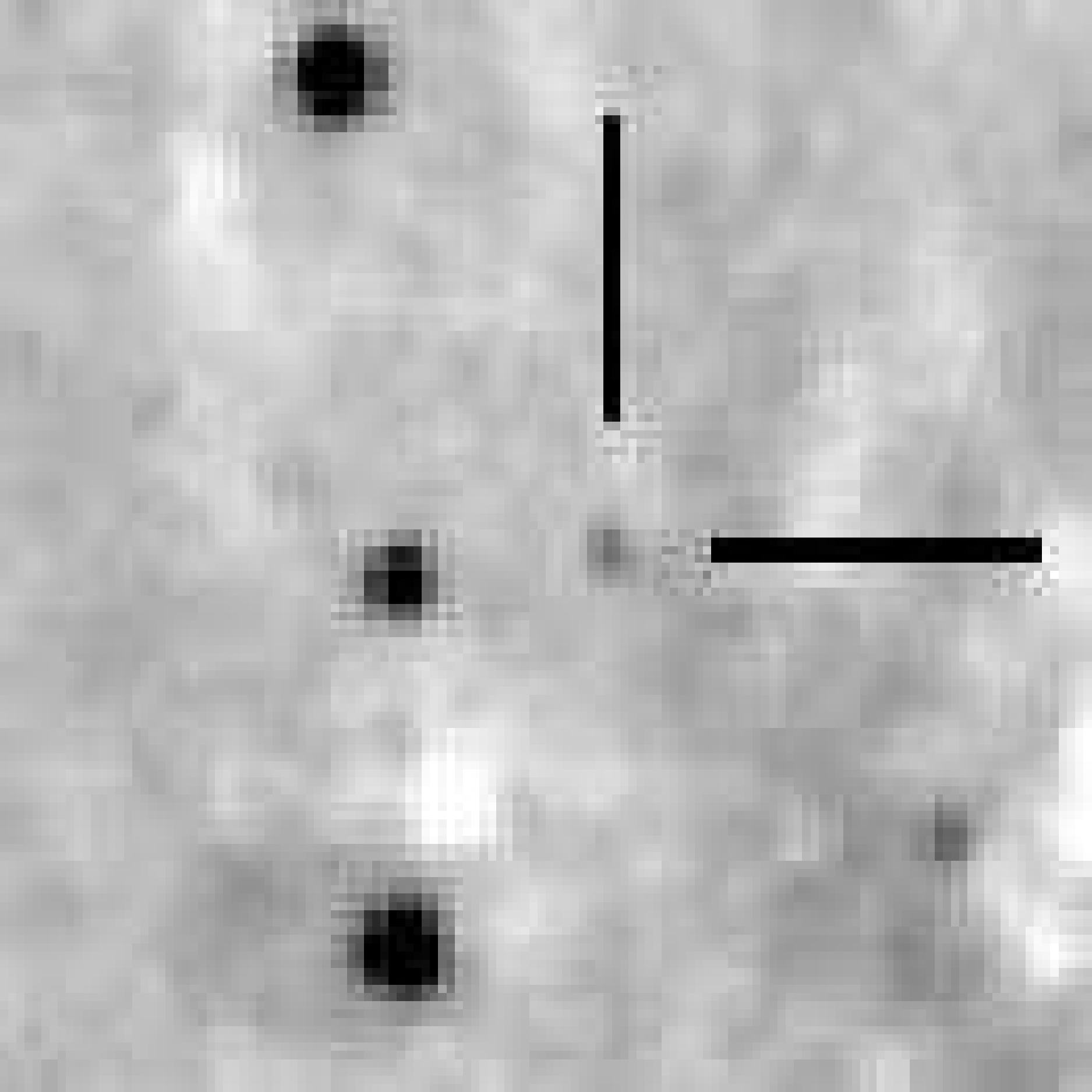}
\includegraphics[angle=0,scale=.28]{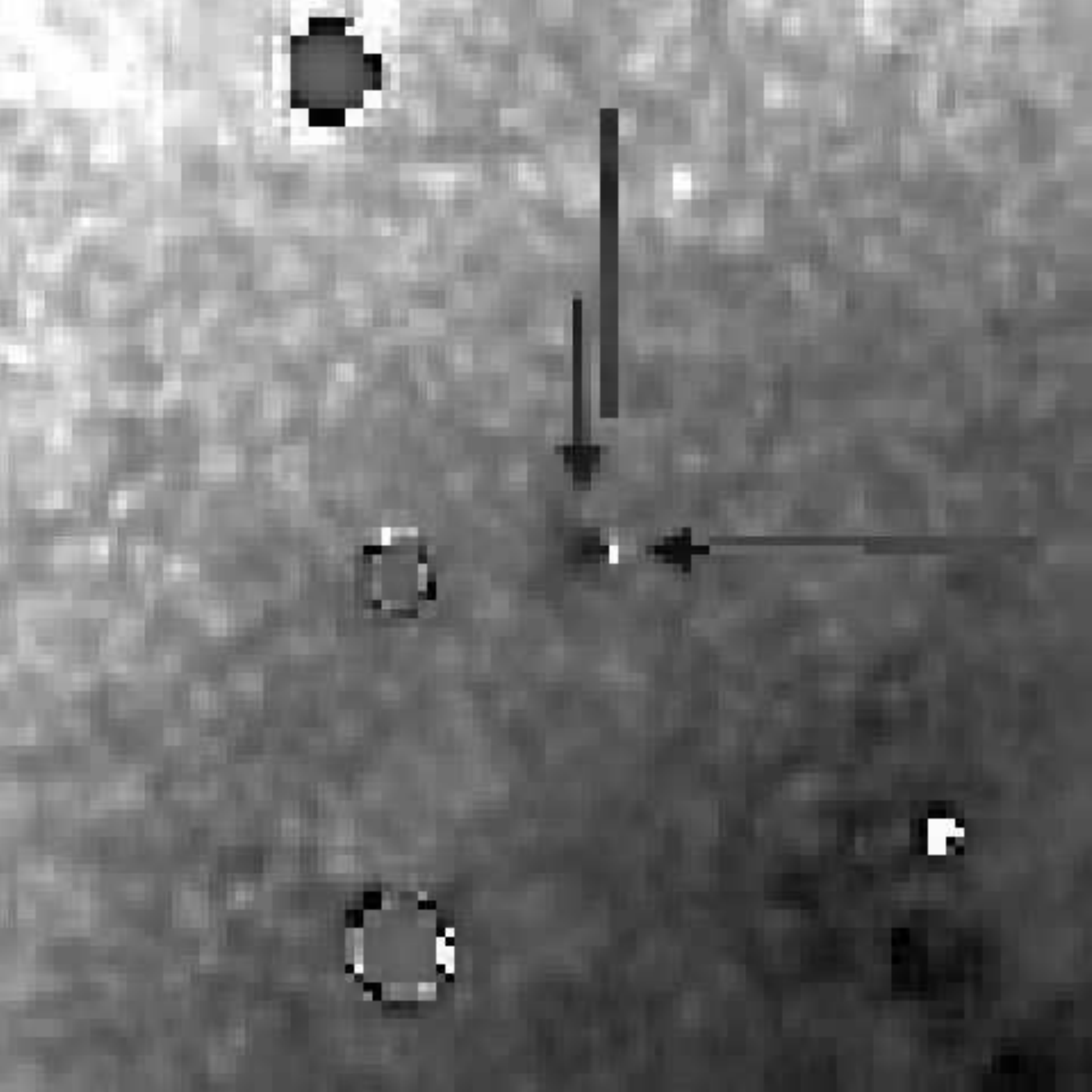}
\caption{Images of M31N 1967-12a, 1923-02a, and their comparison
(top left, center, and right),
M31N 1967-12a, 1993-11c and their comparison
(middle left, center, and right), and
M31N 1967-12a, 2013-08b, and their comparison
(bottom left, center, and right, respectively).
M31N 1923-02a, 1967-12a, 1993-11c and 2013-08b are clearly not coincident
with one another;
however, M31N 1967-12a and 2013-08b appear quite close. Despite their proximity,
precise astrometry and the comparison image
reveal that the novae are not coincident, with
M31N 2013-08b (in white) being $\sim2''$ West of the position of 1967-12a.
The image for M31N1923-02a was produced from a photograph of plate A47
in the Carnegie archives taken on 1923 Feb 15.
The charts for M31N 1967-12a and 1993-11c are taken from unpublished
data from the surveys of \citet{ros73} and \citet{sha01}, with the
chart for 2013-08b taken from \citet{hor13}.
North is up and East to the left, with a scale of $\sim1'$ on a side.
\label{fig21}}
\end{figure}

M31N 1923-02a, nova \#22 in the survey of \citet{hub29},
erupted close to the nucleus of M31 ($a\simeq1.74'$).
Since
then, there have been three potential recurrences
M31N 1967-12a \citep{ros73}, 1993-11c \citep{sha01} and
2013-08b \citep{hor13}. Finding charts for the novae are shown
in Figure~\ref{fig21}.
It is clear that M31N 1923-02a, 1967-12a and
1993-11c are different novae. It is less
obvious that M31N 2013-08b is distinct from 1967-12a, but the comparison
image clearly shows that the novae are not spatially coincident. Revised
astrometry of M31N 1923-02a and
1967-12a are given in Table~\ref{revcoord}, confirming that
all four novae are distinct objects.

\subsubsection{M31N 1924-02b}

M31N 1924-02b was flagged as a potential RN candidate because
nova 1995-09d was observed to lie $5.2''$ from its reported position.
M31N 1924-02b was a faint ($m_{pg}=19$) transient discovered
on 1924 February 03 UT by \citet{hub29} who noted that the
object varied between $m_{pg}=19.0$ and 19.5 for several years.
Given its long-term photometric behavior,
it seems clear that M31N 1924-02b is not a nova in M31, but likely a fainter
long-period variable.

\subsubsection{M31N 1924-08a}

\begin{figure}
\includegraphics[angle=0,scale=.28]{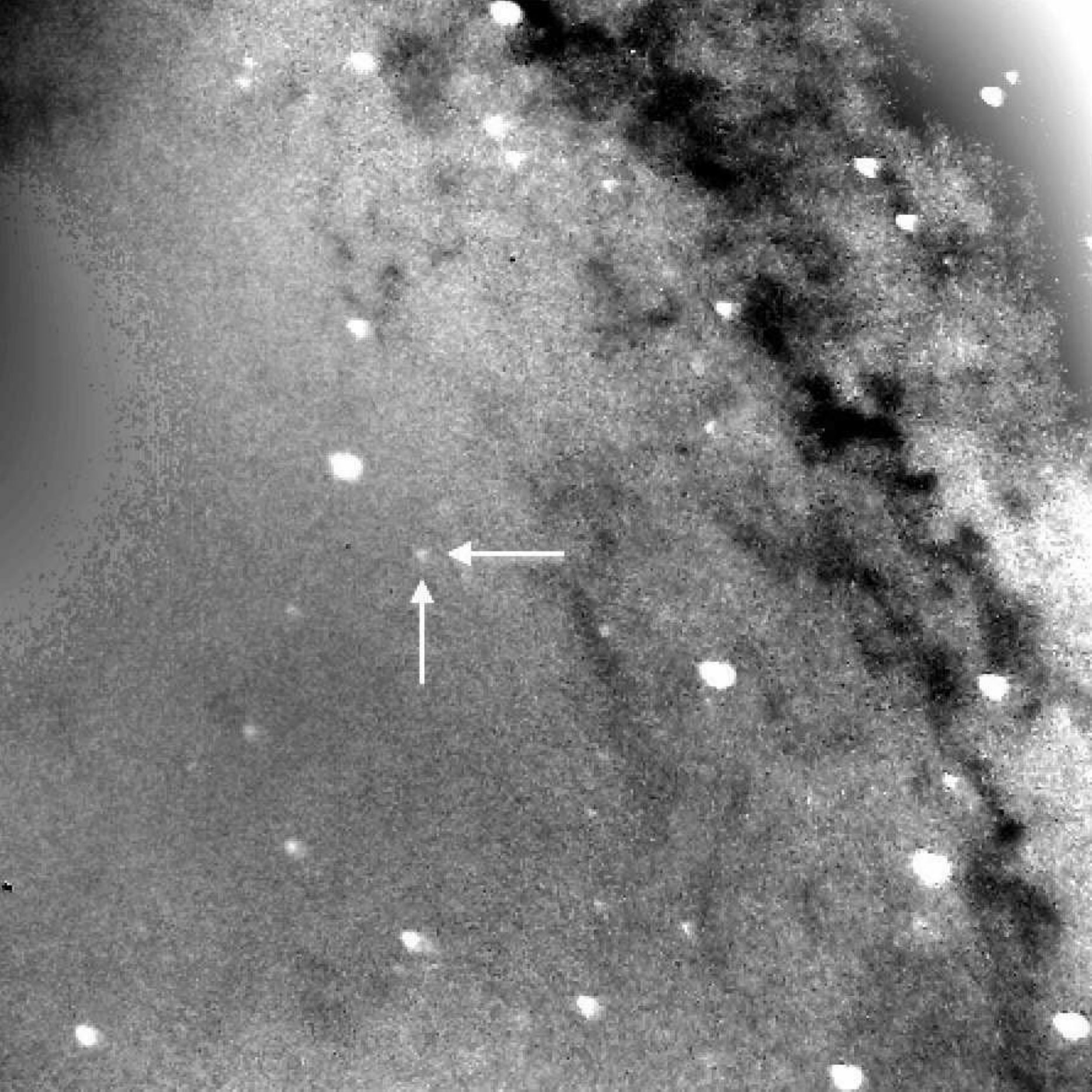}
\includegraphics[angle=0,scale=.28]{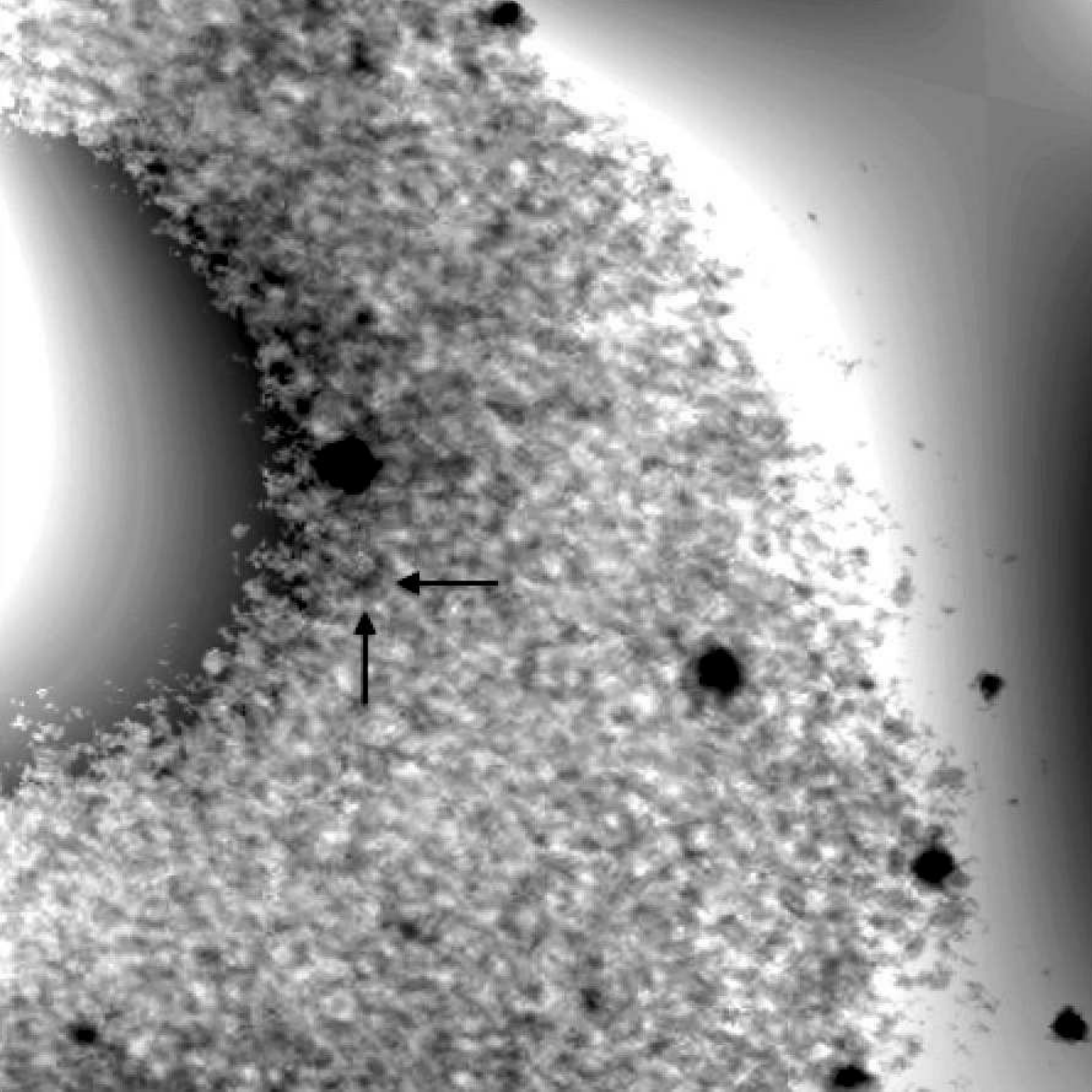}
\includegraphics[angle=0,scale=.28]{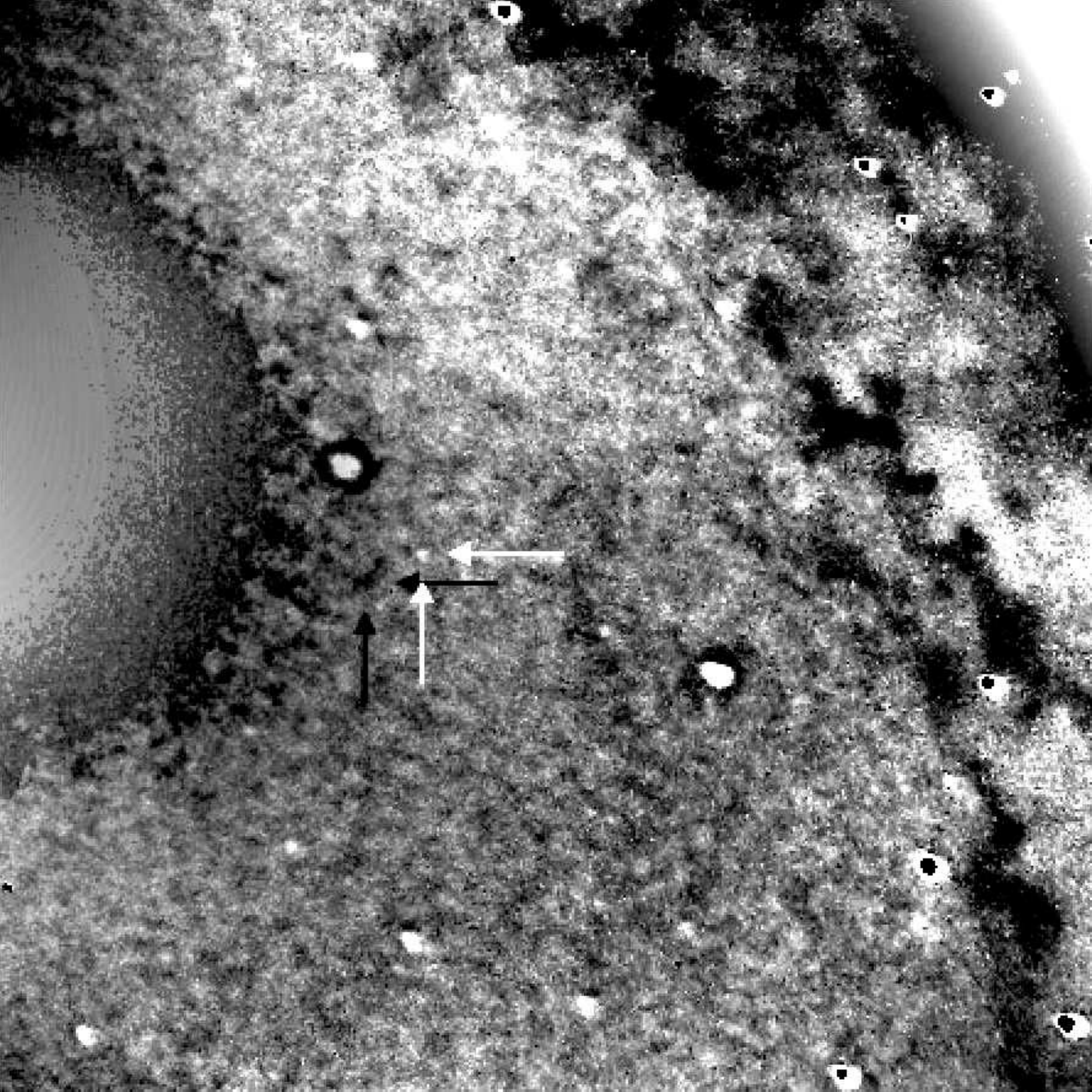}
\caption{Images of M31N 1924-08a, M31N 1987-12a, and their comparison
(left, center, and right, respectively).
The comparison image reveals that
the two novae are clearly not spatially coincident, with
M31N 1987-12a (in black) being $\sim20''$ SE of the position of 1924-08a.
The image for M31N 1924-08a is a reproduction from plate H246D from the
Carnegie archives, while that of 1987-12a is from a 35mm negative
obtained by one of us (JB) on 1987 Dec. 20.13 UT \citep{bry87}.
North is up and East to the left, with a scale of $\sim5'$ on a side.
\label{fig22}}
\end{figure}

M31N 1987-12a was discovered at $m_{pg} = 16.8$ on 1987 December 20.13 UT
\citep{bry87}. Our screen revealed that the nova erupted
$3.1''$ from the reported position of M31N 1924-08a. At an
isophotal radius of just $a\sim2.61'$, the probability of a chance coincidence
is quite high ($P_C\simeq0.900$). Fortunately, we were
able to recover both the original 35-mm negative from the 1987 observation
as well as the original Mt Wilson plates, H241D and H246D, from 1924
August 26 and 28, respectively.
Figure~\ref{fig22} shows the field of M31N 1924-08a taken from plate H246D
compared with the position of 1987-12a from a 35mm photographic negative
taken on 1987 December 20.13 UT.
The two novae are clearly not spatially coincident. Updated astrometry
has been performed, and
revised coordinates measured, for both M31N 1924-08a and 1987-12a,
and are given in Table~\ref{revcoord}.

\subsubsection{M31N 1925-07c}

\begin{figure}
\includegraphics[angle=0,scale=.28]{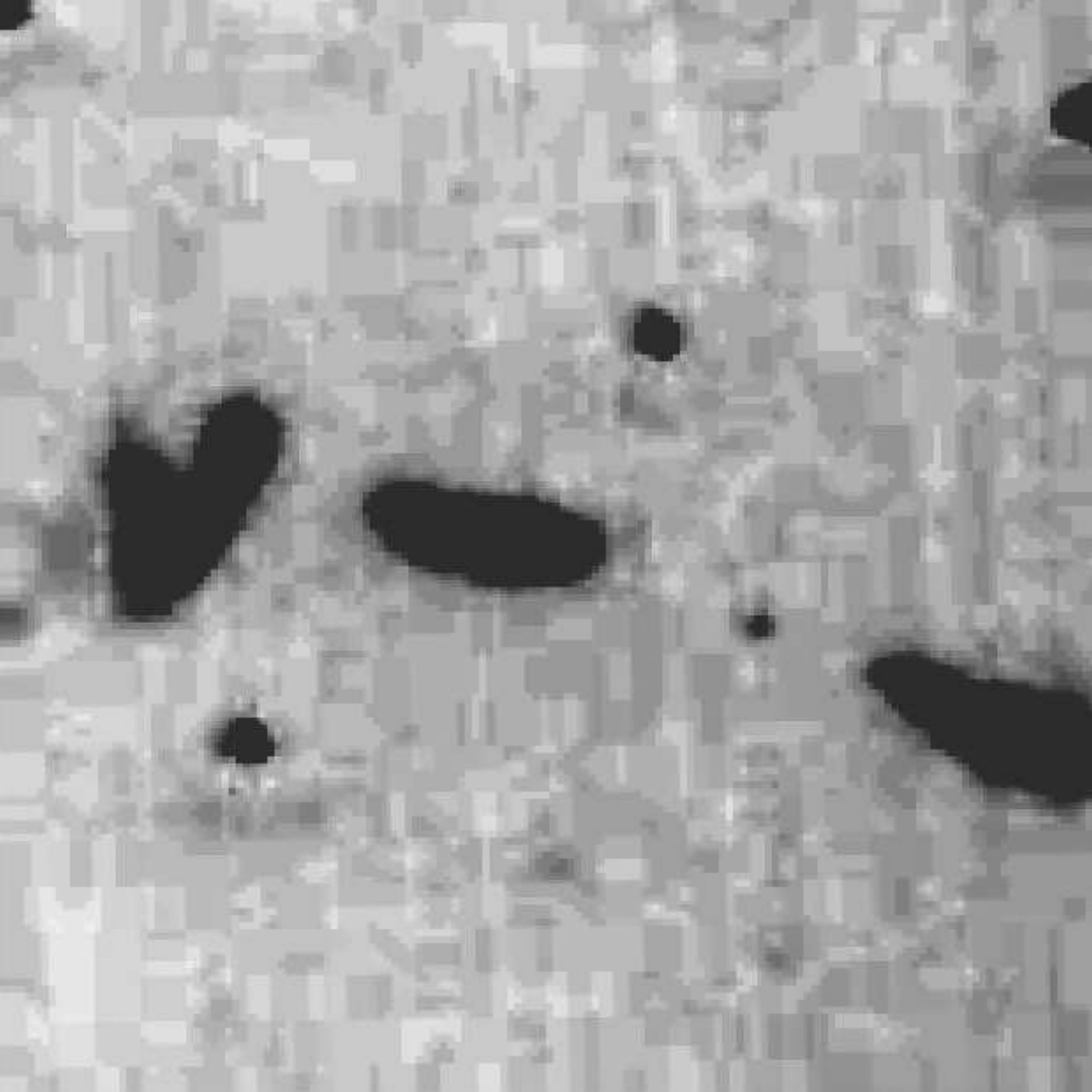}
\includegraphics[angle=0,scale=.28]{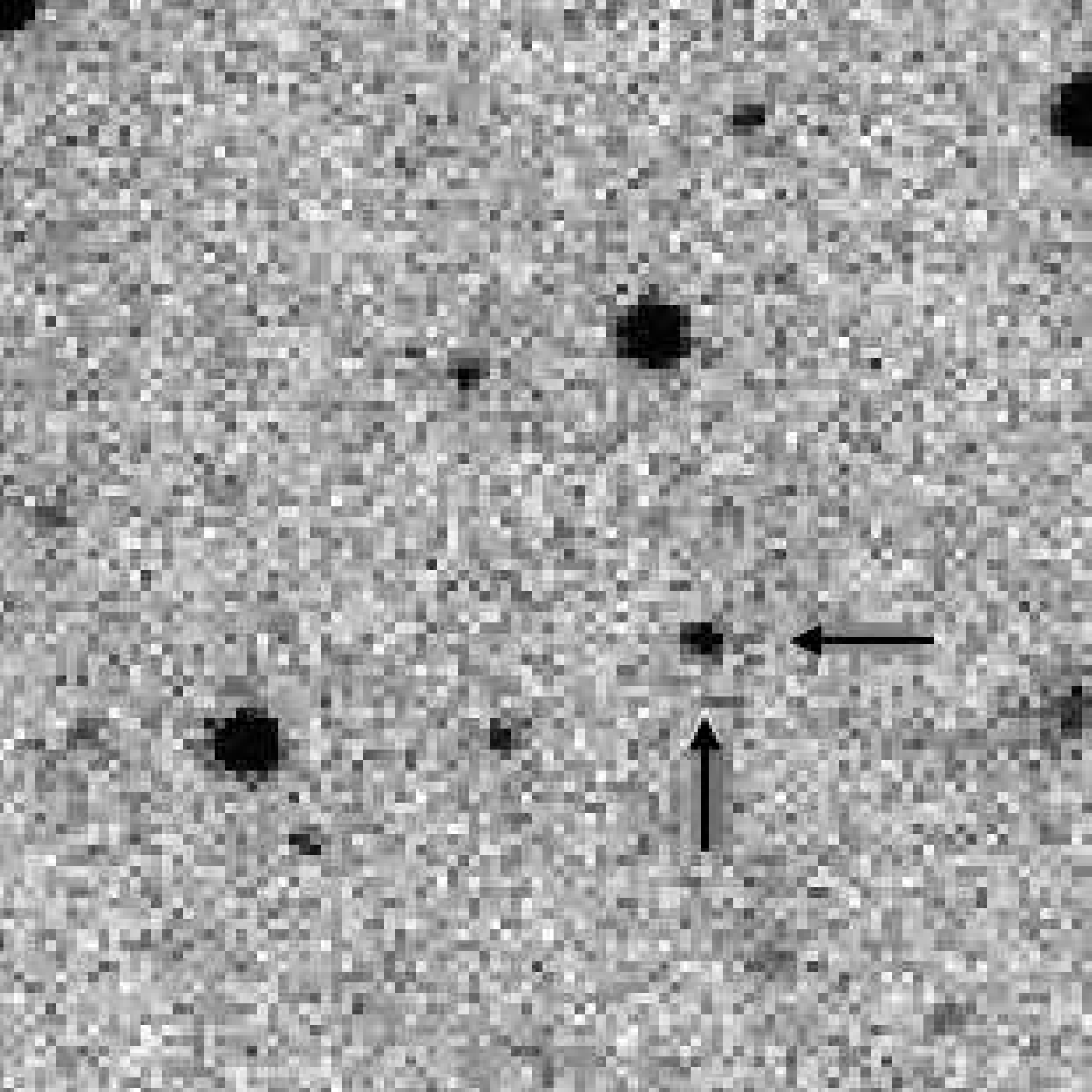}
\includegraphics[angle=0,scale=.28]{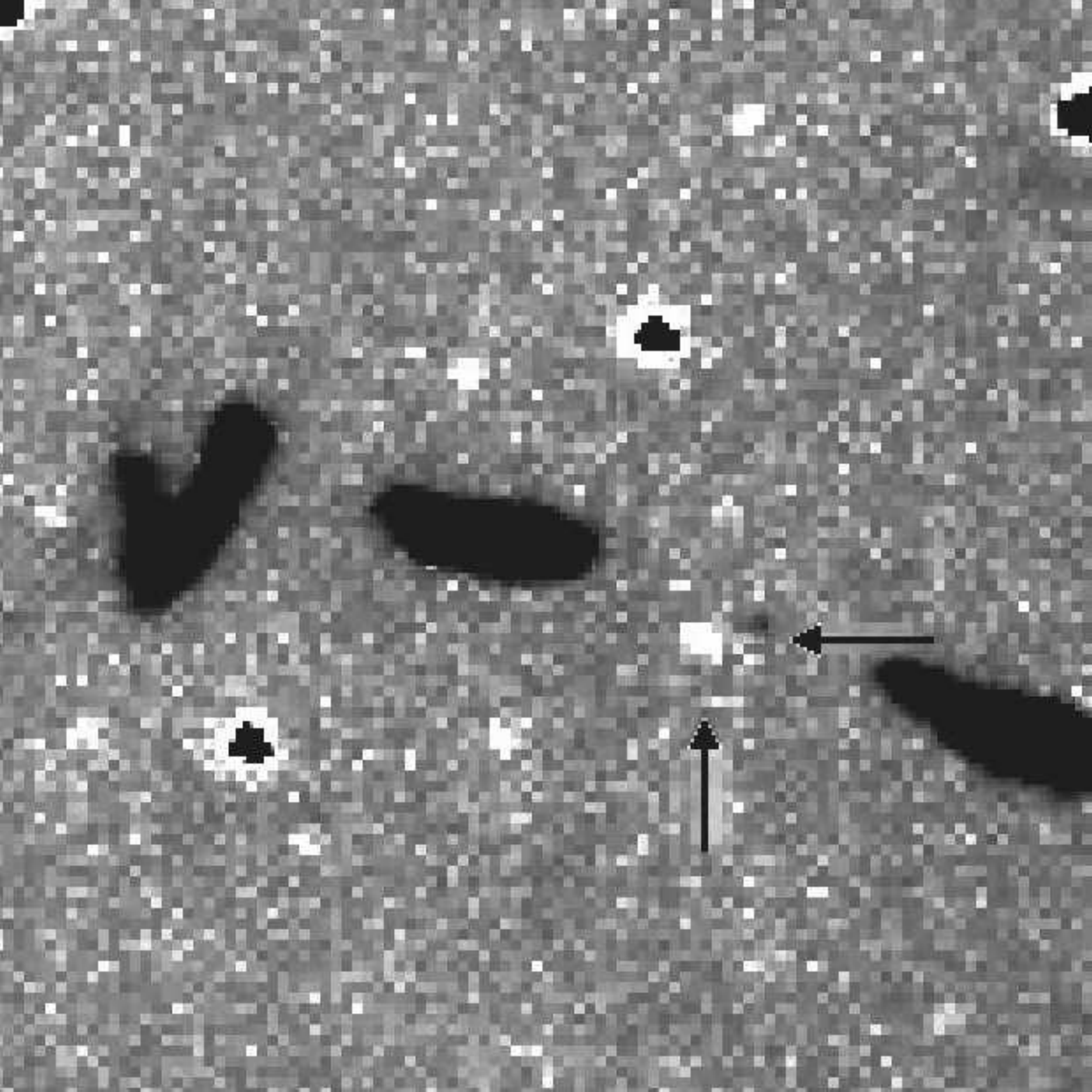}
\caption{Images of M31N 1925-07c, 2011-12b, and their comparison
(left, center, and right, respectively).
As is clear from the comparison image,
the recent outburst M31N 2011-12b (in white) erupted
$\sim6.6''$ ESE of the position of 1925-07c, and thus is not a recurrence.
The image of M31N 1925-07c has been reproduced from plate H580H in
the Carnegie archives, while the image for 2011-12b is courtesy
of K. Nishiyama of the Miyaki-Argenteus Observatory.
North is up and East to the left, with a scale of $\sim2.2'$ on a side.
\label{fig23}}
\end{figure}

M31N 2011-12b (PNV J00435583+4121265), which was discovered by
K. Nishiyama and F. Kabashima on 2011 December 27.448 UT
at $m = 17.4$ (unfiltered),
was found to lie $\sim7.5''$ from the position of Nova \#49 in
Hubble's survey \citep{hub29}. Despite the relatively large reported separation,
our expanded initial screen for early photographic surveys flagged
this pair of novae for closer inspection.
Figure~\ref{fig23} shows a comparison of the positions of M31N 1925-07c
and 2011-12b, revealing that the novae to be distinct objects.
Revised coordinates for M31N 1925-07c are given in Table~\ref{revcoord}.

\subsubsection{M31N 1925-09a}

\begin{figure}
\includegraphics[angle=0,scale=.28]{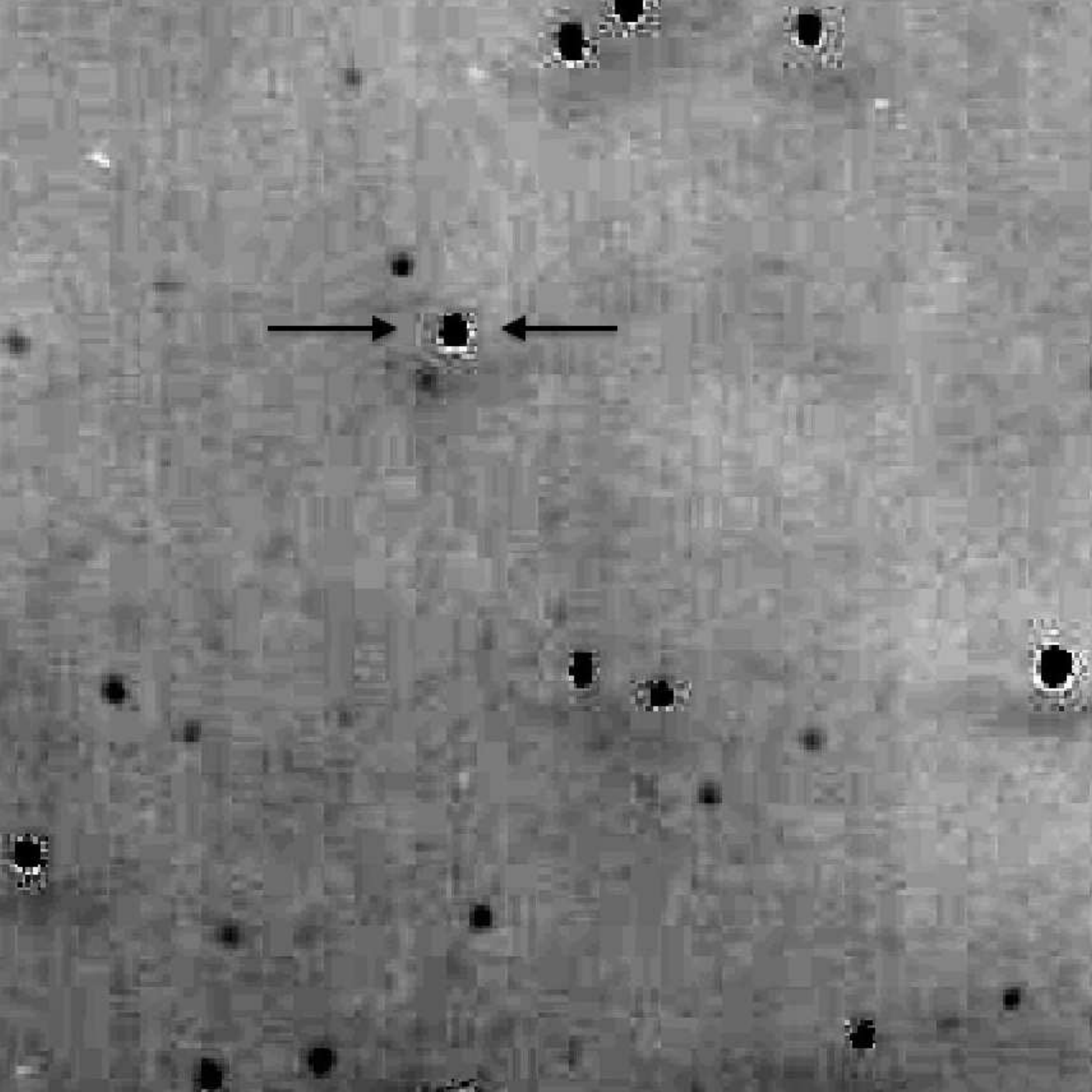}
\includegraphics[angle=0,scale=.28]{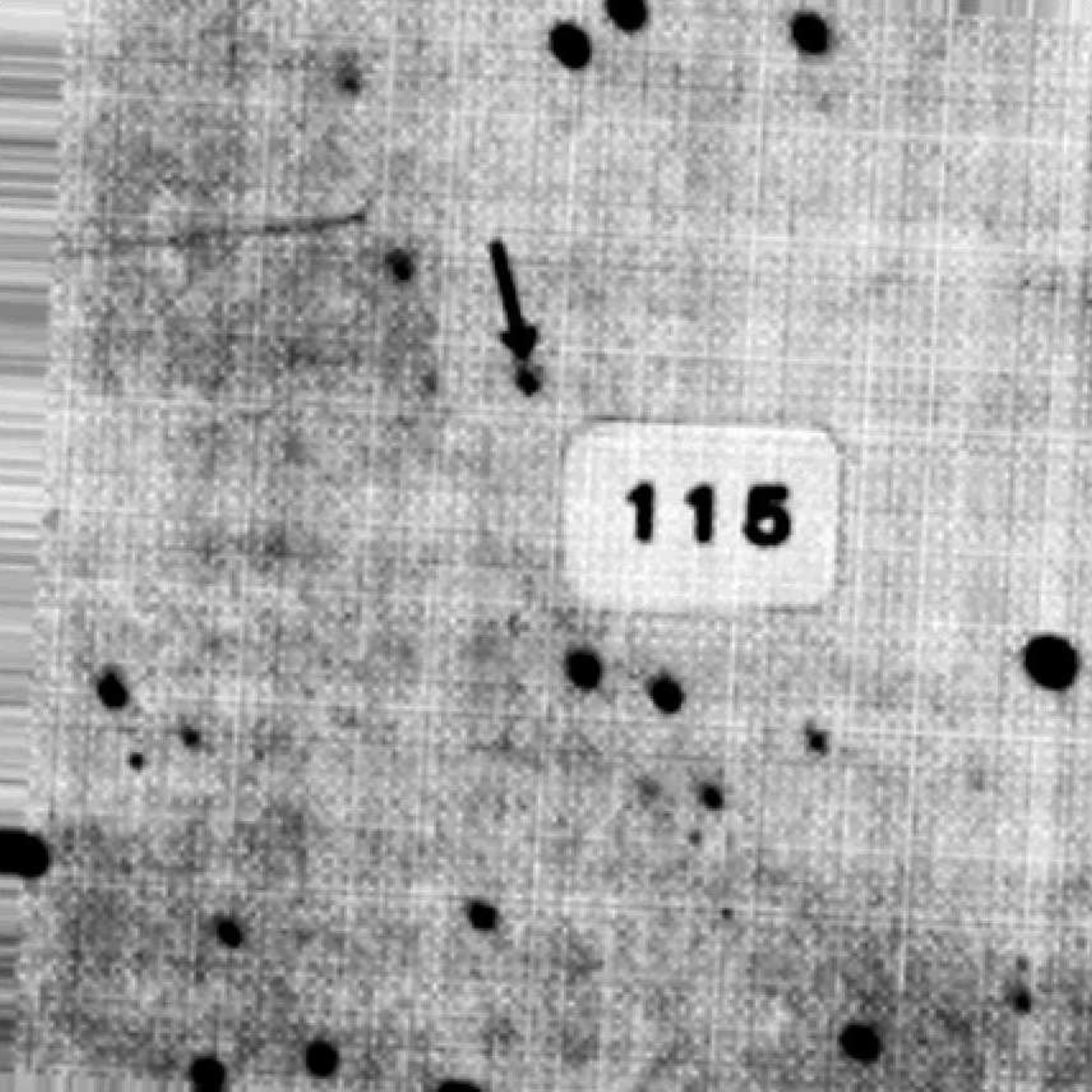}
\includegraphics[angle=0,scale=.28]{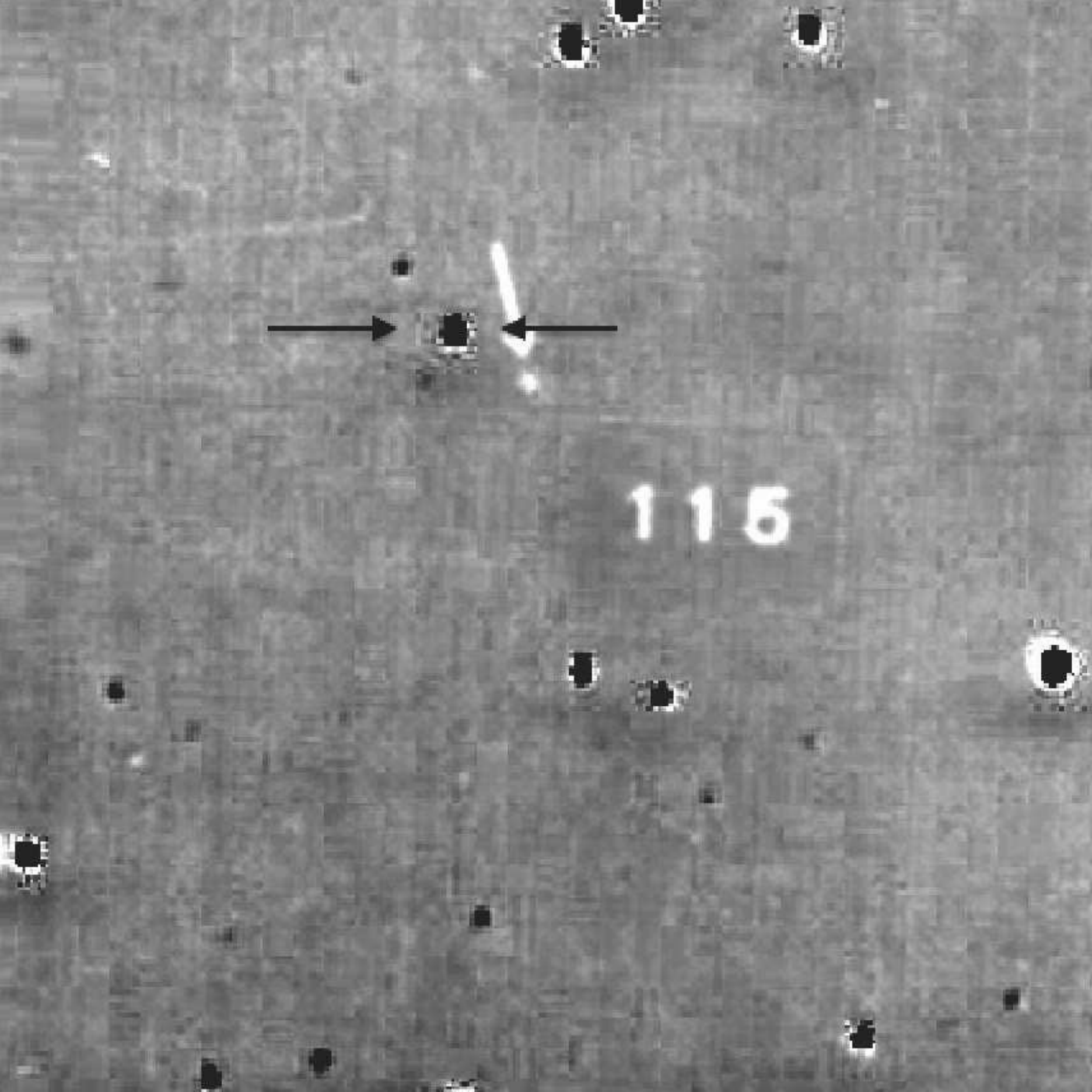}
\caption{Images of M31N 1925-09a, 1976-12a and their comparison
(left, center, and right, respectively).
The comparison image shows that
M31N 1976-12a is clearly not a recurrence of 1925-09a (shown in white), with
the former nova being located $\sim15''$ to the SW of 1925-09a.
The finding chart for M31N 1925-09a has been reproduced from plate
H609H from the
Carnegie archives, while the chart for 1976-12a is taken from \citet{ros89}.
North is up and East to the left, with a scale of $\sim3.2'$ on a side.
\label{fig24}}
\end{figure}

M31N 1976-12a was discovered at $B=18.3$ by \citet{ros89} as the 115th
nova in his survey. Our coarse screen
reveals that Rosino \#115 lies within $9.7''$ of a very bright
nova ($m_{pg}=15.3$), which is nova \#54 in the survey of \citet{hub29}.
Given the disparity in peak brightness, and the relatively large apparent
separation, we considered it unlikely that the 1976 eruption was
a recurrence of Hubble \#54. Nevertheless, given that the novae are relatively
far from the nucleus of M31 ($a\simeq20'$), the
probability for a chance coincidence is relatively low ($P_C\simeq0.155$),
and we decided to
take a closer look at the positions of these two novae. The positions
are compared in Figure~\ref{fig24}, confirming
that the novae are in fact distinct objects.
Revised coordinates for M31N 1925-07c are given in Table~\ref{revcoord}.

\subsubsection{M31N 1927-08a}

\begin{figure}
\includegraphics[angle=0,scale=.28]{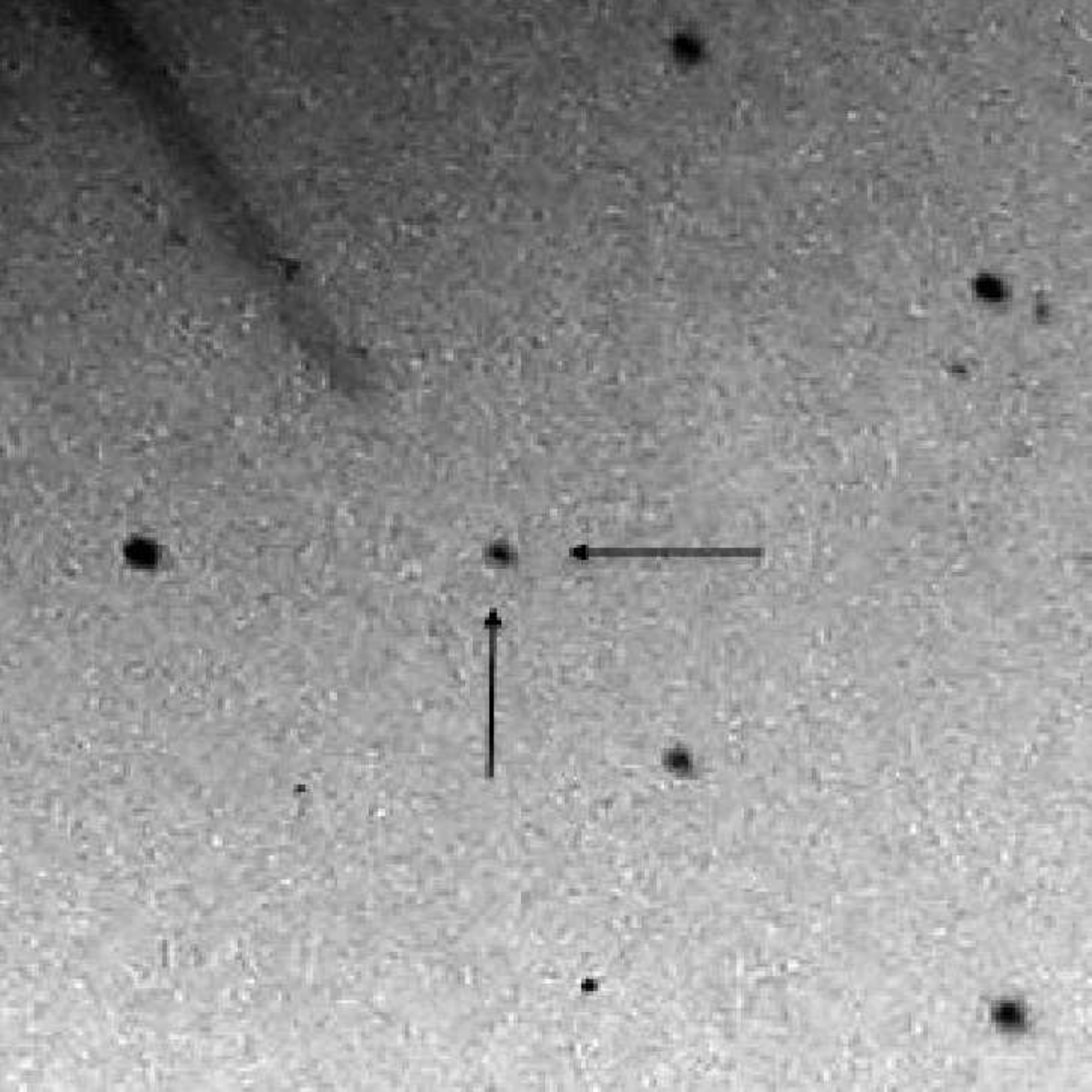}
\includegraphics[angle=0,scale=.28]{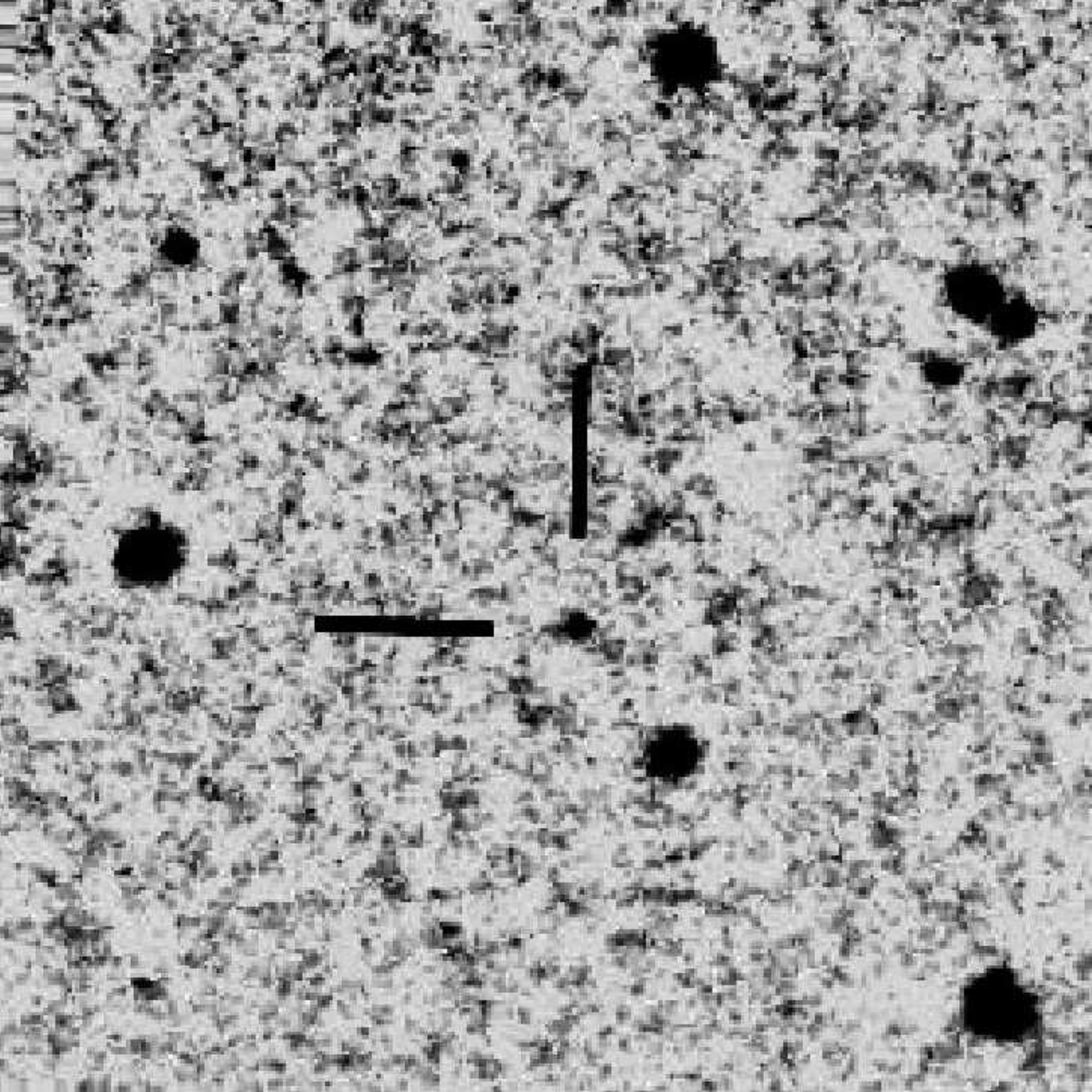}
\includegraphics[angle=0,scale=.28]{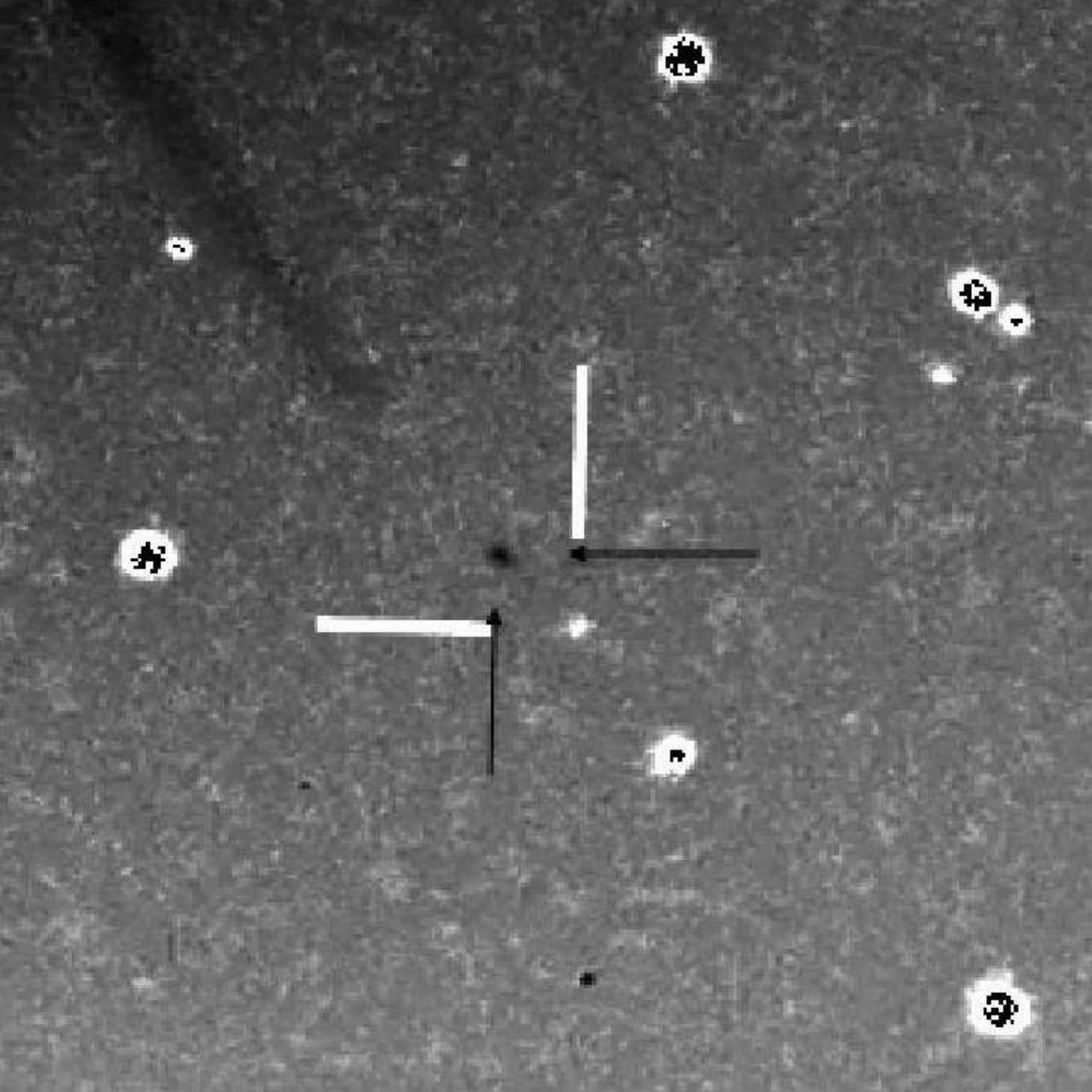}
\caption{Images of M31N 1927-08a, 2009-11e and their comparison
(left, center, and right, respectively).
The comparison image shows that
M31N 2009-11e (shown in white) is clearly not a recurrence of 1927-08a, with
the former nova being located $\sim20''$ to the SW of 1927-08a.
The finding chart for M31N 1927-08a was reproduced from plate
H828H from the
Carnegie archives, while the chart for 2009-11e is taken from
SuperLOTIS project \citep{bur10}.
North is up and East to the left, with a scale of $\sim2.3'$ on a side.
\label{fig25}}
\end{figure}

M31N 1927-08a was discovered as nova \#76 in the survey of \cite{hub29},
reaching a magnitude of $m_{pg}=17.1$ on 1927 August 23 UT. A possible,
but unlikely recurrence with a reported position
$\sim8''$ from 1927-08a, the $R=17.4$ nova M31N 2009-11e \citep{pie09b},
was flagged by our screen. Figure~\ref{fig25} shows finding charts for the
two novae, and clearly establishes that they are in fact distinct objects.

\subsubsection{M31N 1930-06b}

M31N 1930-06b was reported by \citet{may31}. The object was detected on
Mount Wilson plates taken on 1929 Nov 30 and 1930 June 19. Both times the
apparent nova was seen at $m\sim18$. A possible recurrence,
M31N 2008-08e, was discovered by D. Balam on 2008 August 31. Unpublished
observations by
one of us (KH) at the Ondrejov Observatory, establish that the object is
visible on multiple images taken over a two-year period from March 2007 to
August 2008. Given its photometric behavior, the object appears to be a
long period variable star and not a nova in M31.

\subsubsection{M31N 1930-06c}

\begin{figure}
\includegraphics[angle=0,scale=.28]{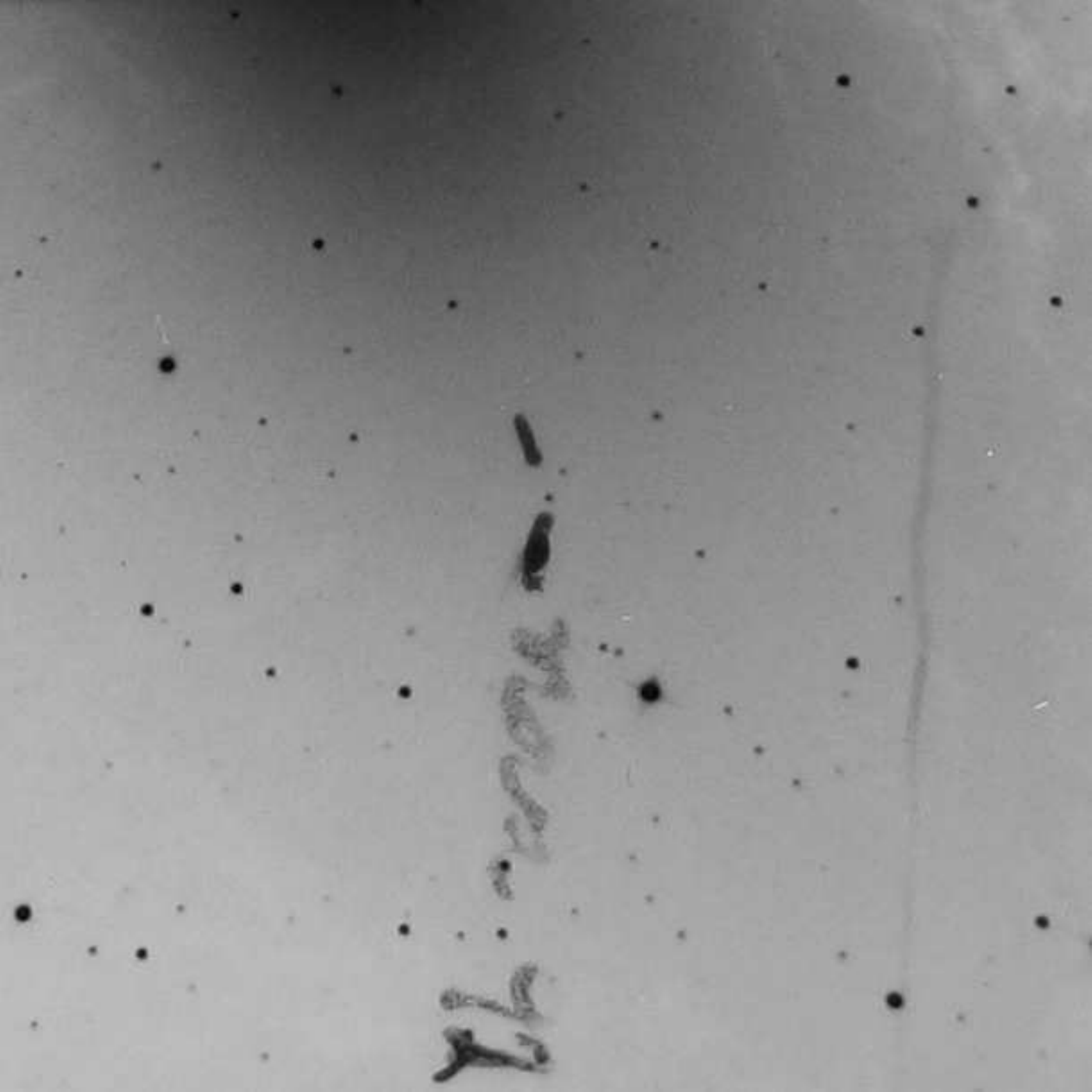}
\includegraphics[angle=0,scale=.28]{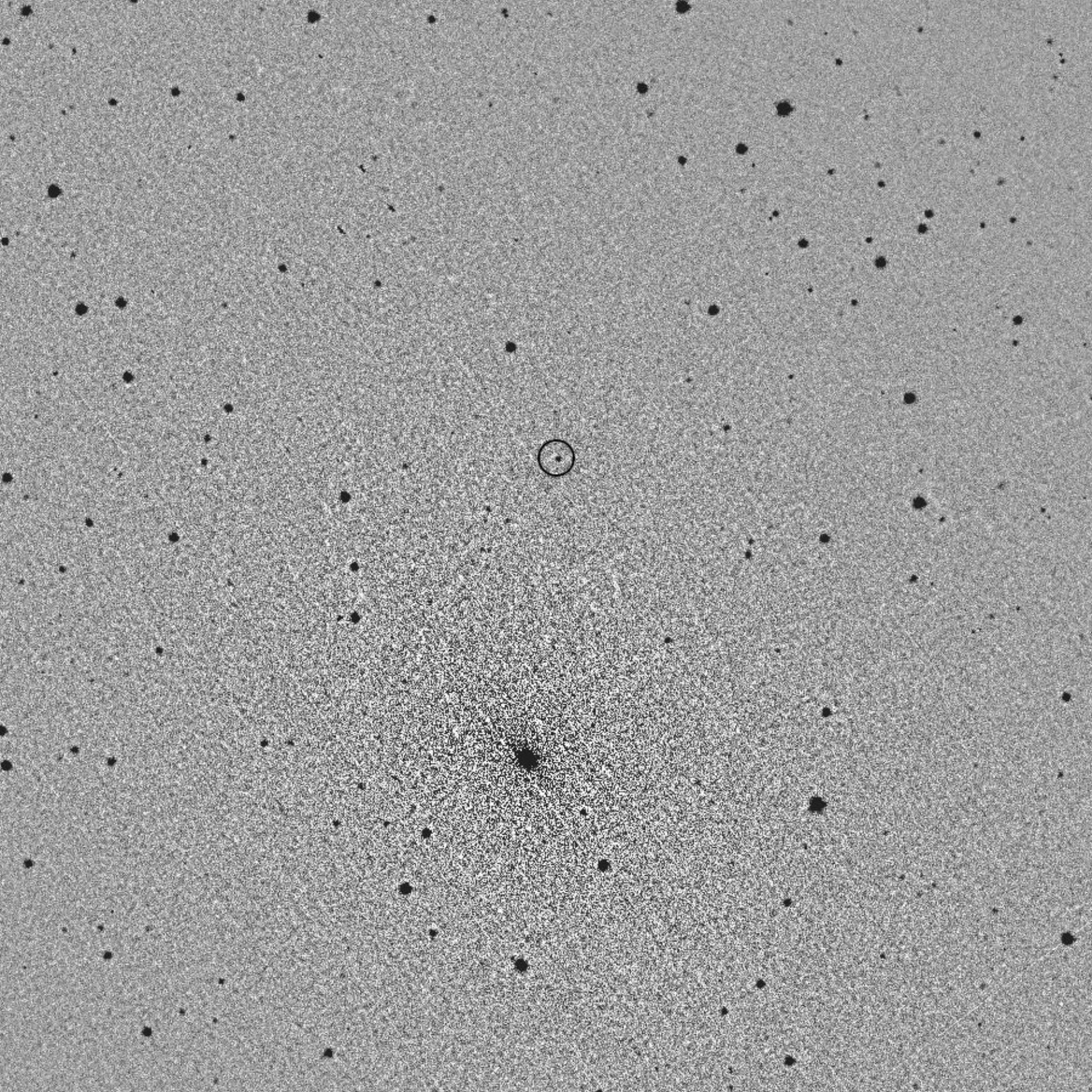}
\caption{Images of M31N 1930-06c and 1996-08a
(left and right, respectively).
A comparison of the position of M31N 1930-06c (Hubble's nova \#92) with that
of 1996-08a shows that the novae
lie on opposite sides of the nucleus of M31, and are clearly distinct objects.
North is up and East to the left, with a scale of $\sim8'$ on a side.
\label{fig26}}
\end{figure}

M31N 1996-08a \citep{sha01} was observed to erupt within $s\sim7.2''$
of the reported position of
M31N 1930-06c \citep[nova \#92 in][]{may31} and made it through our coarse screen
for RN candidates.
A review of plate H1155H in
the Carnegie archives taken on 1930 June 28 fails to
reveal a source at the reported position of M31N 1930-06c. Instead,
an object labeled ``nova 92'' is found far from the reported position of
M31N 1930-06c on the opposite (South) side of the nucleus of M31.
A comparison of the position of Hubble's nova \#92 with that
of M31N 1996-08a is shown in Figure~\ref{fig26}.
We conclude that an error was made in the conversion of Hubble's original
X and Y position of the nova to the equatorial coordinates for M31N 1930-06c
given in the online M31 nova catalog.
Revised coordinates for both novae are given in Table~\ref{revcoord}.

\subsubsection{M31N 1953-11a}

\begin{figure}
\includegraphics[angle=0,scale=.28]{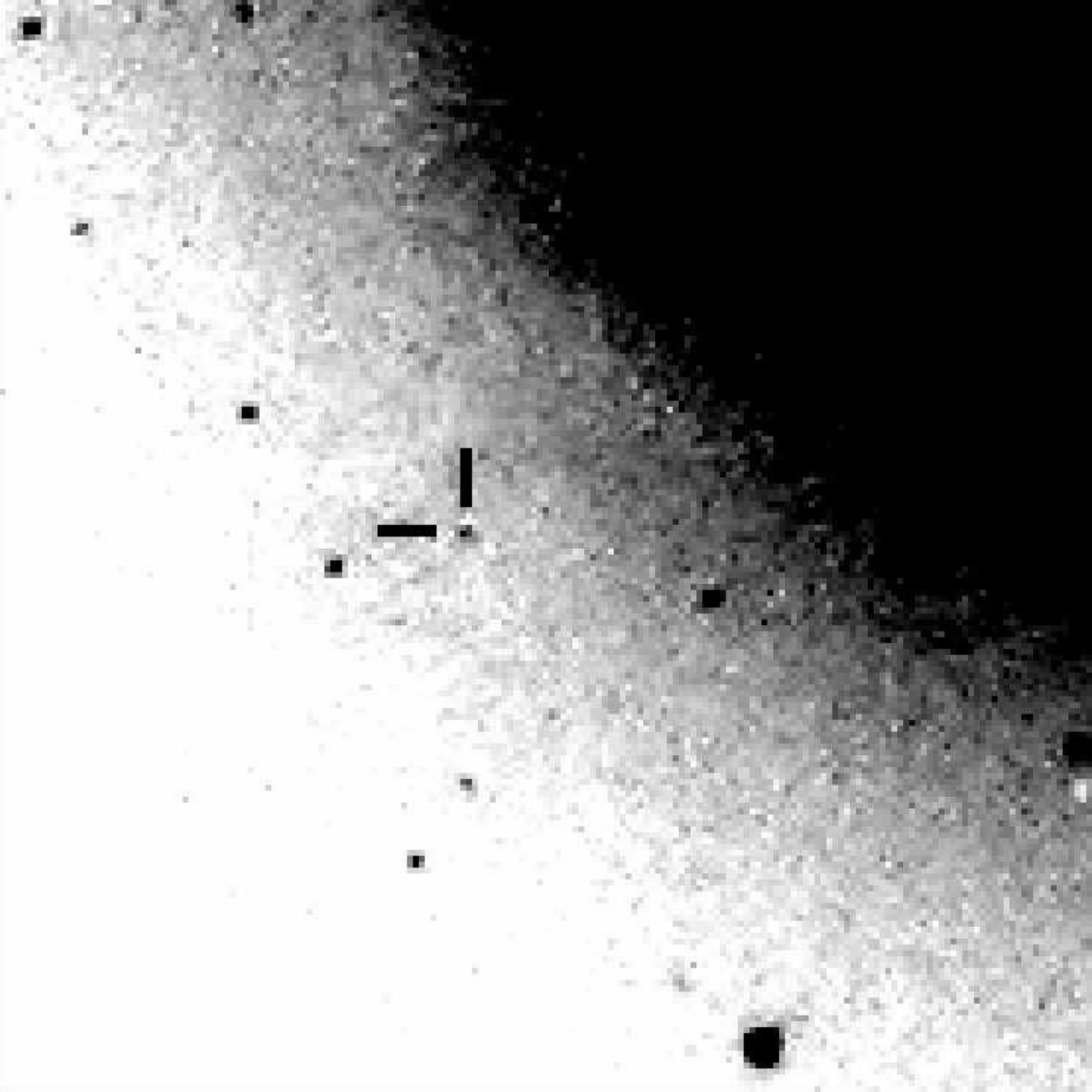}
\includegraphics[angle=0,scale=.28]{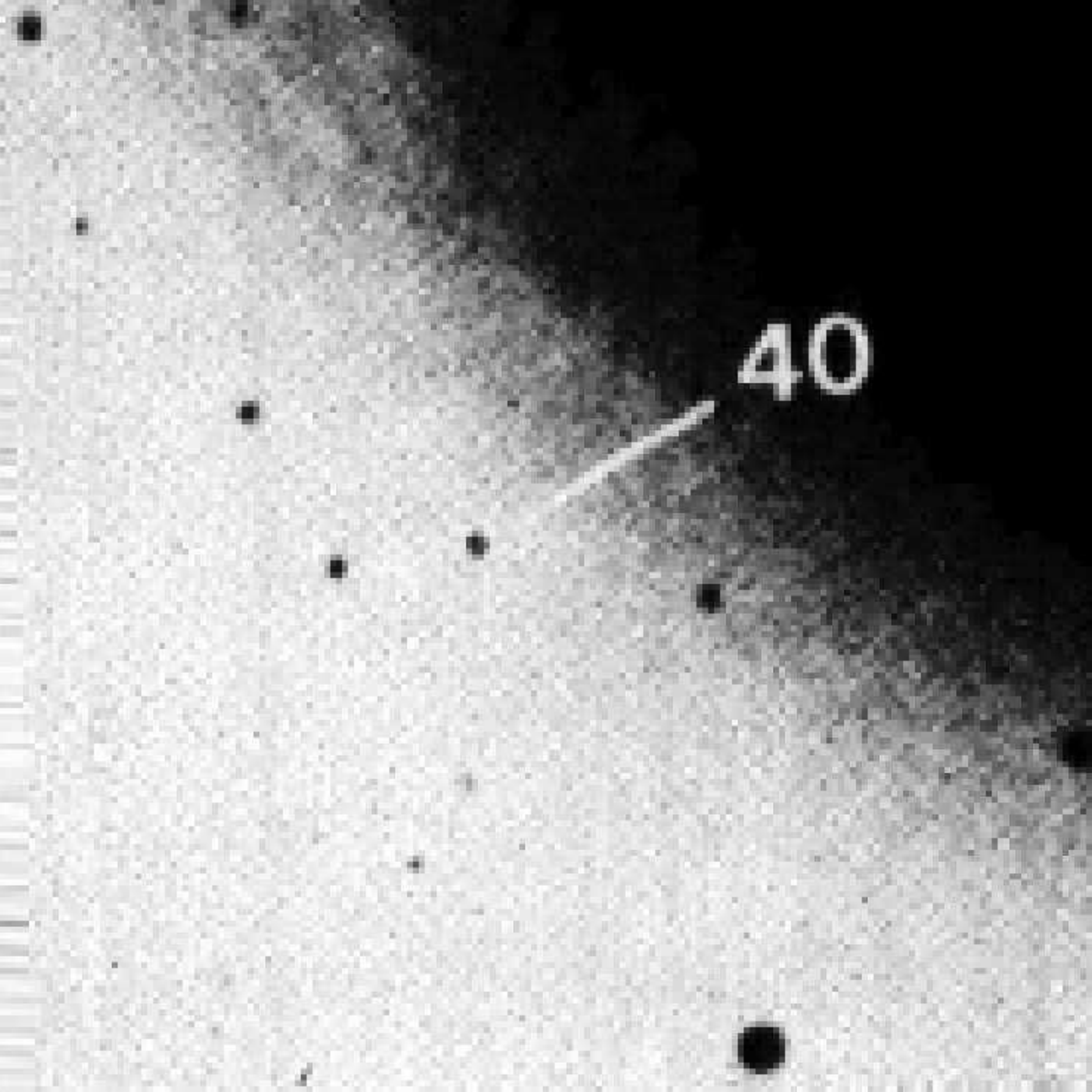}
\includegraphics[angle=0,scale=.28]{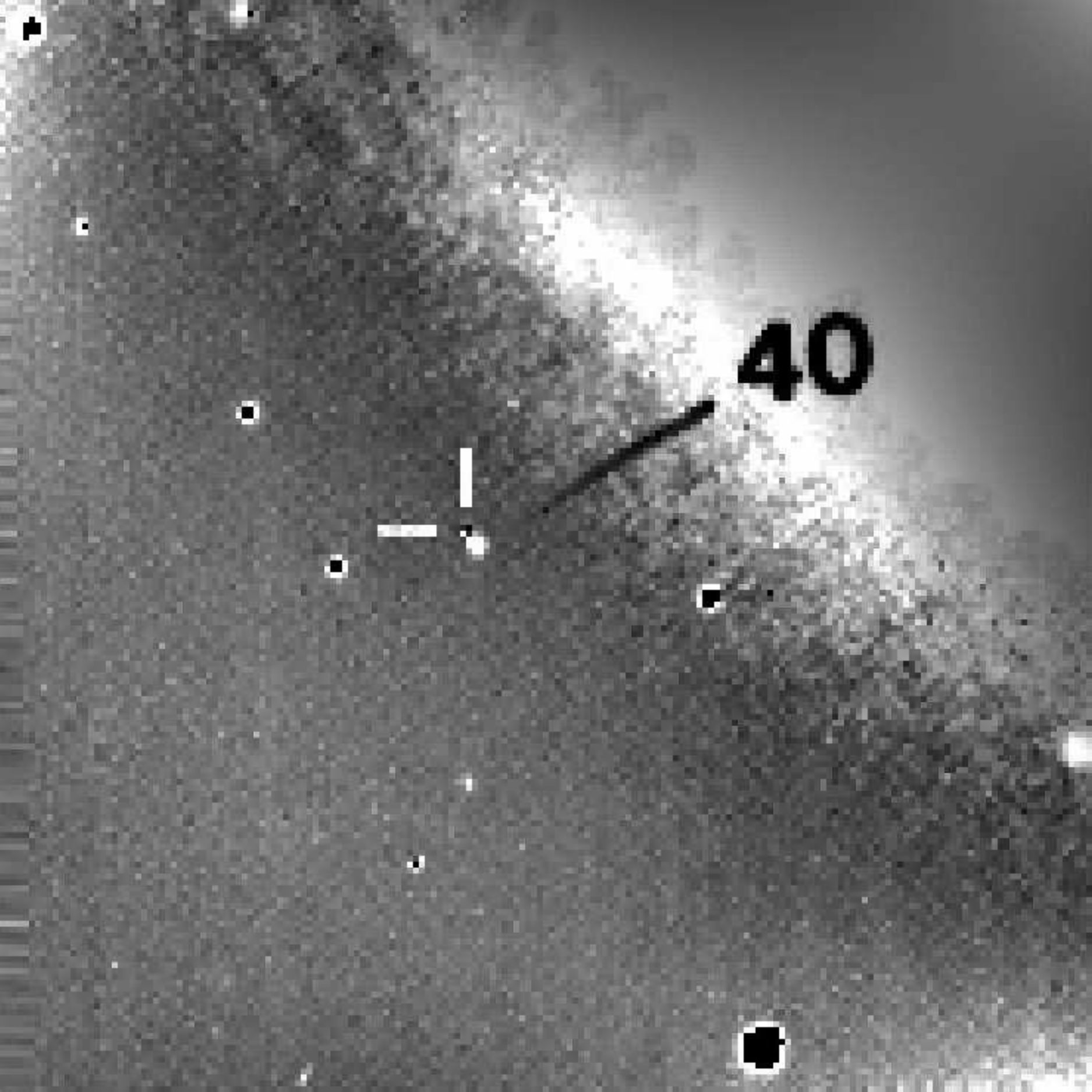}
\caption{Images of M31N 1953-11a, 1962-11b and their comparison
(left, center, and right, respectively).
A careful comparison afforded by the comparison image shows that
M31N 1962-11b (in white) is located slightly SW of the position of 1953-11a,
establishing that that the former nova is not a recurrence of the latter.
The image of M31N 1953-11a is a scan of the plate S1137A from \citet{arp56},
while that of 1962-11b is from \cite{ros73}.
North is up and East to the left, with a scale of $\sim3.5'$ on a side.
\label{fig27}}
\end{figure}

M31N 1953-11a was discovered by \citet{arp56} as nova \#19 in his survey.
Three subsequent novae were seen to erupt nearby and passed our coarse screen:
M31N 1960-12a, 1962-11b, and 2013-05b. 
Of these, 
M31N 1962-11b \citep{ros64} was the closest to 1953-11a,
with a reported position lying just $\sim2.9''$ away.
These novae are all relatively close to the
nucleus of M31 ($a\sim4.4'$), making
the probability of a chance positional coincidence between M31N 1953-11a and
1962-11b relatively high ($P_C\simeq0.622$).
As with M31N 1953-09b, we were able to locate the
original Arp plate (S1137A) in the Carnegie archives. A reproduction
of the nova field, along with that of M31N 1962-11b,
is shown in Figure~\ref{fig27}.
The novae are clearly not spatially coincident. Furthermore, as expected,
a comparison of Figure~\ref{fig27}
with Figure~\ref{fig8} shows that both M31N 1953-11a and 1962-11a
are also not coincident with either M31N 1960-12a or 2013-05b.
However, as discussed earlier and shown in Figure~\ref{fig8},
the latter two novae are in fact spatially coincident
to within observational uncertainties, strongly suggesting that
M31N 1960-12a is a RN.
Revised coordinates for M31N 1953-11a have been included, along
with those of 1960-12a and 1961-11b, in Table~\ref{revcoord}.

\subsubsection{M31N 1954-06c}

\begin{figure}
\includegraphics[angle=0,scale=.28]{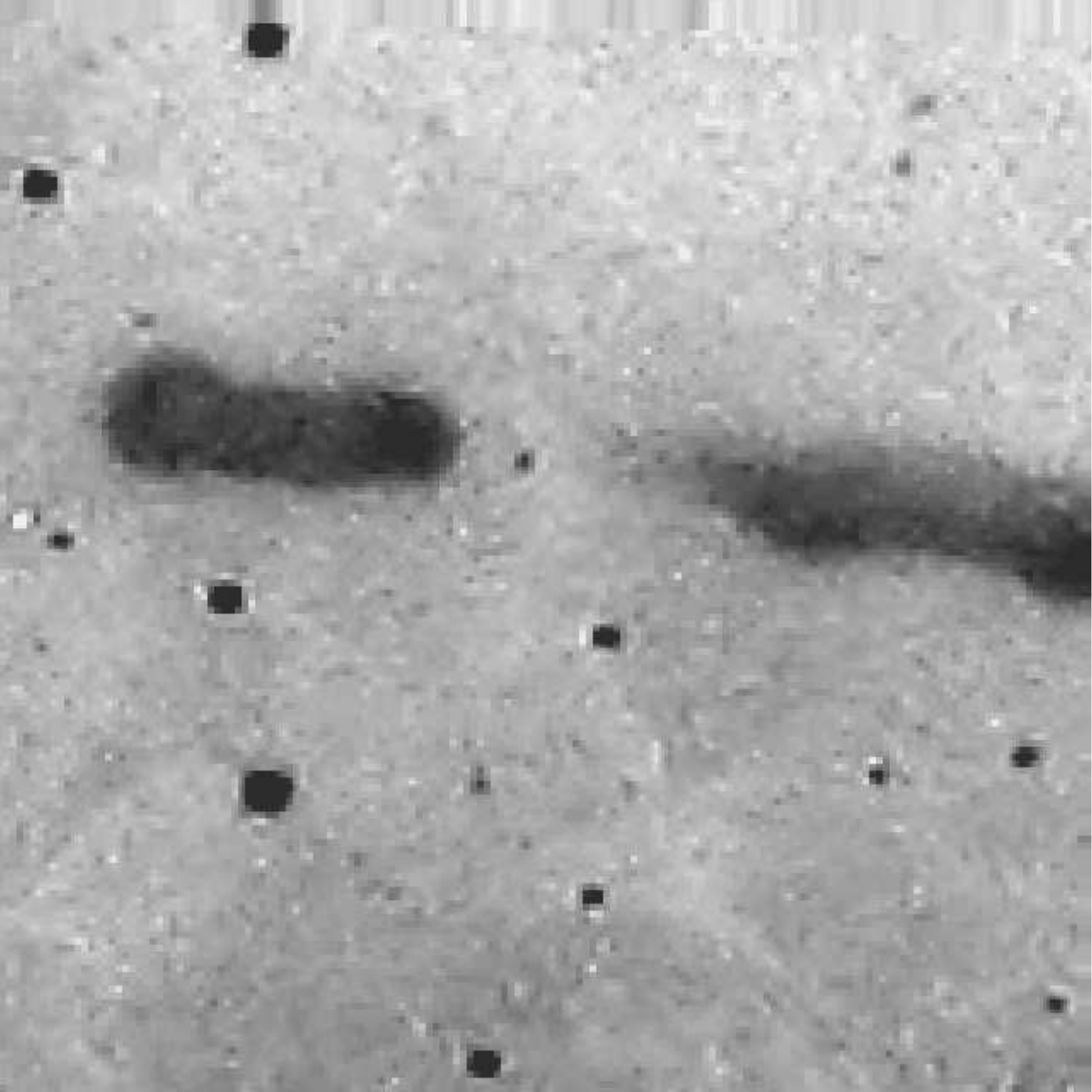}
\includegraphics[angle=0,scale=.28]{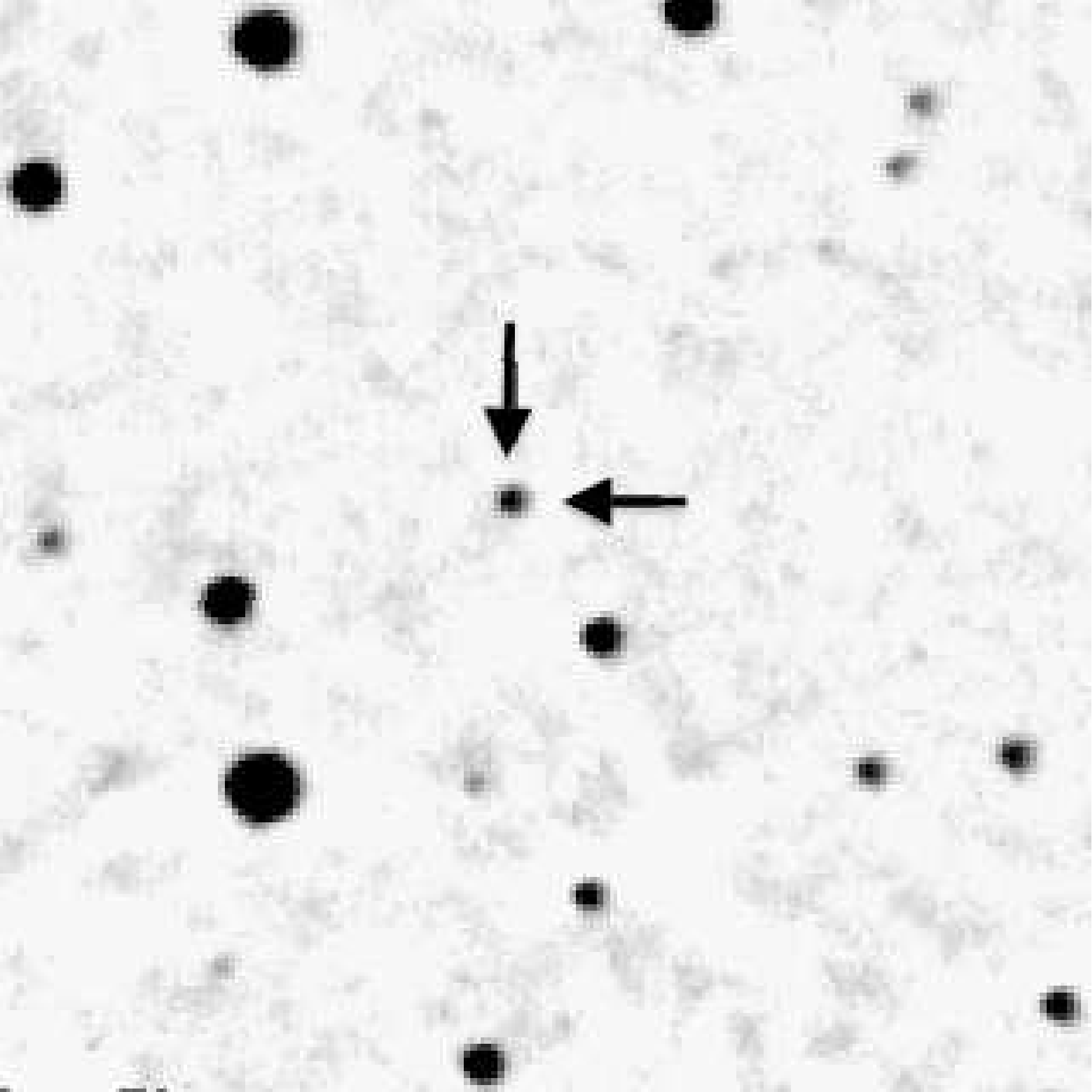}
\includegraphics[angle=0,scale=.28]{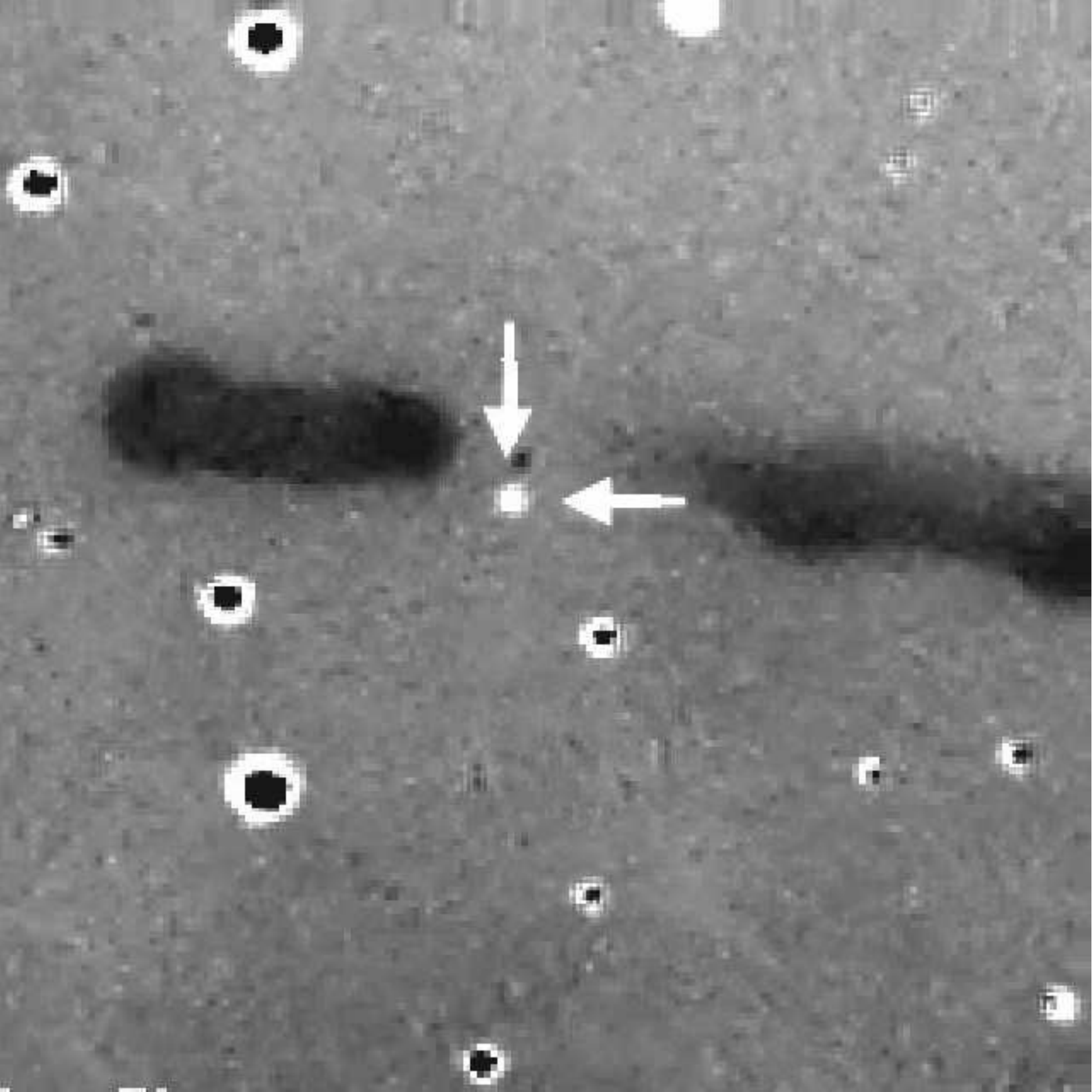}
\caption{Images of M31N 1954-06c, 2010-01b and their comparison
(left, center, and right, respectively).
M31N 2010-01b (in white) is located several arcsec to the SSE of
1954-06c and is clearly not spatially coincident with it.
The image of M31N 1954-06c is a scan of the plate S1335A from \citet{arp56},
while that of 2010-01b is a previously unpublished finding chart based
on data obtained by V. Burwitz using the TOU/OAM PIRATE
Schmidt-Cassegrain telescope at Costitx, Mallorca.
North is up and East to the left, with a scale of $\sim3'$ on a side.
\label{fig28}}
\end{figure}

M31N 2010-01b was discovered by \citet{nis10} on 2010
January 17.438 UT at $m=18.1$. The position of the nova lies $\sim8.8''$
(just inside our coarse screen) from the position of M31N 1954-06c,
which is Nova \#28 in Arp's nova survey \citep{arp56}. The light curve
of M31N 1954-06c is very slow, with \citet{arp56} finding a ``duration"
(the number of days where the nova is brighter than $m=20$) of 115 days.
In contrast, light curve information on 2010-01b is much more sparse;
however, photometry by \cite{pie10a} shows that the nova remained near
$m=17.7$ between 2010 January 25 and February 02.
As with the previous two novae from Arp's survey, we located
the discovery plate (S1335A) in the Carnegie archives, and re-measured the
coordinates of the nova (see Table~\ref{revcoord}).
The field of M31N 1954-06c is compared with that of
2010-01b in Figure~\ref{fig28}, clearly showing that the two novae are
not spatially coincident.

\subsubsection{M31N 1955-09b}

\begin{figure}
\includegraphics[angle=0,scale=.28]{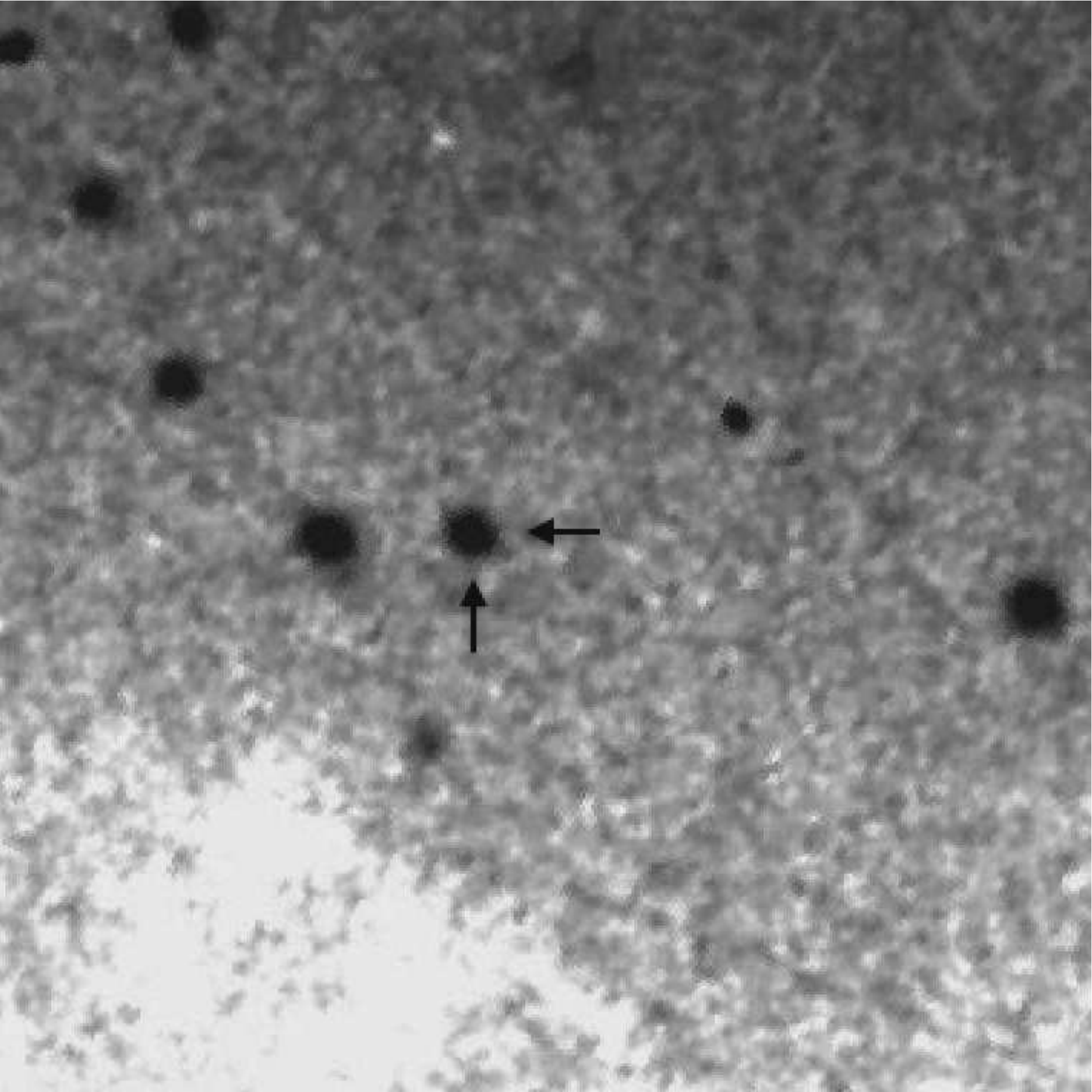}
\includegraphics[angle=0,scale=.28]{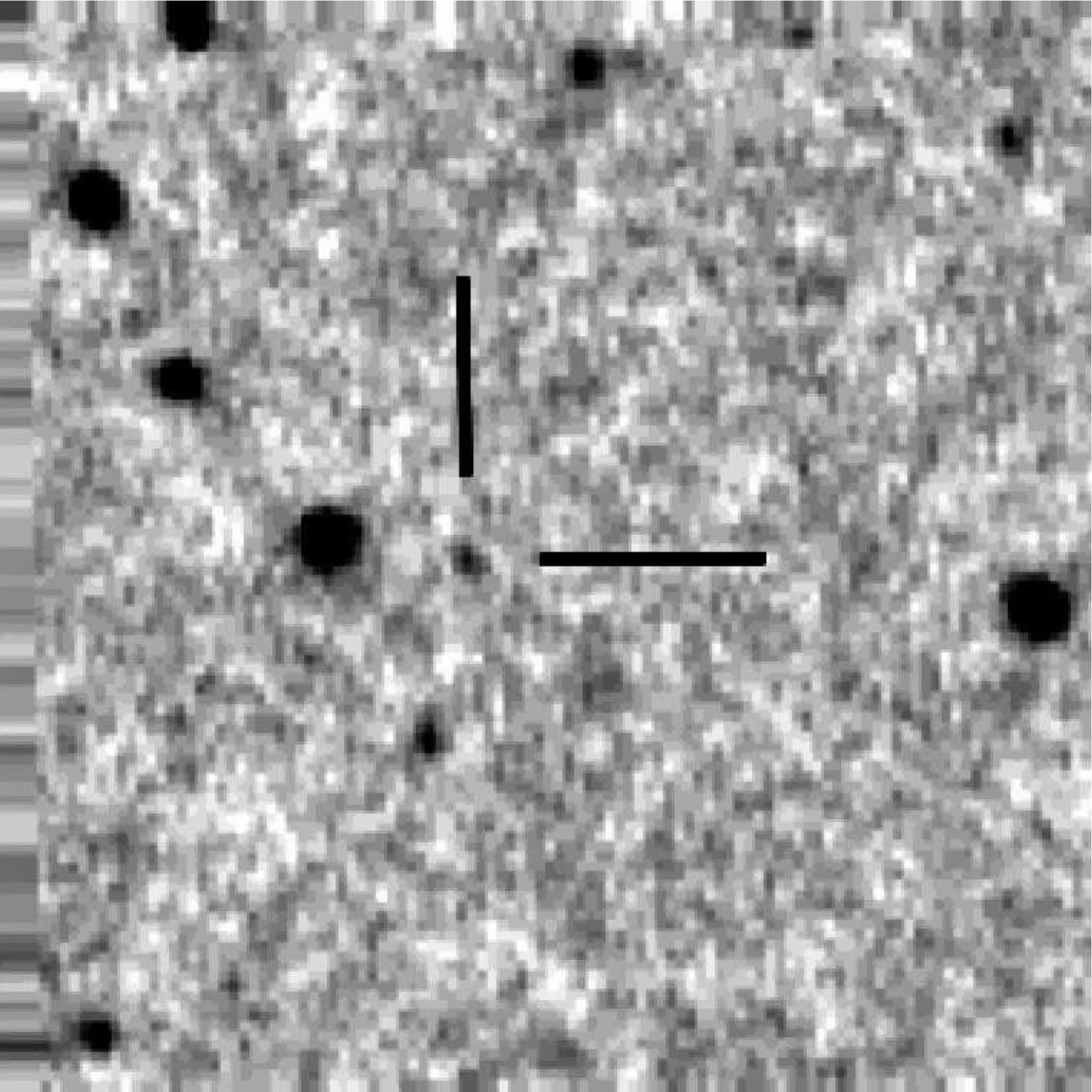}
\includegraphics[angle=0,scale=.28]{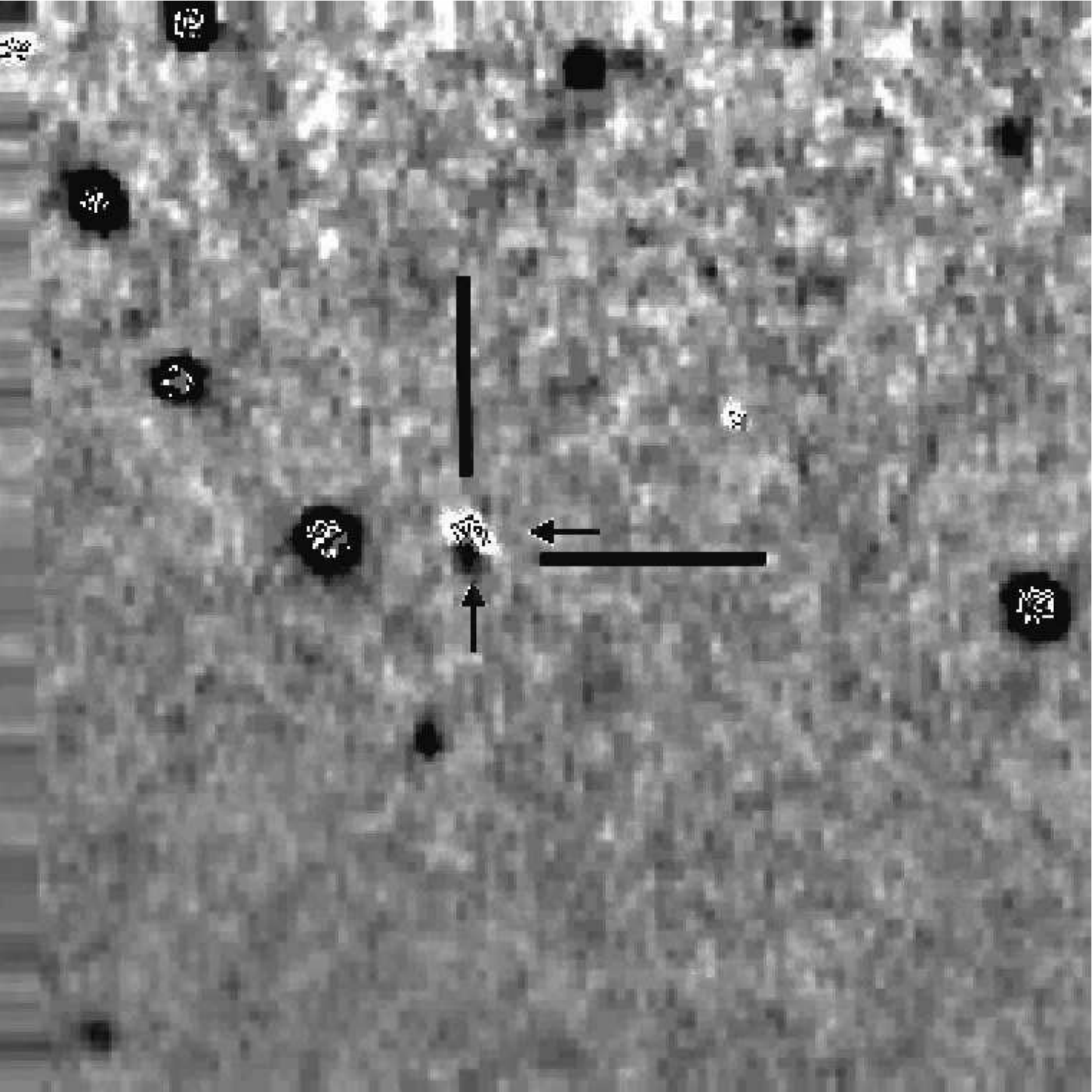}
\caption{Images of M31N 1955-09b, M31N 2012-03b and their comparison
(left, center, and right, respectively).
As can be seen from the comparison image,
M31N 2012-03b is clearly not coincident with 1955-09b (in white),
and we conclude that the former nova is not a recurrence of the latter.
The image of M31N 1955-09b is from \citet{ros64}, while that of
2012-03b is from \cite{hor12b}.
North is up and East to the left, with a scale of $\sim2'$ on a side.
\label{fig29}}
\end{figure}

M31N 2012-03b erupted $2.5''$ from the reported position of
M31N 1955-09b, nova \#1 in the extensive survey
of \citet{ros64}. The pair of novae are located at an
isophotal radius of $a\sim10.5'$ from the center of M31, resulting
in a relatively low probability of chance coincidence, $P_C\simeq0.032$.
Finding charts for the two novae are shown in Figure~\ref{fig29}. The
two novae are clearly not spatially coincident, and thus
M31N 2012-03b is not a recurrence of 1955-09b.

\subsubsection{M31N 1964-12b}

M31N 1964-12b \citep{ros73}
erupted very close to the nucleus of M31 ($a\simeq0.71'$).
A possible recurrence, M31N 1998-07b \citep{rec99},
lying $3.74''$ from that of 1964-12b was flagged by our screen.
After re-analyzing the original
data from the RBSE project, we have determined that M31N 1998-07b
was a spurious detection, and the nova does not exist.

\subsubsection{M31N 1967-11a}

\begin{figure}
\includegraphics[angle=0,scale=.28]{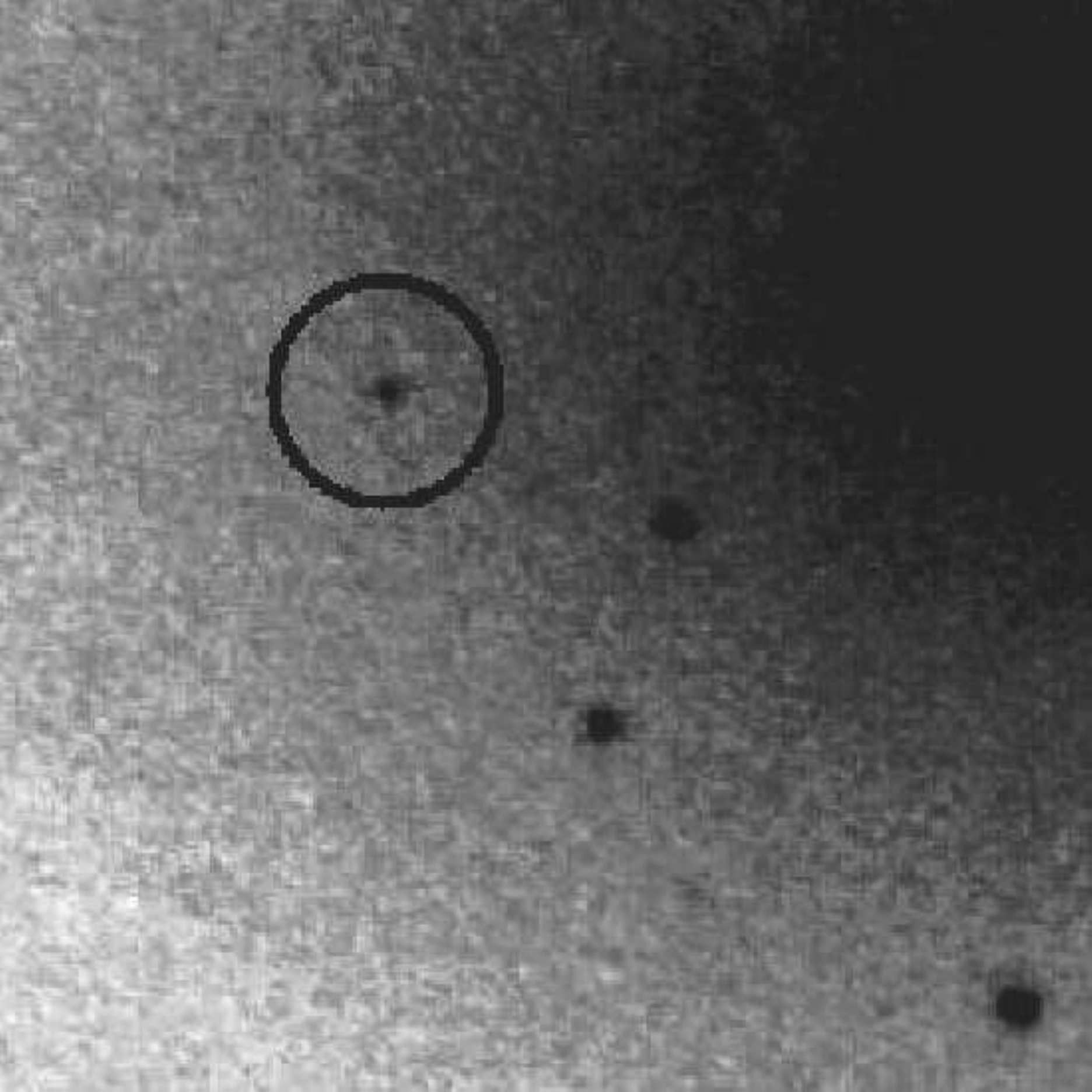}
\includegraphics[angle=0,scale=.28]{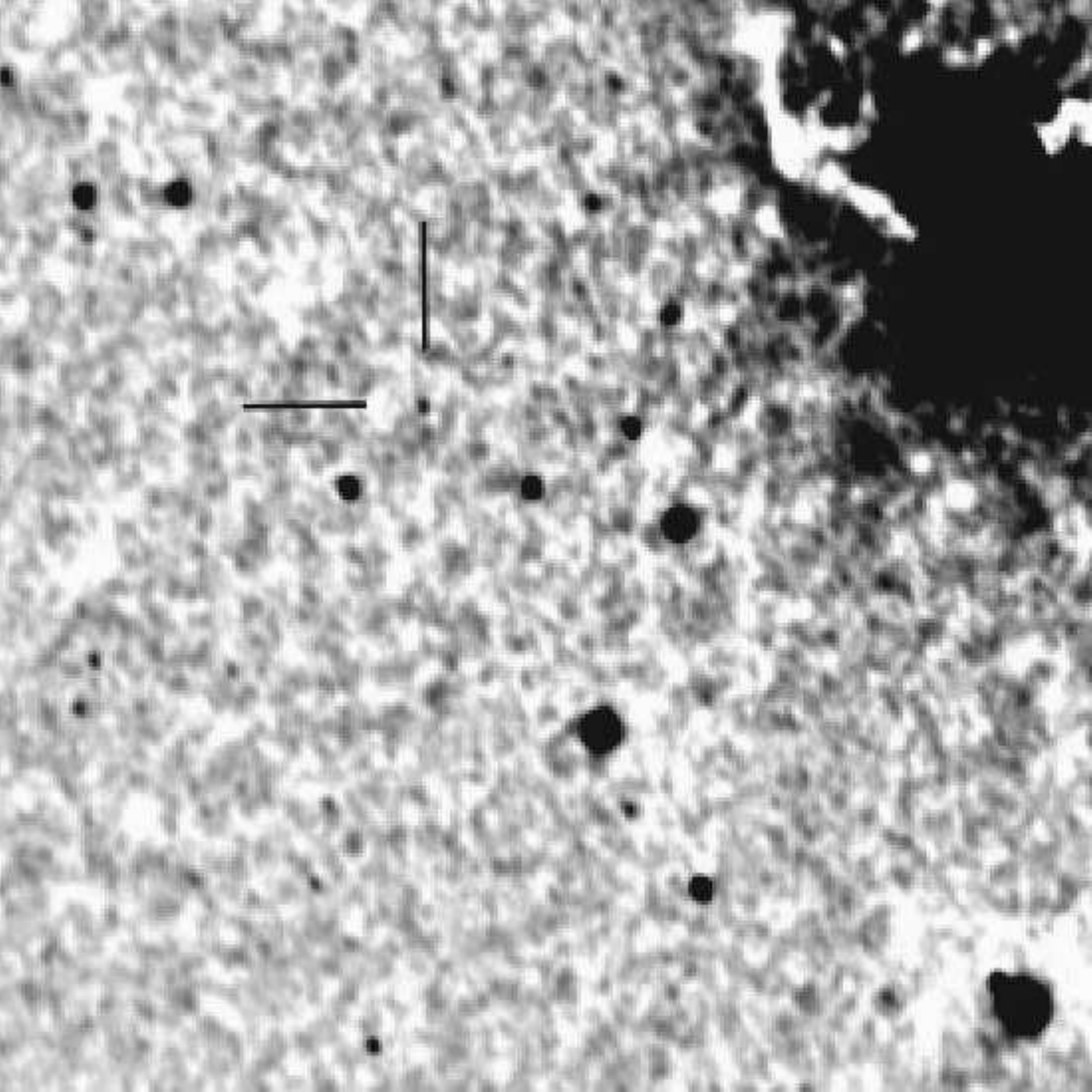}
\includegraphics[angle=0,scale=.28]{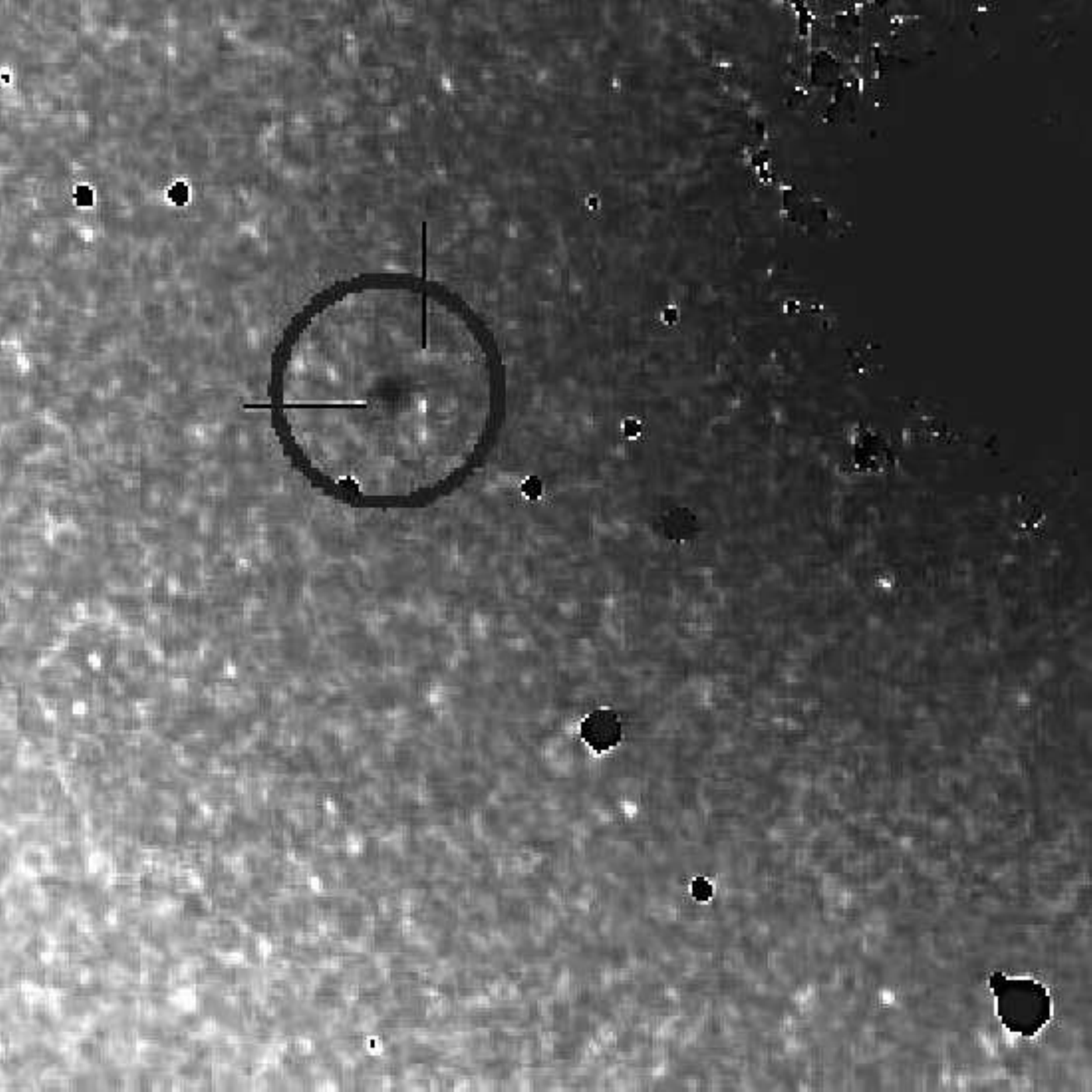}
\caption{Images of M31N 1967-11a, 2006-02a, and their comparison
(left, center, and right, respectively).
A careful comparison of the images, 1967-11a from \citet{hen08a} and
2006-02a from Hornoch (unpublished), shows that the novae are not coincident,
with M31N 2006-02a (in white) being $\sim 4''$ WSW of M31N 1967-11a.
North is up and East to the left, with a scale of $\sim2'$ on a side.
\label{fig30}}
\end{figure}

M31N 1967-11a was discovered on archival plates by \citet{hen08a}.
A possible recurrence, M31N 2006-02a, was later discovered by
K. Hornoch \citep{pie07a}. Charts for the two novae are shown
in Figure~\ref{fig30}. Although the positions of the novae appear close,
a careful inspection reveals that the positions differ, with
M31N 2006-02a located $\sim4''$ WSW of the position of 1967-11a, in
agreement with their reported positions.

\subsubsection{M31N 1967-12b}

\begin{figure}
\includegraphics[angle=0,scale=.28]{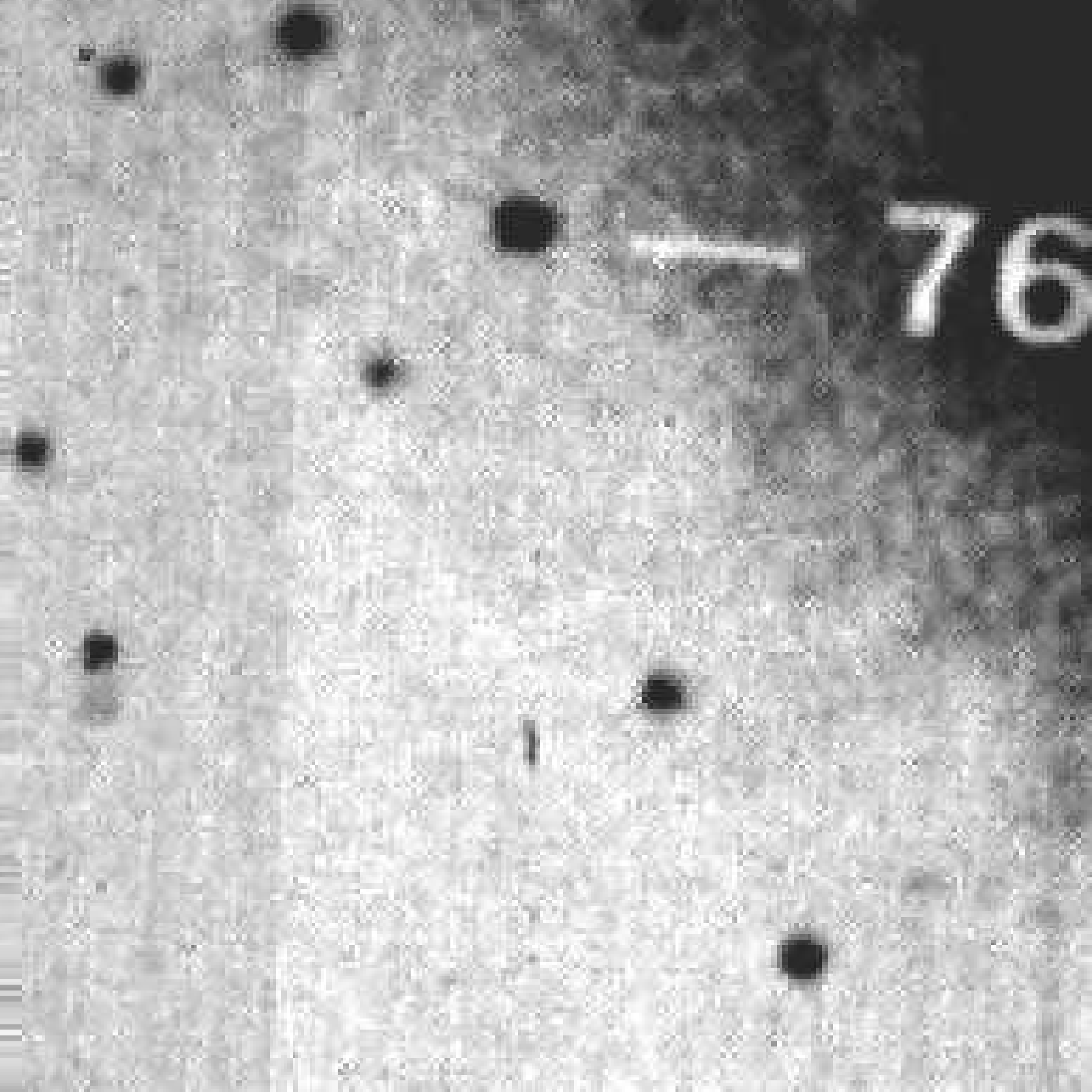}
\includegraphics[angle=0,scale=.28]{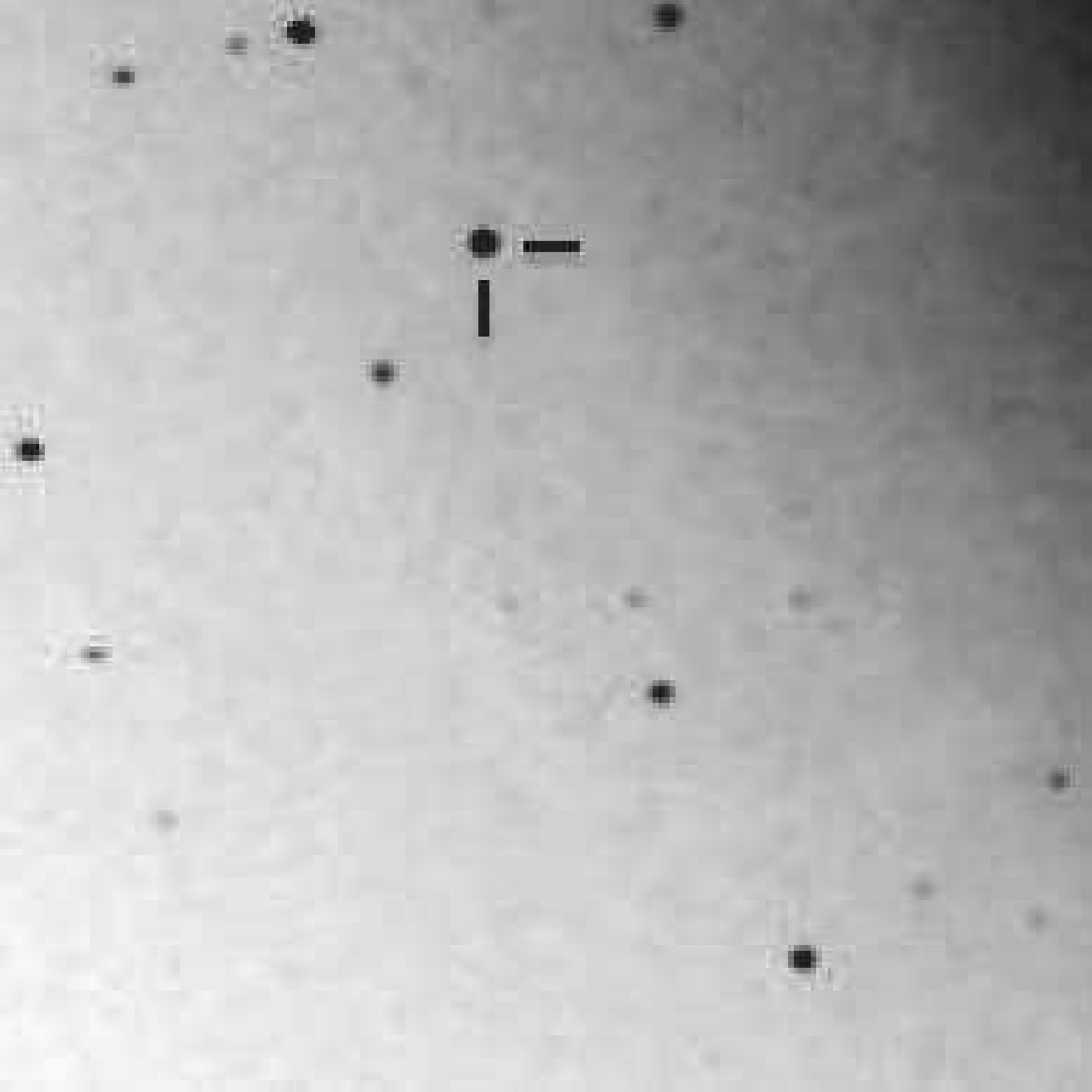}
\includegraphics[angle=0,scale=.28]{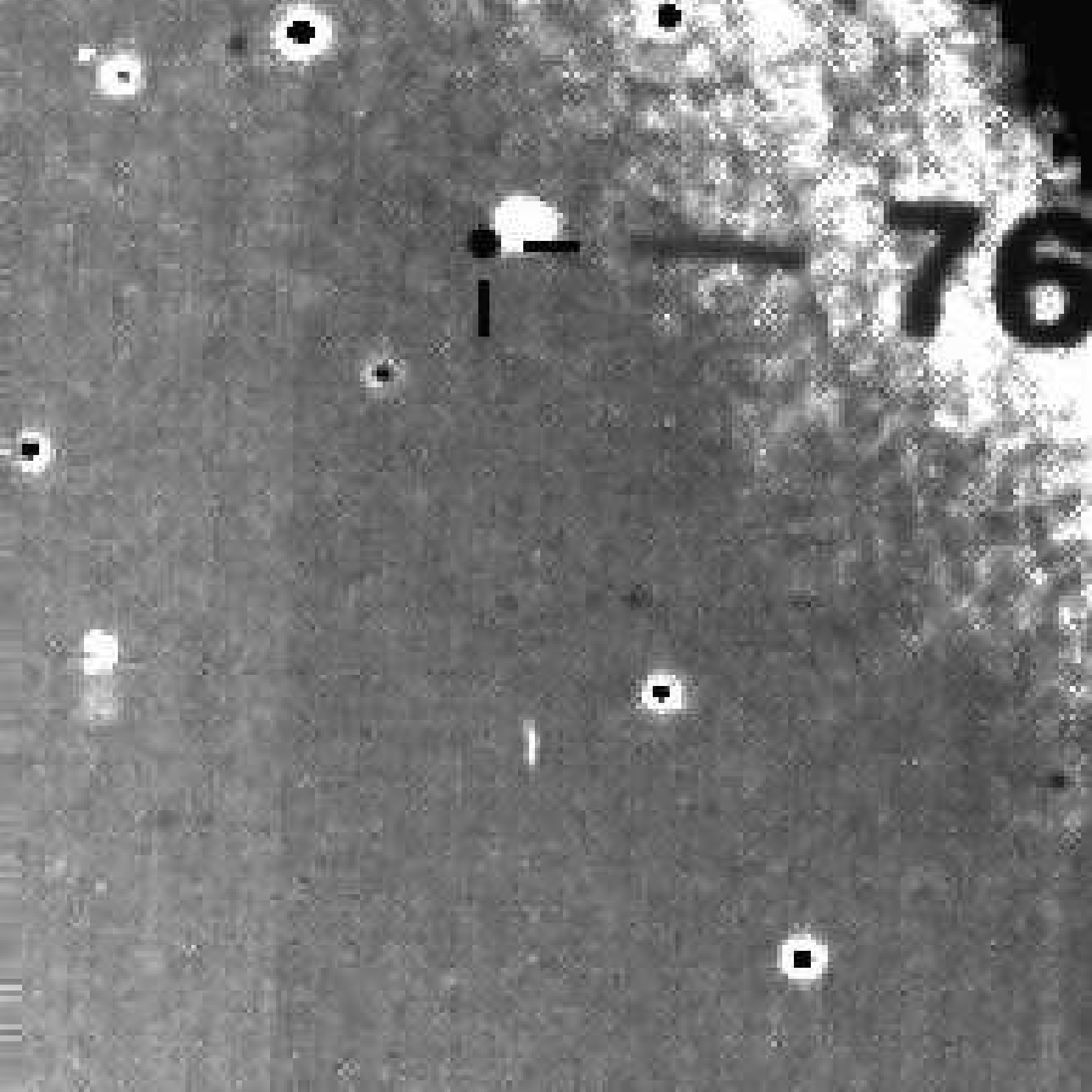}
\caption{Images of M31N 1967-12b, 1999-06b, and their comparison
(left, center, and right, respectively).
These novae are clearly not coincident, with
M31N 1999-06b (black) being located $\sim6''$ ESE of the position of 1967-12b.
The chart for 1967-12b is taken from \citet{ros73} and that for
1999-06b from the RBSE project on M31 novae \citep{rec99}.
North is up and East to the left, with a scale of $\sim2.2'$ on a side.
\label{fig31}}
\end{figure}

M31N 1967-12b is another nova discovered by \citet{ros73}. Its reported
position is close to that of M31N 1999-06b discovered later as part of
the RBSE program
\citep{rec99}. Charts for the two novae are shown in Figure~\ref{fig31}. The
novae are clearly not spatially coincident, with M31N 1999-06b
lying $\sim6''$ ESE
of 1967-12b. Refined coordinates for M31N 1967-12b
are given in Table~\ref{revcoord}.

\subsubsection{M31N 1969-08a}

M31N 2007-12b erupted very close to the reported position of M31N 1969-08a.
Nevertheless, as discussed by \cite{bod09}, the objects have been shown to be
distinct novae.
Interestingly however, \citet{bod09} were able to show that M31N 2007-12b
has a red giant companion star. This discovery, combined with its
He/N spectroscopic type, makes it likely that M31N 2007-12b has a short
recurrence time. Thus, it is possible that it will be revealed to be a RN
in the not too distant future. We also note that \citet{pie11} found a
1110 s coherent periodicity in the X-ray light curve of M31N 2007-12b, which
they interpreted as the rotation period of a magnetized white dwarf in
the system, and
suggested that the nova erupted from a intermediate polar progenitor.

\subsubsection{M31N 1975-09a}

\begin{figure}
\includegraphics[angle=0,scale=.28]{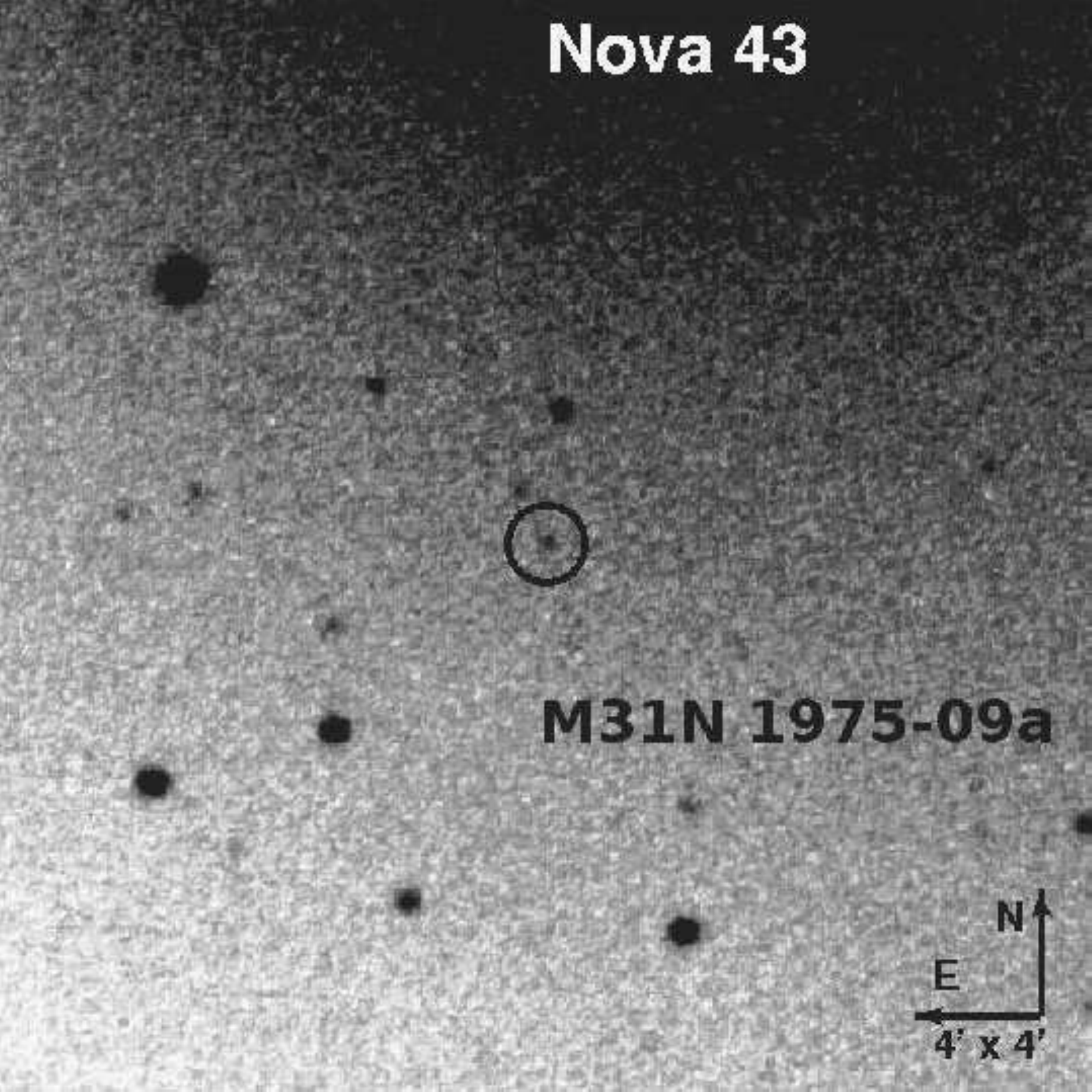}
\includegraphics[angle=0,scale=.28]{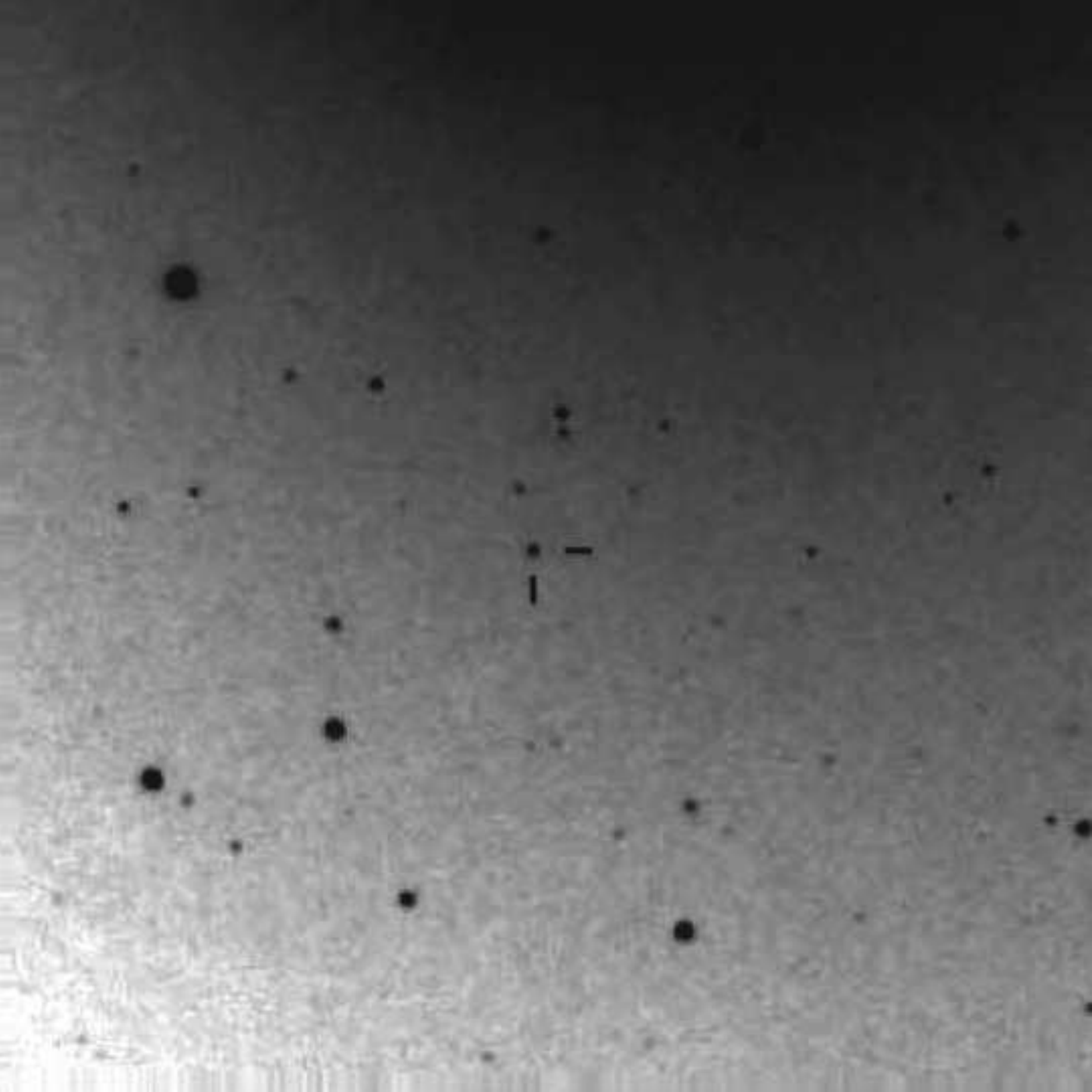}
\includegraphics[angle=0,scale=.28]{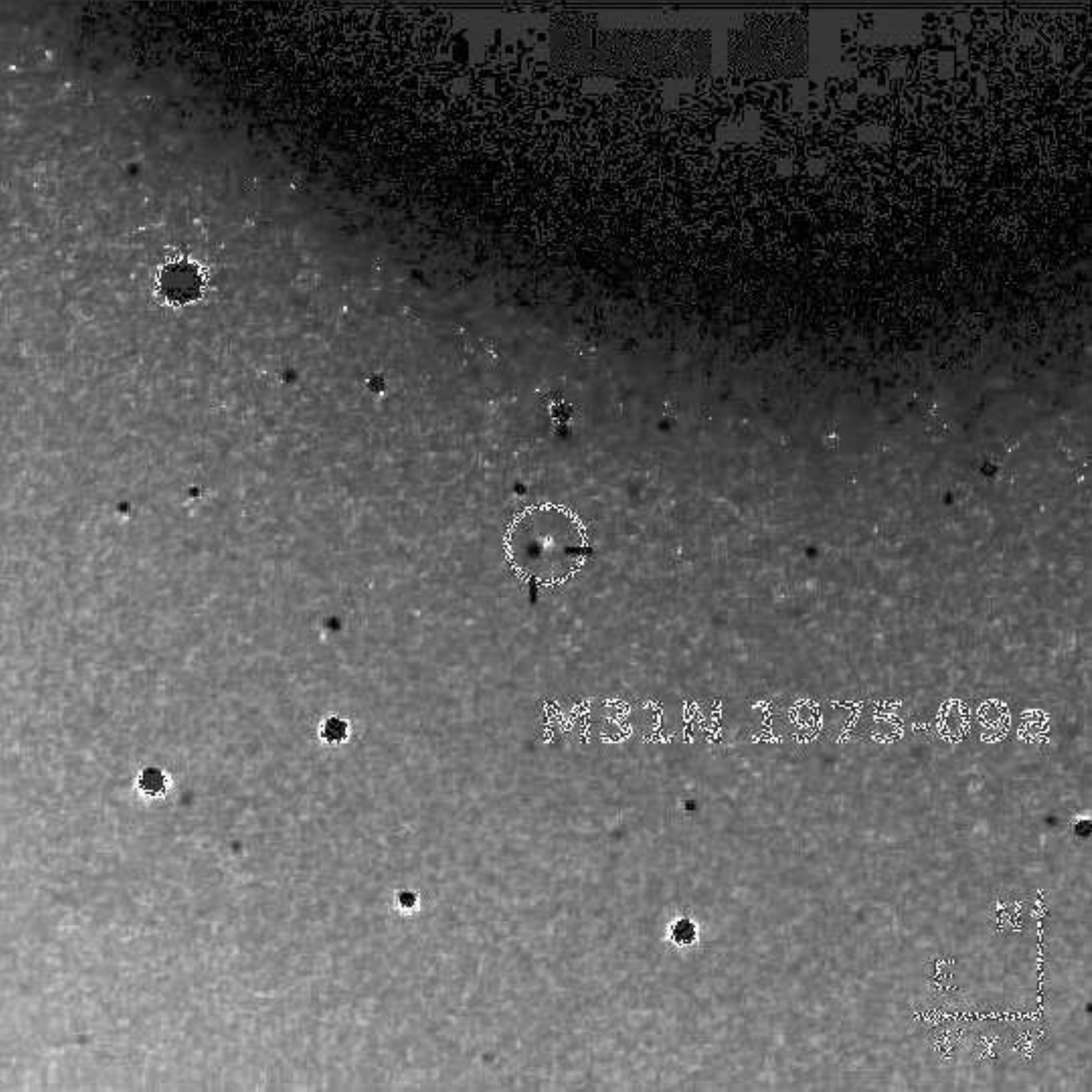}
\caption{Images of M31N 1975-09a, M31N 1999-01a, and their comparison
(left, center, and right, respectively).
As revealed by the comparison image, although close,
the novae are not spatially coincident with
M31N 1999-01a (in white) being located
$\sim3.3''$ SE of the position of 1975-09a.
The chart for 1975-09a is from \citet{hen08a}, with that for
1999-01a taken from the RBSE project \citet{rec99}.
North is up and East to the left, with a scale of $\sim4'$ on a side.
\label{fig32}}
\end{figure}

M31N 1975-09a was discovered by \citet{hen08a}
from an analysis of the Tautenburg Schmidt plates.
\cite{hen08a} explored the possibility that M31N 1999-01a was a
recurrence of 1975-09a and determined that the novae were
not spatially coincident. We confirm their conclusion as can
be seen in Figure~\ref{fig32}.

\subsubsection{M31N 1977-12a}

\begin{figure}
\includegraphics[angle=0,scale=.28]{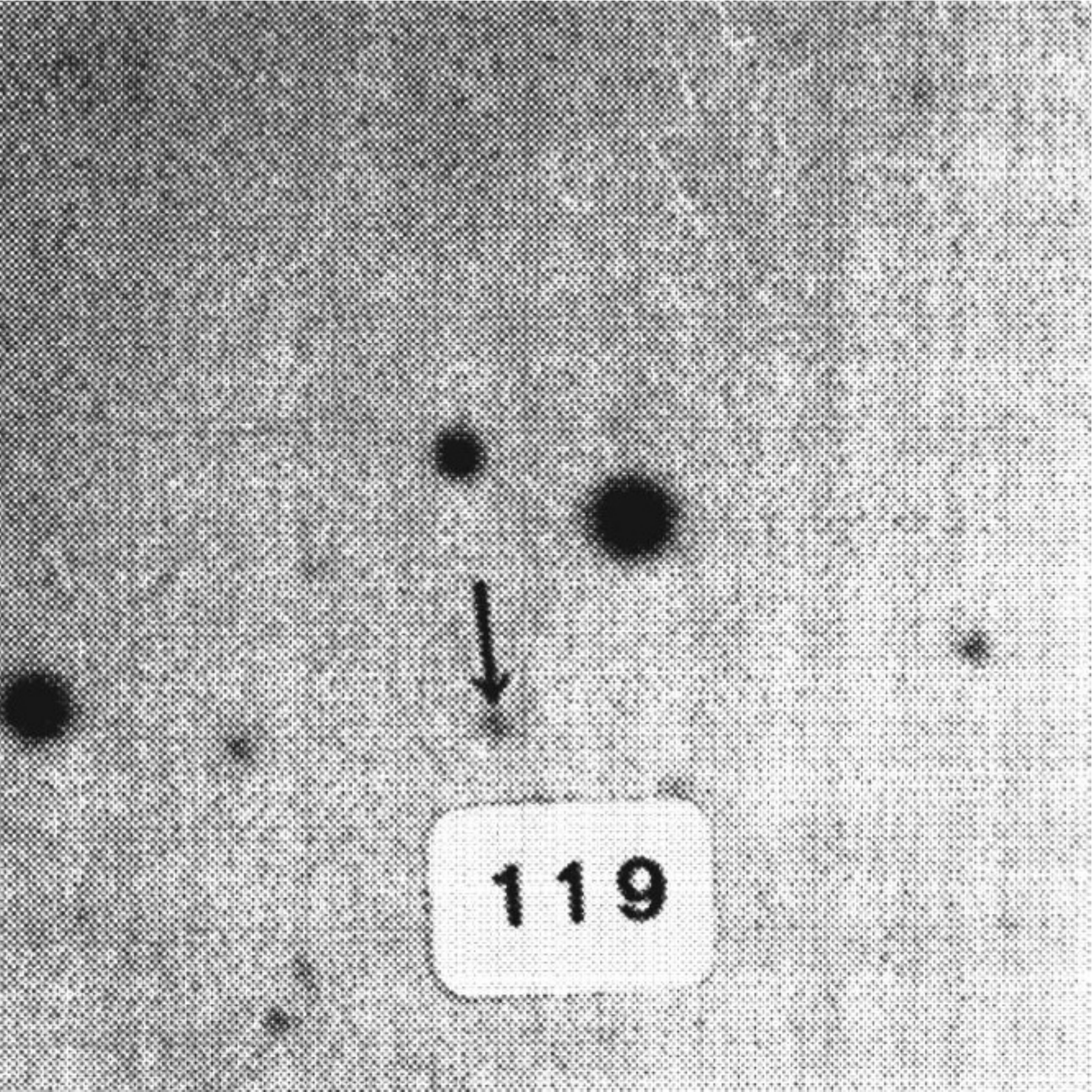}
\includegraphics[angle=0,scale=.28]{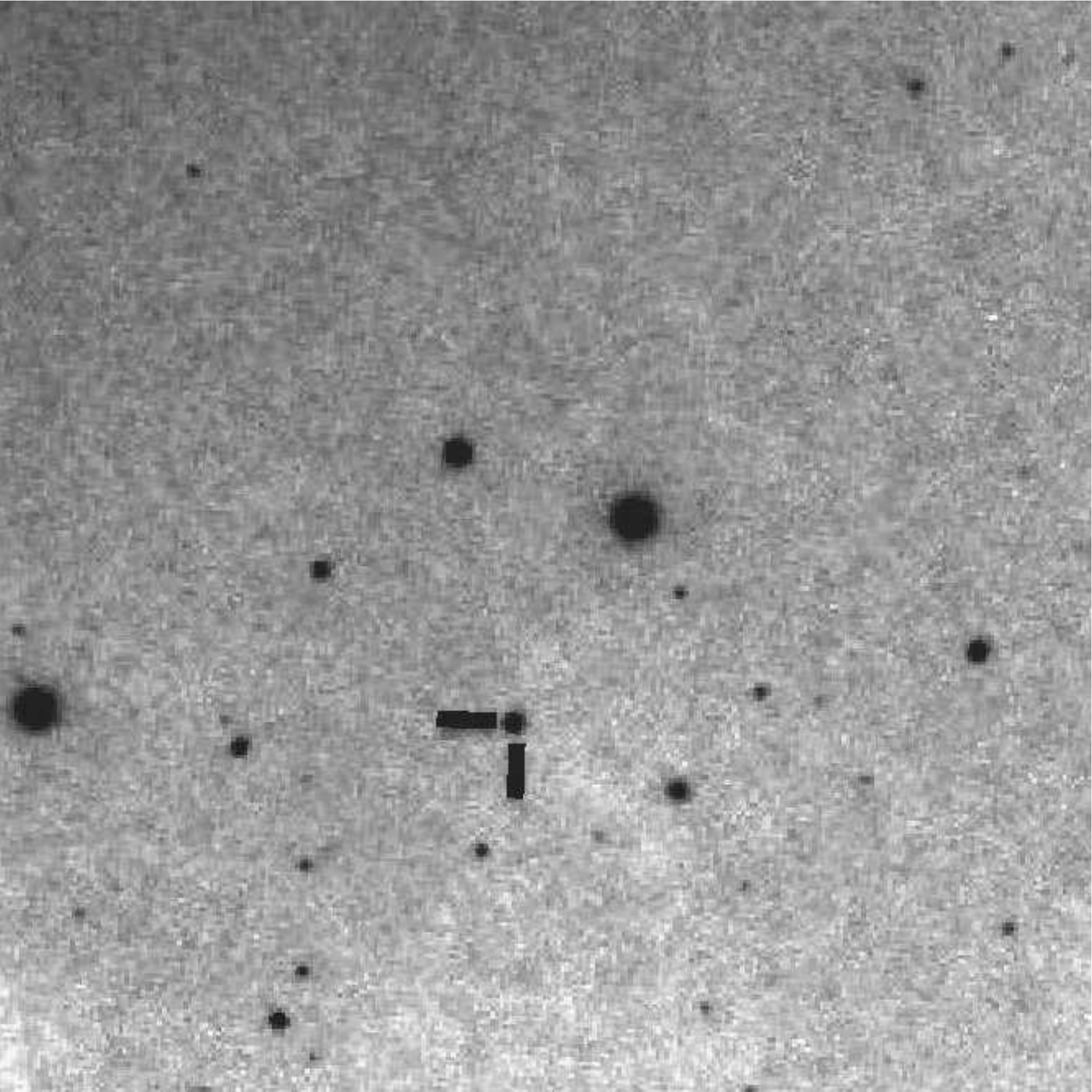}
\includegraphics[angle=0,scale=.28]{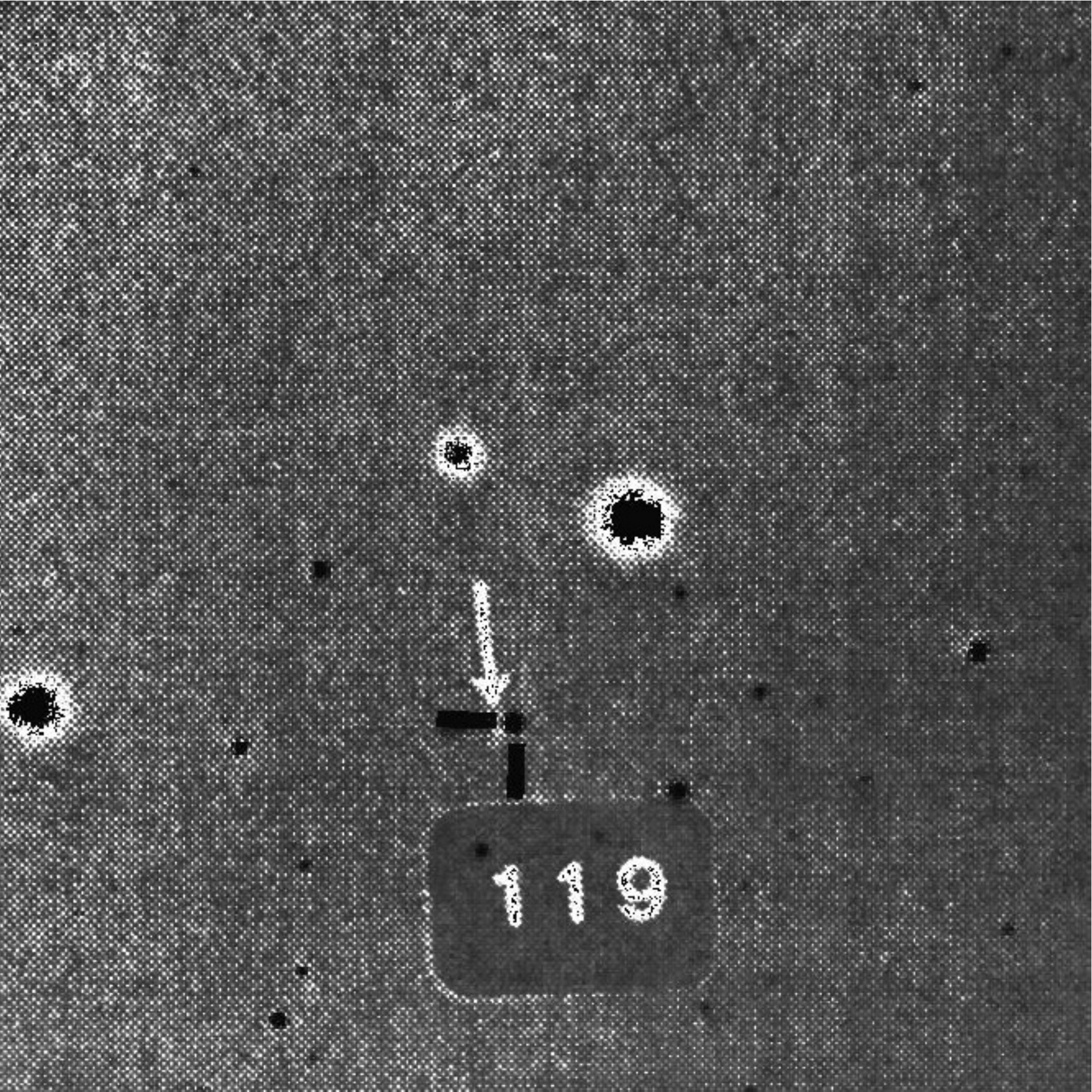}
\caption{Images of M31N 1977-12a, 1998-08a and their comparison
(left, center, and right, respectively).
A careful inspection of the comparison image reveals that
the novae are not coincident, with
M31N 1977-12a (in white) being $\sim2.9''$ ESE of the position of 1998-08a.
The chart for M31N 1977-12a is from \citet{ros89}, with that for
1998-08a from the RBSE at KPNO \citep{rec99}.
North is up and East to the left, with a scale of $\sim2.8'$ on a side.
\label{fig33}}
\end{figure}

M31N 1977-12a is nova \#119 from \citet{ros89}. A possible recurrence
is M31N 1998-08a, which was reported to erupt only $2.1''$ from the nominal
position of 1977-12a. Given that the novae were observed at a
relatively large galactocentric radius ($a=10.8'$), the probability
of a chance positional coincidence is small ($P_C\simeq0.054$).
Despite the likelihood that the novae are related, comparison of the
finding charts (Fig.~\ref{fig33}) shows that the novae are in fact distinct.
Revised astrometry of M31N 1977-12a (see Table~\ref{revcoord}) reveals that
the separation between the eruptions is slightly larger, with
1977-12a lying $\sim 2.9''$ ESE of 1998-08a.

\subsubsection{M31N 1992-12b}

\begin{figure}
\includegraphics[angle=0,scale=.28]{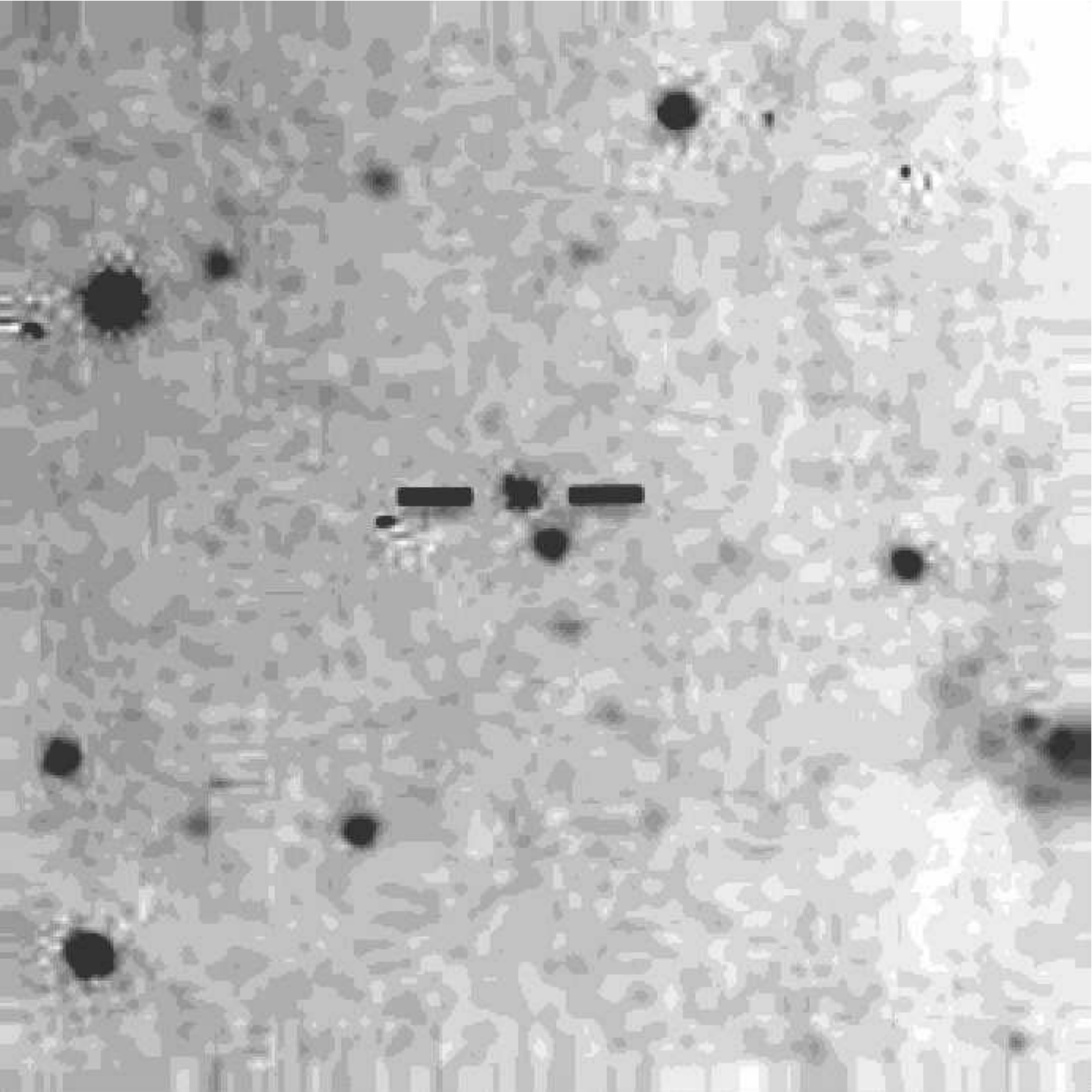}
\includegraphics[angle=0,scale=.28]{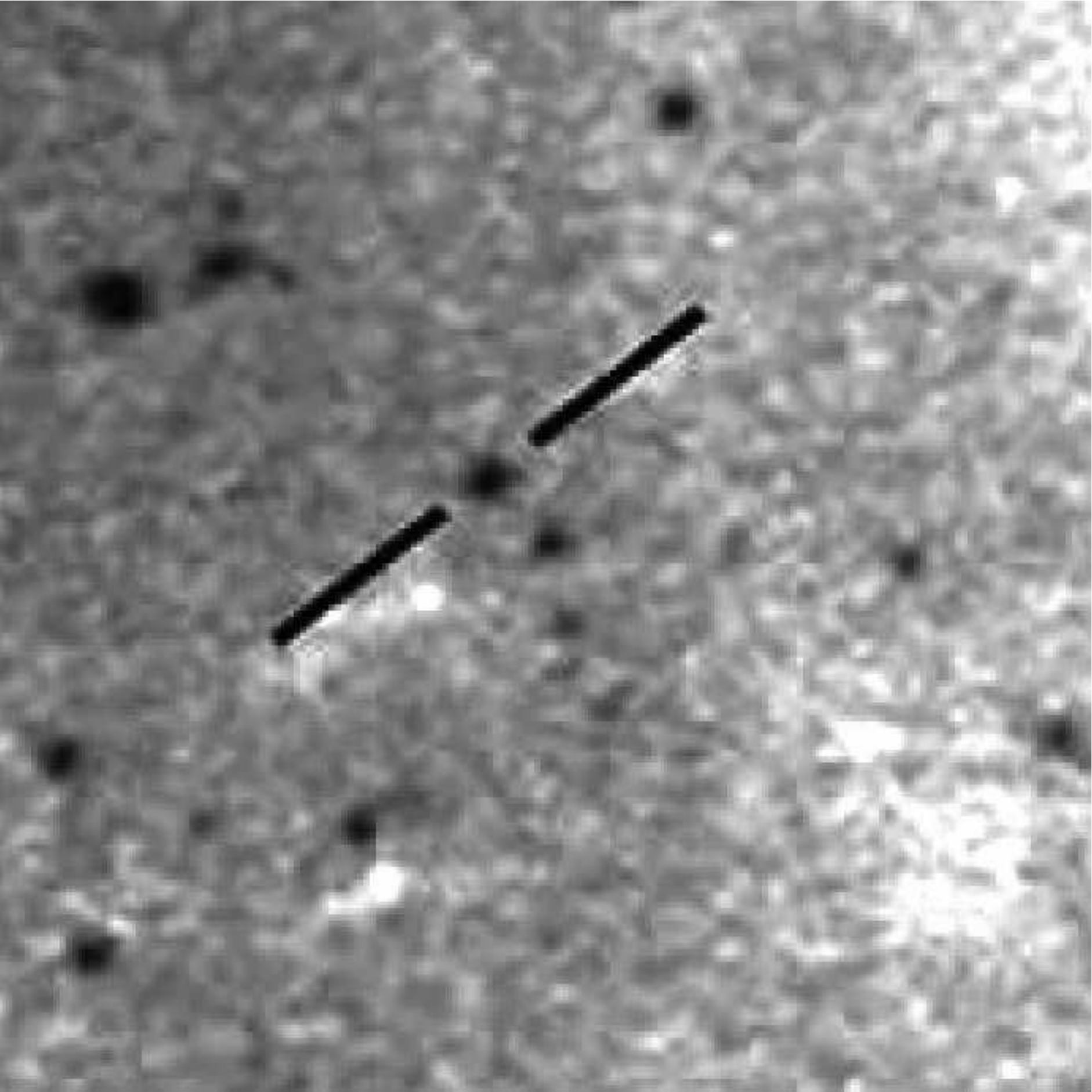}
\includegraphics[angle=0,scale=.28]{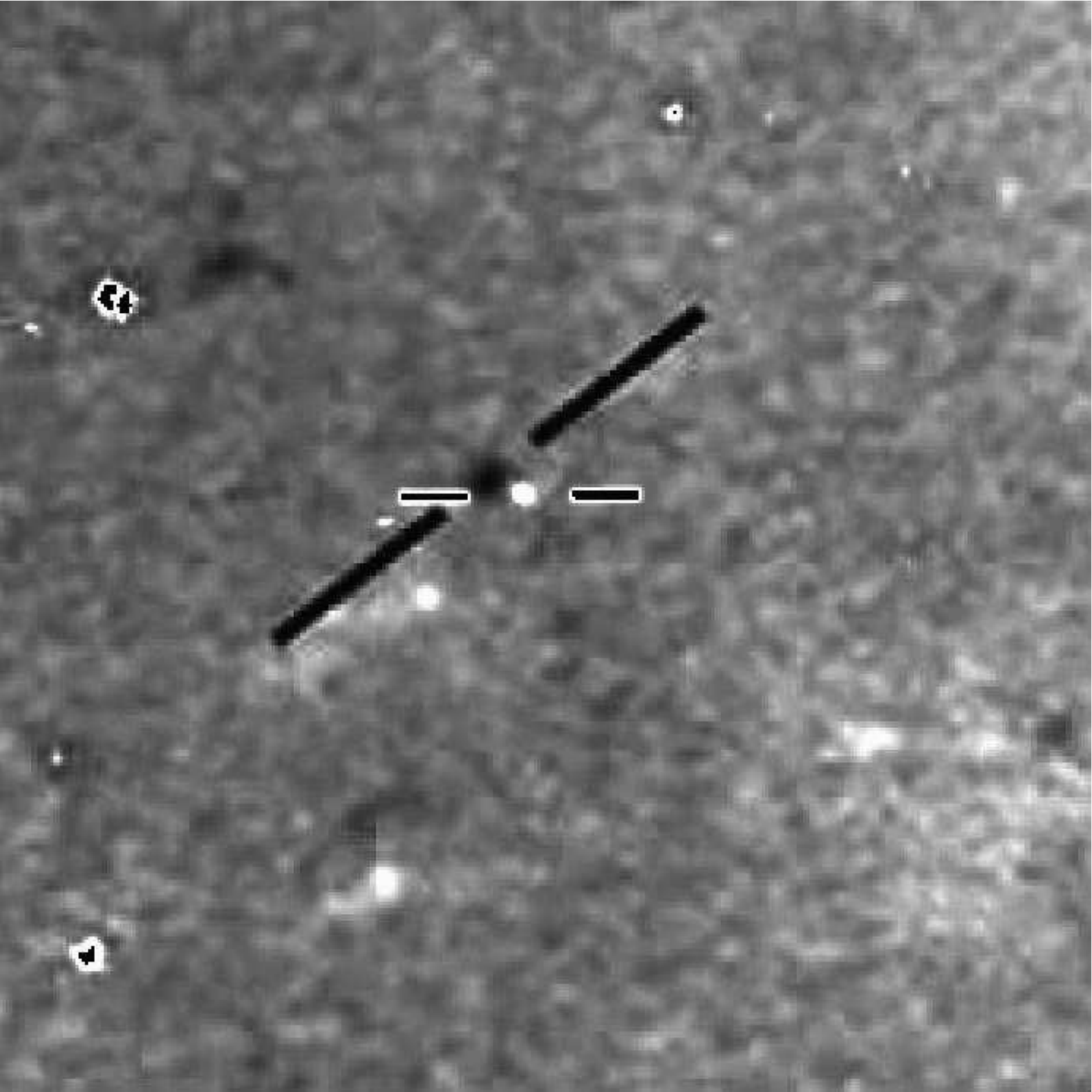}
\caption{Images of M31N 1992-12b, M31N 2001-10f and their comparison
(left, center, and right, respectively).
As demonstrated by the comparison image,
the novae are clearly not coincident, with
M31N 2001-10f (black) being $\sim5.3''$ ENE of the position of 1992-12b.
The chart for 1992-12b is from the survey of \citet{sha01}, and that
of 2001-10f from \citet{alk08}.
North is up and East to the left, with a scale of $\sim2.5'$ on a side.
\label{fig34}}
\end{figure}

M31N 1992-12b was discovered by \citet{sha01} at an isophotal radius of
$a\simeq12.91'$ from the center of M31. A possible recurrence,
M31N 2001-10f was seen to erupt $\sim5.31''$ from the position of 1992-12b.
Given this large separation, it
is unlikely that M31N 2001-10f is related to 1992-12b, despite the modest
probability of a chance positional coincidence ($P_C\simeq0.256$). Revised
astrometry of M31N 1992-12b given in Table~\ref{revcoord}
shows that the position is
within $\sim0.8''$ of that given in \citet{sha01}. The positions of
M31N 1992-12b and 2001-10f are shown in Figure~\ref{fig34}, making it clear that
the two novae are distinct objects.

\subsubsection{M31N 1993-09b}

\begin{figure}
\includegraphics[angle=0,scale=.28]{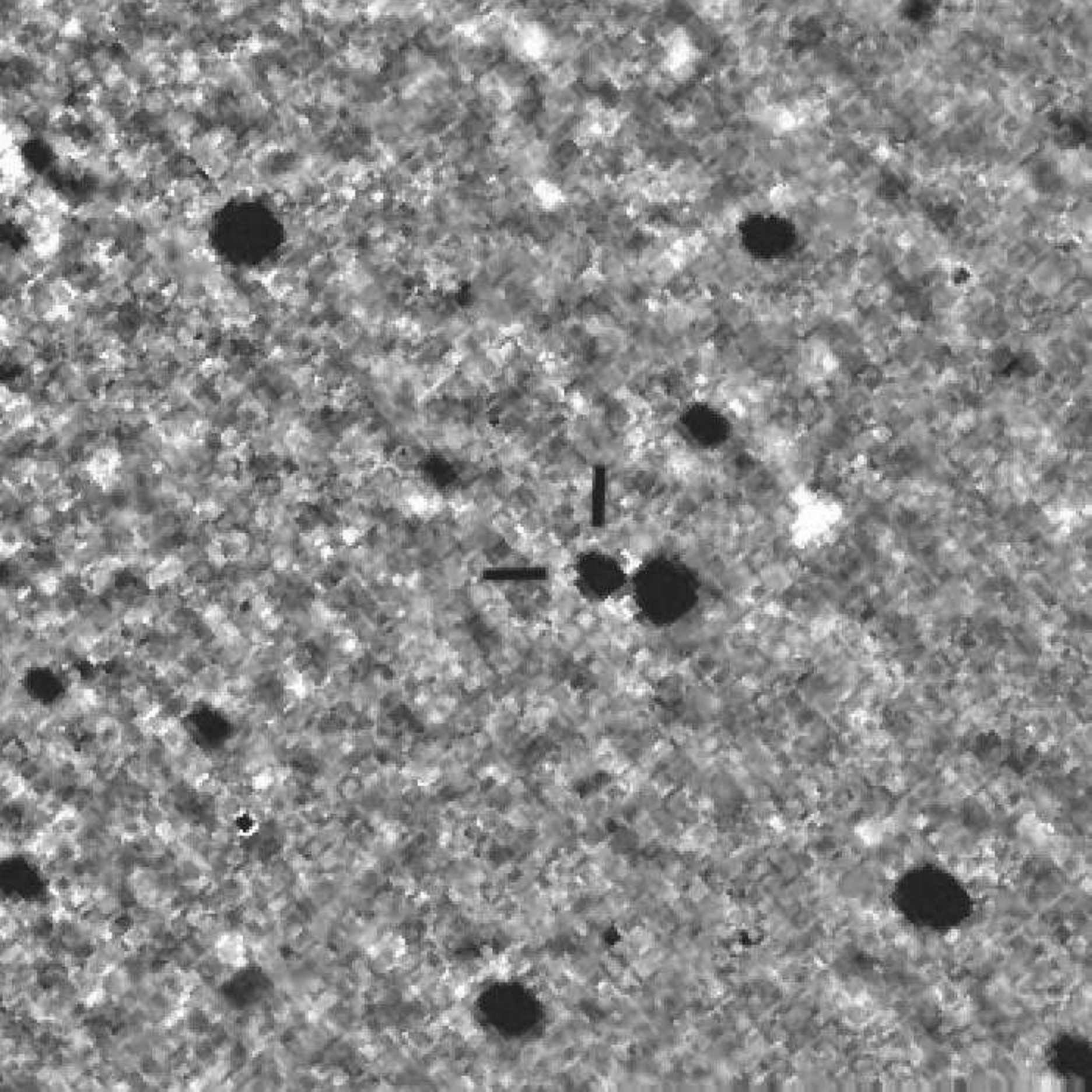}
\includegraphics[angle=0,scale=.28]{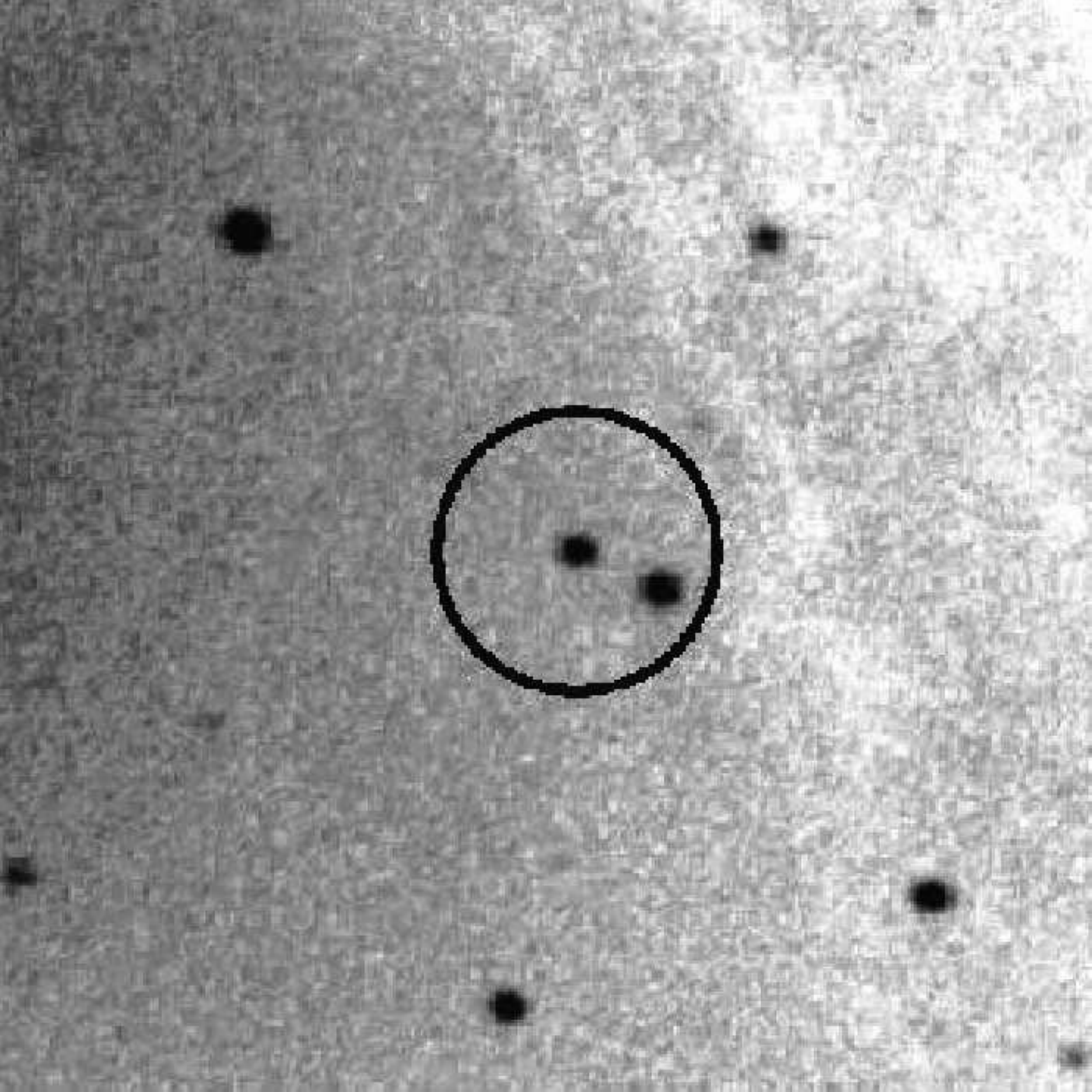}
\includegraphics[angle=0,scale=.28]{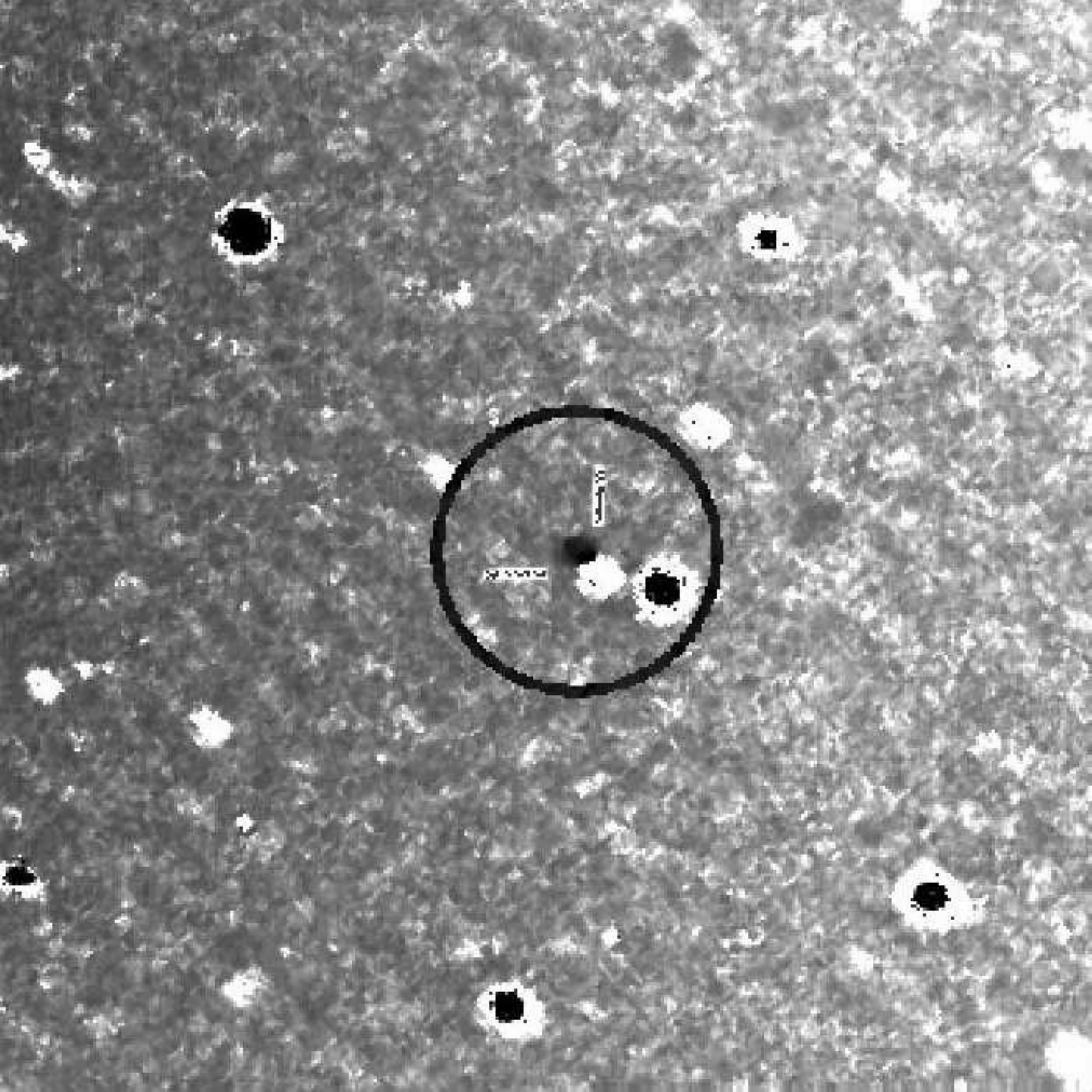}
\caption{Images of M31N 1993-09b, M31N 1996-08g, and their comparison
(left, center, and right, respectively).
The novae are clearly not coincident, with
M31N 1996-08g (in black) being $\sim5.5''$ NE of the position of 1993-09b.
The chart for 1993-09b is taken from the survey of \citet{sha01}, with
that for 1996-08g from \citet{hen08a}.
North is up and East to the left, with a scale of $\sim2.6'$ on a side.
\label{fig35}}
\end{figure}

M31N 1993-09b was another nova discovered in the H$\alpha$ survey of
\citet{sha01}. Approximately 3 years later, a possible recurrence,
M31N 1996-08g \citep{sha97, hen08a}, was observed
$\sim5.47''$ away from the position of 1993-09b. The probability of
a chance positional coincidence is relatively high with $P_C\simeq0.843$.
Figure~\ref{fig35}
clearly establishes that the two novae are distinct objects.
Revised coordinates for M31N 1993-09b are given in Table~\ref{revcoord}.

\subsubsection{M31N 2001-08d}

\begin{figure}
\includegraphics[angle=0,scale=.28]{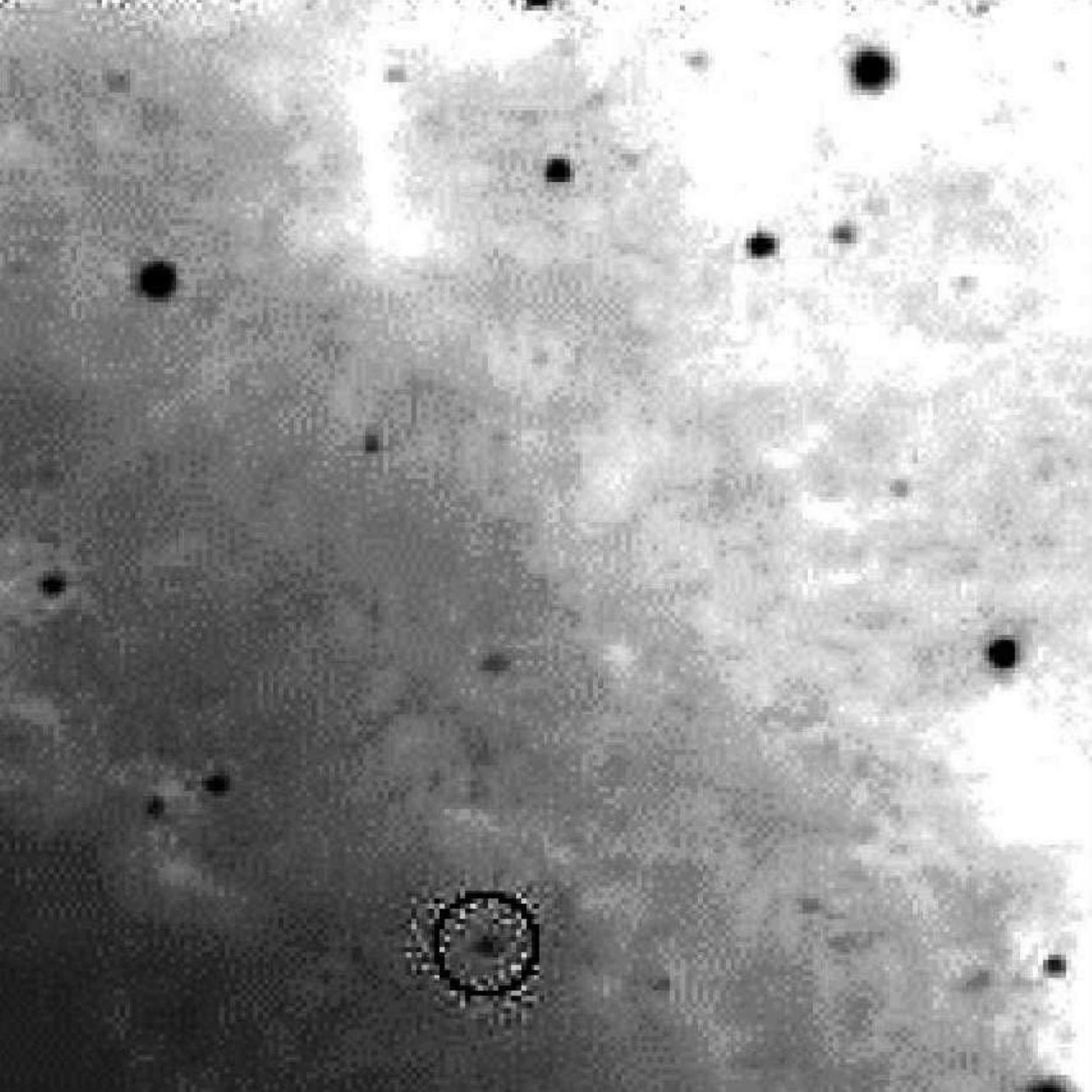}
\includegraphics[angle=0,scale=.28]{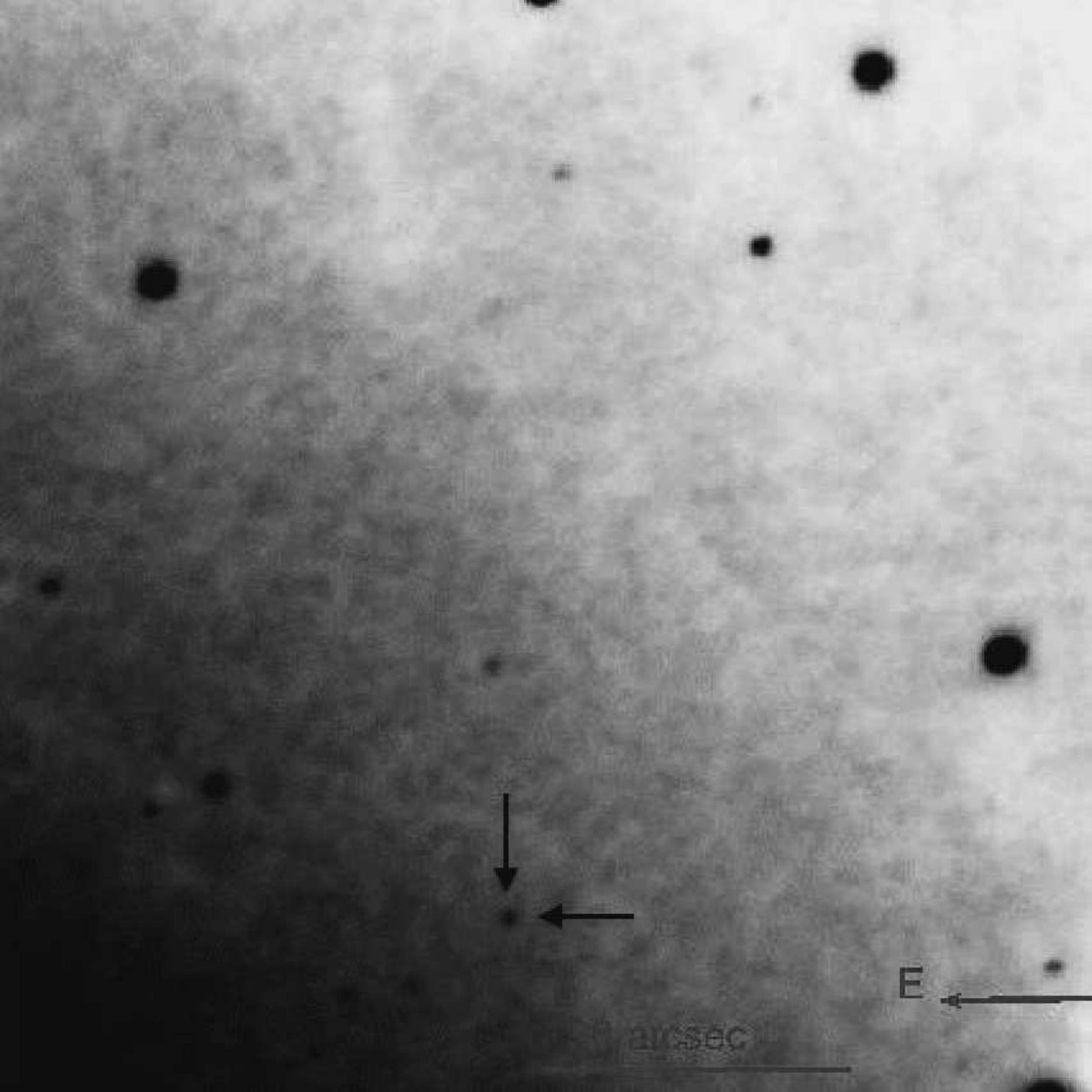}
\includegraphics[angle=0,scale=.28]{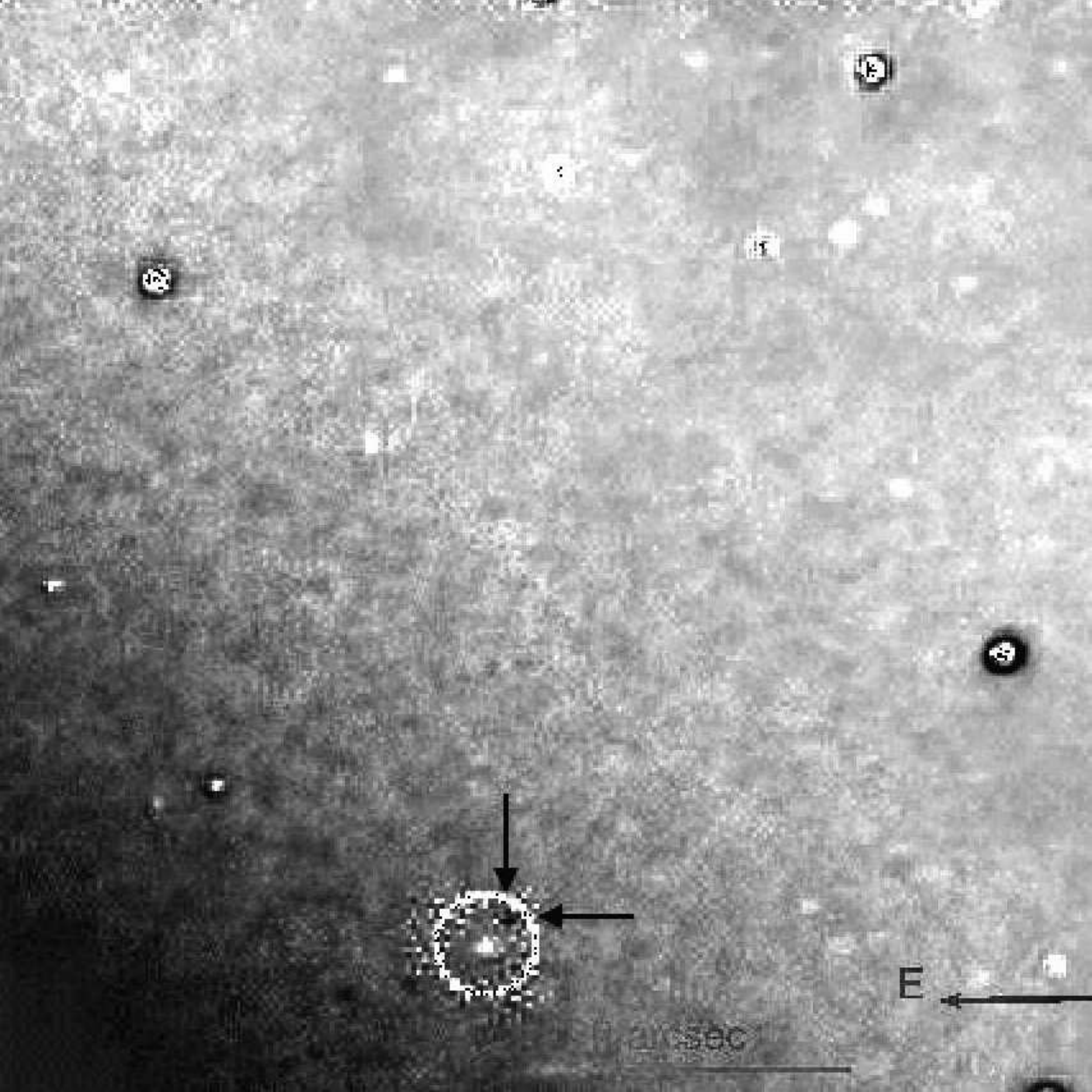}
\caption{Images of M31N 2001-08d \citet{mas06},
M31N 2008-07a \citet{hen08c}, and their comparison
(left, center, and right, respectively).
A careful comparison shows that the novae are not coincident, with
M31N 2008-07a (in black) being $\sim3.5''$ NW of the position of 2001-08d.
North is up and East to the left, with a scale of $\sim1.6'$ on a side.
\label{fig36}}
\end{figure}

M31N 2001-08d was discovered in the bulge of M31 ($a\simeq4.34'$)
by \citet{fia01} on an H$\alpha$ image
of M31 taken on 2001 September 02.93 UT. A possible recurrence, M31N 2008-07a
\citep{hen08b},
was observed within $\sim3.5''$ of the reported position of 2001-08d
with an estimated probability of change positional coincidence given
by $P_C\simeq0.774$. A comparison of finding charts for the two novae
(Fig.~\ref{fig36}) shows that the two novae are not positionally coincident.

\subsubsection{M31N 2004-11b}

\begin{figure}
\includegraphics[angle=0,scale=.28]{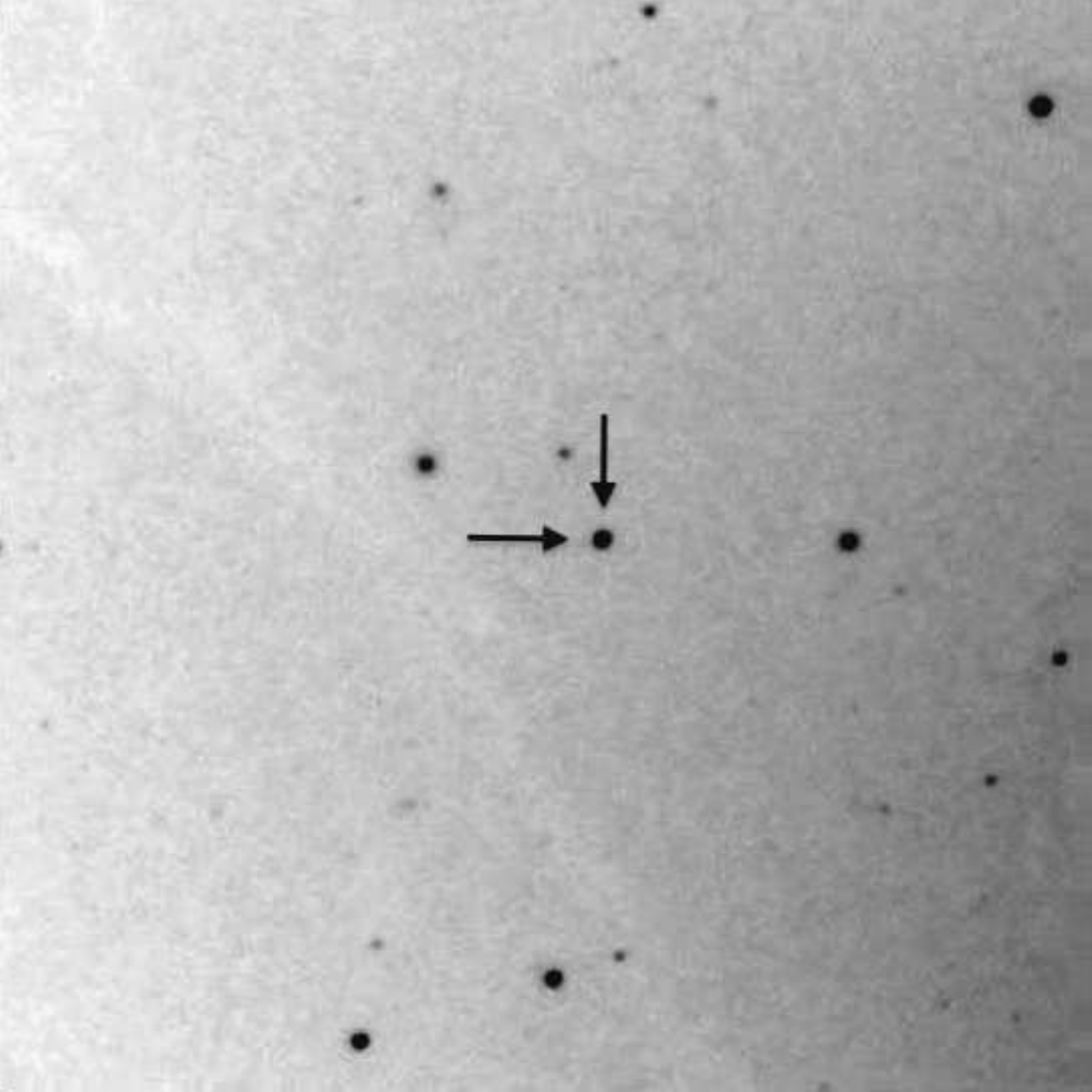}
\includegraphics[angle=0,scale=.28]{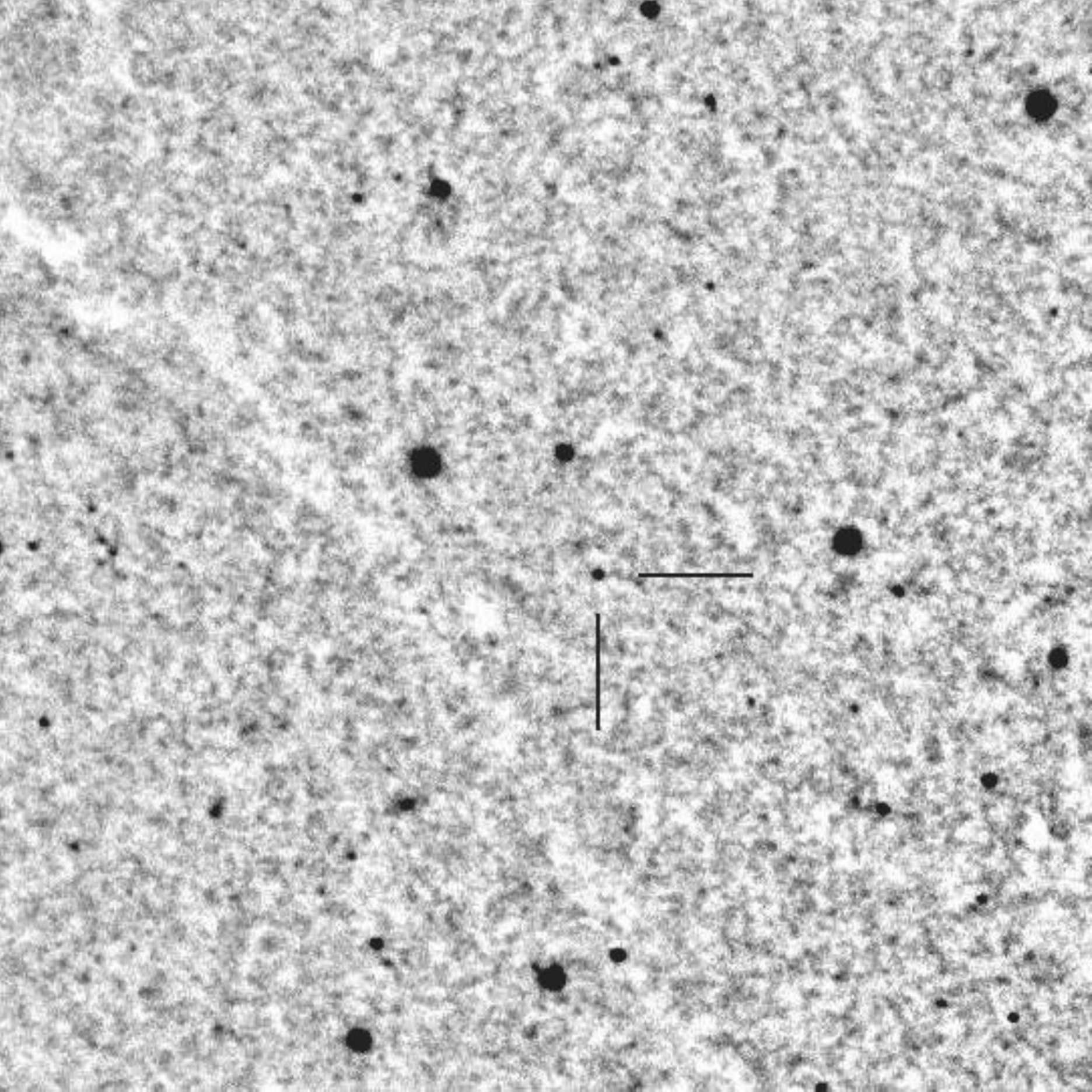}
\includegraphics[angle=0,scale=.28]{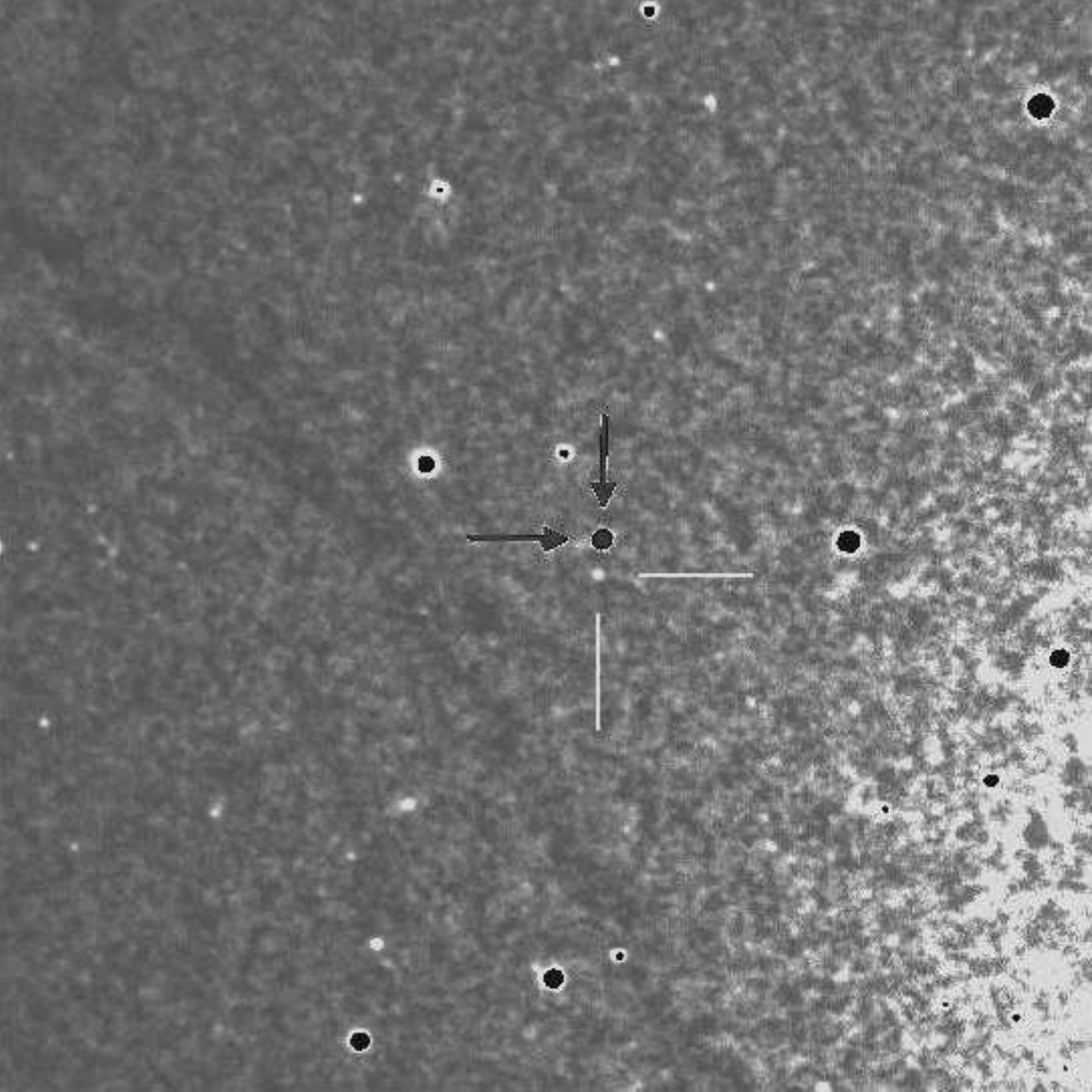}
\caption{Images of M31N 2004-11b,  M31N 2010-07b, and their comparison
(left, center, and right, respectively). As shown by the comparison image,
the novae are clearly not coincident, with
M31N 2010-07b (in white) being $\sim5.9''$ S of the position of 2004-11b.
The images for 2004-11b and 2010-07b were taken by one of us (K.H.)
as part of an ongoing patrol of M31 \citep{pie07a,hor10c}.
North is up and East to the left, with a scale of $\sim3'$ on a side.
\label{fig37}}
\end{figure}

M31N 2004-11b \citep{lee12} and 2010-07b \citep{hor10c}
represent two nova outbursts in the bulge of M31 
($a\simeq5'$) with reported positions within $\sim5.9''$ of one another.
Not surprisingly, given that the coordinates are unlikely to be in error
by more than an arcsec, Figure~\ref{fig37} confirms that the two novae are not
spatially coincident.

\subsubsection{M31N 2005-05b}

\begin{figure}
\includegraphics[angle=0,scale=.28]{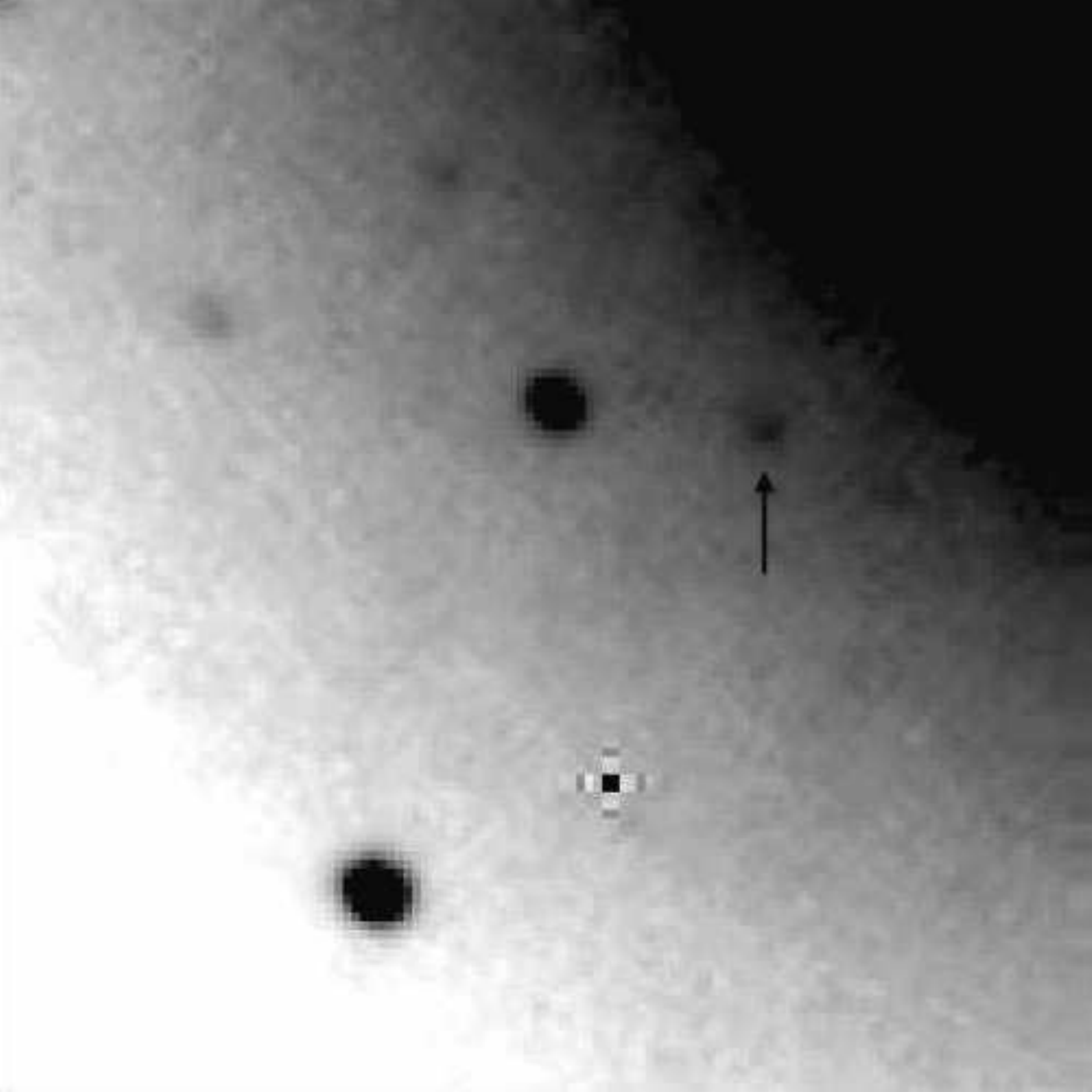}
\includegraphics[angle=0,scale=.28]{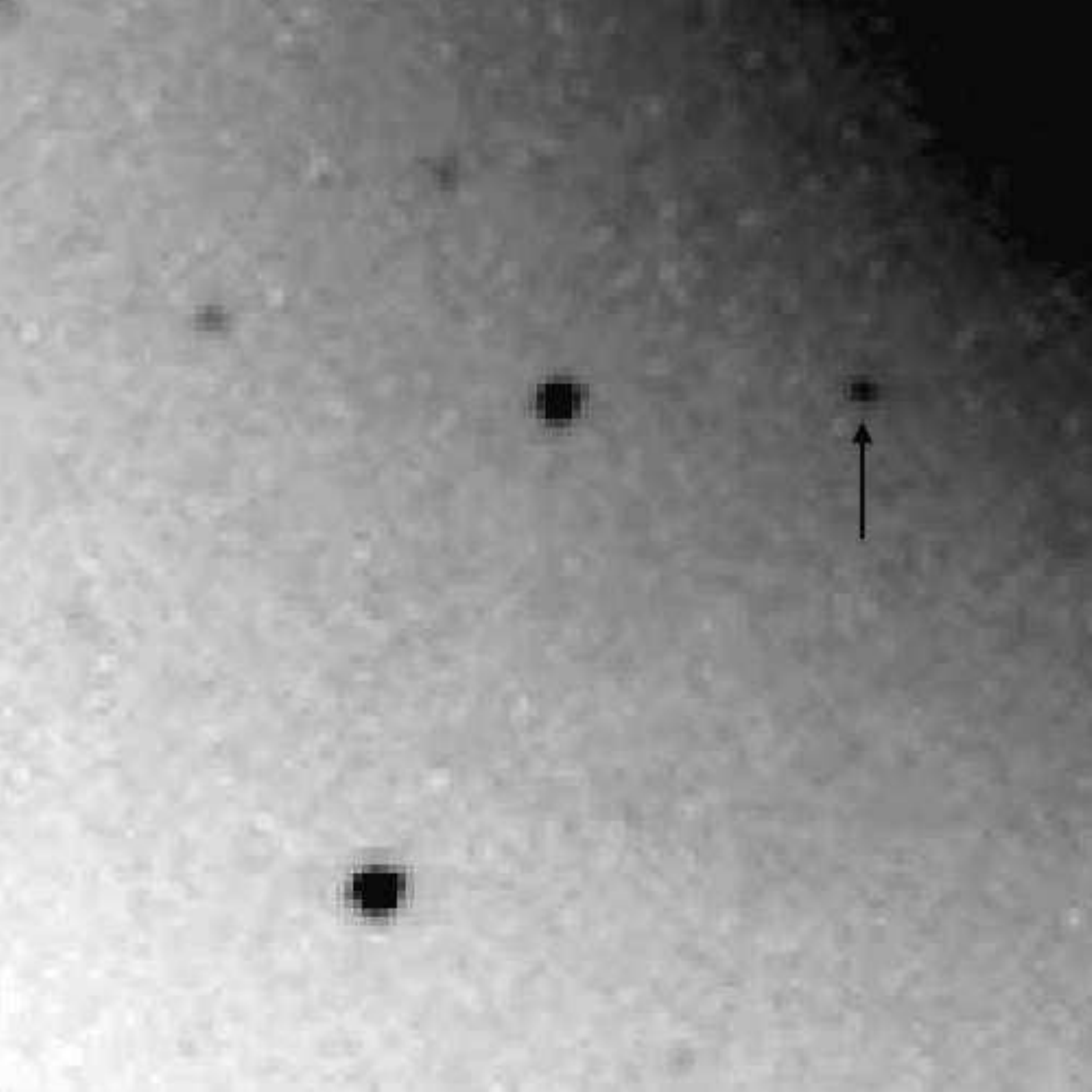}
\includegraphics[angle=0,scale=.28]{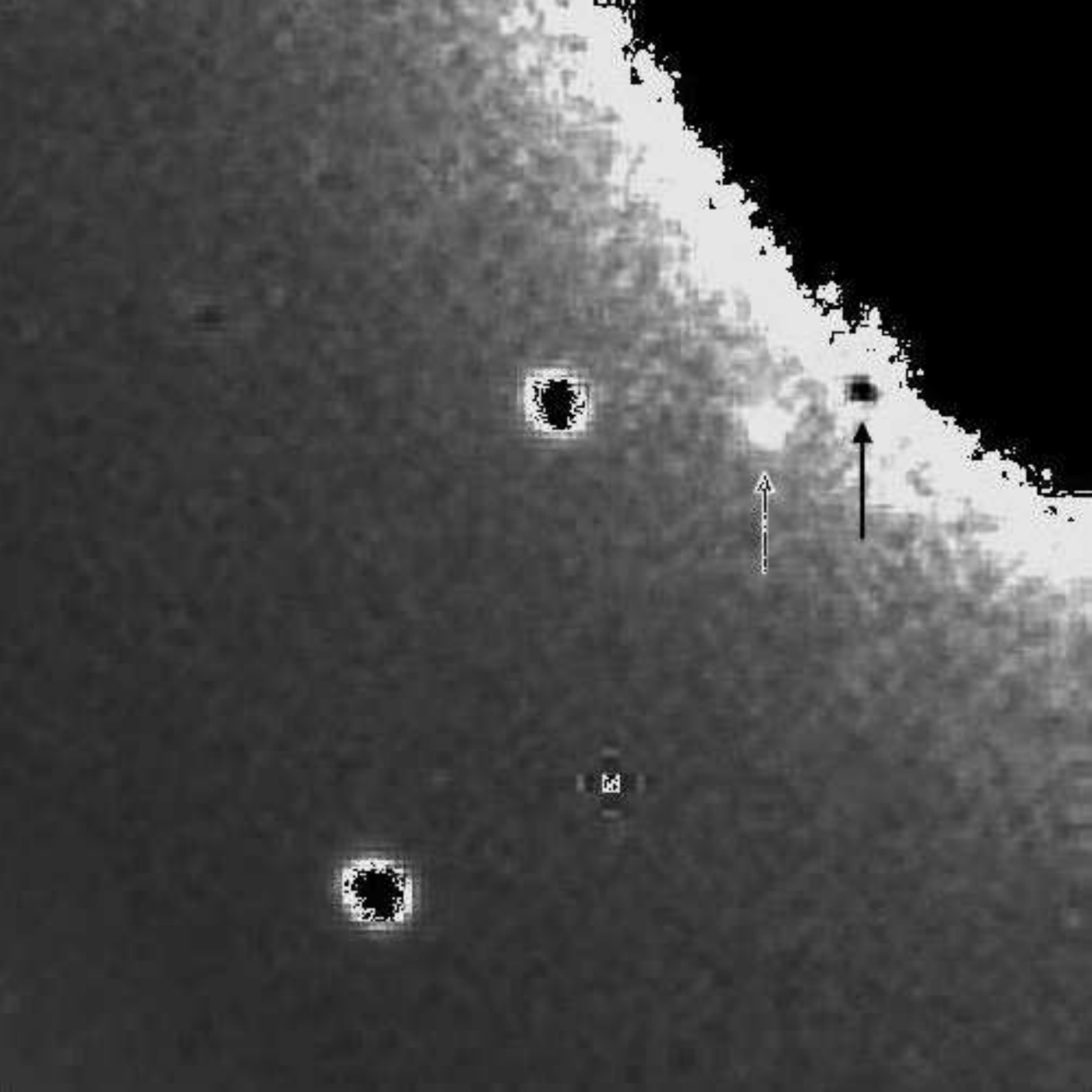}
\caption{Images of M31N 2005-05b, M31N 2009-08d, and their comparison
(left, center, and right, respectively).
These novae are clearly not coincident, with
M31N 2009-08d (in white) being $\sim5''$ WNW of the position of 2005-05b.
Images for both novae come from the RBSE M31 nova patrol.
North is up and East to the left, with a scale of $\sim0.8'$ on a side.
\label{fig38}}
\end{figure}

M31N 2005-05b \citep{lee12} and M31N 2009-08d \citep{hen09a}
erupted
within $\sim5''$ of one another, and
very close to the nucleus ($a\sim0.8'$), resulting in a relatively
high probability of a chance positional coincidence ($P_C\simeq0.998$).
The finding charts from the RBSE data shown in Figure~\ref{fig38}
reveal, as expected, that the objects are distinct systems.

\subsubsection{M31N 2006-11b}

M31N 2006-11b and M31N 2006-12d \citep{lee12}
are two cataloged novae observed near the nucleus
of M31 with reported positions within $0.34''$ of one another. The probability
of a chance coincidence is small ($P_C\simeq0.029$).
The objects appear to be spatially coincident, but it seems clear that
M31N 2006-12d, observed just 38 days later,
is a re-brightening of 2006-11b, and not a distinct eruption.

\subsubsection{M31N 2006-12c}

\begin{figure}
\includegraphics[angle=0,scale=.28]{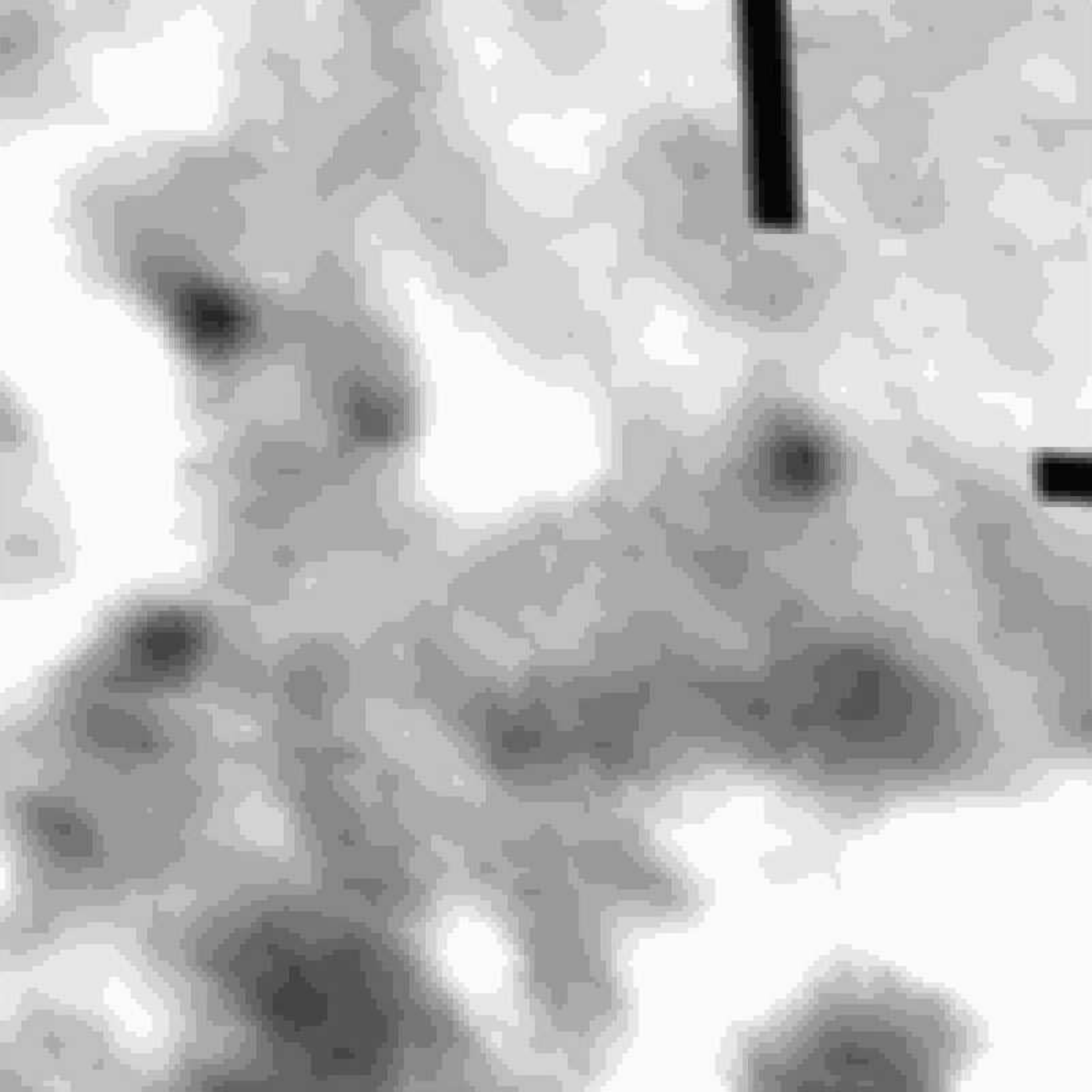}
\includegraphics[angle=0,scale=.28]{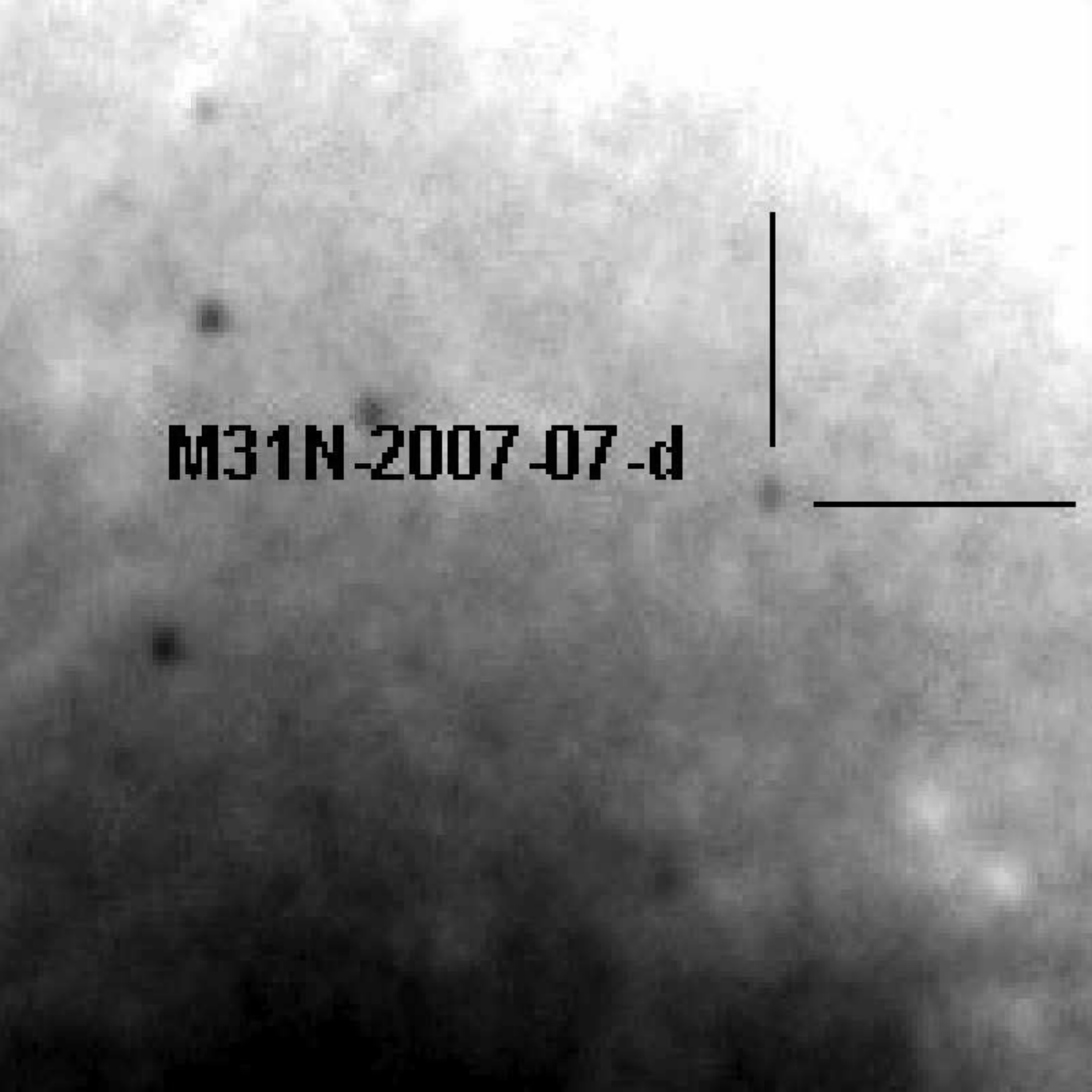}
\includegraphics[angle=0,scale=.28]{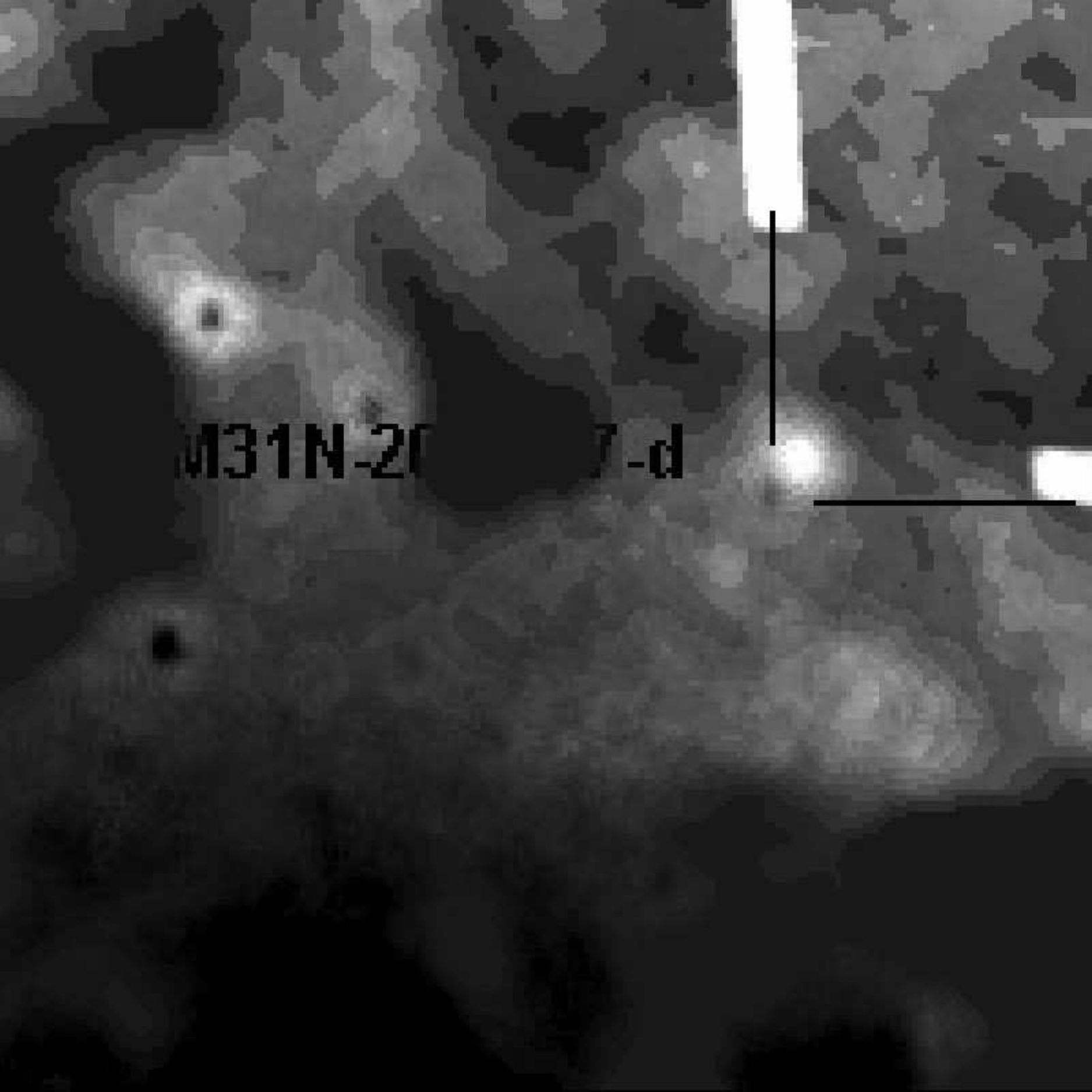}
\caption{Images of M31N 2006-12c, M31N 2007-07e, and their comparison
(left, center, and right, respectively).
The novae are clearly not coincident, with
M31N 2007-07e (in black) being located
$\sim4.3''$ SE of the position of 2006-12c.
The image for M31N 2006-12c was taken from the nova patrol of K.H., while
that of M31N 2007-07e is from the survey of \citet{hat07b}
(Note that the chart for 2007-07e is incorrectly labeled as 2007-07d).
North is up and East to the left, with a scale of $\sim1'$ on a side.
\label{fig39}}
\end{figure}

M31N 2006-12c and M31N 2007-07e \citep{lee12}
erupted close to the nucleus ($a\sim2'$), and
within $\sim4.4''$ of one another resulting in a high probability of a chance
positional coincidence. Once again, finding charts for the objects
(see Fig.~\ref{fig39}) reveal that they are distinct systems.

\subsubsection{M31N 2010-01a}

\begin{figure}
\includegraphics[angle=0,scale=.28]{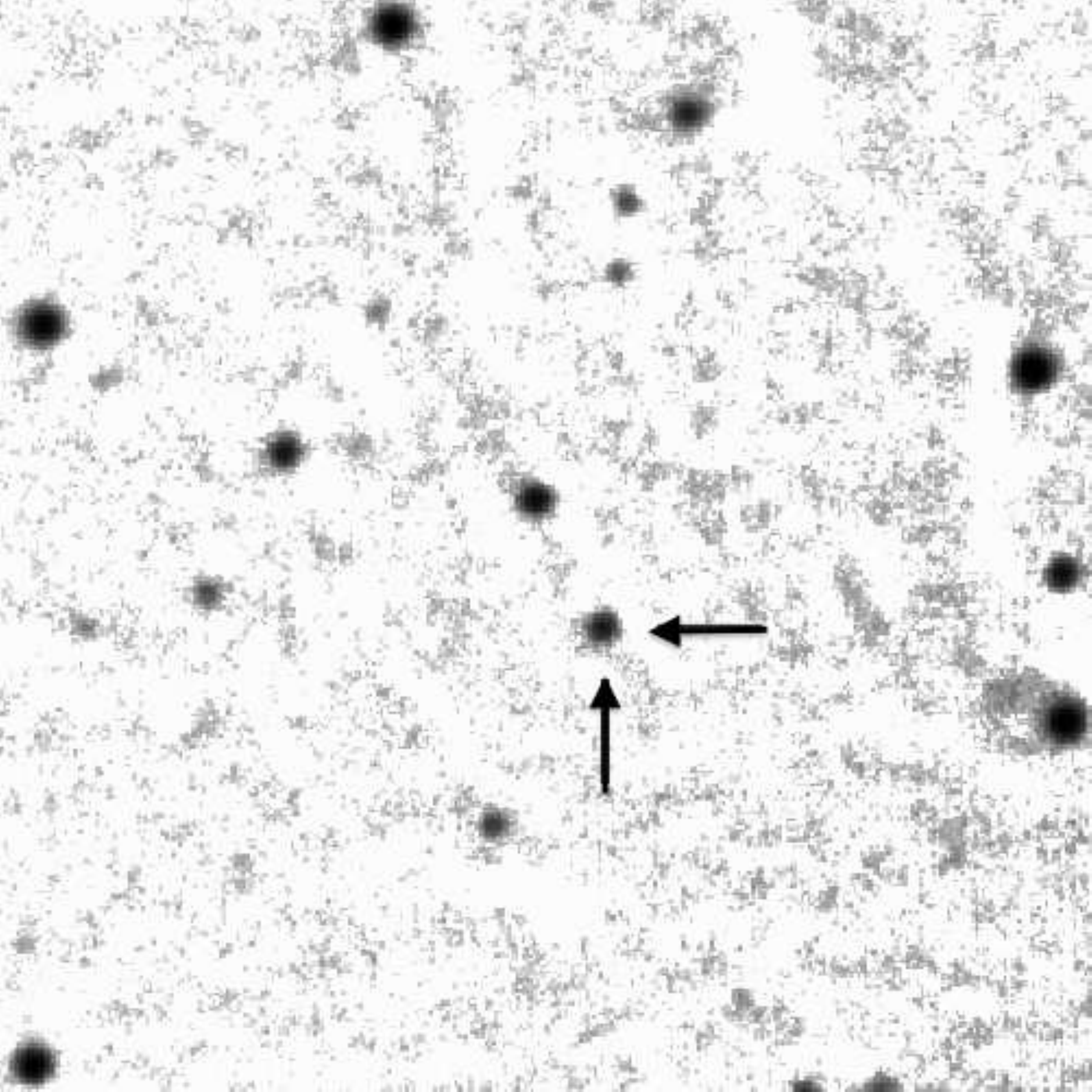}
\includegraphics[angle=0,scale=.28]{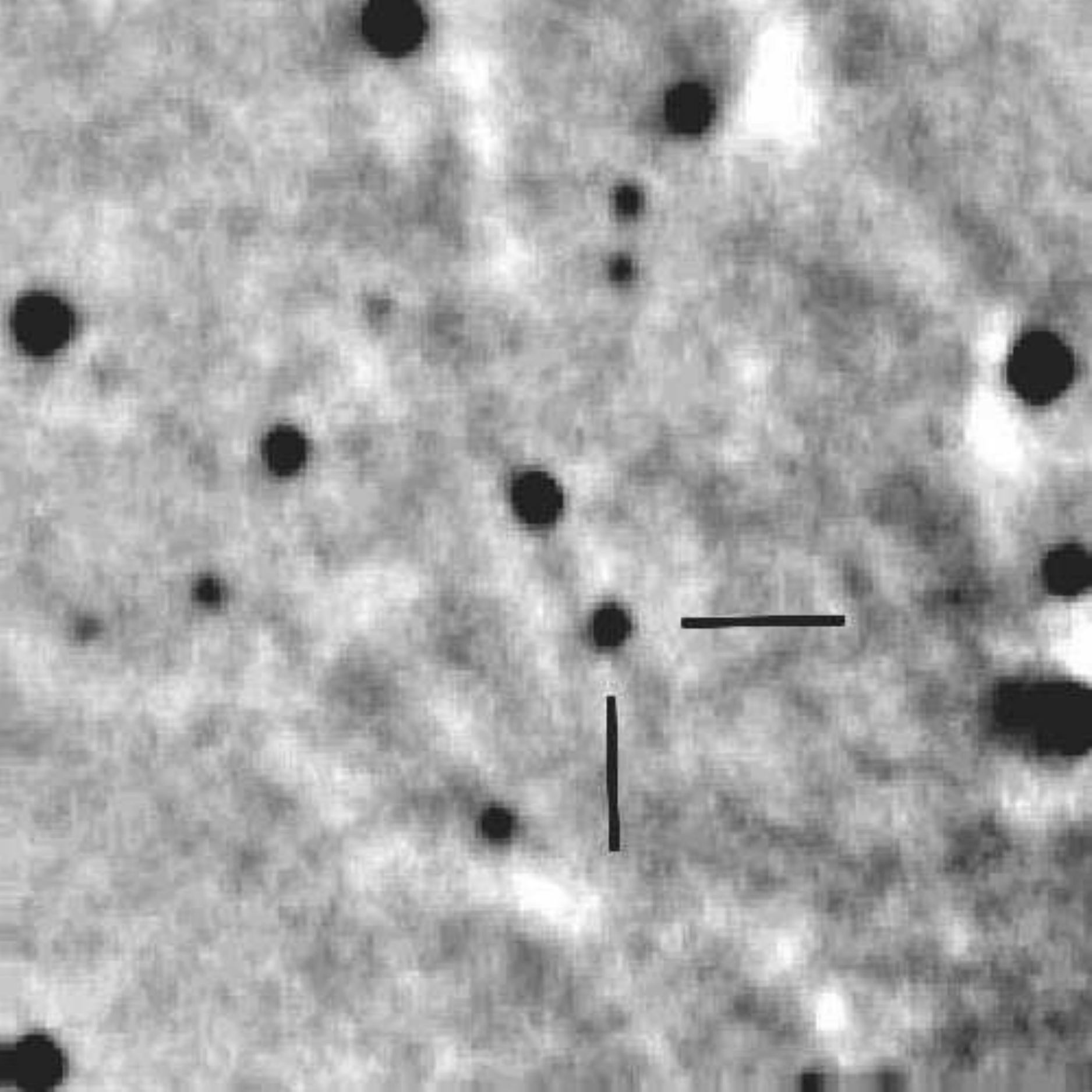}
\includegraphics[angle=0,scale=.28]{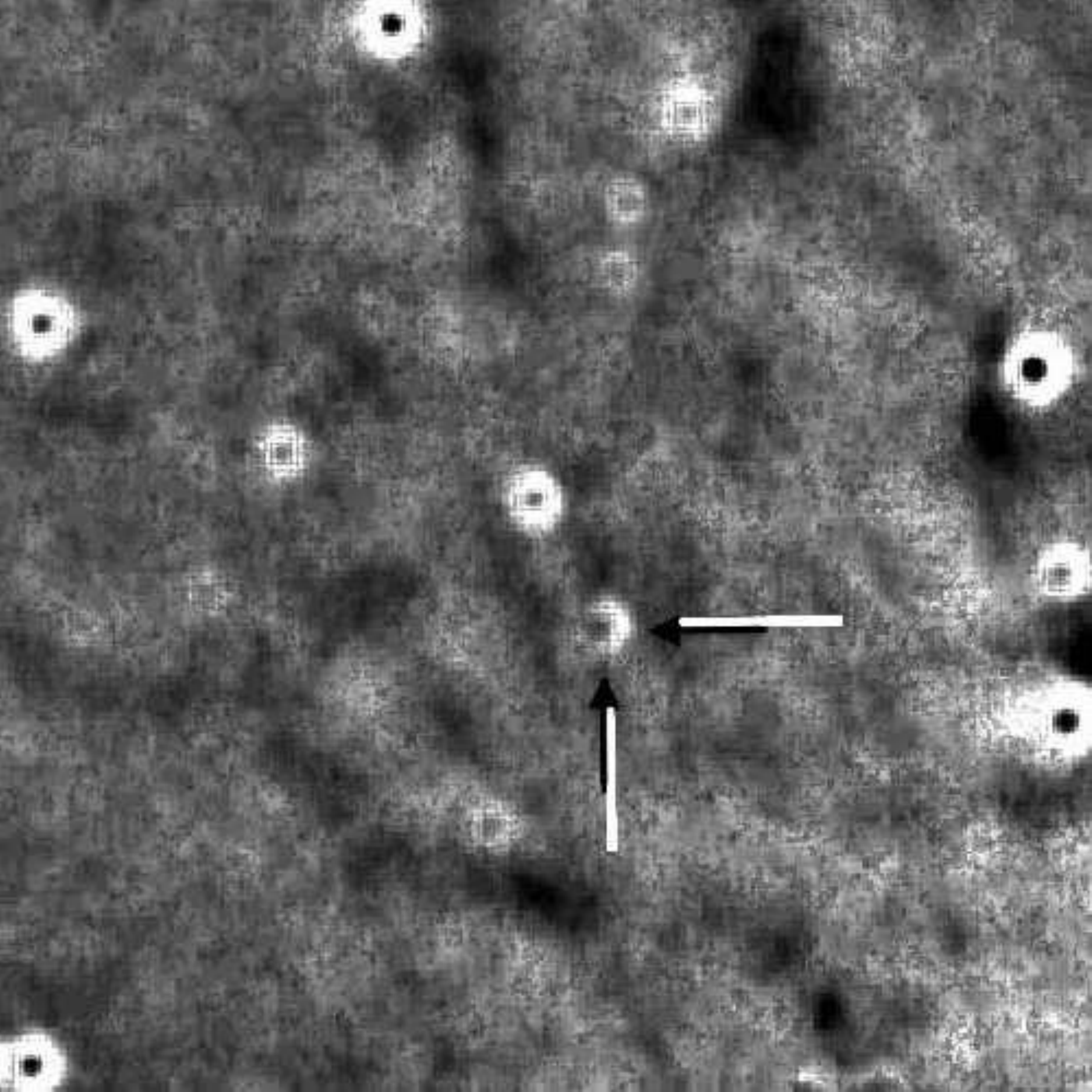}
\caption{Images of M31N 2010-01a, M31N 2010-12c, and their comparison
(left, center, and right, respectively). As can be seen from the comparison
image,
the positions of these novae differ very
slightly with M31N 2010-12c (in white) lying just $\sim0.7''$ NW
of the position of M31N 2010-01a. This result is in agreement with the
findings of \citet{hor10a}.
The image for 2010-01a is from the SuperLOTIS project \citep{bur10},
while that of 2010-12c is from the nova patrol of K.H.
North is up and East to the left, with a scale of $\sim2.5'$ on a side.
\label{fig40}}
\end{figure}

M31N 2010-01a \citep{bur10}
was discovered close to the nucleus of M31 ($a\simeq2.7'$),
and observed to have a possible recurrence less than a year later,
M31N 2010-12c \citep{hor10a,rua10b}, which was reported just $0.82''$ away.
Despite a probability
of a chance positional coincidence of just $P_C\simeq0.141$,
a careful inspection of the images for each nova shows that they
are in fact distinct objects (see Figure~\ref{fig40}).
This conclusion is supported
by the work of \citet{hor10a}, who performed
careful astrometry of both novae and also came to the conclusion that
they are in fact distinct systems.
Revised coordinates based on astrometry performed on the charts
shown in Figure~\ref{fig40} are given in Table~\ref{revcoord}.
We conclude that M31N 2010-01a is not a RN.

\section{The Recurrent Nova Sample in M31} 

Based on the analysis of the RN candidates from Table~\ref{rntable}
as described in the preceding section,
we have determined that there are a total of at least 12 RNe
that have been observed in M31 over approximately the past century.
Four additional systems, M31N 1953-09b, 1961-11a, 1966-08a,
and 1990-10a are considered possible RNe.
These 16 RN systems, representing 39 independent nova eruptions,
are summarized in Table~\ref{rntable_conf}, where we have
recalculated the observed separation, $s$, and updated the
probabilities for chance positional coincidence, $P_C$, to
reflect the revised coordinates given in Table~\ref{revcoord}.
We note that
since not all of the RNe identified above have been spectroscopically confirmed,
we cannot completely eliminate the possibility that one or more LPVs could
be lurking in our RN sample.

\subsection{Spatial Distribution}

\begin{figure}
\includegraphics[angle=0,scale=.80]{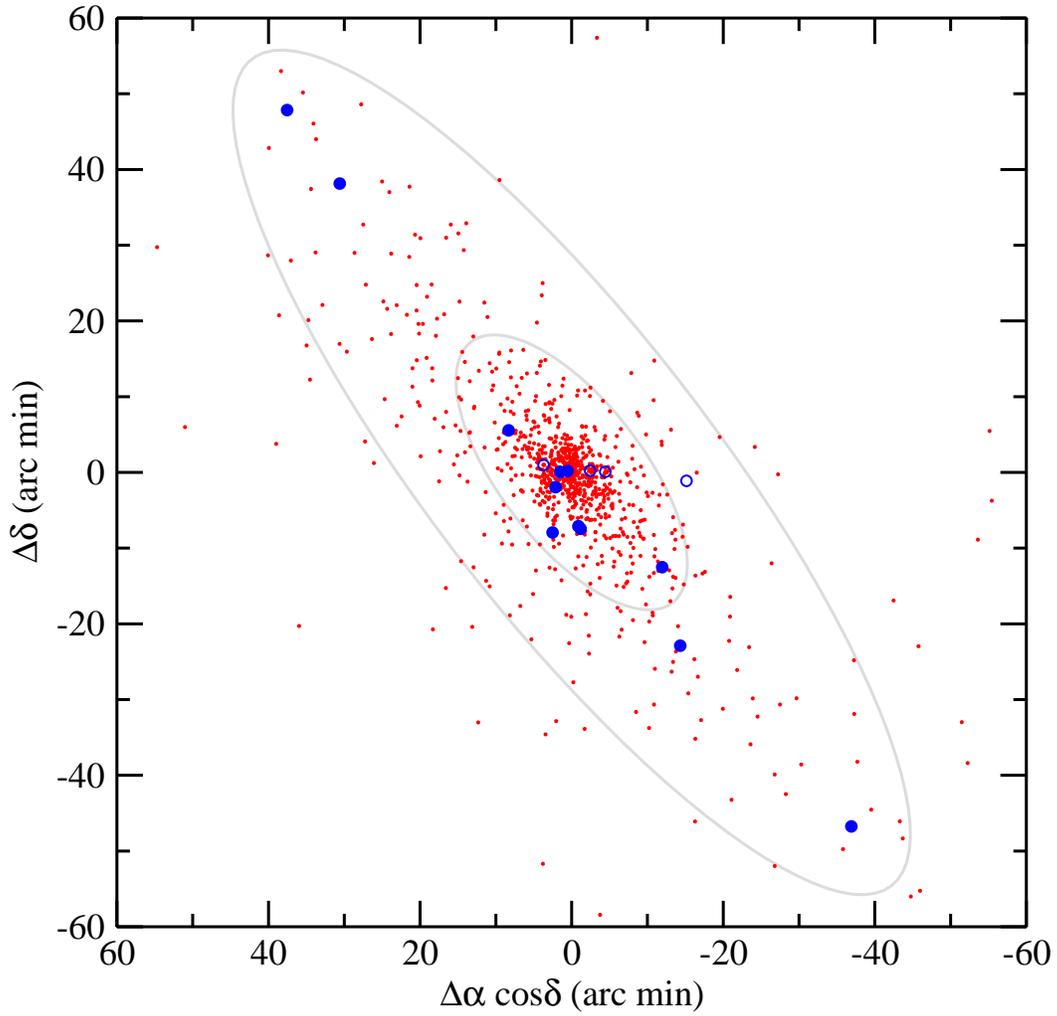}
\caption{The spatial distribution of RNe compared with all novae
in M31 (small red points). The filled blue circles represent
the confirmed RNe, with the
blue open circles showing the positions of the possible RNe.
Two $B$-band isophotes are shown to illustrate the orientation
of M31 on the sky.
\label{fig41}}
\end{figure}

\begin{figure}
\includegraphics[angle=90,scale=.70]{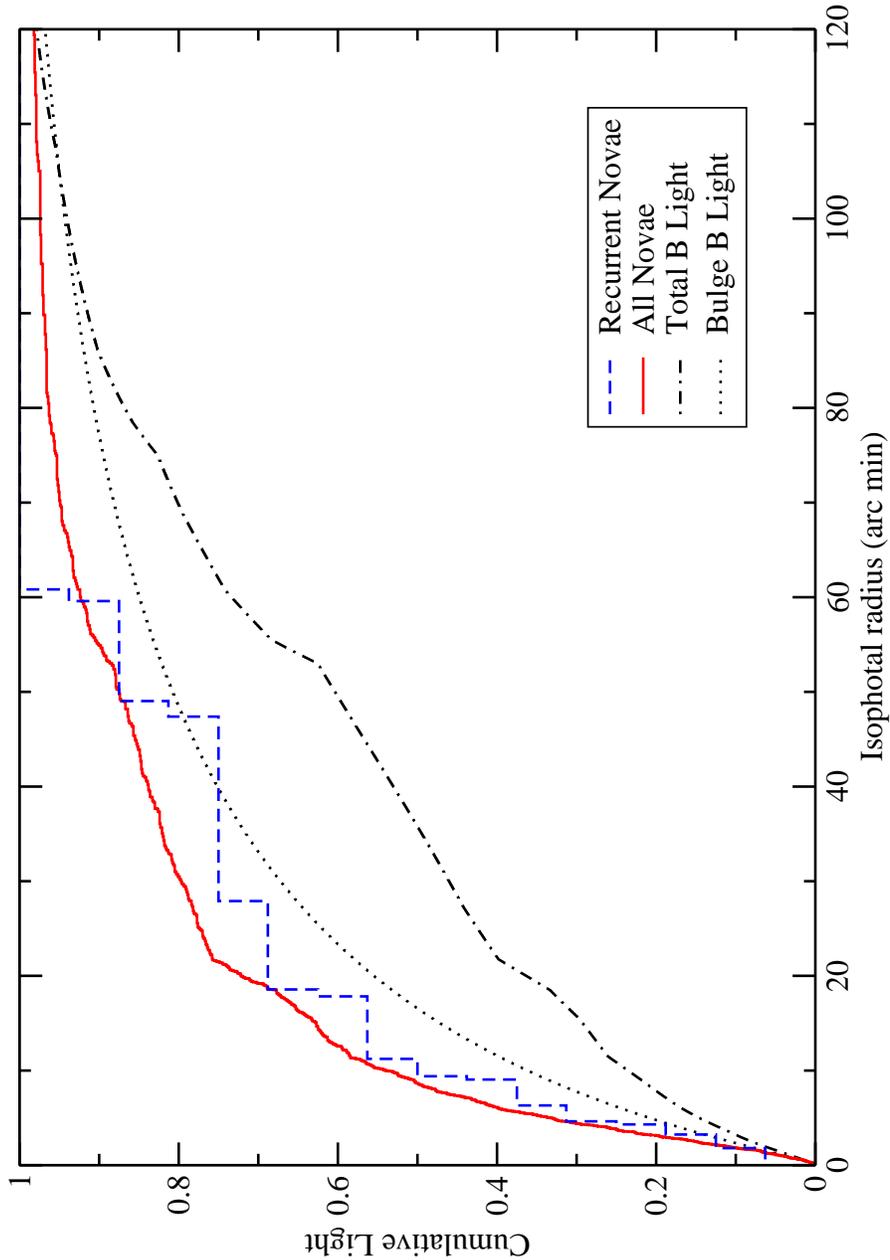}
\caption{The cumulative RN distribution (including the 4 possible RNe)
compared with the cumulative distribution for all novae and with
the background $B$-band bulge and total light. There is no significant
difference between the cumulative distributions for RNe and novae in general
(KS = 0.95). The nova distributions fall off faster then the background
light, as expected given that the nova surveys are not spatially complete,
particularly in the outer regions of M31.
\label{fig42}}
\end{figure}

Figure~\ref{fig41} shows the spatial distribution of the 12 RNe and
the 4 possible RNe compared with all novae in our sample. There is no obvious
difference in the overall
spatial position of the RNe compared with that of
novae in general. To explore the distributions more quantitatively, we
have plotted the cumulative distributions of the RNe and the remaining
nova sample as a function of isophotal radius in Figure~\ref{fig42}.
For comparison, we have included the cumulative $B$-band light for M31 as
well as an estimate of the cumulative bulge light. The cumulative
$B$-band light was computed from the surface photometry of \citet{dev58},
with the cumulative bulge light computed from a standard $r^{1/4}$ law

\begin{equation}
\mu(r) = \mu_e + 8.33 [({r/r_e})^{1/4} -1]
\end{equation}

\noindent
parameterized by $\mu_e=22.9$ mag arcsec$^{-2}$ and $r_e=18'$ for $B$-band
light as found by \citet{cia87}.

It is clear from Figure~\ref{fig42} that there is no discernible difference
between the cumulative distributions for RNe and novae in general (KS = 0.95),
however both distributions differ appreciably from the integrated bulge
and (especially) total light. The poor agreement between the nova
distributions and the light distributions is not unexpected, and can be
explained by the fact that the combined nova sample is far from being spatially
complete. The Andromeda galaxy covers a relatively large angular area
with the major axis extending more than 3 degrees on the sky.
Most M31 nova surveys over the years have concentrated on the inner
(bulge dominated)
regions of the galaxy, with extended coverage being far more sporadic.
As a result, the cumulative nova samples are observed to fall off faster
than the cumulative background light, with the bulge light providing a
better fit than the total (bulge + disk) light. In cases where the
nova distribution from a given spatially-defined survey is compared with
the cumulative light included in that particular survey, the nova distributions
have been shown to agree quite well with the integrated bulge (but not the
total) light \citep[e.g., see][]{cia87,sha01,dar06}.

\subsection{Optical and X-Ray Outburst Properties}

The optical and X-ray outburst properties of the full sample of RNe
are summarized in Table~\ref{rntable_op}. The optical properties include the
outburst discovery magnitude, $m_{\rm dis}$, and when known, the time for
the nova to fade by two magnitudes ($t_2$) from maximum light. X-ray
properties have been measured for five RNe, including
the turn-on and turn-off times ($t_{\rm on}$, $t_{\rm off}$) for the SSS
phase of the eruption, and the best-fit blackbody temperature, $T_{\rm bb}$,
for three of the systems. When known, we also have included the
spectroscopic class of the nova.
Unfortunately, very little photometric
or spectroscopic data are available for most of the initial RN outbursts,
mostly because the first recorded outbursts occurred prior to the
time that routine
imaging and spectroscopic observations became feasible.
In no cases do we have spectroscopic classes available for
the initial outburst, and in only two cases
(M31N 1963-09c and 1997-10f) do we have estimates of
the rate of decline from maximum light. The
$t_2$ times for
M31N 1963-09c and 1997-10f are given by $t_2\grtsim17$~d and $t_2\sim10$~d,
respectively. Despite the uncertain $t_2$ value of M31N 1963-09c,
the estimates of $t_2$ for the three recurrences
are consistent with one another, and establish that the nova
declined rapidly. In addition, \citet{sha10} established that 2010-10e
belonged to the He/N spectroscopic class. Thus, both the photometric and
spectroscopic properties are consistent with the identification of
M31N 1963-09c as a RN. In the case of M31N 1997-10f and its
likely recurrence 2008-08b, we have no estimate of the $t_2$ value
of the latter eruption, however an optical spectrum by \citet{dim10}
shows that the object is a likely member of the He/N class.

The remarkable RN system M31N 2008-12a displayed 4 observed recurrences
through the end of 2013:
2009-12b, 2011-10e, 2012-10a, and 2013-11f
\citep[e.g., see][]{dar14a,hen14b,tan14}.
Available photometry
for two of the outbursts (2011-10e and 2013-11f)
is sufficient to establish
that the nova faded extremely rapidly ($t_2\simeq2$~d), while spectroscopic
observations of the last two eruptions establish that the nova is a
member of the He/N spectroscopic class. Such systems comprise approximately
15\% of novae observed in M31, with the remaining 85\% mostly belonging
to the Fe~II class \citep{sha11b}. The former novae are characterized
by relatively high ejection velocities
(FWHM of H$\alpha$ typically $\grtsim2500$ km~s$^{-1}$) and rapid
photometric development as expected from a nova arising from a massive
white dwarf. As explained in \citet{hen14b},
the relatively modest peak optical absolute magnitude of $M_{\rm max}\sim-6$
is consistent with the extremely short recurrence period of this RN.

Of the remaining RN systems in M31, a total of 4 more have
spectroscopic classifications. Two of these 4 have been classified
as Fe~II systems (M31N 2007-08d and 2009-11b),
with the other two belonging to the
He/N \citep[2012-01b,][]{sha12c} or
broad-lined Fe~II \citep[2012-09a,][]{sha12e} classes.
Thus, among the 7 RN systems with known spectroscopic types, 5 are
classified as either He/N or Fe~IIb.

Of the 12 RNe and the 4 possible RNe systems,
five were detected in X-rays as SSS:
M31N 1919-09a (as M31N 1998-06a), M31N 1953-09b (as M31N 2004-08a),
M31N 1963-09c (as M31N 2010-10e), M31N 1984-07a (as M31N 2004-11f),
and M31N 2008-12a (as M31N 2013-11f).
Four of these sources displayed very short SSS phases ($<100$ days)
with high effective X-ray temperatures (only measured for three of them).
Short SSS time scales and high effective temperatures have been found
to be correlated in M31 novae \citep{hen14a}
and are believed to indicate a high white dwarf mass \citep{hac06}.
Therefore, their X-ray properties are consistent with these objects
being RNe, which require a high white dwarf mass and a high accretion rate
to produce frequent eruptions.
The notable exception from this picture is M31N 1919-09a,
which was visible as a SSS between 1000 and 2000 days after its 1998 outburst.
Although this property points towards a significantly lower white dwarf mass
than for the other RNe, the available X-ray data is insufficient
to challenge seriously the classification of M31N 1919-09a as a RN.

\section{The Recurrent Nova Fraction in M31}

The 12 highly probable RN systems identified above
have produced a total of $n_{\rm out}({\rm RNe})=32$ nova outbursts out
of the total of 964 outbursts observed in M31 up to the end
of calendar year 2013. From these data we find that the {\it observed\/}
ratio of RN to CN outbursts in M31 is given by:

\begin{equation}
n_{\rm out}({\rm RNe})/n_{\rm out}({\rm CNe}) = n_{\rm out}({\rm RNe})/[964-n_{\rm out}({\rm RNe})] = 0.0343.
\end{equation}

\noindent
If we consider all plausible RNe, including the 4 systems (8 outbursts)
classified as uncertain, $n_{\rm out}({\rm RNe})=40$, and we have:

\begin{equation}
n_{\rm out}({\rm RNe})/n_{\rm out}({\rm CNe}) = 0.0433.
\end{equation}

\noindent
Given the uncertainties, in the analysis to follow we adopt a
mean value of

\begin{equation}
n_{\rm out}({\rm RNe})/n_{\rm out}({\rm CNe})\simeq0.04
\end{equation}

\noindent
(one in 25) as a reasonable estimate of the ratio of recurrent to classical
nova outbursts observed in M31.
It is clear, however, that this ratio must be a lower limit to
the true ratio of recurrent-to-classical nova outbursts in the galaxy given that
the sporadic temporal and uneven spatial
coverage of M31 surveys over the past century
has impacted the discovery
efficiency for RNe, where at least two outbursts must be observed,
considerably more than it has the discovery of systems classified as CNe.

\subsection{Estimating the Discovery Efficiencies for CNe and RNe}

In order to accurately estimate how the discovery efficiencies
of CNe and RNe differ in M31, it would be necessary to know the exact
temporal and spatial coverage as well as the limiting magnitudes of the many
M31 surveys that have taken place over the past century. In addition,
the typical recurrence times and light curve
properties of both RNe and CNe would need to be known. Obviously, the
precise values for all of these parameters are not available. However, we can
estimate the expected ratio of RN to CN outbursts from numerical simulations
that are based on plausible values for these unknowns.
Before describing the simulations, we first turn to a discussion
of the important biases that will affect the analysis.

\subsubsection{Temporal and Spatial Biases of M31 Nova Surveys}

\begin{figure}
\includegraphics[angle=90,scale=.70]{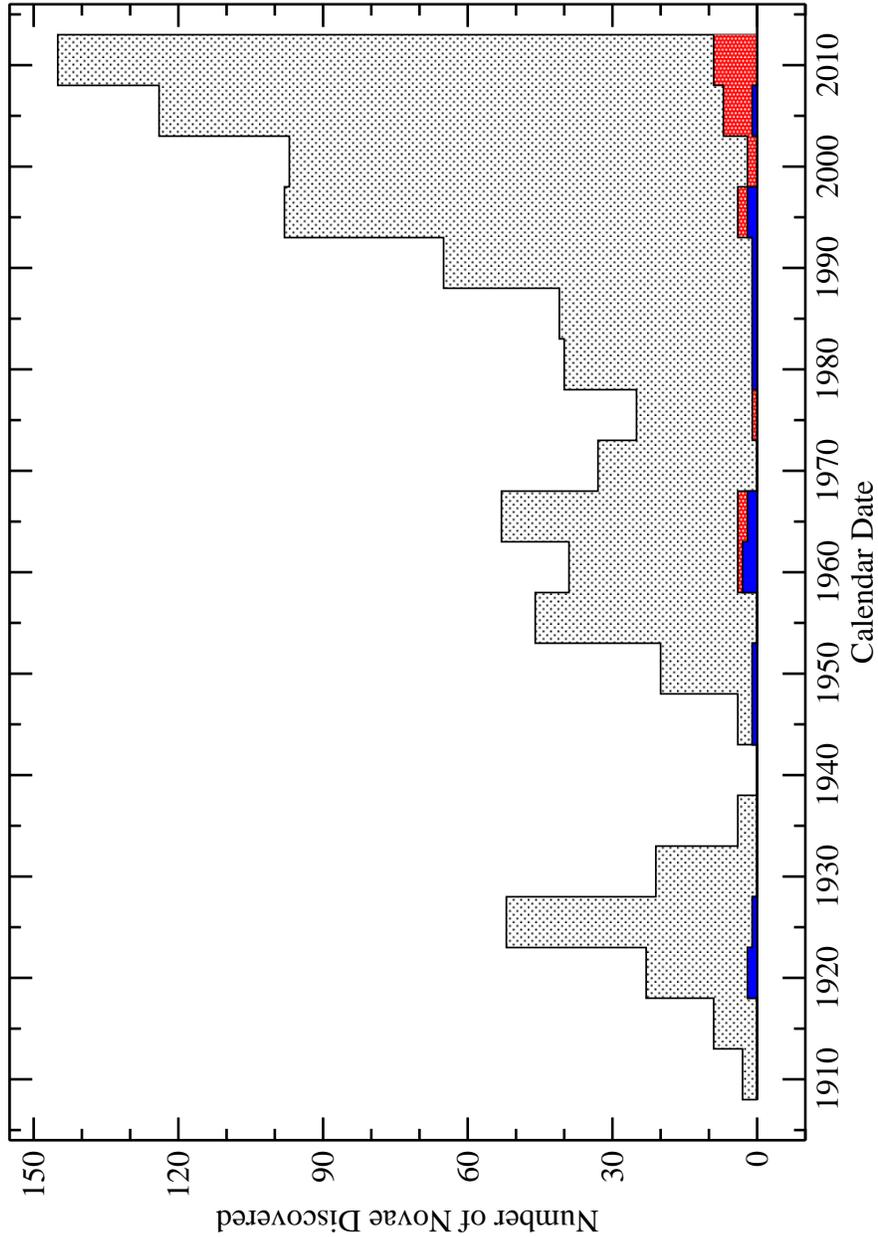}
\caption{The history of nova discoveries in M31. The discovery dates
of nova candidates in M31 have been divided into 5 year bins and plotted
as a function of time (lightly shaded grey region).
The shaded red area shows the overall
number of RN outbursts
observed in a given time interval, while the dark blue region shows
the number of outbursts for newly discovered RNe as a function of time.
\label{fig43}}
\end{figure}

The history
of nova discoveries in M31 is shown in Figure~\ref{fig43}. A few notable
features are apparent. The first clump of discoveries centered
in the 1920s is due largely to Hubble's pioneering survey \citep{hub29}. Nova discoveries
tailed off precipitously in the 1930s and essentially dropped off completely during
World War II. Nova discoveries picked up significantly in the 1950s and 1960s
as a result of the Arp \citep{arp56} and Rosino \citep{ros64, ros73}
photographic surveys. Then, after a short lull in the 1970s, nova discoveries
began a steady increase due primarily to the CCD surveys of \citet{cia87},
\citet{sha01}, \citet{dar06}, and others. In recent years the number
of novae discovered in M31 has exploded thanks to the contributions
from automated surveys such as the Palomar Transient Factory \citep{law09}, and
amateur astronomers throughout the world.

In addition to the overall number of novae discovered over the past century,
Figure~\ref{fig43} also shows the distribution of RN outbursts observed over time.
As expected,
the number of {\it outbursts\/} of RN systems
has increased roughly in proportion to the number of nova discoveries. However,
despite the dramatic increase in the overall
number of nova discoveries in recent years,
the first recorded eruptions of our 12 RNe and our 4 possible RNe are
spread relatively evenly over time. This result is not surprising.
If surveys had been uniform over time, with the discovery of novae increasing
at a steady rate, we would have expected the percentage
of novae identified as recurrent to be higher for systems
discovered early on,
where more would be available for the discovery of subsequent outbursts.
The fact that the density of observations has increased
over time has blunted this effect and resulted in what we see, namely
that the identification of new RN systems is approximately evenly
distributed over the past century.

In addition to the sporadic temporal coverage, the uneven spatial coverage
of the various surveys can be expected to
impact the relative discovery efficiencies of RNe and CNe. In particular,
the discovery efficiency of RNe should be biased towards
the bulge of the galaxy where the temporal
coverage is more extensive. From a theoretical standpoint,
it is unclear whether the
recurrence times of novae (and hence the relative populations of RNe and CNe)
should vary with spatial position in M31.
Population synthesis models predict that younger stellar populations should produce
nova progenitors that have on average more massive white dwarfs than do older
stellar populations \citep[e.g.,][]{dek92, pol96}. Since the recurrence times
of novae are strongly dependent on the mass of the white dwarf
\citep[e.g.,][]{yar05, wol13},
younger stellar populations should not only produce a higher observed nova
rate \citep{yun97}, but also a higher fraction of systems classified as RNe.
Despite these predictions, observations have consistently shown
that novae are more prominent in older bulge populations
\citep[e.g.,][]{cia87, sha01, dar06}.
The fact that the observed spatial distribution of RNe does
not differ significantly
from that of novae generally (see Figure~\ref{fig42})
suggests that an unbiased survey (with more complete disk coverage)
may find that RNe are more spatially
extended than novae in general.
If so, such a finding would be consistent with the predictions of
the population synthesis models.

\subsubsection{Limiting Magnitudes of M31 Nova Surveys}

\begin{figure}
\includegraphics[angle=90,scale=.70]{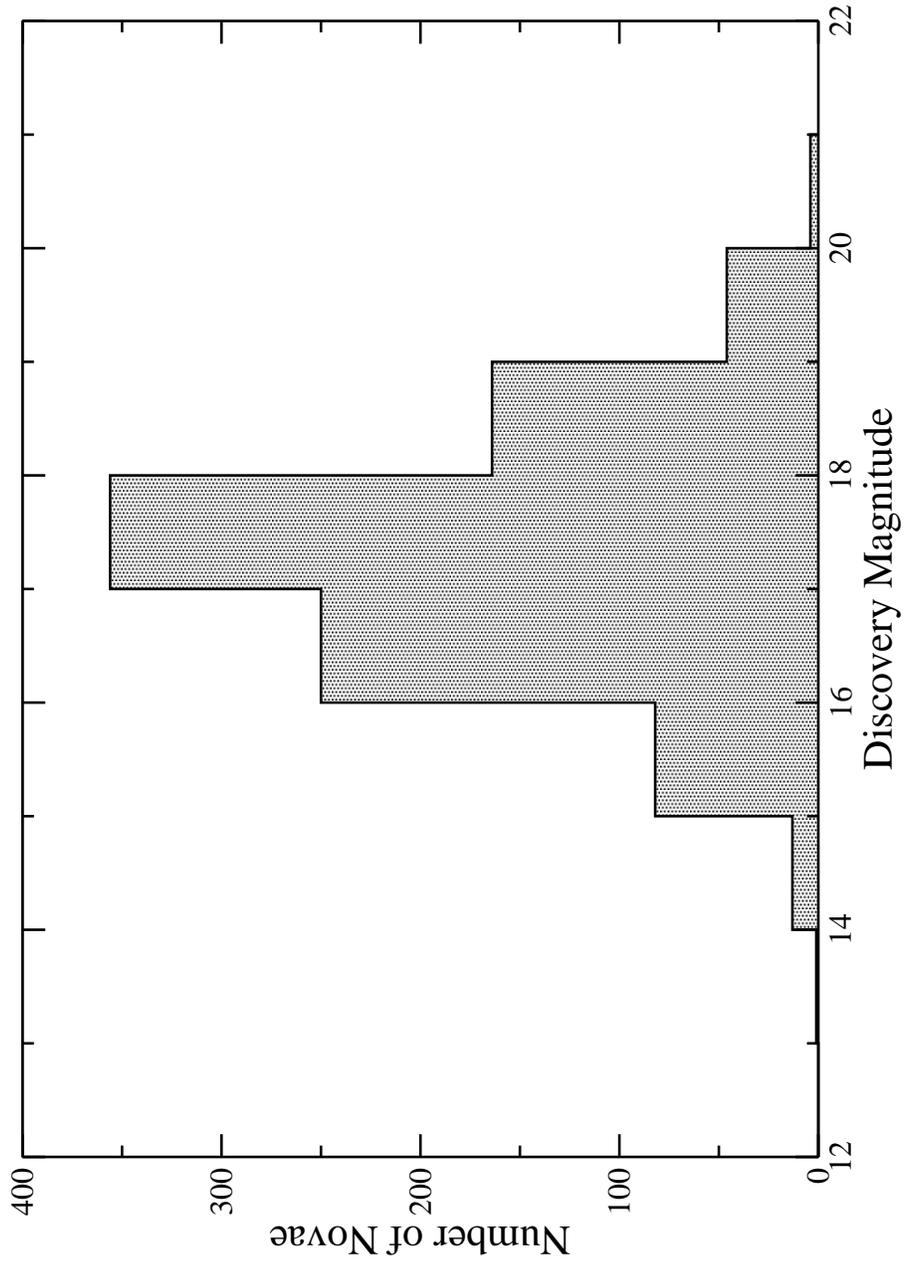}
\caption{The distribution of discovery magnitudes from the online
catalog of \citet{pie07a}. The number of novae discovered starts to
drop off precipitously for $m\grtsim18$.
\label{fig44}}
\end{figure}

Aside from the temporal and spatial uncertainties, we must also consider
how variations in the limiting magnitudes of the observations will
impact the discovery of novae in M31.
Limiting magnitudes for most of the M31 surveys vary significantly, both
between surveys and within the surveys themselves. Thus, it is not possible
to determine any single limiting magnitude representative of the full
M31 nova sample. Nevertheless,
we can get an idea of the
limiting magnitude of the ``average" survey by
referring to Figure~\ref{fig44}, which shows the distribution of discovery
magnitudes for the 964 M31 nova candidates in the \citet{pie07a} online
database.
Here we assume that the average survey reaches a limiting magnitude
down to the point where the discovery magnitude distribution turns over
and begins to drop precipitously.
Based on the distribution of
discovery magnitudes, it appears that a limiting magnitude of
$m_{\rm lim}\sim18$ is representative of the typical M31 nova survey.
Given the considerable uncertainty in $m_{\rm lim}$,
in the simulations described below we have also considered
how our results are affected by adopting limiting magnitudes of $m_{\rm lim}=17$
and 19.

\subsubsection{Monte Carlo Simulations}

We have performed a series of Monte Carlo simulations
to predict the discovery efficiencies of both CN and RN outbursts in M31
under a variety of scenarios.
In the simulations, we produce a total of 6500 synthetic
novae erupting at random times over a 100 year time span\footnote{
We have chosen 6500 eruptions based on an estimated annual nova rate
of 65~yr$^{-1}$
\citep{dar06}, although the predicted ratio of RNe to CNe
outbursts is not sensitive to the assumed nova rate.}. Model nova
light curves are based on the
the maximum magnitudes and rates of decline from
the ``high quality" M31 sample of \citet{cap89} and from the
Galactic RN sample from \citet{sch10} for CNe and RNe, respectively.
Table~\ref{lightcurve}
gives the values of $m_B(\rm{max})$ and $\nu_B$~(mag~d$^{-1}$) assuming
a distance and foreground reddening for M31 of $(m-M)_o = 24.38$ and
$A_B=0.25$ \citep{fre01, sch98}. 
For CNe, an eruption
reaches a maximum brightness and then fades at a rate that is chosen
at random from the CN light curve parameters given in Table~\ref{lightcurve}.
A model nova
is considered ``detected" if it remains brighter than the
adopted limiting magnitude ($m_{\rm lim}$=17, 18, or 19)
on the nearest survey date immediately following the date of the simulated
eruption. The fraction
of the 6500 novae that are detected is reported as the discovery efficiency
for CNe.

As a lower limit on the temporal
coverage of M31 we have simply taken the discovery dates of the nova
candidates that
have been discovered in M31 over the past century.
Then, to explore the sensitivity of
our simulations to the temporal coverage, we have also considered the
possibility that the actual survey coverage extended one or two days
on either side of each of the dates were a nova was reported to be discovered.
We refer
to the temporal coverage estimates either as ``Clump~1" (discovery dates only),
``Clump~3" (discovery dates plus the day
immediately preceding and following each discovery date), or
``Clump~5" (discovery date plus the 2 days preceding and the 2 days
following all discovery days).
It is also possible that there were more extended intervals of coverage where
no novae were discovered, but we have no way of accurately accounting
for this possibility.

In the case of RNe, our simulations must produce
multiple outbursts, and the RN discovery efficiencies will
clearly depend on the assumed recurrence time, or the distribution
of recurrence times, adopted in the simulations.
To explore this dependence, we have initially
considered several trial recurrence
times ranging from as short as 1 year (the shortest observed RN recurrence time)
to a maximum of 90 years (longer
recurrence times have a negligible chance of being detected in our simulations
that span just 100 years).
As for CNe, the light curve parameters for RNe have also
been chosen randomly, this time
from the RN sample given given in \citet{sch10} and
summarized here in Table~\ref{lightcurve}.

To qualify as a RN outburst in our simulations,
a particular system must be detected {\it at least twice\/} over the
100 year time span of the simulation.
The RN discovery efficiency is computed as the ratio of the
total number of multiple outbursts detected to the total number
of outbursts generated in the RN simulation (i.e., 6500 simulated nova
outbursts plus all recurrences in the 100 year interval).
The entire simulation for both CNe and RNe has been
repeated a total of 100,000 times, with the CN outburst fraction,
the RN outburst fraction recorded as the respective discovery efficiencies.

\begin{figure}
\includegraphics[angle=90,scale=.70]{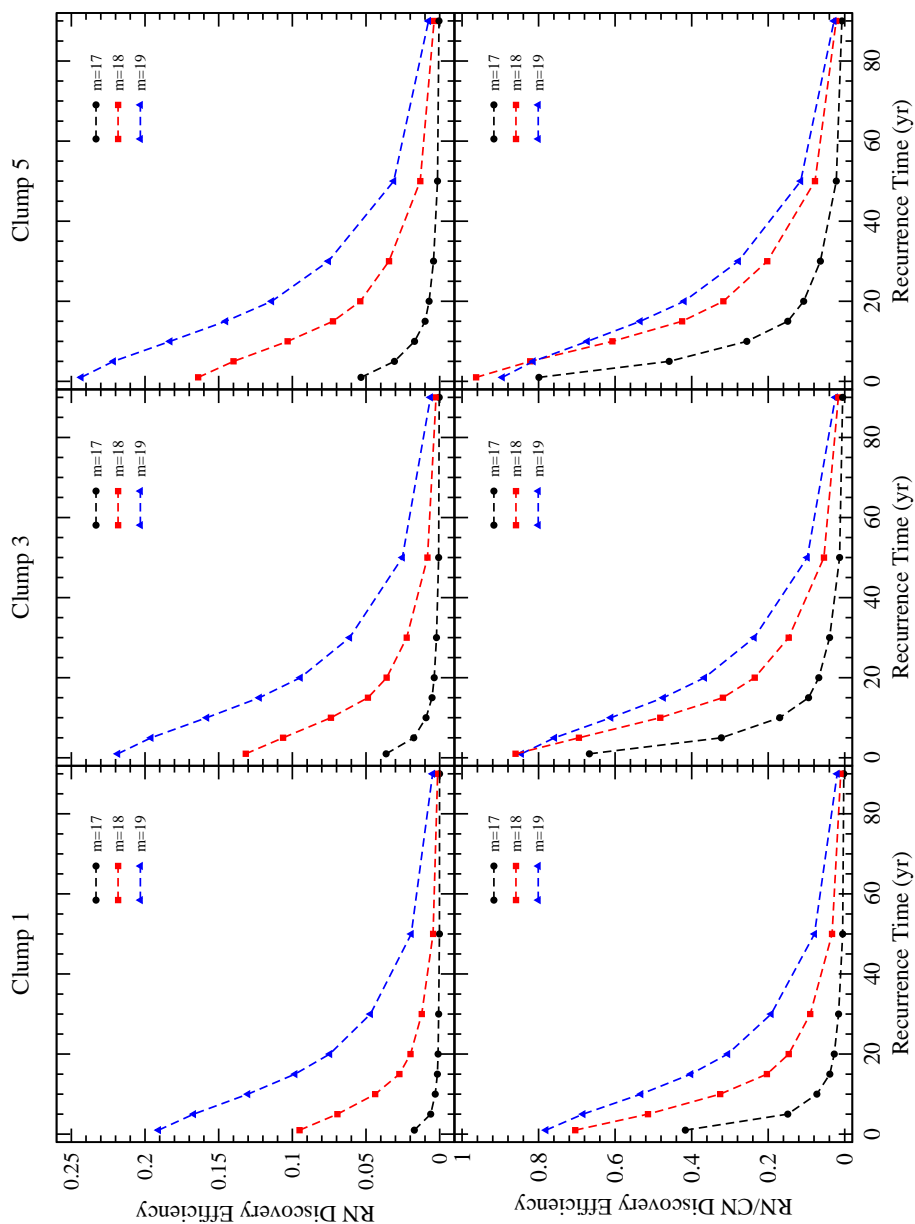}
\caption{Discovery efficiencies for RNe, and for RNe relative to CNe,
are plotted as a function of RN recurrence time for the three
different assumed limiting magnitudes and temporal coverages
($m_{\rm lim}=17$, black circles;
$m_{\rm lim}=18$, red squares; $m_{\rm lim}=19$,
blue triangles).
As expected, the discovery efficiency for RNe is strongly
dependent on the recurrence time and on the limiting magnitude
of the surveys.
\label{fig45}}
\end{figure}

The results of our Monte Carlo
simulations of the RN outburst discovery efficiencies and their ratio to that
of CNe are shown in Figure~\ref{fig45}.
Not surprisingly, the discovery efficiencies are strongly dependent on
the assumed limiting magnitude of the surveys, with the deeper surveys
resulting in higher RN discovery efficiencies. The assumed density of the
temporal coverage is also important, with the discovery efficiencies
increasing with increasing temporal coverage, as expected.
It is interesting to note that at the shortest recurrence times,
the discovery efficiencies for RNe relative to CNe for
$m_{\rm lim}=18$ slightly exceeds that for $m_{\rm lim}=19$
for the ``Clump 3" and ``Clump 5" temporal distributions. This behavior
results from the fact that the CN light curves have generally slower rates of
decline when compared with the RN light curve sample. A fainter limiting
magnitude thus increases the CN discovery efficiency more than it increases
the RN discovery efficiency. This effect is most pronounced for the
shorter recurrence times and denser temporal coverage where the
RN discovery efficiency is already quite high.

\begin{figure}
\includegraphics[angle=90,scale=.70]{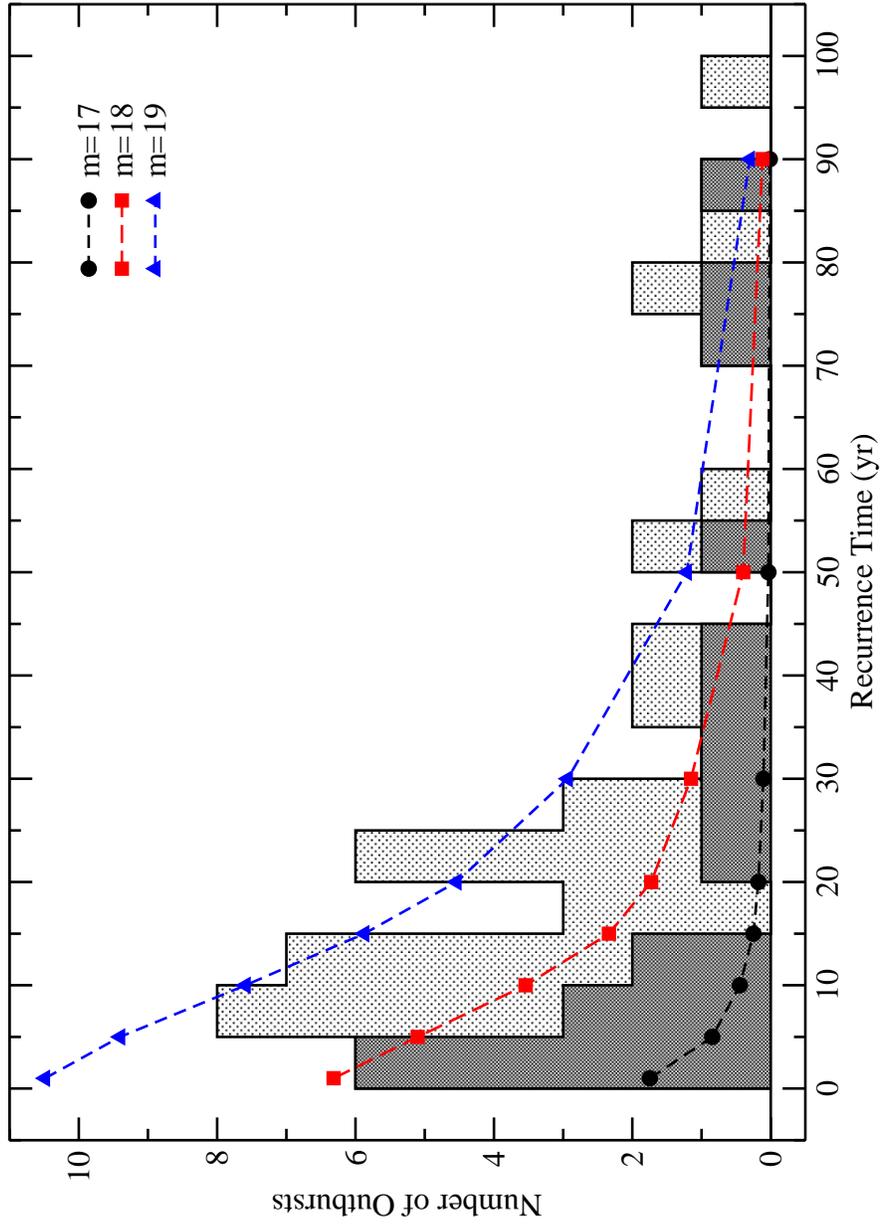}
\caption{The observed RN recurrence time distribution for known RNe in
the Galaxy (light shaded region) and M31 (dark shaded region) compared with
our Clump 3 model prediction for how the RN discovery efficiencies
should vary with recurrence time.
\label{fig46}}
\end{figure}

Regardless of the limiting survey magnitude or the temporal coverage,
these simulations confirm that the observed
RN recurrence time distribution for known RNe should strongly favor the
discovery of RNe having shorter recurrence times. To verify this prediction,
in Figure~\ref{fig46} we have plotted the observed outburst
distribution for Galactic and M31 RNe, and compared it with the $m_{\rm lim}=18$,
``Clump 3" discovery efficiencies from Figure~\ref{fig45}. Since we
simulated the same number of RNe across all recurrence times,
the qualitative agreement shown in Figure~\ref{fig46} is consistent with the hypothesis
that the observed bias toward the
discovery of short recurrence time RNe can be explained solely by
selection bias from an intrinsically
flat distribution of RN recurrence times.

Under the assumption that the intrinsic RN recurrence time distribution is
flat over the range of $1-100$ years,
we have conducted two composite recurrence time RN simulations. In the first
composite simulation we choose recurrence times at random from
the full range of $t_{\rm rec}=1-100$ years.
Then, we consider a restricted range with
recurrence times ranging from $t_{\rm rec}=5-100$
years that would be more appropriate
for a population of RNe where ultra-short recurrence time RNe such as
M31N 2008-12a are extremely rare.

\begin{figure}
\includegraphics[angle=0,scale=.70]{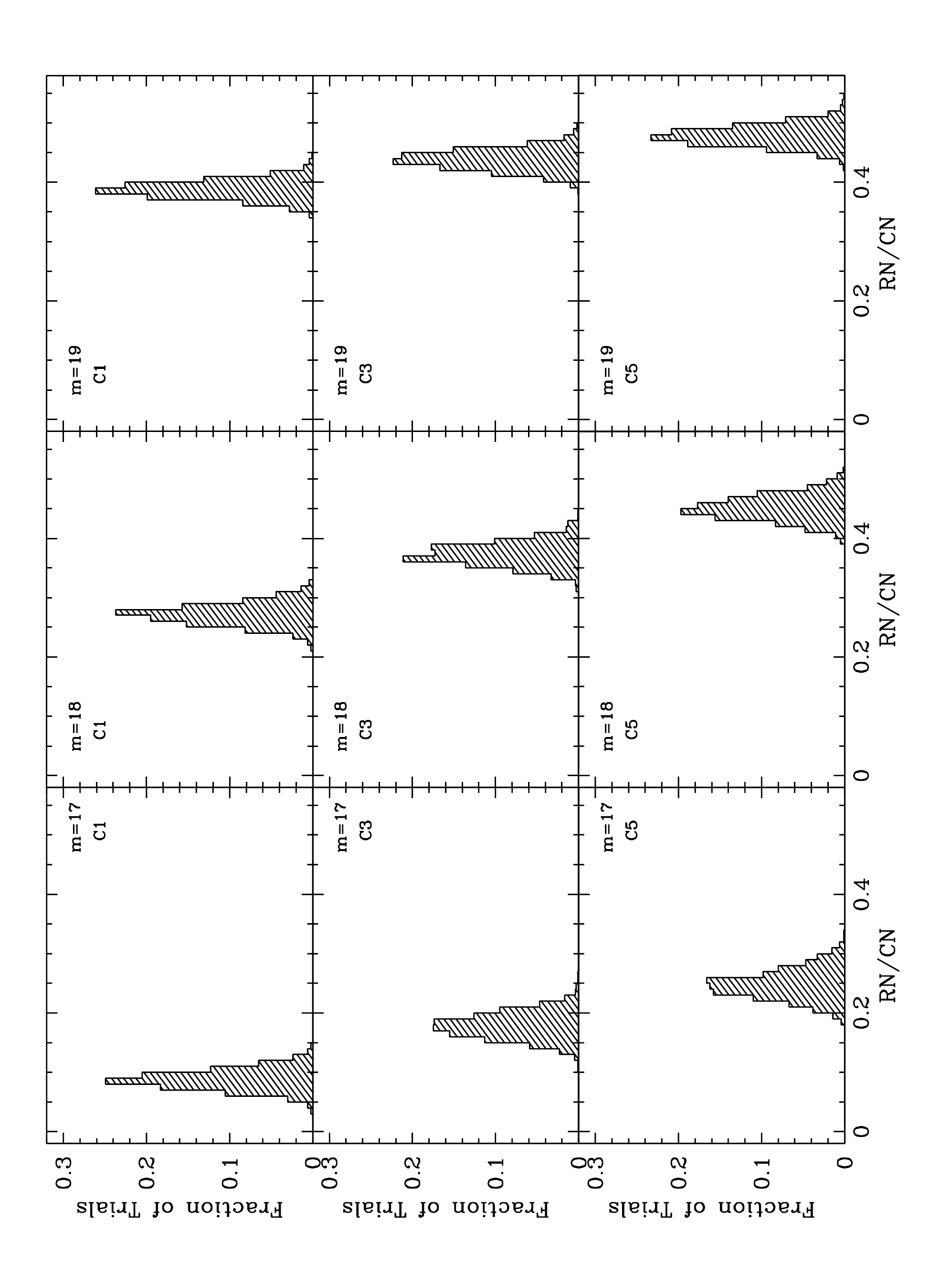}
\caption{The ratio of RN to CN outbursts detected in our Monte Carlo simulation
for RN recurrence times chosen at random from a range of 1 to 100 years.
Results are shown for assumed limiting magnitudes $m=17$, $m=18$ and $m=19$
(columns left to right), and temporal coverages Clump 1, Clump 3, and Clump 5
(rows top to bottom).
\label{fig47}}
\end{figure}

\begin{figure}
\includegraphics[angle=0,scale=.70]{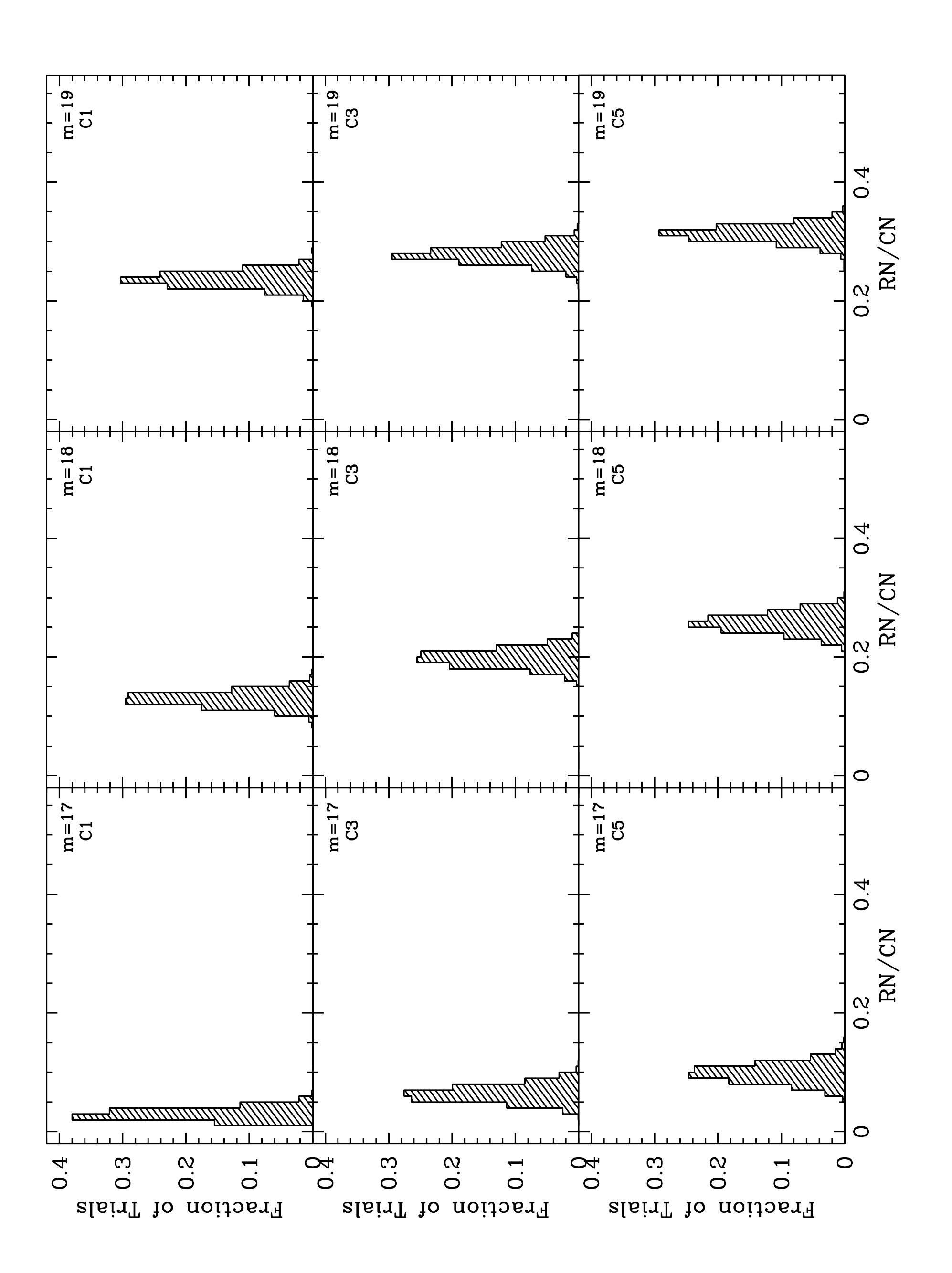}
\caption{Same as Figure~47, except for RN recurrence times chosen randomly
from a range of 5 to 100 years.
\label{fig48}}
\end{figure}

The results of the Monte Carlo simulation are
reported in Table~\ref{monte} and shown in Figures~\ref{fig47} and~\ref{fig48}
for the various input parameters.
Columns 3 and 4 of Table~\ref{monte}
give the fractions of CN and RN outbursts that are detected in the
simulation, with column 5 giving the principal result of our simulation,
the ratio of RN to CN outbursts detected (see Figs.~\ref{fig47},~\ref{fig48}).
Column 6 gives our estimate of the true ratio of RN to CN outbursts,
$N_{\rm out}({\rm RNe})/N_{\rm out}({\rm CNe})$,
after the observed outburst ratio
given in Equation (5) is corrected for the
relative outburst discovery efficiencies. Columns 7 and 8 then give
our estimates of the expected numbers
of RN and CN outbursts produced in M31 each year
assuming a combined annual nova rate of
65~yr$^{-1}$ \citep{dar06}.

As a check on our simulations, we note that the CN discovery efficiency predicted by
our Monte Carlo simulations is in qualitative agreement with
observations. Given
an annual nova rate of 65~yr$^{-1}$, we estimate that
approximately 6500 novae have erupted in M31 over the past century.
Our standard model ($m_{\rm lim}=18$, ``Clump 3") produces
a CN discovery fraction of 0.15 and predicts that approximately
975 novae will have been discovered over the past century, in very
good agreement with the number of novae discovered in M31.
It is also worth noting that our values for the true ratio
of RN to CN outbursts,
$N_{\rm out}({\rm RNe})/N_{\rm out}({\rm CNe})$, for our
$m_{\rm lim} = 18, 19$ models are consistent
with the ratio estimated by \citet{del96} in their study of the RN populations
in the Galaxy, the LMC, and M31. 

Under the assumption that the recurrence time distribution
is flat in the range of $1-100$ years, we predict that RN outbursts make
up $\sim$10\% of nova eruptions generally. For an annual nova rate
of 65~yr$^{-1}$, as many as $5-10$ outbursts per year are expected to
arise from RN progenitors.
If ultra-short recurrence
time ($t_{\rm rec}<5$ yr) RNe are relatively rare,
our $5<t_{\rm rec}/{\rm yr}<100$ models suggest that
as many as one out of three ($\sim20$) nova eruptions
observed annually in M31 could arise from RN systems. This result is
consistent with that found recently by \citet{pag14} for that fraction
of Galactic novae that are recurrent, and \citet{wil14,wil15} for the
fraction of M31 novae with evolved secondary stars.

\section{RNe as a Channel for the Production of Type Ia Supernovae}

SNe~Ia are thought to arise from the deflagration of an
accreting white dwarf in a binary system \citep{mao12}. There are
two primary models for the progenitor systems (1) a single degenerate
channel, where the white dwarf accretes from a companion star in a
semi-detached binary, and (2) a double-degenerate channel, where
a pair of orbiting white dwarfs merge as a result of the loss
of orbital energy and angular momentum from the emission of gravitational
waves. RNe are sometimes proposed as potential contributors to
the single degenerate channel. However, the consensus
has generally been that RNe do not exist in sufficient numbers to
explain the observed rate of SNe~Ia production in a galaxy like M31.

Based on our estimates for the RN rate in M31, we can explore whether
these systems might make a significant contribution to
the production of SNe~Ia in a galaxy like M31.
If we assume RNe have a mean mass accretion rate,
$\dot M = 10^{-7}$~M$_{\odot}$~yr$^{-1}$, which is typical for
Galactic RN systems \citep{sch10}, and that each RN system
must accrete a total of 0.3 M$_{\odot}$ of material before reaching the
Chandrashakhar limit, it would take approximately 3
million years on average for a given RN to
become a SNe~Ia. Assuming a RN rate of as many as 20 per year and a mean recurrence
time of 30 years between outbursts, we estimate that there could be as many as
600 active RN systems in M31. This number likely represents an upper limit
to the number of active RN systems in M31 given that most models
predict fewer than 20 RN outbursts per year and the mean recurrence time
is likely shorter than 30 years. Thus, we arrive at
an estimated upper limit to the RN ``death rate"
of $\sim2\times10^{-4}$~yr$^{-1}$, which could produce
perhaps as many as one SNe~Ia every 5000 yr in this galaxy.

\citet{man05} has estimated that for typical S0a/b and Sbc/d spirals,
we can expect SNe~Ia birthrates
of $\sim$0.04 and 0.08 SNe~Ia per century per $10^{10}~L_{\odot,K}$,
respectively.
Adopting $L_{K} \simeq 16\times10^{10}~L_{\odot,K}$ for M31 \citep{sha13},
gives expected SNe~Ia rates of $\sim0.64$ and $\sim1.3$ per century
depending on Hubble type.
In the case of M31, with an intermediate Hubble type of Sbc, we
estimate a rate of $\sim1.0\pm0.4$ SNe~1a per century. Thus, it appears that
RNe might be able to supply only $\sim$2\% of the SNe~Ia in a galaxy like M31.
Given the uncertainties in the assumptions,
this estimate clearly constitutes only a very rough estimate of the
contribution that RNe might make to the production of SNe~Ia in M31.
Nevertheless, based on our analysis of the RN population in M31,
it appears very unlikely that RNe could play a major role
as SNe~Ia progenitors.

\section{Conclusions}

We have undertaken a program to identify the fraction of RNe in the
nearby spiral galaxy, M31. We have identified RN candidates by
cross-correlating the published coordinates for 964 nova candidates
recorded in M31 from 1909 September through the end of 2013. Allowing for
a positional uncertainties $s<6''$
($9''$ for early photographic plate surveys), we have identified a total of
51 potential RN systems responsible for 118 observed eruptions.
In order to determine which of these potential RNe are in fact RNe,
we have made an attempt to locate and analyze original plates, films,
CCD images, and finding charts for each of the 118 eruptions. We were
successful in locating a number of images that have allowed us to
identify a total of 12 RN systems, with another 4 likely RNe,
representing a total of 40 eruptions.
In addition, we have been able to rule out a total of 27 potential
RN systems representing a total of 61 individual eruptions.
The remaining 8 systems, representing 17 eruptions, are unlikely to
be RNe, but we have not been able to definitively rule out their RN nature.

Taking the 12 RN systems representing 32 eruptions, and assuming
that 2 of the 4 possible RNe representing 4 eruptions
are in fact recurrent,
we find that approximately 1 in 25 of the {\it observed\/}
nova outbursts in M31 arise from RN progenitors.
To correct for observational
bias against the detection of RN systems,
we conducted a simple Monte Carlo simulation to estimate the
relative discovery probability for CN and RN outbursts.
For plausible estimates of the M31 observational coverage, we estimate
that as many as one in three nova outbursts observed in M31 could
arise from a population of RN progenitors having
recurrence times less than a century. Despite the rudimentary nature
of our simulation, which is unable to accurately account for the inhomogeneous spatial
and temporal coverage of the M31 data set, this number is consistent with the
estimate for Galactic RNe found by \citet{pag14}, and with the relatively high
fraction of M31 novae found to harbor evolved secondary stars
\citep{wil14,wil15}.
Finally, based on our rough estimates for the number of RNe active in M31 and their
expected white dwarf masses and accretion rates, we conclude that
it appears unlikely that such systems can provide
a significant channel for the production of Type Ia supernovae.

\acknowledgments

We would like
to thank an anonymous referee for valuable comments and suggestions
that have led to an improved presentation. We also thank
S. M. Lauber for her assistance in searching for RN candidates in the
RBSE data and for proof reading the manuscript.
A.W.S. is grateful to C.T. Daub for discussions regarding the
probabilities of chance positional coincidences of novae, and to
J. Rice for discussions during the early stages of related work
on RN in M31.
This work is based in part on observations made with the Isaac Newton
Telescope operated on the island of La Palma by the Isaac Newton Group
in the Spanish Observatorio del Roque de los Muchachos of the Instituto
de Astrof\'isica de Canarias.
M.H. acknowledges support from an ESA fellowship.
T.A.R. was funded for research on this project through NSF grants
DUE-0920293 and DUE-0618849.
K.H. was supported by the project RVO:67985815.
A.W.S. acknowledges financial support through NSF grant AST-1009566.




\clearpage

\clearpage

\begin{planotable}{lcrcrrcccr}
\tabletypesize{\scriptsize}
\tablenum{1}
\tablewidth{0pt}
\tablecolumns{10}
\tablecaption{M31 Recurrent Nova Candidates\label{rntable}}
\tablehead{\colhead{Nova} & \colhead{Possible Recurrence} & \colhead{$\Delta t$ (yr)} & \colhead{$s$ ($''$)} & \colhead{$a$ ($'$)} & $N$ & \colhead{$P_C$} & \colhead{Type\tablenotemark{a}} & \colhead{Charts?} & \colhead{Comments}
}
\startdata
M31N 1909-09b & M31N 2009-02b &99.4& 4.20 &  3.96 & 64 & 0.8552 &     &Y & Not RN \cr
M31N 1918-02b & M31N 2013-10g &95.7&12.18 &  9.54 & 17 & 0.4326 &Fe II&Y & Not RN \cr
M31N 1919-09a & M31N 1998-06a &78.7& 1.80 & 10.50 & 27 & 0.0296 &     &Y & RN \cr
\cr
M31N 1923-02a & M31N 1967-12a &44.8&  6.55 &  1.74 & 69 &  0.9999 & &Y & Not RN\cr
\dots        & M31N 1993-06a &70.3&  6.70 &  1.64 & 63 &  0.9999 & &N & Not RN\tablenotemark{b}\cr
\dots        & M31N 1993-11c &70.7&  6.48 &  1.73 & 68 &  0.9999 & &Y & Not RN\cr
\dots        & M31N 2013-08b &90.5&  6.64 &  1.73 & 68 &  0.9999 & &Y & Not RN\cr
M31N 1967-12a & M31N 1993-11c &25.9&  3.30 &  1.73 & 68 &  0.9326 & &Y & Not RN\cr
\dots        & M31N 2013-08b &45.6&  1.41 &  1.73 & 68 &  0.3813 & &Y & Not RN\cr
M31N 1993-11c & M31N 2013-08b &19.9&  4.66 &  1.73 & 68 &  0.9960 & &Y & Not RN\cr
\cr
M31N 1923-12c & M31N 2012-01b &88.1& 0.83 & 10.99 & 36 & 0.0112 & He/N&Y & RN \cr
M31N 1924-02b & M31N 1995-09d &71.6& 5.20 & 13.57 & 14 & 0.0512 &     &N & Not Nova \cr
M31N 1924-08a & M31N 1987-12a &63.3& 3.10 &  2.61 & 81 & 0.9008 &     &Y & Not RN \cr
M31N 1925-07c & M31N 2011-12b &86.4& 7.55 & 20.30 & 19 & 0.1477 &Fe II&Y & Not RN \cr
M31N 1925-09a & M31N 1976-12a &51.3& 9.70 & 19.57 & 15 & 0.1548 &     &Y & Not RN \cr
M31N 1926-06a & M31N 1962-11a &36.5& 6.27 & 17.45 & 24 & 0.1870 &     &Y & RN \cr
M31N 1926-07c & M31N 1980-09d &54.2& 6.95 &  2.01 & 72 & 0.9999 &     &Y & Not RN\tablenotemark{c} \cr
M31N 1927-08a & M31N 2009-11e &82.2& 8.02 &  3.86 & 62 & 0.9990 &     &Y & Not RN \cr
M31N 1930-06b & M31N 2008-08e &78.2& 6.29 & 12.95 & 29 & 0.3035 &     &N & Not Nova \cr
M31N 1930-06c & M31N 1996-08a &66.1& 7.23 &  2.67 & 82 & 0.9999 & &Y & Not RN\cr
M31N 1932-09d & M31N 2001-07d &68.9& 3.63 &  3.03 & 80 & 0.9390 &     &N & Unlikely RN\cr
M31N 1945-09c & M31N 1975-11a & 27:& 0.88 & 29.19 &  9 & 0.0004 &     &Y & RN \cr
M31N 1953-09b & M31N 2004-08a &50.8& 4.80 &  5.77 & 48 & 0.6399 &     &Y & RN: \cr
\cr
M31N 1953-11a & M31N 1960-12a &7.1 &  7.85 &  4.40 & 70 &  0.9996 & &Y & Not RN\cr
\dots        & M31N 1962-11b &9.1 &  2.90 &  4.38 & 69 &  0.6215 & &Y & Not RN\cr
\dots        & M31N 2013-05b &59.6&  6.67 &  4.40 & 70 &  0.9959 & &Y & Not RN\cr
M31N 1960-12a & M31N 1962-11b &1.9 &  5.31 &  4.38 & 69 &  0.9645 & &Y & Not RN\cr
\dots         & M31N 2013-05b &52.4&  1.20 &  4.40 & 70 &  0.1551 & &Y & RN\cr
M31N 1962-11b & M31N 2013-05b &50.5&  4.12 &  4.40 & 70 &  0.8684 & &Y & Not RN\cr
\cr
M31N 1954-06c & M31N 2010-01b &55.6& 8.78 & 12.39 & 28 & 0.4995 &     &Y & Not RN\cr
M31N 1955-09b & M31N 2012-03b &56.5& 1.94 & 10.46 & 26 & 0.0319 &     &Y & Not RN \cr
M31N 1957-10b & M31N 2010-12a &53.1& 1.10 & 51.43 &  5 & 0.0001 &     &N & Unlikely RN \cr
M31N 1961-11a & M31N 2005-06c &43.6& 2.37 &  3.03 & 80 & 0.6910 &     &Y & RN: \cr
\cr
M31N 1963-09c & M31N 1968-09a &5.0 & 0.89 & 17.60 & 24 & 0.0041 &     &Y & RN \cr
\dots         & M31N 2001-07b &37.8& 1.23 & 17.60 & 24 &0.0079 &     &N & RN \cr
\dots         & M31N 2010-10e &47.1& 1.32 & 17.60 & 24 &0.0090 & He/N&Y & RN \cr
M31N 1968-09a & M31N 2001-07b &32.8& 0.45 & 17.60 & 24 &  0.0011 & &Y & RN \cr
\dots         & M31N 2010-10e &42.1& 0.49 & 17.60 & 24 &  0.0013 &He/N&Y & RN \cr
M31N 2001-07b & M31N 2010-10e &9.3 & 0.11 & 17.60 & 24 &  0.0001 &He/N&N & RN\cr
\cr
M31N 1964-12b & M31N 1998-07b\tablenotemark{d} &33.6&  4.51 &  0.71 & 51 & 0.9981 & &N & Not RN\cr
M31N 1966-08a & M31N 1968-10c &2.2 & 0.00 & 31.18 &  3 & 0.0000 &     &Y & RN: \cr
M31N 1966-09e & M31N 2007-08d &40.9& 0.36 & 59.56 &  7 & 0.0000 &Fe II& Y& RN \cr
M31N 1967-11a & M31N 2006-02a &38.2& 4.18 &  1.46 & 62 & 0.9859 &     & Y& Not RN\cr
M31N 1967-12b & M31N 1999-06b &31.5& 5.98 &  4.43 & 67 & 0.9811 &     & Y& Not RN\cr
M31N 1969-08a & M31N 2007-12b &38.3& 4.86 & 14.68 & 11 & 0.0264 & He/N& N& Not RN \cr
M31N 1975-09a & M31N 1999-01a &23.3& 3.34 &  6.51 & 46 & 0.3337 &     & Y& Not RN \cr
M31N 1977-12a & M31N 1998-08a &20.7& 2.07 & 10.77 & 32 & 0.0536 &     & Y& Not RN \cr
M31N 1982-08b & M31N 1996-08c\tablenotemark{e} &14.0& 2.99 & 60.77 &  7 &0.0014 &  &Y    & RN\cr
M31N 1982-09a & M31N 2011-02b &28.5& 5.46 &  1.08 & 50 &0.9980 &  &N    & Unlikely RN\cr
M31N 1983-09c & M31N 1997-11c &14.0& 5.99 &  8.03 & 35 & 0.4932 &     &N & Unlikely RN\cr
\tablebreak
\cr
M31N 1984-07a & M31N 2001-10c &17.2& 1.29 &  0.57 & 40 & 0.2981 &     &N & Not RN \cr
\dots         & M31N 2004-02a &19.5& 1.61 &  0.60 & 45 &0.4896 &     &N & Not RN \cr
\dots         & M31N 2004-11f &20.3& 0.49 &  0.57 & 40 &0.0496 &     &Y & RN \cr
\dots         & M31N 2012-09a &28.1& 0.51 &  0.58 & 40 &0.0520 &Fe IIb&Y& RN  \cr
M31N 2001-10c & M31N 2004-02a &2.3 &  2.81 &  0.60 & 45 &  0.8764 & &N & Not RN\cr
\dots         & M31N 2004-11f &3.1 &  1.19 &  0.57 & 40 &  0.2588 & &N & Not RN\cr
\dots         & M31N 2012-09a &10.9&  0.80 &  0.58 & 40 &  0.1260 &Fe II& N & Not RN\cr
M31N 2004-02a & M31N 2004-11f &0.7 &  2.02 &  0.57 & 40 &  0.5827 & &N & Not RN\cr
\dots         & M31N 2012-09a &8.6 &  2.10 &  0.58 & 40 &  0.6051 &Fe II& N & Not RN\cr
M31N 2004-11f & M31N 2012-09a &7.9 &  0.45 &  0.58 & 40 &  0.0418 &Fe II& Y & RN\cr
\cr
M31N 1984-09b & M31N 2012-12a &28.2& 1.52 &  1.27 & 62 & 0.4498 & Fe II&N& Unlikely RN \cr
M31N 1985-09d & M31N 2009-08d &23.9& 4.12 &  0.81 & 53 & 0.9931 &     &N & Unlikely RN \cr
\cr
M31N 1985-10c & M31N 1995-12a &10.2& 2.40 &  0.31 & 29 & 0.7077 &     &N & Unlikely RN \cr
\dots         & M31N 2003-10b &18.0& 1.60 &  0.30 & 29 & 0.4216 &     &N & Unlikely RN\cr
M31N 1995-12a & M31N 2003-10b &7.9 & 3.66 &  0.30 & 29 & 0.9529 &     &N & Unlikely RN\cr
\cr
M31N 1986-09a & M31N 2006-09b &20.0& 4.80 &  1.63 & 61 & 0.9943 &     &N & Unlikely RN \cr
\cr
M31N 1990-10a & M31N 1997-10b\tablenotemark{f} &7.0 & 2.26 &  4.24 & 73 & 0.4877 &     &N & Not RN\cr
\dots         & M31N 2007-07a &16.7& 0.80 &  4.28 &\dots&0.0726 &     &Y & RN: \cr
\cr
M31N 1992-12b & M31N 2001-10f &9.0 & 5.31 & 12.91 & 31 & 0.2563 &     &Y & Not RN \cr
M31N 1993-09b & M31N 1996-08g &2.9 & 5.47 &  5.33 & 55 & 0.8427 &     &Y & Not RN \cr
M31N 1997-10f & M31N 2008-08b &10.8& 0.45 &  1.69 & 65 & 0.0452 &He/N?&Y & RN\cr
\cr
M31N 1997-11k & M31N 2001-12b &4.2 & 0.00 & 10.57 & 30 & 0.0000 &    & Y & RN \cr
\dots         & M31N 2009-11b &12.1& 0.38 & 10.56 &\dots&0.0016 &Fe II&Y & RN \cr
M31N 2001-12b & M31N 2009-11b &7.9 &  0.38 & 10.56 & 30 &  0.0016 & &Y & RN \cr
\cr
M31N 2001-08d & M31N 2008-07a &6.8 & 3.51 &  4.34 & 70 & 0.7739 &Fe II&Y & Not RN \cr
M31N 2004-11b & M31N 2010-07b &5.6 & 5.91 &  4.97 & 63 & 0.9539 &     &Y & Not RN \cr
M31N 2005-05b & M31N 2009-08d &4.2 & 4.59 &  0.81 & 53 & 0.9981 &Fe II&Y & Not RN \cr
M31N 2006-11b & M31N 2006-12d &0.1 & 0.34 &  1.27 & 62 & 0.0292 &     &Y & Not RN \cr
M31N 2006-12c & M31N 2007-07e &  1 & 4.39 &  1.99 & 74 & 0.9928 & Fe II&Y& Not RN \cr
\cr
M31N 2008-12a & M31N 2009-12b\tablenotemark{g} &1.0 &\dots & 48.92 &\dots& \dots  &     & Y& RN \cr
\dots         & M31N 2011-10e &2.8 & 0.30 & 48.92 & 11 & 0.0000 &     & Y& RN \cr
\dots         & M31N 2012-10a &3.8 & 0.74 & 48.92 & 11 &0.0003 & He/N& Y& RN \cr
\dots         & M31N 2013-11f &4.9 & 1.01 & 48.93 & 11 &0.0005 &     & Y& RN \cr
M31N 2011-10e & M31N 2012-10a &1.0 &  0.46 & 48.92 & 11 &  0.0001 &He/N& Y& RN \cr
\dots         & M31N 2013-11f &2.1 &  0.84 & 48.93 & 11 &  0.0004 &    & Y& RN \cr
M31N 2012-10a & M31N 2013-11f &1.1 &  0.89 & 48.93 & 11 &  0.0004 &    & Y& RN \cr
\cr
M31N 2010-01a & M31N 2010-12c &0.9 & 0.82 &  2.69 & 80 & 0.1410 &Fe II &Y& Not RN \cr
\enddata
\tablenotetext{a}{Spectroscopic type of the possible recurrence.}
\tablenotetext{b}{Although a chart for M31N 1993-06a does not exist,
 our revised coordinates for 1923-02a differ from those of 1993-06a by 11.6$''$
 making it extremely unlikely that the two novae are coincident.}
\tablenotetext{c}{Unexpectedly, M31N 1926-07c was found to be coincident with 1997-10f and 2008-08b.}
\tablenotetext{d}{M31N 1998-07b is not a nova}
\tablenotetext{e}{The nova was discovered on 1997 August 01; not 1996 August 12 as reported in \citet{sha01}.}
\tablenotetext{f}{M31N 1997-10b is not a nova.}
\tablenotetext{g}{We have designated the outburst just announced in \citet{tan14} as M31N 2009-12b.}
\end{planotable}

\clearpage

\begin{deluxetable}{lcc}
\tabletypesize{\scriptsize}
\tablenum{2}
\tablewidth{0pt}
\tablecolumns{3}
\tablecaption{Revised Coordinates\tablenotemark{a}\label{revcoord}}
\tablehead{\colhead{} & \colhead{R.A.} & \colhead{Decl.} \\
\colhead{Nova} & \colhead{(2000.0)} & \colhead{(2000.0)} }
\startdata
M31N 1909-09b  &  00 42 27.48 & 41 13 32.4  \\
M31N 1918-02b  &  00 43 23.76 & 41 21 34.5  \\
M31N 1919-09a  &  00 43 28.65 & 41 21 42.1  \\
M31N 1923-02a  &  00 42 50.03 & 41 17 20.9  \\
M31N 1923-12c  &  00 42 38.07 & 41 08 41.4  \\
M31N 1924-08a  &  00 42 31.49 & 41 15 25.4  \\
M31N 1925-07c  &  00 43 55.28 & 41 21 28.4  \\
M31N 1925-09a  &  00 43 36.56 & 41 31 56.2  \\
M31N 1926-06a  &  00 41 40.66 & 41 03 33.7  \\
M31N 1926-07c  &  00 42 52.37 & 41 16 12.8  \\
M31N 1927-08a  &  00 42 36.08 & 41 13 06.4  \\
M31N 1930-06c  &  00 42 35.26 & 41 11 37.7  \\
M31N 1945-09c  &  00 41 28.58 & 40 53 13.7  \\
M31N 1953-09b  &  00 42 20.69 & 41 16 07.9  \\
M31N 1953-11a  &  00 42 56.10 & 41 14 18.4  \\
M31N 1954-06c  &  00 42 59.97 & 41 25 24.2  \\
M31N 1960-12a  &  00 42 55.71 & 41 14 12.5  \\
M31N 1961-11a  &  00 42 31.36 & 41 16 21.0  \\
M31N 1962-11b  &  00 42 55.92 & 41 14 16.6  \\
M31N 1966-09e  &  00 39 30.33 & 40 29 14.0  \\
M31N 1977-12a  &  00 42 13.70 & 41 07 44.4  \\
M31N 1982-08b  &  00 46 06.68 & 42 03 49.3  \\
M31N 1984-07a  &  00 42 47.15 & 41 16 19.6  \\
M31N 1987-12a  &  00 42 32.91 & 41 15 14.2  \\
M31N 1990-10a  &  00 43 04.00 & 41 17 08.1  \\
M31N 1992-12b  &  00 41 53.77 & 41 07 21.9  \\
M31N 1993-09b  &  00 42 19.53 & 41 14 04.2  \\
M31N 1993-11c  &  00 42 50.10 & 41 17 29.4  \\
M31N 1996-08a  &  00 42 43.23 & 41 18 14.8  \\
M31N 1996-08c  &  00 46 06.60 & 42 03 49.4  \\
M31N 1997-10f  &  00 42 52.34 & 41 16 12.8  \\
M31N 1998-08a  &  00 42 13.45 & 41 07 44.9  \\
M31N 2010-01a  &  00 42 56.68 & 41 17 21.2  \\
M31N 2010-12c  &  00 42 56.65 & 41 17 22.3  \\
\enddata
\tablenotetext{a}{Uncertainties estimated to be 0.1s in R.A. and $1''$ in Decl.}
\end{deluxetable}

\clearpage

\begin{deluxetable}{lcrcrclc}
\tabletypesize{\scriptsize}
\tablenum{3}
\tablewidth{0pt}
\tablecolumns{8}
\tablecaption{M31 Recurrent Novae \label{rntable_conf}}
\tablehead{\colhead{Nova} & \colhead{Recurrence} & \colhead{$\Delta t$ (yr)} & \colhead{$s$ ($''$)} & \colhead{$a$ ($'$)} & \colhead{$P_C$} & \colhead{Type} & \colhead{Comments}
}
\startdata
M31N 1919-09a & M31N 1998-06a &78.7& 1.33 & 10.50 &  0.0163 &        & RN \cr
M31N 1923-12c & M31N 2012-01b &88.1& 0.45 & 10.99 &  0.0033 & He/N   & RN \cr
M31N 1926-06a & M31N 1962-11a &36.5& 1.40 & 17.45 &  0.0102 &        & RN \cr
M31N 1926-07c & M31N 1997-10f &71.3& 0.34 &  1.68 &  0.0240  &       & RN \cr
              & M31N 2008-08b &10.8& 0.15 & 1.69  & 0.0050 & He/N?  & RN \cr
M31N 1945-09c & M31N 1975-11a & 27:& 0.41 & 29.19 &  0.0001 &        & RN \cr
M31N 1953-09b & M31N 2004-08a &50.8& 1.79 &  5.77 &  0.1310 &        & RN: \cr
M31N 1960-12a & M31N 2013-05b &52.4& 0.90 & 4.40 &  0.0909 &        & RN \cr
M31N 1961-11a & M31N 2005-06c &43.6& 0.45 & 3.03 &  0.0417 &        & RN: \cr
M31N 1963-09c & M31N 1968-09a &5.0 & 0.89 & 17.61 & 0.0041 &        & RN \cr
\dots         & M31N 2001-07b &32.8& 1.23 & 17.60 & 0.0079 &        & RN \cr
\dots         & M31N 2010-10e &9.3& 1.32 & 17.60 & 0.0090 & He/N   & RN \cr
M31N 1966-08a & M31N 1968-10c &2.2 & 0.00 & 31.18 & 0.0000 &          & RN: \cr
M31N 1966-09e & M31N 2007-08d &40.9& 0.71 & 59.56 & 0.0001 & Fe II&   RN \cr
M31N 1982-08b & M31N 1996-08c\tablenotemark{a} &14.0& 0.90 & 60.77 &  0.0001 &       & RN\cr
M31N 1984-07a & M31N 2004-11f &20.3& 0.31 &  0.57 & 0.0197 &        & RN \cr
\dots         & M31N 2012-09a &7.8& 0.57 &  0.58 & 0.0660 &Fe IIb&   RN \cr
M31N 1990-10a & M31N 2007-07a &16.7& 0.59 & 4.28  & 0.0411 &        & RN: \cr
M31N 1997-11k & M31N 2001-12b &4.2 & 0.00 & 10.57  & 0.0000 &        & RN \cr
\dots         & M31N 2009-11b &7.9& 0.38 & 10.56 &  0.0016 & Fe II  & RN \cr
M31N 2008-12a & M31N 2009-12b\tablenotemark{b} &  1 &\dots & 48.92 & \dots  &         & RN \cr
\dots         & M31N 2011-10e &2.8 & 0.30 & 48.92 & 0.0001 &         & RN \cr
\dots         & M31N 2012-10a &1.0 & 0.74 & 48.92 & 0.0002 & He/N    & RN \cr
\dots         & M31N 2013-11f &1.1 & 1.01 & 48.93 & 0.0005 & He/N    & RN \cr
\enddata
\tablenotetext{a}{M31N 1996-08c was discovered on 1997 August 01;
not 1996 August 12 as reported in \citet{sha01}.}
\tablenotetext{b}{Coordinates for 2009-12b were not provided by \citet{tan14}.}
\end{deluxetable}

\clearpage

\begin{deluxetable}{lcccccccccccr}
\rotate
\tabletypesize{\scriptsize}
\tablenum{4}
\tablewidth{0pt}
\tablecolumns{13}
\tablecaption{Recurrent Nova Optical and X-Ray Outburst Properties\label{rntable_op}}
\tablehead{\colhead{Nova} & \colhead{$m_{\rm dis}$} & \colhead{Filter} & \colhead{$t_2$~(d)} & \colhead{Recurrence} & \colhead{$m_{\rm dis}$} & \colhead{Filter} & \colhead{$t_2$~(d)} & \colhead{$t_{\rm on}$ (d)} & \colhead{$t_{\rm off}$ (d)} & \colhead{$T_{\rm bb}$} & \colhead {Type} & \colhead{References\tablenotemark{a}}
}
\startdata
M31N 1919-09a & 17.6 &$pg$& \dots & 1998-06a & 16.3 & H$\alpha$ & \dots & $1028\pm92$ & $1773\pm463$ & \dots & \dots & 1, 2, 3 \cr
M31N 1923-12c & 17.5 &$pg$& \dots & 2012-01b & 17.1 & \dots       & \dots & \dots & \dots & \dots& He/N & 1, 4  \cr
M31N 1926-06a & 17.4 &$pg$& \dots & 1962-11a & 17.2 & $U$        & $\grtsim5$  &  \dots & \dots & \dots&\dots& 1, 5  \cr
M31N 1945-09c & \dots&\dots& \dots &1975-11a & 18.6 & $V$        & \dots & \dots & \dots & \dots& \dots& 5, 6  \cr
M31N 1953-09b & 17.3 &$pg$& \dots &2004-08a & 17.4 & $R$         & \dots & $48\pm16$  & $77\pm13$ & $77_{-24}^{+21}$ & \dots & 3, 7 \cr
M31N 1957-10b & 16.3 &$pg$& \dots & 2010-12a & 15.6 & $R$        & $\grtsim13$ & \dots & \dots & \dots& \dots& 8, 9, 10  \cr
M31N 1960-12a & 16.9 &$pg$& \dots & 2013-05b & 15.2 & UV        & $1:$  & \dots & \dots & \dots& \dots& 8, 11  \cr
M31N 1961-11a & 17.0 &$pg$& \dots & 2005-06c & 16.5 & $R$        & 6.3   & \dots & \dots & \dots& \dots& 7, 8 \cr
M31N 1963-09c & 17.8 &$pg$& $\grtsim17$ & 1968-09a & 17.3 & $pg$       & $\grtsim4$  & \dots & \dots & \dots& \dots& 12  \cr
\dots         &\dots &\dots& \dots &2001-07b & 18.1 & $R$        & $5:$  & \dots & \dots & \dots& \dots& 13  \cr
\dots         &\dots &\dots& \dots &2010-10e & 17.8 & $R$        & $\grtsim5$  &$10\pm1$ & $92\pm5$ & $61_{-3}^{+6}$ & He/N & 3, 10, 14  \cr
M31N 1966-09e & 18.8 &$U$ & \dots & 2007-08d & 18.7 & $R$        & $81:$ & \dots & \dots & \dots& Fe II& 5, 15  \cr
M31N 1982-08b & 15.6 &$B$ & \dots & 1996-08c & 15.8 & H$\alpha$ & \dots & \dots & \dots & \dots& \dots& 2, 16, 17  \cr
M31N 1984-07a & 17.6 &$B$ & \dots & 2004-11f & 17.9 & $R$        & $28.4$& $17\pm17$ & $45\pm10$ & \dots & \dots& 3, 8, 18 \cr
\dots         &\dots &\dots& \dots &2012-09a & 16.3 & $R$        & \dots & \dots & \dots & \dots& Fe IIb& 19 \cr
M31N 1990-10a & 16.7 &$B$ & \dots & 2007-07a & 16.6 & $R$        & $\lessim10$ & \dots & \dots & \dots& \dots& 14, 20  \cr
M31N 1997-10f & 18.1 &$R$       & $10:$ & 2008-08b & 16.4 & $R$  & \dots & \dots & \dots & \dots& He/N?& 14, 21  \cr
M31N 1997-11k &\dots &\dots& \dots &2001-12b & 18.7 & $R$        & $\grtsim100$& \dots & \dots & \dots & \dots& 13  \cr
\dots         &\dots &\dots& \dots &2009-11b & 18.3 & $R$        & $88$  & \dots & \dots & \dots & Fe II& 14  \cr
M31N 2008-12a & 18.7 &\dots  & \dots &2009-12b & 18.8: &$R$        &\dots  & $6\pm1$ & $19\pm1$ & $97_{-4}^{+5}$ & \dots&  3, 22 \cr
\dots         & \dots&\dots& \dots &2011-10e & 18.2 & $R$      & 2.0   & \dots & \dots & \dots & \dots  &  22, 23 \cr
\dots         & \dots&\dots& \dots &2012-10a & 18.4 & $R$       & \dots & \dots & \dots & \dots & He/N &  24 \cr
\dots         & \dots&\dots& \dots &2013-11f & 18.3 & $R$        & 2.1   & \dots & \dots & \dots & He/N &  22, 25, 26 \cr
\enddata
\tablenotetext{a}{(1) \citet{hub29}; (2) \citet{rec99}; (3) \citet{hen14a}; (4) \citet{sha12d}; (5) \citet{hen08a}; (6) \cite{baa64}; (7) \citet{pie07a}; (8) \citet{ros64}; (9) \citet{rua10a}; (10) \citet{cao12}; (11) \citet{bar13}; (12) \citet{ros73}; (13) \citet{lee12}; (14) \citet{sha10}; (15) \citet{sha11b}; (16) \citet{sha92}; (17) \citet{sha01}; (18) \citet{ros89}; (19) \cite{hor12a}; (20) \citet{bry90}; (21) \citet{dim10}; (22) \citet{tan14}; (23) \citet{bar11}; (24) \citet{sha12a}; (25) \citet{dar14a}; (26) \citet{hen14b}}
\end{deluxetable}

\clearpage

\begin{deluxetable}{lcc}
\tabletypesize{\scriptsize}
\tablenum{5}
\tablewidth{0pt}
\tablecolumns{3}
\tablecaption{Model Light Curve Properties\label{lightcurve}}
\tablehead{ \colhead{Nova} & \colhead{$m_B$(max)}  & \colhead{$\nu_B$~(mag~d$^{-1}$)}
}
\startdata

\cutinhead{Classical Novae\tablenotemark{a}}

A04 & 18.2&0.200 \cr
A06 & 16.0&0.230 \cr
A07 & 15.9&0.150 \cr
A12 & 16.1&0.180 \cr
A13 & 17.0&0.077 \cr
A19 & 17.6&0.070 \cr
A20 & 17.2&0.060 \cr
A21 & 17.4&0.075 \cr
A24 & 17.8&0.059 \cr
A25 & 17.6&0.061 \cr
A26 & 18.0&0.043 \cr
A29 & 18.0&0.017 \cr
R06 & 16.5&0.140 \cr
R12 & 17.6&0.050 \cr
R18 & 17.1&0.090 \cr
R20 & 16.9&0.090 \cr
R28 & 14.9&0.250 \cr
R30 & 16.2&0.170 \cr
R38 & 16.9&0.110 \cr
R43 & 16.3&0.220 \cr
R52 & 16.4&0.125 \cr
R53 & 17.0&0.040 \cr
R57 & 15.0&0.220 \cr
R67 & 16.2&0.130 \cr
R77 & 15.9&0.220 \cr
R80 & 17.4&0.034 \cr
R85 & 15.7&0.120 \cr

\cutinhead{Recurrent Novae\tablenotemark{b}}

T Pyx & 17.6&0.055 \cr
IM Nor & 17.9&0.039 \cr
CI Aql & 17.0&0.087 \cr
V2487 Oph & 17.3&0.340 \cr
U Sco & 16.2&1.410 \cr
V394 CrA & 15.8&0.705 \cr
T CrB & 17.2&0.500 \cr
RS Oph & 16.4\tablenotemark{c}&0.254 \cr
V745 Sco & 16.7&0.328 \cr
V3890 Sgr & 15.9&0.260 \cr

\enddata
\tablenotetext{a}{From \citet{cap89}.}
\tablenotetext{b}{From \citet{sch10}. Apparent magnitudes at the distance of M31
have been computed assuming $(m-M)_o = 24.38$ \citep{fre01} and a foreground extinction of
$A_B=0.25$ \citep{sch98}.}
\tablenotetext{c}{Peak brightness for RS Oph
is based on the distance ($d=1.4$ kpc) and reddening ($E(B-V)=0.7$) given in \citet{dar12}}
\end{deluxetable}

\clearpage

\begin{deluxetable}{lccccccc}
\tabletypesize{\scriptsize}
\tablenum{6}
\tablewidth{0pt}
\tablecolumns{8}
\tablecaption{Monte Carlo Results\label{monte}}
\tablehead{\colhead{(1)} & \colhead{(2)} & \colhead{(3)} & \colhead{(4)} & \colhead{(5)} & \colhead{(6)} & \colhead{(7)} & \colhead{(8)}\\
\colhead{Coverage} & \colhead{$m_{\rm lim}$} & \colhead{CN Fraction} & \colhead{RN Fraction} & \colhead{RN/CN Dis. Efficiency} & \colhead{$N_{\rm out}({\rm RNe})/N_{\rm out}({\rm CNe})$} & \colhead{CN (yr$^{-1}$)} & \colhead{RN (yr$^{-1}$)}
}
\startdata
\cutinhead{$t_{\rm rec}=1-100$ yr}
 Clump1 & 17 & 0.0415 & 0.0034 & 0.0825 & 0.485 & 44 & 21\cr
 \dots  & 18 & 0.1335 & 0.0360 & 0.2665 & 0.150 & 57 &  8\cr
 \dots  & 19 & 0.2457 & 0.0936 & 0.3825 & 0.105 & 59 &  6\cr
\\
 Clump3 & 17 & 0.0547 & 0.0095 & 0.1736 & 0.230 & 53 & 12\cr
 \dots  & 18 & 0.1532 & 0.0563 & 0.3680 & 0.109 & 59 &  6\cr
 \dots  & 19 & 0.2586 & 0.1119 & 0.4328 & 0.092 & 60 &  5\cr
\\
 Clump5 & 17 & 0.0670 & 0.0162 & 0.2427 & 0.165 & 56 & 9\cr
 \dots  & 18 & 0.1702 & 0.0758 & 0.4459 & 0.090 & 60 & 5\cr
 \dots  & 19 & 0.2721 & 0.1287 & 0.4732 & 0.085 & 60 & 5\cr
\cutinhead{$t_{\rm rec}=5-100$ yr}
 Clump1 & 17 & \dots  & 0.0010 & 0.0244 & 1.640 & 25 & 40\cr
 \dots  & 18 & \dots  & 0.0167 & 0.1237 & 0.323 & 49 & 16\cr
 \dots  & 19 & \dots  & 0.0564 & 0.2311 & 0.173 & 55 & 10\cr
\\
 Clump3 & 17 & \dots  & 0.0032 & 0.0592 & 0.676 & 39 & 26\cr
 \dots  & 18 & \dots  & 0.0295 & 0.1926 & 0.208 & 54 & 11\cr
 \dots  & 19 & \dots  & 0.0704 & 0.2726 & 0.147 & 57 &  8\cr
\\
 Clump5 & 17 & \dots  & 0.0062 & 0.0931 & 0.430 & 45 & 20\cr
 \dots  & 18 & \dots  & 0.0428 & 0.2518 & 0.159 & 56 & 9\cr
 \dots  & 19 & \dots  & 0.0839 & 0.3083 & 0.130 & 58 & 7\cr
\enddata
\end{deluxetable}

\end{document}